\pdfoutput=1
\documentclass[12pt]{article}
\usepackage[margin=1in]{geometry}
\usepackage{amsmath}
\usepackage{amssymb}
\usepackage{bm}
\usepackage{mathtools} 
\usepackage{graphicx}
\usepackage{caption}
\usepackage{subcaption}
\usepackage{multirow}
\usepackage{algpseudocode}
\usepackage{enumitem}
\usepackage{booktabs}
\usepackage{makecell}
\usepackage{algorithmicx}
\usepackage{algorithm}
\usepackage{float}
\usepackage[normalem]{ulem}
\usepackage{setspace}
\usepackage[round]{natbib}
\usepackage{xcolor}
\usepackage{xr}
\usepackage{hyperref}
\makeatletter
\def\prep@gobbleopt[#1]{}
\def\prep@nolnopt{\@ifnextchar[{\prep@gobbleopt}{}}
\def\prep@nolinenumbers{\@ifstar{\prep@nolnopt}{\prep@nolnopt}}
\AtBeginDocument{\@ifundefined{nolinenumbers}{}{%
  \nolinenumbers\let\linenumbers\prep@nolinenumbers
  \typeout{prep_arxiv: line numbers switched off}}}
\makeatother
\begin{document}

\title{\textbf{Time-Specific Multinomial Logistic Regression for Competing Risks}}

\author{
\textbf{Haoxuan (Charlie) Zhou}$^{1}$,
\textbf{Xiaoqiong(Joan) Hu}$^{1}$,
\textbf{Yi Xiong}$^{\S 2}$ \textbf{and}
\textbf{Yan Yuan}$^{3}$
}
 
\date{}
\maketitle

\renewcommand{\thefootnote}{\fnsymbol{footnote}}
\footnotetext[4]{email: yxiong3@buffalo.edu}
\renewcommand{\thefootnote}{\arabic{footnote}}
 
\begin{center}
$^{1}$Department of Statistics and Actuarial Science, Simon Fraser University,\\
Burnaby, Canada\\[6pt]
$^{2}$Department of Biostatistics, University at Buffalo,\\
Buffalo, USA\\[6pt]
$^{3}$School of Public Health, University of Alberta,\\
Edmonton, Canada
\end{center}
 
\medskip

\begin{abstract}
Motivated by the project on premature menopause among childhood cancer survivors in the Childhood Cancer Survivor Study (CCSS, \citealp{im2023development}), we develop methods for estimating time-specific factor effects   using doubly censored event time data with competing risks. We begin with fitting separate binary logistic regression models for different causes. However, the approach does not ensure the sum of the resulting estimated cumulative incidence rates, a natural estimator of the cumulative distribution function of the event time, remains bounded by one. To overcome this limitation, we consider a multinomial logistic regression formulation to jointly model all cause-specific risks. Two inverse-probability-of-censoring-weight (IPCW) based estimating functions are proposed accordingly. To handle the estimation instability caused by data separation, we incorporate the Firth-type penalization \citep{firth1993bias} in those estimating functions to obtain more stable coefficient estimators. The sandwich variance estimator is applied to estimate the standard errors of the parameter estimators. A weighted quasi-likelihood information criterion, paired with a backward elimination, is further adapted to the proposed estimating functions for selecting covariates at a given age. Simulation studies are conducted to numerically examine the proposed approaches regarding consistency, efficiency, and robustness of the estimators. The Firth correction substantially reduces the bias induced by separation and provides valid standard error estimation. We illustrate the proposed approach with  the CCSS data viewing premature ovarian insufficiency and surgical premature menopause as the two potential causes of the primary event, premature menopause. The data analysis reveals that the effects of the cancer treatment-related covariates are significant on premature ovarian insufficiency but not on surgical premature menopause. The variable selection procedure, illustrated on an extended covariate set, yields estimated risks close to those under the full model.
\end{abstract}

\noindent\textbf{Keywords:} Cumulative Incidence Rate, Doubly Censored Event Time, Estimating Function, Firth Correction, Inverse Probability of Censoring Weight (IPCW), Variable Selection

\section{Introduction}\label{sec1: Introduction}
The Childhood Cancer Survivor Study (CCSS) is a multi-institutional cohort study of long-term survivors of childhood cancer \citep{robison2002study, robison2009childhood}. It records cumulative doses of key chemotherapeutic agents as well as detailed information on radiation dose and treatment fields, allowing the identification of risk factors for premature menopause (PM). Assessing the risk of PM at a given age in this population can inform family planning, fertility-preservation decisions, and long-term survivorship care. Such an assessment requires accurately estimating time-varying covariate effects on the distinct causes of this condition. PM leads to the early and often unexpected loss of reproductive potential and the cessation of ovarian sex hormone production, placing women at an increased risk of osteoporosis, cardiovascular disease mortality, and psychosexual dysfunction \citep{SklarCharlesA.2006PMiS, ShusterLynneT.2010Pmoe, okeke2013premature, HamodaHaitham2024Poie}. Two competing causes of PM among female childhood cancer survivors are premature ovarian insufficiency (POI) and surgical premature menopause (SPM). Recently, \cite{im2023development} developed a time-specific model for assessing the risk of POI in this cohort, treating SPM as part of the censoring time under the assumption that POI and SPM are independent events. This assumption is supported by the observation that the cumulative incidence functions (i.e., the cumulative incidence rate functions in this paper) of SPM were similar between female childhood cancer survivors and their siblings, whereas the cumulative incidence function of POI differed substantially between the two groups \citep{im2023development, SklarCharlesA.2006PMiS}. However, when PM is the event of interest, POI and SPM act as competing causes of the outcome. Consequently, analyzing data within a competing risks framework is necessary.

Statistically, assessing the risk of PM at a given age requires accurately estimating the cumulative incidence functions of the two causes conditional on available covariates, as well as characterizing how these covariate effects vary with age. Two standard regression models for this purpose are the cause-specific hazard (CSH) model \citep{prentice1978analysis} and the subdistribution hazard (Fine--Gray) model \citep{fine1999proportional}. The CSH model is easy to implement and provides a direct interpretation of covariate effects on the CSH, but not on the cumulative incidence function. Furthermore, covariate effects on a CSH can differ from their effects on the corresponding cumulative incidence function \citep{haller2013applying, fine1999proportional, scheike2008predicting}. The Fine--Gray model directly assesses covariate effects on the cumulative incidence function and handles right-censored data using inverse probability of censoring weight (IPCW) \citep{robins1992recovery, fine1999analysing}, alongside various extensions \citep{scheike2008flexible, he2016proportional, sankaran2016prop}. However, while the Fine--Gray model indicates the direction of covariate effects, the estimated effect sizes are not always easy to interpret on the cumulative incidence function scale \citep{austin2017practical}.

Beyond the interpretability concerns of these standard models, modeling the incidence of PM using CCSS data presents an additional challenge: the event times are doubly censored. The data are subject to left censoring because we assume cancer diagnosis precedes PM, which we handle by introducing a risk-set indicator. Right censoring is addressed using IPCW. The cumulative incidence function evaluated at a given age $t_{0}$ is a probability, which we refer to as the cumulative incidence rate (CIR). To model the CIR of PM directly, one can use the logistic regression considered by \cite{zhou202xagespecific}, which includes two IPCW approaches: one weighting the estimating function and the other weighting the response variable. These two approaches under logistic regression can be extended to model CIRs under competing risks. The extended approach that weights the estimating function is similar to that of \citet{blanche2023logistic}, and the extended approach that weights the response variable is similar to those of \citet{scheike2008predicting} and \citet{blanche2023logistic}. However, fitting a logistic regression model for different causes separately has a known limitation: the sum of the estimated CIRs may exceed one \citep{gerds2012absolute}. To resolve this issue, we consider a multinomial logistic regression for competing risks, which constrains the total estimated CIR, a natural estimator of the cumulative distribution function (CDF) of the event time at a given $t_{0}$, to at most one by construction. While multinomial logistic regression has been applied in discrete-time survival analysis, prior applications have generally modeled the discrete time cause-specific hazard rather than the cumulative incidence function \citep{tutz1995competing, ambrogi2009estimating}. Although \cite{shuai2018multinomial} adapted the multinomial model for interval-censored competing risks data, this model has not yet been developed for the doubly censored setting.

Despite its advantages, multinomial logistic regression for estimating CIRs can suffer from severe estimation instability due to data separation. Separation occurs when one or more covariates perfectly or nearly perfectly predict the outcome \citep{albert1984existence}. This can arise when the sample size is small, or when the sample size is large but the event is rare or strong predictors are present. This is another challenge when using the CCSS data, in which SPM events are sparse among subgroups such as African American survivors and bone marrow transplant recipients. \citet{firth1993bias} proposed a bias-reduction method that removes the first-order bias of the maximum likelihood estimator by solving a modified score equation, which is equivalent to penalizing the likelihood with Jeffreys' prior. This method has been applied to logistic regression \citep{firth1993bias, heinze2002solution} and extended to multinomial logistic regression \citep{bull2002modified, bull2007confidence}. \citet{mondol2019bias} introduced a penalized generalized estimating equation (GEE) that incorporates the Firth correction term to address separation in correlated binary data with a logit link.

In this paper, we consider a multinomial logistic regression for estimating CIRs in doubly censored data with competing risks. We propose two IPCW-based approaches and incorporate Firth's correction to ensure stable estimation in the presence of data separation. We also adapt a weighted quasi-likelihood information criterion to the proposed estimating functions, so that covariates can be selected at a given age by backward elimination. The remainder of this paper is organized as follows: Section~\ref{sec2: estimation procedure} describes the model, the proposed estimation procedures, and an adapted variable selection procedure for the multinomial logistic regression model. Section~\ref{sec4: simulation} reports the simulation studies to evaluate the performance of the proposed estimation procedure for CIR under settings designed to mimic the CCSS data. In Section~\ref{sec3: real data analysis}, the proposed approaches are applied to the CCSS data that motivated this research. We conclude in Section~\ref{sec5: disscusion} with final remarks.
\section{Time-Specific Competing Risk Regressions}\label{sec2: estimation procedure}

\subsection{Notation and Modeling}
Consider a study on the risk of a particular event subject to competing risks. Let the primary event time $T_{i}$ be the age at the event for subject $i$, with cause indicator $\eta_{i} \in \{1,2\}$. In the CCSS application, $\eta_{i}=1$ and $\eta_{i}=2$ correspond to POI and SPM, respectively. The event time is subject to double censoring: left censoring at age $V_i$ and right censoring at age $C_i$, where both $V_i$ and $C_i$ are observed. For subject $i$, let $U_i = \min\{T_i,C_i\}$ denote the observed age at study exit, $\delta_i = \mathbb{I}(T_i \le C_i)$ the event indicator, and $\mathbf{Z}_i$ a $p$-dimensional covariate vector. We assume that $(T_{i}, \eta_{i})$ is independent of $C_{i}$ conditional on $\mathbf{Z}_i$. The observed data consist of $n$ independent and identically distributed copies of $O_{i} = (U_{i}, \delta_{i}, \eta_{i}\delta_{i}, V_{i}, \mathbf{Z}_{i})$.

Our aim is to estimate the conditional risk of experiencing the event by age $t_{0}$ due to a specific cause, known as the CIR for cause $k \in \{1, 2\}$, defined as $P(T \le t_{0},\, \eta = k \mid \mathbf{Z},\, T > V,\, V < t_{0})$. Considering only a single cause of failure, \citet{zhou202xagespecific} proposed the following time-specific logistic regression model for the conditional probability of failure by $t_0$:
\begin{align*} 
    \log\Bigg[
      \frac{P(T \le t_{0} \mid \mathbf{Z},\, T > V,\, V < t_{0})}
           {1 - P(T \le t_{0} \mid \mathbf{Z},\, T > V,\, V < t_{0})}
    \Bigg]
    = \kappa(t_{0}) + \bm{\tau}(t_{0})^{\top} \mathbf{Z},
\end{align*}
where $\kappa(t_0)$ is the intercept and $\bm{\tau}(t_0)$ is a $p$-dimensional vector of regression coefficients at $t_0$. This model can be extended to accommodate competing risks by modeling the conditional cumulative incidence rate of each cause $k$ through a time-specific logistic regression:
\begin{align} \label{model: project2 model 1}
    \log\Bigg[
      \frac{P(T \le t_{0},\, \eta = k \mid \mathbf{Z},\, T > V,\, V < t_{0})}
           {1 - P(T \le t_{0},\, \eta = k \mid \mathbf{Z},\, T > V,\, V < t_{0})}
    \Bigg]
    = \kappa_{k}(t_{0}) + \bm{\tau}_{k}(t_{0})^{\top} \mathbf{Z},
\end{align}
where $\kappa_k(t_0)$ is the intercept and $\bm{\tau}_k(t_0)$ is a $p$-dimensional vector of covariate effects at $t_0$, for cause $k = 1, 2$. When there is only a single cause of failure, or when each cause is modeled marginally by treating the other causes as additional censoring, model~\eqref{model: project2 model 1} reduces to the model in \cite{zhou202xagespecific}.

Although model~\eqref{model: project2 model 1} estimates the cause-specific risks directly, it specifies a logistic regression for different causes separately, leaving the resulting probabilities unconstrained across causes. As a consequence, a natural estimator for the CDF of the event time at a given $t_{0}$, $\hat{P}(T \le t_0 \mid \mathbf{Z},\, T > V,\, V < t_0) = \sum_{k=1}^{2} \hat{P}(T \le t_0,\, \eta = k \mid \mathbf{Z},\, T > V,\, V < t_0)$, may exceed one for certain covariate values and time points.
 
To overcome this limitation, we consider the following multinomial logistic regression model, which jointly models both cause-specific risks:
\begin{align}\label{model: project2 model 2}
    \log\Bigg[
      \frac{P(T \le t_0,\, \eta = k \mid \mathbf{Z},\, T > V,\, V < t_0)}
           {P(T > t_0 \mid \mathbf{Z},\, T > V,\, V < t_0)}
    \Bigg]
    = \alpha_k(t_0) + \bm{\beta}_k(t_0)^{\top} \mathbf{Z},
    \quad k = 1, 2,
\end{align}
where $\alpha_k(t_0)$ is the cause-specific intercept and $\bm{\beta}_k(t_0)$ is a $p$-dimensional vector of regression coefficients for cause $k$ at age $t_0$. By construction, this model ensures that the estimated CIRs are nonnegative and that their sum across all causes lies in $[0,1]$.

\subsection{Proposed Estimation Procedures}
Since $T_i$ is subject to censoring, we extend the IPCW approaches from \citet{zhou202xagespecific}. We assume that the support of the conditional survival function of the censoring time $G(c \mid \mathbf{Z}) = P(C > c \mid \mathbf{Z})$ contains $[0, t_0]$. The two IPCW approaches for model~\eqref{model: project2 model 1} can be straightforwardly extended from \citet{zhou202xagespecific}, which are reported in Supplementary Materials~\ref{supp: logistic regression}.

\subsubsection{Weighted GEE}
We now construct the estimating functions for model~\eqref{model: project2 model 2}. For the case of two competing causes, define the outcome vector 
$\mathbf{Y}_{i}
= \bigl(\mathbb{I}(T_{i} \le t_0,\, \eta_{i} = 1),\,
        \mathbb{I}(T_{i} \le t_0,\, \eta_{i} = 2)\bigr)^{\top}$, the parameter vector 
$\bm{\theta}(t_{0}) = (\alpha_{1}(t_{0}), \bm{\beta}_{1}(t_{0})^{\top}, 
\alpha_{2}(t_{0}), \bm{\beta}_{2}(t_{0})^{\top})^{\top}$, 
and the probability vector 
$\bm{\pi}_{i}(t_{0},\bm{Z}_{i})
= \bigl(\pi_{1i}(t_{0},\bm{Z}_{i}),\, \pi_{2i}(t_{0},\bm{Z}_{i})\bigr)^{\top}$, where 
\[
\pi_{ki}(t_{0}, \bm{Z}_{i}) = \frac{\exp\{\alpha_{k}(t_{0}) +
\bm{\beta}_{k}(t_{0})^{\top}\mathbf{Z}_{i}\}}{1 +
\exp\{\alpha_{1}(t_{0}) + \bm{\beta}_{1}(t_{0})^{\top}\mathbf{Z}_{i}\} +
\exp\{\alpha_{2}(t_{0}) + \bm{\beta}_{2}(t_{0})^{\top}\mathbf{Z}_{i}\}}.
\]
For notational simplicity, we write $\bm{\theta}$ for $\bm{\theta}(t_{0})$ hereafter.
Let
\[
\mathbf{\tilde{Z}}_i =
\begin{pmatrix}
1 & Z_{i1} & \dots & Z_{ip} & 0 & 0 & \dots & 0 \\
0 & 0 & \dots & 0 & 1 & Z_{i1} & \dots & Z_{ip}
\end{pmatrix}
\]
denote the block-diagonal design matrix, with $Z_{ij}$ denoting the $j$-th covariate for subject $i$, and let
\begin{align}\label{eq: IPCW}
    W_i(t_{0};G)=
    \frac{\mathbb{I}(U_i \le t_{0})\delta_i}{G(U_i \mid\mathbf{Z}_i)}+
    \frac{\mathbb{I}(U_i > t_{0})}{G(t_{0}\mid\mathbf{Z}_i)}
\end{align}
denote the IPCW. The estimating function under Approach A is
\begin{align} \label{eq: proj2 multinomial estimating equation a} 
    \bm{U}_{2A}(\bm{\theta};t_{0}\mid G) 
    = \sum_{i=1}^{n} \mathbb{I}(t_0 \ge V_i)\, W_i(t_0; G)\, 
    \mathbf{\tilde{Z}}_i^{\top} \Bigg[ \mathbf{Y}_i - 
    \bm{\pi}_{i}(t_{0},\bm{Z}_{i})\Bigg].
\end{align}
Note that the weight $W_i(t_0;G)$ in~\eqref{eq: IPCW} assigns zero weight to subjects censored before $t_0$, which excludes them from the estimating function. Appropriately incorporating the available information from these excluded subjects is likely to yield more efficient estimation, particularly at values of $t_0$ where censoring is heavy. This motivates Approach B, whose estimating function is
\begin{align} \label{eq: proj2 multinomial estimating equation b}
    \bm{U}_{2B}(\bm{\theta};t_{0}\mid G) 
    = \sum_{i=1}^{n} \mathbb{I}(t_0 \ge V_i)\, 
    \mathbf{\tilde{Z}}_i^{\top} \Bigg[ \mathbf{Y}_i\, W_i(t_0; G) - 
    \bm{\pi}_{i}(t_{0},\bm{Z}_{i})\Bigg].
\end{align}
Both estimating functions are unbiased under the conditional independence of $(T_{i},\eta_{i})$ and $C_{i}$ given $\mathbf{Z}_{i}$, and the resulting estimators are consistent and asymptotically normal at a fixed $t_0 > 0$ under standard regularity conditions.

\subsubsection{Weighted GEE With Firth Correction}
Separation can occur when the event for a particular cause is rare at certain ages, leading to unstable or non-existent estimates. To address this issue, we apply a Firth-type penalization \citep{firth1993bias} to the estimating equations \eqref{eq: proj2 multinomial estimating equation a} and \eqref{eq: proj2 multinomial estimating equation b} introduced above, following the extension to GEE developed by \citet{mondol2019bias}. We apply the correction for both of its purposes. It delivers finite estimates where a covariate level carries no events at an analysis age, and it reduces the sparse-data bias where such a level carries only a few events \citep{heinze2002solution, joshi2022solutions}.

The derivation of Firth-corrected estimating functions under both approaches when using logistic regression is given in Supplementary Materials~\ref{supp: deriv_firth_ee_model1}. When using multinomial logistic regression, let $\bm{U}(\bm{\theta};\, t_{0} \mid G)$ denote a generic estimating function of either Approach~A or~B. The total number of parameters is $P = 2(p + 1)$. The $r$-th component of the Firth-corrected estimating function is
\begin{align}\label{eq:firth_score}
U^{*}_{r}(\bm{\theta};\, t_{0} \mid G)
= U_{r}(\bm{\theta};\, t_{0} \mid G)
+ A_{r}(\bm{\theta};\, t_{0} \mid G),
\quad r = 1, \ldots, P,
\end{align}
where the Firth penalty term is
\begin{align}\label{eq:firth_penalty}
A_{r}(\bm{\theta};\, t_{0} \mid G)
= \frac{1}{2}\,\mathrm{trace}\!\left[
  \mathcal{I}(\bm{\theta};\, t_0 \mid G)^{-1}\,
  \frac{\partial\, \mathcal{I}(\bm{\theta};\, t_0 \mid G)}
  {\partial \theta_{r}}
  \right]
= \frac{\partial}{\partial \theta_{r}}\left[
  \frac{1}{2}\log\bigl|\mathcal{I}(\bm{\theta};\, t_0 \mid G)\bigr|
    \right],
\end{align}
$\theta_r$ is the $r$-th element of $\bm{\theta}$ where $1 \le r\le P$, and $\mathcal{I}(\bm{\theta};\, t_0 \mid G) = -\partial \bm{U}(\bm{\theta};\, t_{0} \mid G) \big/ \partial \bm{\theta}^{\top}$.

The Firth-corrected versions of $\bm{U}_{2A}$ and $\bm{U}_{2B}$ are
\begin{align}\label{eq: u2_firth}
\begin{split}
\bm{U}_{2A}^{*}(\bm{\theta};\, t_0 \mid G)
&= \sum_{i=1}^{n}
\mathbb{I}(t_0 \ge V_{i})\,
W_{i}(t_0; G)\,
\mathbf{\tilde{Z}}_i^{\top}
\Bigg[
  \mathbf{Y}_{i}
  - \bm{\pi}_{i}(t_{0},\bm{Z}_{i})
  \Bigg]
+ \bm{A}_{2A}(\bm{\theta};\, t_0 \mid G),\\
\bm{U}_{2B}^{*}(\bm{\theta};\, t_0 \mid G)
&= \sum_{i=1}^{n}
\mathbb{I}(t_0 \ge V_{i})\,
\mathbf{\tilde{Z}}_i^{\top}
\Bigg[
  \mathbf{Y}_{i}\, W_{i}(t_0; G)
  - \bm{\pi}_{i}(t_{0},\bm{Z}_{i})
  \Bigg]
+ \bm{A}_{2B}(\bm{\theta};\, t_0 \mid G),
\end{split}
\end{align}
where $\bm{A}_{2A}(\bm{\theta};\, t_0 \mid G)$ and $\bm{A}_{2B}(\bm{\theta};\, t_0 \mid G)$ are vectors with $r$-th components
\begin{align}\label{eq:A_2A}
A_{2A,r}(\bm{\theta};\, t_0 \mid G)
= \frac{1}{2}
\sum_{i=1}^{n}
\mathbb{I}(t_0 \ge V_{i})\, W_{i}(t_0; G)\,
\mathrm{trace}\!\left[
  \frac{\partial \mathbf{V}_{i}(\bm{\theta})}{\partial \theta_{r}}\,
  \mathbf{H}_{2A,i}
  \right],
\end{align}
\begin{align}\label{eq:A_2B}
A_{2B,r}(\bm{\theta};\, t_0 \mid G)
= \frac{1}{2}
\sum_{i=1}^{n}
\mathbb{I}(t_0 \ge V_{i})\,
\mathrm{trace}\!\left[
  \frac{\partial \mathbf{V}_{i}(\bm{\theta})}{\partial \theta_{r}}\,
  \mathbf{H}_{2B,i}
  \right].
\end{align}
Here, $\mathcal{I}_{2m}(\bm{\theta};\, t_0 \mid G)
= -\partial \bm{U}_{2m} / \partial \bm{\theta}^{\top}$ for $m \in \{A, B\}$,
$\mathbf{V}_{i}(\bm{\theta})
= \mathrm{diag}\bigl(\bm{\pi}_{i}(t_{0},\bm{Z}_{i})\bigr)
- \bm{\pi}_{i}(t_{0},\bm{Z}_{i})\, \bm{\pi}_{i}(t_{0},\bm{Z}_{i})^{\top}$
is the $2 \times 2$ variance matrix of $\mathbf{Y}_{i}$, and $\mathbf{H}_{2m,i} = \tilde{\mathbf{Z}}_{i}\, \mathcal{I}_{2m}(\bm{\theta};\, t_0 \mid G)^{-1}\, \tilde{\mathbf{Z}}_{i}^{\top}$.

Under model~\eqref{model: project2 model 1}, the Firth penalty for each parameter depends only on the fitted probability of its own cause (see equations~\eqref{eq:A_1A_final} and~\eqref{eq:A_1B_final} in Supplementary Materials~\ref{supp: deriv_firth_ee_model1}). In contrast, the $2 \times 2$ structure of $\mathbf{V}_{i}(\bm{\theta})$ in model~\eqref{model: project2 model 2} couples the two causes through the off-diagonal term $-\pi_{1i}\, \pi_{2i}$, so the penalty for each parameter depends on the fitted probabilities of both causes. The explicit form of $\partial \mathbf{V}_{i}(\bm{\theta}) / \partial \theta_{r}$ and the full derivation regarding multinomial logistic regression are given in Supplementary Materials~\ref{supp: deriv_firth_ee_model2}. 

\subsubsection{Asymptotic Properties and Variance Estimation}
We examine the asymptotic properties of the proposed estimators without Firth correction at a fixed time $t_0$. For model~\eqref{model: project2 model 1}, the asymptotic properties of $\bm{U}_{1A}(\kappa_{k}, \bm{\tau}_{k}; t_{0} \mid G)$ and $\bm{U}_{1B}(\kappa_{k}, \bm{\tau}_{k}; t_{0} \mid G)$ follow from the theory developed in \citet{zhou202xagespecific}. We now present the corresponding results for model~\eqref{model: project2 model 2}. Let $\hat{\bm{\theta}}_A$ and $\hat{\bm{\theta}}_B$ denote the estimators obtained by solving
$\bm{U}_{2A}(\bm{\theta};\, t_0 \mid G) = \bm{0}$ and $\bm{U}_{2B}(\bm{\theta};\, t_0 \mid G) = \bm{0}$, respectively.
Write $\bm{U}_{2A}(\bm{\theta};\, t_0 \mid G) = \sum_{i=1}^{n} \bm{A}_i$ and $\bm{U}_{2B}(\bm{\theta};\, t_0 \mid G) = \sum_{i=1}^{n} \bm{B}_i$, where the $\bm{A}_i$ are i.i.d.\ with mean zero and the $\bm{B}_i$ are i.i.d.\ with mean zero. These subject-level contributions are distinct from the Firth penalty vectors $\bm{A}_{2A}$ and $\bm{A}_{2B}$ defined in~\eqref{eq:A_2A} and~\eqref{eq:A_2B}. Under standard regularity conditions and provided that the number of subjects with an observed event, $n^{*} = \sum_{i=1}^{n} \mathbb{I}(V_i < t_0)\, \delta_i \to \infty$ as $n \to \infty$, we have
\[
  \sqrt{n}\,(\hat{\bm{\theta}}_{A} - \bm{\theta})
  \xrightarrow{d} N\!\bigl(0,\, AV_{A}(\bm{\theta}; G)\bigr)
  \quad \text{and} \quad
  \sqrt{n}\,(\hat{\bm{\theta}}_{B} - \bm{\theta})
  \xrightarrow{d} N\!\bigl(0,\, AV_{B}(\bm{\theta}; G)\bigr),
  \]
where
\begin{align*}
AV_{A}(\bm{\theta}; G)
&= \Gamma_{A}^{-1}(\bm{\theta}; G)\,
\Sigma_{A}(\bm{\theta}; G)\,
\bigl[\Gamma_{A}^{-1}(\bm{\theta}; G)\bigr]^{\top}, \\
AV_{B}(\bm{\theta}; G)
&= \Gamma_{B}^{-1}(\bm{\theta}; G)\,
\Sigma_{B}(\bm{\theta}; G)\,
\bigl[\Gamma_{B}^{-1}(\bm{\theta}; G)\bigr]^{\top}.
\end{align*}
Here, $\Gamma_A(\bm{\theta}; G)$ and $\Gamma_B(\bm{\theta}; G)$ are the limits of $-n^{-1}\,\partial \bm{U}_{2A}(\bm{\theta}; t_0 \mid G) \big/ \partial \bm{\theta}^{\top}$ and $-n^{-1}\,\partial \bm{U}_{2B}(\bm{\theta}; t_0 \mid G) \big/ \partial \bm{\theta}^{\top}$, respectively. Similarly, $\Sigma_{A}(\bm{\theta}; G)$ and $\Sigma_{B}(\bm{\theta}; G)$ are the asymptotic variances of $n^{-1/2}\bm{U}_{2A}(\bm{\theta}; t_0 \mid G)$ and $n^{-1/2}\bm{U}_{2B}(\bm{\theta}; t_0 \mid G)$, respectively. The detailed derivation is provided in Supplementary Materials~\ref{supp: derv_asy_var}.

In practice, $\Gamma_A(\bm{\theta};G)$ and $\Gamma_B(\bm{\theta};G)$ can be estimated by 
$-n^{-1}\,\partial \bm{U}_{2A}(\bm{\theta}; t_0 \mid G) \big/ \partial \bm{\theta}^{\top}$ and $-n^{-1}\,\partial \bm{U}_{2B}(\bm{\theta}; t_0 \mid G) \big/ \partial \bm{\theta}^{\top}$, evaluated at $\hat{\bm{\theta}}_A$ and $\hat{\bm{\theta}}_B$, respectively. Similarly, $\Sigma_{A}(\bm{\theta};G)$ and $\Sigma_{B}(\bm{\theta};G)$ can be estimated using 
$n^{-1}\sum_{i=1}^n \bm{A}_i \bm{A}_i^{\top}$ and $n^{-1}\sum_{i=1}^n \bm{B}_i \bm{B}_i^{\top}$, where $\bm{\theta}$ is replaced by $\hat{\bm{\theta}}$.

We now show that the Firth-corrected estimators share the same asymptotic distribution as their uncorrected counterparts. Recall from~\eqref{eq:firth_score} that the $r$-th component of the corrected estimating function is
$U^{*}_{r}(\bm{\theta};\, t_0 \mid G)
= U_{r}(\bm{\theta};\, t_0 \mid G)
+ A_{r}(\bm{\theta};\, t_0 \mid G)$.
Since
$\mathcal{I}(\bm{\theta};\, t_0 \mid G)
= -\partial \bm{U}(\bm{\theta};\, t_0 \mid G)
\big/ \partial \bm{\theta}^{\top}$
is a sum of $n$ subject-level contributions, each component of the Firth penalty~\eqref{eq:firth_penalty} satisfies $A_{r}(\bm{\theta};\, t_0 \mid G) = (1/2)\,\mathrm{trace}\bigl[\mathcal{I}(\bm{\theta};\, t_0 \mid G)^{-1}\, \partial \mathcal{I}(\bm{\theta};\, t_0 \mid G) \big/ \partial \theta_{r}\bigr] = O(1)$, since $\mathcal{I} = O(n)$, $\mathcal{I}^{-1} = O(n^{-1})$, and $\partial \mathcal{I} / \partial \theta_r = O(n)$. Therefore, $n^{-1/2}\, U^{*}_{r}(\bm{\theta};\, t_0 \mid G) = n^{-1/2}\, U_{r}(\bm{\theta};\, t_0 \mid G) + n^{-1/2}\, A_{r}(\bm{\theta};\, t_0 \mid G)$. The second term is $O(n^{-1/2}) \to 0$ as $n \to \infty$, so the Firth-corrected and uncorrected estimating functions yield the same limiting distribution for $\sqrt{n}\,(\hat{\bm{\theta}} - \bm{\theta})$. Consequently, the variance of all Firth-corrected estimators can be estimated using the same sandwich form described above.

\subsection{Practical Implementation}\label{sec2: practical implementation}
The conditional censoring distribution $G(\cdot \mid \mathbf{Z})$ is usually unknown in practice. To implement the proposed approaches, we replace $G(\cdot \mid \mathbf{Z})$ in~\eqref{eq: IPCW} with an estimator $\hat{G}(\cdot \mid \mathbf{Z})$ obtained from the observed data. If $\hat{G}(\cdot \mid \mathbf{Z})$ is uniformly consistent for $G(\cdot \mid \mathbf{Z})$, one can establish the consistency and asymptotic normality of the resulting estimator $\tilde{\bm{\theta}}$ obtained by solving $\bm{U}(\bm{\theta};\, t_0 \mid \hat{G}) = \bm{0}$.

For model~\eqref{model: project2 model 1}, the variance estimation procedure for approaches~A and~B can be found in \citet{zhou202xagespecific}. We now describe the procedure for model~\eqref{model: project2 model 2}.

Let $\tilde{\bm{\theta}}_A$ and $\tilde{\bm{\theta}}_B$ denote the estimators obtained by solving
$\bm{U}_{2A}(\bm{\theta};\, t_0 \mid \hat{G}) = \bm{0}$ and
$\bm{U}_{2B}(\bm{\theta};\, t_0 \mid \hat{G}) = \bm{0}$,
respectively. We first consider variance estimation for Approach~A. Write
$\bm{U}_{2A}(\bm{\theta};\, t_0 \mid \hat{G}) = \sum_{i=1}^{n} \tilde{\bm{A}}_i$,
where the $\tilde{\bm{A}}_i$'s are correlated through their shared dependence on $\hat{G}(\cdot \mid \mathbf{Z})$ but satisfy
$\mathrm{E}(\tilde{\bm{A}}_i \mid \mathbf{Z}_i, V_i) = \bm{0}$.
This conditional mean-zero property ensures that the standard sandwich variance estimator remains valid:
\begin{equation}\label{eq: varestmAapp}
  \widehat{\mathrm{Var}}\bigl\{\tilde{\bm{\theta}}_A\bigr\}
  = \mathcal{I}_{2A}^{-1}\!\left(\bm{\theta};\, t_0 \mid \hat{G}
    \right)\,
    \hat{V}\bigl\{\bm{U}_{2A}(\bm{\theta};\, t_0 \mid \hat{G})
    \bigr\}\,
    \left[\mathcal{I}_{2A}^{-1}\!\left(\bm{\theta};\, t_0 \mid
    \hat{G}\right)\right]^{\top},
\end{equation}
where
$\hat{V}\bigl\{\bm{U}_{2A}(\bm{\theta};\, t_0 \mid \hat{G})\bigr\}
= \sum_{i=1}^{n}
  \bigl(\tilde{\bm{A}}_i - \bar{\tilde{\bm{A}}}\bigr)
  \bigl(\tilde{\bm{A}}_i - \bar{\tilde{\bm{A}}}\bigr)^{\top}$
with
$\bar{\tilde{\bm{A}}}
= n^{-1} \sum_{i=1}^{n} \tilde{\bm{A}}_i$.

Variance estimation for Approach~B is more involved. Write $\bm{U}_{2B}(\bm{\theta};\, t_0 \mid \hat{G}) = \sum_{i=1}^{n} \tilde{\bm{B}}_i$, where the $\tilde{\bm{B}}_i$'s are again correlated through
$\hat{G}(\cdot \mid \mathbf{Z})$. Unlike Approach~A, $\mathrm{E}(\tilde{\bm{B}}_i \mid \mathbf{Z}_i, V_i) \neq \bm{0}$ in general. To see the consequences, consider the variance decomposition
\[
  \mathrm{Var}\bigl\{\bm{U}_{2B}(\bm{\theta};\, t_0 \mid \hat{G})
    \bigr\}
  = \mathrm{E}\Bigl\{\mathrm{Var}\Bigl[
    \textstyle\sum_{i=1}^{n} \tilde{\bm{B}}_i
    \,\Big|\, \mathbf{Z}_i, V_i
    \Bigr]\Bigr\}
  + \mathrm{Var}\Bigl\{\mathrm{E}\Bigl[
    \textstyle\sum_{i=1}^{n} \tilde{\bm{B}}_i
    \,\Big|\, \mathbf{Z}_i, V_i
    \Bigr]\Bigr\}.
  \]
The first term can be estimated in the same manner as Approach~A by
$\hat{V}_{\mathrm{app}}\bigl\{
  \bm{U}_{2B}(\bm{\theta};\, t_0 \mid \hat{G})
  \bigr\}
= \sum_{i=1}^{n}
\bigl(\tilde{\bm{B}}_i - \bar{\tilde{\bm{B}}}\bigr)
\bigl(\tilde{\bm{B}}_i - \bar{\tilde{\bm{B}}}\bigr)^{\top}$
  with
$\bar{\tilde{\bm{B}}}
= n^{-1} \sum_{i=1}^{n} \tilde{\bm{B}}_i$.
The second term arises from estimating $G(\cdot \mid \mathbf{Z})$ and is generally nonnegative. Since
\begin{align}\label{eq:cond_mean_B}
\mathrm{E}(\tilde{\bm{B}}_i \mid \mathbf{Z}_i, V_i)
= \mathbb{I}(t_0 \ge V_i)\,
\tilde{\mathbf{Z}}_{i}^{\top}\,
\mathrm{E}\!\left\{
  \mathbf{Y}_i \left[
    \frac{G(T_i \mid \mathbf{Z}_i)}
    {\hat{G}(T_i \mid \mathbf{Z}_i)} - 1
    \right]
  \,\bigg|\, U_{i},\mathbf{Z}_i, V_i
  \right\},
\end{align}
estimating this term depends heavily on $\hat{G}(\cdot \mid \mathbf{Z})$. One may use a bootstrap variance estimator. Alternatively, we assume that $\hat{G}(\cdot \mid \mathbf{Z})$ is uniformly consistent and hence sufficiently close to $G(\cdot \mid \mathbf{Z})$, so that the conditional mean in~\eqref{eq:cond_mean_B} is close to zero. By omitting the second term, we obtain the following approximate sandwich variance estimator:
  \begin{equation}\label{eq: varestmBapp}
\widehat{\mathrm{Var}}\bigl\{\tilde{\bm{\theta}}_B\bigr\}
= \mathcal{I}_{2B}^{-1}\!\left(\bm{\theta};\, t_0 \mid \hat{G}
                               \right)\,
\hat{V}_{\mathrm{app}}\bigl\{
  \bm{U}_{2B}(\bm{\theta};\, t_0 \mid \hat{G})
  \bigr\}\,
\left[\mathcal{I}_{2B}^{-1}\!\left(\bm{\theta};\, t_0 \mid
                                   \hat{G}\right)\right]^{\top}.
\end{equation}

Let $\tilde{\bm{\theta}}_A^{*}$ and $\tilde{\bm{\theta}}_B^{*}$ denote the estimators obtained by solving the Firth-corrected
estimating functions $\bm{U}^{*}(\bm{\theta};\, t_0 \mid \hat{G}) = \bm{0}$ under approaches~A and~B, respectively. Following \citet{mondol2019bias}, the same sandwich forms~\eqref{eq: varestmAapp} and~\eqref{eq: varestmBapp} can be used to estimate the variance of $\tilde{\bm{\theta}}_A^{*}$ and $\tilde{\bm{\theta}}_B^{*}$.

To implement both approaches, we consider four estimators for the censoring survival function $G(\cdot \mid \mathbf{Z})$: (i) the survival random forest (SRF); (ii) a stratified empirical cumulative distribution function
(ECDF), where the continuous covariate $Z_1$ is partitioned into four levels ($[0, 0.25)$, $[0.25, 0.5)$, $[0.5, 0.75)$, and
$[0.75, 1]$); (iii) a Cox proportional hazards model for the censoring time $C$ (referred to as the Cox PH model on $C$); and
(iv) a Cox proportional hazards model for the gap time $C^{*} = C - (V + 5)$ (referred to as the Cox PH model on $C^{*}$). The gap-time formulation accounts for the CCSS cohort structure, in which all subjects survived at least five years after cancer diagnosis. Section~\ref{sec4: simulation} provides extensive simulations, and Section~\ref{sec3: real data analysis} illustrates the application of approaches~A and~B to CCSS data.

\subsection{Variable Selection}\label{sec2: variable selection}

Model~\eqref{model: project2 model 2} is fitted at a fixed analysis age $t_{0}$, so the relevant covariate set may differ across $t_{0}$. We therefore adapt an information criterion to the estimating functions~\eqref{eq: proj2 multinomial estimating equation a} and~\eqref{eq: proj2 multinomial estimating equation b}, and use it to drive a backward elimination at each $t_{0}$. The criterion of \citet{akaike1998} is not directly available here, since these estimating functions are not the score functions of a likelihood. \citet{pan2001} modified the AIC for GEE by replacing the likelihood with the quasi-likelihood under the independence working correlation, yielding the penalty $2\,\mathrm{trace}(\bm{\Omega}_{I}\bm{V}_{R})$, in which $\bm{\Omega}_{I}$ is the negative second derivative of the quasi-loglikelihood and $\bm{V}_{R}$ is the robust variance of the estimator; this penalty reduces to $2P$ when all modeling specifications are correct. Several proposals extend such criteria to estimating functions carrying inverse probability weights \citep{shenchen2012, gosho2016, kojima2023}. Our development adapts the criterion of \citet{gosho2016}, whose structure matches~\eqref{eq: proj2 multinomial estimating equation a} and~\eqref{eq: proj2 multinomial estimating equation b} most closely, and follows \citet{nagashima2017} for the Firth-corrected case.

Write $Y_{ki} = \mathbb{I}(T_{i} \le t_{0},\, \eta_{i} = k)$ for $k = 1, 2$ and $Y_{0i} = 1 - Y_{1i} - Y_{2i}$, and let $\pi_{0i} = 1 - \pi_{1i} - \pi_{2i}$ denote the probability of the reference category, with the arguments $(t_{0}, \bm{Z}_{i})$ suppressed. Each of $\bm{U}_{2A}$ and $\bm{U}_{2B}$ is the derivative of a weighted quasi-loglikelihood, in the sense that $\partial Q_{w}^{m}(\bm{\theta};\, t_{0} \mid \hat{G}) \big/ \partial \bm{\theta} = \bm{U}_{2m}(\bm{\theta};\, t_{0} \mid \hat{G})$ for $m \in \{A, B\}$, where
\begin{align}\label{eq: Qw}
\begin{split}
Q_{w}^{A}(\bm{\theta};\, t_{0} \mid \hat{G})
&= \sum_{i=1}^{n} \mathbb{I}(t_{0} \ge V_{i})\, W_{i}(t_{0}; \hat{G})
\Bigg[ \sum_{k=1}^{2} Y_{ki} \log \frac{\pi_{ki}}{\pi_{0i}} + \log \pi_{0i} \Bigg], \\
Q_{w}^{B}(\bm{\theta};\, t_{0} \mid \hat{G})
&= \sum_{i=1}^{n} \mathbb{I}(t_{0} \ge V_{i})
\Bigg[ W_{i}(t_{0}; \hat{G}) \sum_{k=1}^{2} Y_{ki} \log \frac{\pi_{ki}}{\pi_{0i}} + \log \pi_{0i} \Bigg].
\end{split}
\end{align}
The two differ in the placement of the IPCW, mirroring the distinction between~\eqref{eq: proj2 multinomial estimating equation a} and~\eqref{eq: proj2 multinomial estimating equation b}. Following \citet{pan2001} and \citet{gosho2016}, the criterion at analysis age $t_{0}$ is
\begin{align}\label{eq: qicw}
\mathrm{QICW}^{m}(t_{0})
= -2\, Q_{w}^{m}(\tilde{\bm{\theta}}_{m};\, t_{0} \mid \hat{G})
+ 2\, \mathrm{trace}\Bigl( \hat{\mathcal{I}}_{2m}\, \widehat{\mathrm{Var}}\bigl\{ \tilde{\bm{\theta}}_{m} \bigr\} \Bigr),
\quad m \in \{A, B\},
\end{align}
where $\hat{\mathcal{I}}_{2m} = \mathcal{I}_{2m}(\bm{\theta};\, t_{0} \mid \hat{G}) \big|_{\bm{\theta} = \tilde{\bm{\theta}}_{m}}$ plays the role of $\bm{\Omega}_{I}$, and $\widehat{\mathrm{Var}}\{\tilde{\bm{\theta}}_{m}\}$ is the sandwich variance estimator~\eqref{eq: varestmAapp} or~\eqref{eq: varestmBapp} and plays the role of $\bm{V}_{R}$. Replacing the trace penalty by $2P$, where $P = 2(p+1)$ is the number of coefficients estimated in the fitted model, gives a second form $\mathrm{QICW}_{u}^{m}(t_{0})$, which relies on correct specification of the censoring model as well as of the outcome model; it is reported in Supplementary Materials~\ref{supp: qicwu}, and~\eqref{eq: qicw} is used throughout the main text. Detailed derivations are given in Supplementary Materials~\ref{supp: qicw_deriv}.

The term $(1/2)\log|\mathcal{I}_{2m}|$ has derivative $\bm{A}_{2m}$ by~\eqref{eq:firth_penalty}, so the Firth-corrected estimating function~\eqref{eq: u2_firth} is the derivative of $Q_{w}^{m,*}(\bm{\theta};\, t_{0} \mid \hat{G}) = Q_{w}^{m}(\bm{\theta};\, t_{0} \mid \hat{G}) + (1/2) \log | \mathcal{I}_{2m}(\bm{\theta};\, t_{0} \mid \hat{G}) |$, so that $\tilde{\bm{\theta}}_{m}^{*}$ of Section~\ref{sec2: practical implementation} is the corresponding estimator. Following \citet{nagashima2017}, the Firth-corrected criterion replaces $Q_{w}^{m}$ by $Q_{w}^{m,*}$ and evaluates every quantity at $\tilde{\bm{\theta}}_{m}^{*}$, leaving the form of the penalty unchanged:
\begin{align}\label{eq: qicw_firth}
\mathrm{QICW}^{m,*}(t_{0})
= -2\, Q_{w}^{m,*}(\tilde{\bm{\theta}}_{m}^{*};\, t_{0} \mid \hat{G})
+ 2\, \mathrm{trace}\Bigl( \hat{\mathcal{I}}_{2m}^{*}\, \widehat{\mathrm{Var}}\bigl\{ \tilde{\bm{\theta}}_{m}^{*} \bigr\} \Bigr),
\end{align}
where $\hat{\mathcal{I}}_{2m}^{*} = \mathcal{I}_{2m}(\bm{\theta};\, t_{0} \mid \hat{G}) \big|_{\bm{\theta} = \tilde{\bm{\theta}}_{m}^{*}}$, and $\widehat{\mathrm{Var}}\{\tilde{\bm{\theta}}_{m}^{*}\}$ is obtained from the same sandwich forms. Expressions~\eqref{eq: qicw} and~\eqref{eq: qicw_firth} give one criterion for each approach with and without the Firth correction, four in total.

Each criterion drives a backward elimination, summarized in Algorithm~\ref{alg: backward}. Write $\mathcal{M}$ for a set of terms and $\mathcal{M}_{\mathrm{full}}$ for the full term set. A covariate is treated as a single term. Removing it drops its coefficients in both $\bm{\beta}_{1}(t_{0})$ and $\bm{\beta}_{2}(t_{0})$ at once, and for a categorical covariate all of its indicators are dropped at once. Marginality is enforced, so an interaction is removed before its main effects. We assume that the censoring model and the event time model share the same set of covariates. The estimator $\hat{G}(\cdot \mid \mathbf{Z})$ depends on the covariate set used to fit it, and is therefore re-estimated at every step: once for the current model $\mathcal{M}$, and once for each candidate model $\mathcal{M} \setminus \{z\}$. Every criterion value compared within a step is thus computed under the weights implied by its own candidate model. Any of the estimators of $G(\cdot \mid \mathbf{Z})$ described in Section~\ref{sec2: practical implementation} may be used for this purpose. The tolerance $\varepsilon$ in Algorithm~\ref{alg: backward} is set to zero, which accepts a removal only when it strictly lowers the criterion. Ways of calibrating a positive $\varepsilon$ are described in Supplementary Materials~\ref{supp: tolerance}.

\begin{algorithm}[ht]
    \small
    \singlespacing
    \caption{Backward elimination driven by $\mathrm{QICW}^{\bullet}$ at analysis age $t_{0}$}\label{alg: backward}
    \raggedright
    \vspace{-10pt}
    \textbf{Fixed throughout the search}: the analysis age $t_{0}$; the criterion $\mathrm{QICW}^{\bullet}$; the SRF configuration; the tolerance $\varepsilon = 0$.\\[2pt]
    \textbf{Re-estimated at every step}: the censoring distribution estimator $\hat{G}_{\mathcal{M}}$, and with it the IPCW $W_{i}(t_{0}; \hat{G}_{\mathcal{M}})$.\\[2pt]
    \textbf{Removal rules}: a covariate is removed as a single term, together with all of its indicators and its coefficients in both $\bm{\beta}_{1}(t_{0})$ and $\bm{\beta}_{2}(t_{0})$; an interaction is removed before its main effects.\\[3pt]
    \textbf{Input}: the full term set $\mathcal{M}_{\mathrm{full}}$.
    \vspace{-5pt}
        \begin{enumerate}[itemsep=2pt, topsep=2pt, parsep=0pt]
            \item \textbf{initialize} $\mathcal{M} \gets \mathcal{M}_{\mathrm{full}}$
            \begin{itemize}[itemsep=1pt, topsep=1pt, parsep=0pt]
                \item \textbf{estimate} $\hat{G}_{\mathcal{M}}$ on the covariates of $\mathcal{M}$, and form the weights $W_{i}(t_{0}; \hat{G}_{\mathcal{M}})$
                \item fit the model on $\mathcal{M}$ and set $c \gets \mathrm{QICW}^{\bullet}(\mathcal{M};\, t_{0} \mid \hat{G}_{\mathcal{M}})$
            \end{itemize}
            \item \textbf{while} $|\mathcal{M}| > 1$ \textbf{do}:
            \begin{itemize}[itemsep=1pt, topsep=1pt, parsep=0pt]
                \item let $\mathcal{R}$ be the set of terms of $\mathcal{M}$ removable under marginality
                \item \textbf{for} each $z \in \mathcal{R}$ \textbf{do}:
                \begin{itemize}[itemsep=1pt, topsep=1pt, parsep=0pt]
                    \item \textbf{re-estimate} $\hat{G}_{\mathcal{M} \setminus \{z\}}$ on the covariates of $\mathcal{M} \setminus \{z\}$, and form the weights $W_{i}(t_{0}; \hat{G}_{\mathcal{M} \setminus \{z\}})$
                    \item fit the model on $\mathcal{M} \setminus \{z\}$ and set $q_{z} \gets \mathrm{QICW}^{\bullet}(\mathcal{M} \setminus \{z\};\, t_{0} \mid \hat{G}_{\mathcal{M} \setminus \{z\}})$
                \end{itemize}
                \item let $z^{\star}$ be the term of $\mathcal{R}$ with the smallest $q_{z}$, and set $d^{\star} \gets q_{z^{\star}} - c$
                \item \textbf{if} $d^{\star} < \varepsilon$ \textbf{then} set $\mathcal{M} \gets \mathcal{M} \setminus \{z^{\star}\}$ and $c \gets q_{z^{\star}}$ \\
                \hspace*{1em} \textbf{else} \textbf{break}
            \end{itemize}
            \item \textbf{output} the selected model $\mathcal{M}^{\star} \gets \mathcal{M}$
        \end{enumerate}
\end{algorithm}

\section{Simulation}\label{sec4: simulation}
We conducted two simulation studies to evaluate the proposed multinomial logistic regression estimators. We designed both studies to mimic the CCSS project analyzed in Section~\ref{sec3: real data analysis}: the simulated PM event times are doubly censored and subject to the competing risks of POI and SPM, and all subjects survive at least five years after cancer diagnosis. Simulation~1 confirms that the estimated CDF of the event remains bounded within $[0, 1]$, and it assesses the consistency of the coefficient estimates and the validity of the sandwich standard error estimator. It also compares four methods for estimating the censoring distribution $G(\cdot|\mathbf{Z})$ described in Section~\ref{sec2: practical implementation}: the Cox PH model on $C$, the Cox PH model on $C^{*}$, the stratified ECDF, and SRF, with results using the true censoring distribution reported as a benchmark. Simulation~2 extends this evaluation to imbalanced data scenarios, demonstrating the necessity of the Firth correction for maintaining unbiased coefficient and standard error estimates when events are rare.
 
In both studies, each generated dataset consists of $n=7000$ subjects and includes two covariates: a continuous variable $Z_{1}$ and a binary variable $Z_{2}$. All results are based on 1000 replications, and all simulations are conducted in \texttt{R}.
 
\subsection{Data Generation}\label{sec4: simulation data generation}
The data generation procedure for the two simulations is reported in Algorithm~\ref{alg:data_sim}, in which the $n$ subjects are generated independently.
\begin{algorithm}[ht]
    \small
    \singlespacing
    \vspace{-20pt}
    \caption{Simulation Process for one repetition}\label{alg:data_sim}
    \begin{flushleft}
        Generated subjects $i = 1, \ldots, n$ independently.\\[3pt]
        \textbf{For} subject $i = 1, \ldots, n$ \textbf{do}
        \vspace{-5pt}
        \begin{enumerate}[itemsep=2pt, topsep=2pt, parsep=0pt]
            \item \textbf{generate} $Z_{1i} \sim \text{Beta}(a_{1},a_{2})$ and $Z_{2i} \sim \text{Bern}(p)$
            \item \textbf{calculate} $V_{i} = 21 \times Z_{1i}$
            \item \textbf{generate censoring time:}
            \begin{itemize}[itemsep=1pt, topsep=1pt, parsep=0pt]
                \item sample $C^{*}_{i} \sim \text{Weibull}(\psi_{3i}, \psi_{4i})$
                \item set $C_{i} = C^{*}_{i} + V_{i} + 5$
            \end{itemize}
        \end{enumerate}
 
        \vspace{5pt}
        \textbf{For} subject $i = 1, \ldots, n$, let $Y_{i}(t_{0}) \coloneqq \mathbb{I}(T_{i} \le t_{0} \mid \mathbf{Z}_{i}, T_{i} > V_{i}, V_{i} < t_{0})$, and for each $t_{0}$ \textbf{do}:
        \vspace{-5pt}
        \begin{enumerate}[itemsep=2pt, topsep=2pt, parsep=0pt]
            \setcounter{enumi}{3}
            \item \textbf{determine event at $t_0$} \\
            calculate $\pi_{i}(t_{0}) = P(T_{i}\le t_{0} \mid \mathbf{Z}_{i}, T_{i} > V_{i}, V_{i} < t_{0})$ \\
            sample $Y_{i}(t_{0}) \sim \text{Bern}(\pi_{i}(t_{0}))$
            \item \textbf{if} $Y_{i}(t_{0}) = 1$ \textbf{then}
            \begin{itemize}[itemsep=1pt, topsep=1pt, parsep=0pt]
                \item set $q \gets 1$
                \item \textbf{while} $Y_{i}(t_{0}-(q-1)s) = 1$ \textbf{do}
                \begin{itemize}[itemsep=1pt, topsep=1pt, parsep=0pt]
                    \item calculate $\pi_{i}(t_{0}-qs) = P(T_{i}\le t_{0}-qs \mid \mathbf{Z}_{i}, T_{i} > V_{i}, V_{i} < t_{0}-qs)$
                    \item \textbf{check threshold:} \textbf{if} $\pi_{i}(t_{0}-qs) < 0.005$ \textbf{then} \\
                    \hspace*{1em} set $Y_{i}(t_{0}-qs) = 0$ and \textbf{break}
                    \item calculate $\rho_{cond} = \pi_{i}(t_{0}-qs) \big/ \pi_{i}(t_{0}-(q-1)s)$
                    \item sample $Y_{i}(t_{0}-qs) \sim \text{Bern}(\rho_{cond})$
                    \item \textbf{if} $Y_{i}(t_{0}-qs)=1$, set $q \gets q + 1$. \textbf{else} break
                \end{itemize}
                \item set $T_{i} = t_{0}-(q-1)s$
                \item \textbf{calculate} $\pi_{spm,i} = P(\eta=2 \mid \mathbf{Z}_{i}, T_i >V_i, V_i<T_{i}, T_i= t_{0}-(q-1)s)$
                \item \textbf{sample} $\mathbb{I}(\eta_{i}=2) \sim \text{Bern}(\pi_{spm,i})$.
            \end{itemize}
            \item let $\delta_{i} = \mathbb{I}(T_i \le C_i)$ and $U_{i} = \min\{T_{i}, C_{i}\}$
            \item store generated data $\{ Y_{i}(t_{0}), V_{i}, U_{i}, \delta_{i},  C_{i}, \delta_{i}\eta_{i}, \mathbf{Z}_{i} \}$
        \end{enumerate}
    \end{flushleft}
\end{algorithm}
 
In all simulation studies, the time unit $s$ is set to 1/12. Each simulated dataset consists of two covariates: a continuous covariate $Z_1$ drawn from a $\text{Beta}(a_1, a_2)$ distribution, and a binary covariate $Z_2$ drawn from a Bernoulli distribution. The age at diagnosis is defined as $V = 21 \times Z_1$. The distributional parameters differ across simulations: in Simulation~1, $(a_1, a_2) = (2.00, 1.00)$ and $Z_2 \sim \text{Bernoulli}(0.40)$; in Simulation~2, $(a_1, a_2) = (2.00, 2.00)$ and $Z_2 \sim \text{Bernoulli}(0.05)$. The lower probability of $Z_2 = 1$ in Simulation~2 is chosen to induce data separation for the SPM outcome. For each subject with $Y_{i}(t_{0}) = 1$ at a given $t_0$, where $Y_{i}(t_{0})$ is as defined in Algorithm~\ref{alg:data_sim}, the event time $T_i \mid \mathbf{Z}_i, T_i > V_i, V_i < t_0$ is generated from its assumed distribution. The cause of the event is then assigned as SPM with probability $\pi_{\text{spm},i}$, which is computed from the true model for $\mathbb{I}(T_i \le t_0, \eta = k \mid \mathbf{Z}_i, T_i > V_i, V_i < t_0)$. When this indicator follows a multinomial logistic regression, $\pi_{\text{spm},i} = e^{\chi_{2}(t_{0})} / (e^{\chi_{1}(t_{0})} + e^{\chi_{2}(t_{0})})$, where $\chi_{1}(t_{0})$ and $\chi_{2}(t_{0})$ are the linear predictors for POI and SPM at age $t_{0}$, respectively. To reflect the CCSS cohort structure, in which all subjects survived at least five years post-diagnosis, the right-censoring time for subject $i$ is defined as $C_i = C_i^{*} + V_i + 5$. The gap time $C_i^{*}$ follows a Weibull distribution with shape parameter $\psi_{3i}$ and scale parameter $\psi_{4i}$. In Simulation~1, $\psi_{3i} = 3.34 - 0.10*Z_{2i}$ and $\psi_{4i} = 18.00$; in Simulation~2, $\psi_{3i} = 3.34 - 0.01*Z_{2i}$ and $\psi_{4i} = 20.00 - 2.00*Z_{2i}$.

\subsection{Simulation Results}
The performance of the proposed estimators was evaluated using four metrics: the sample mean of the estimates (SMEAN), the sample standard deviation of the estimates (SSD), the sample mean of the estimated standard errors (SMESE), and the root sample mean squared error (RSMSE). Consistency is assessed by comparing the SMEAN to the true parameter values, and the validity of the standard error estimator is assessed by comparing the SMESE to the SSD. In addition, we compared the estimated conditional survival probabilities from the separate logistic regressions and the multinomial logistic regression to examine whether the estimated CDF of the event remained bounded within $[0,1]$.

\subsubsection[Simulation 1: Consistency and Robustness of Proposed Estimators with Different G(.|Z)]{Simulation~1: Consistency and Robustness of Proposed Estimators with Different $\hat{G}(\cdot|\mathbf{Z})$}The indicator $\mathbb{I}(T_i \le t_0, \eta_i = k \mid \mathbf{Z}, T > V)$ is generated from a multinomial logistic regression:
\begin{align}\label{model sim1: true model}
\begin{split}
\log\left[ \frac{P(T \le t, \eta = k|\mathbf{Z}, T > V)}{P(T > t|\mathbf{Z}, T > V)} \right]= \alpha_{k}(t)+\beta_{1k}(t) Z_1 + \beta_{2k}(t) Z_2,
\end{split}
\end{align}
where the intercept is linear in age, $\alpha_{k}(t) = \gamma_{0k} + \gamma_{1k}t$, and the regression coefficients $\beta_{1k}(t)$ and $\beta_{2k}(t)$ are constant over time. For the POI outcome ($k=1$), $\gamma_{01} = -7.00$, $\gamma_{11} = 0.24$, $\beta_{11} = -5.04$, and $\beta_{21} = 0.50$. For the SPM outcome ($k=2$), $\gamma_{02} = -7.40$, $\gamma_{12} = 0.23$, $\beta_{12} = -4.83$, and $\beta_{22} = 1.00$. Descriptive statistics for Simulation~1 are summarized in Table~\ref{table: desc_sim1_data}. All the estimates from the two approaches were evaluated at four distinct time points: 16, 25, 30, and 40 using the simulated data.
 
\begin{figure}[H]
\centering
\scalebox{0.9}{%
\begin{minipage}{\textwidth}
\centering
\begin{subfigure}[b]{0.48\textwidth}
    \centering
    \includegraphics[width=0.6\textwidth]{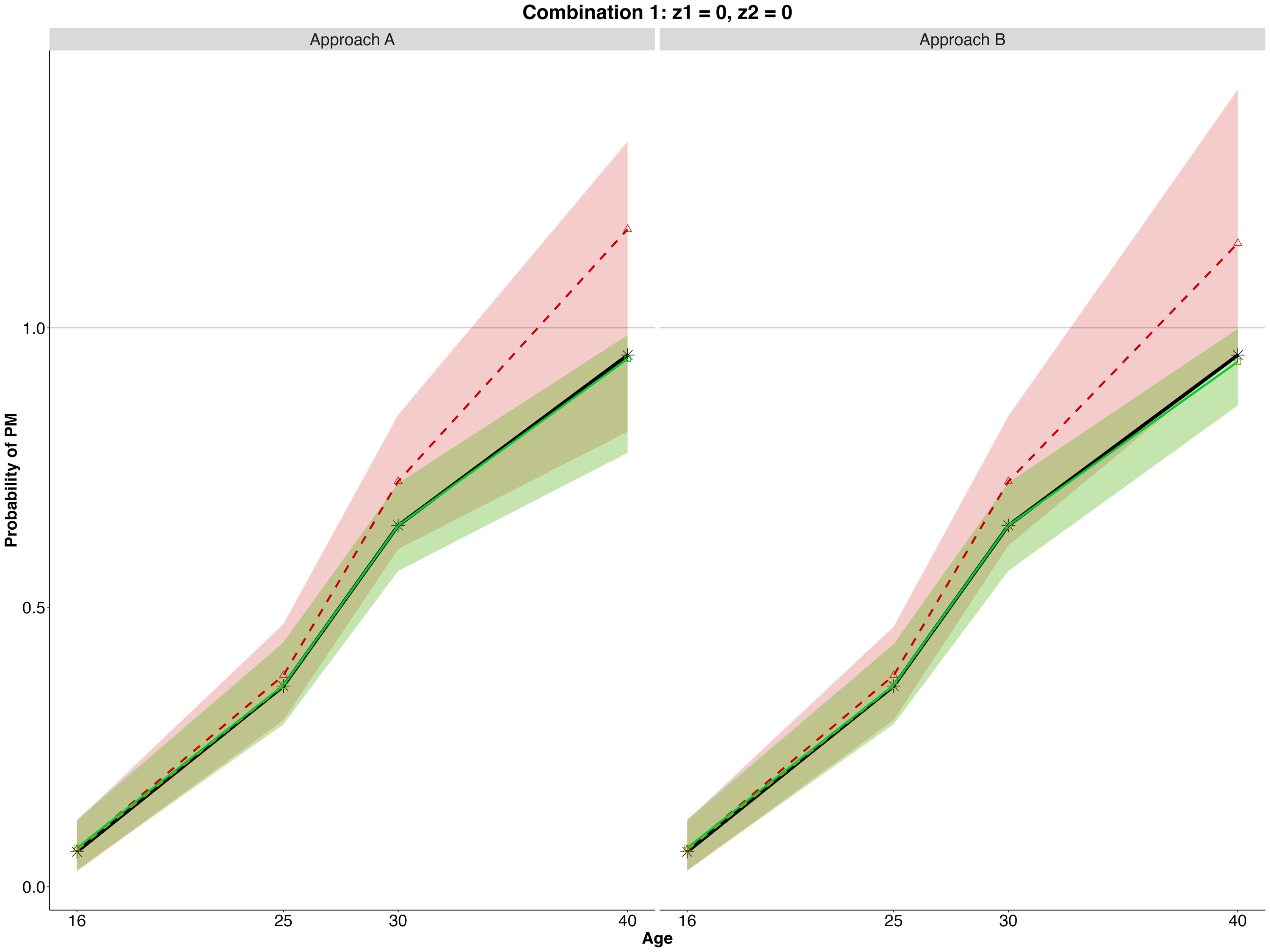}
    \caption{}
\end{subfigure}
\hfill
\begin{subfigure}[b]{0.48\textwidth}
    \centering
    \includegraphics[width=0.6\textwidth]{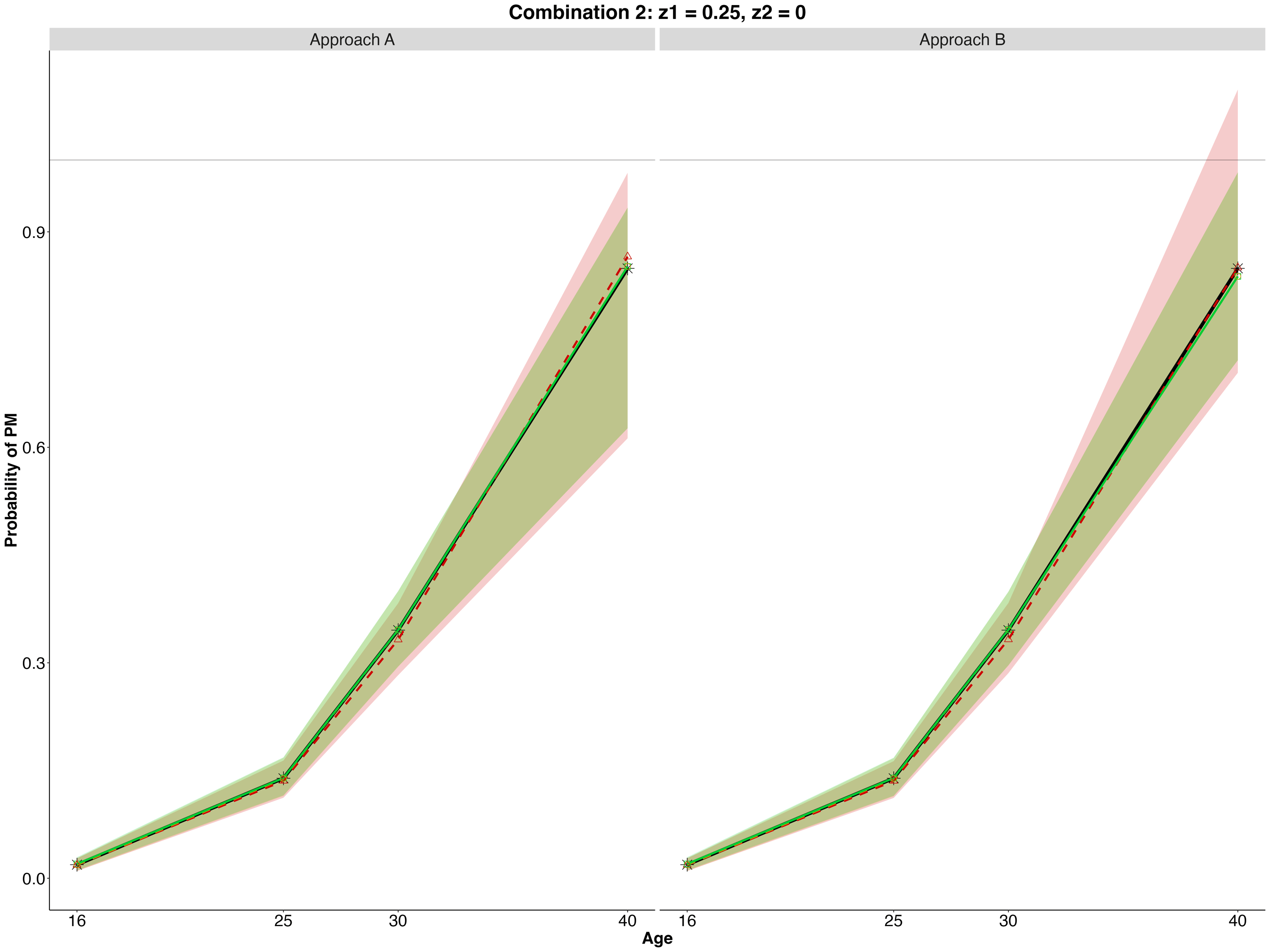}
    \caption{}
\end{subfigure}

\begin{subfigure}[b]{0.48\textwidth}
    \centering
    \includegraphics[width=0.6\textwidth]{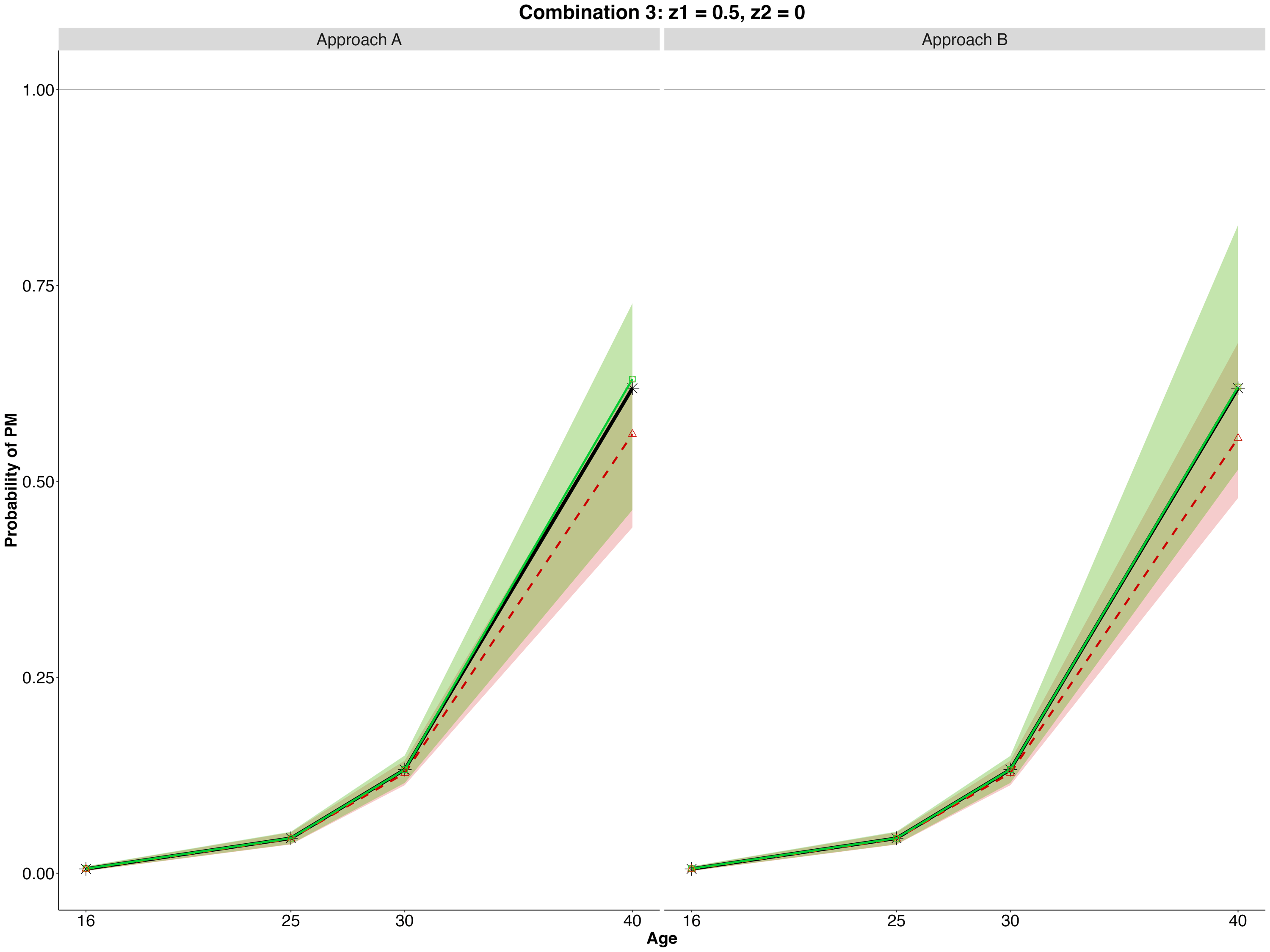}
    \caption{}
\end{subfigure}
\hfill
\begin{subfigure}[b]{0.48\textwidth}
    \centering
    \includegraphics[width=0.6\textwidth]{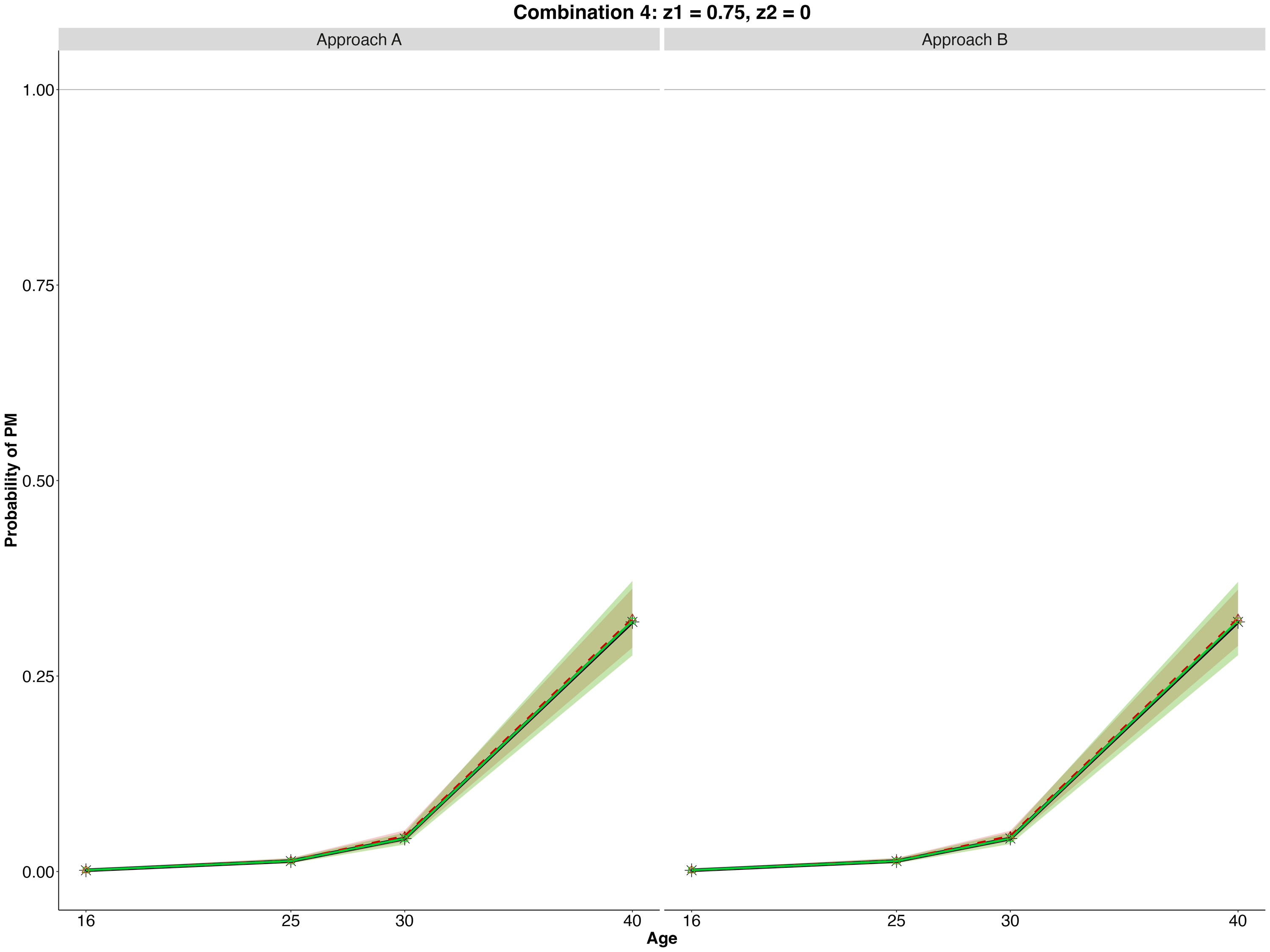}
    \caption{}
\end{subfigure}

\begin{subfigure}[b]{0.48\textwidth}
    \centering
    \includegraphics[width=0.6\textwidth]{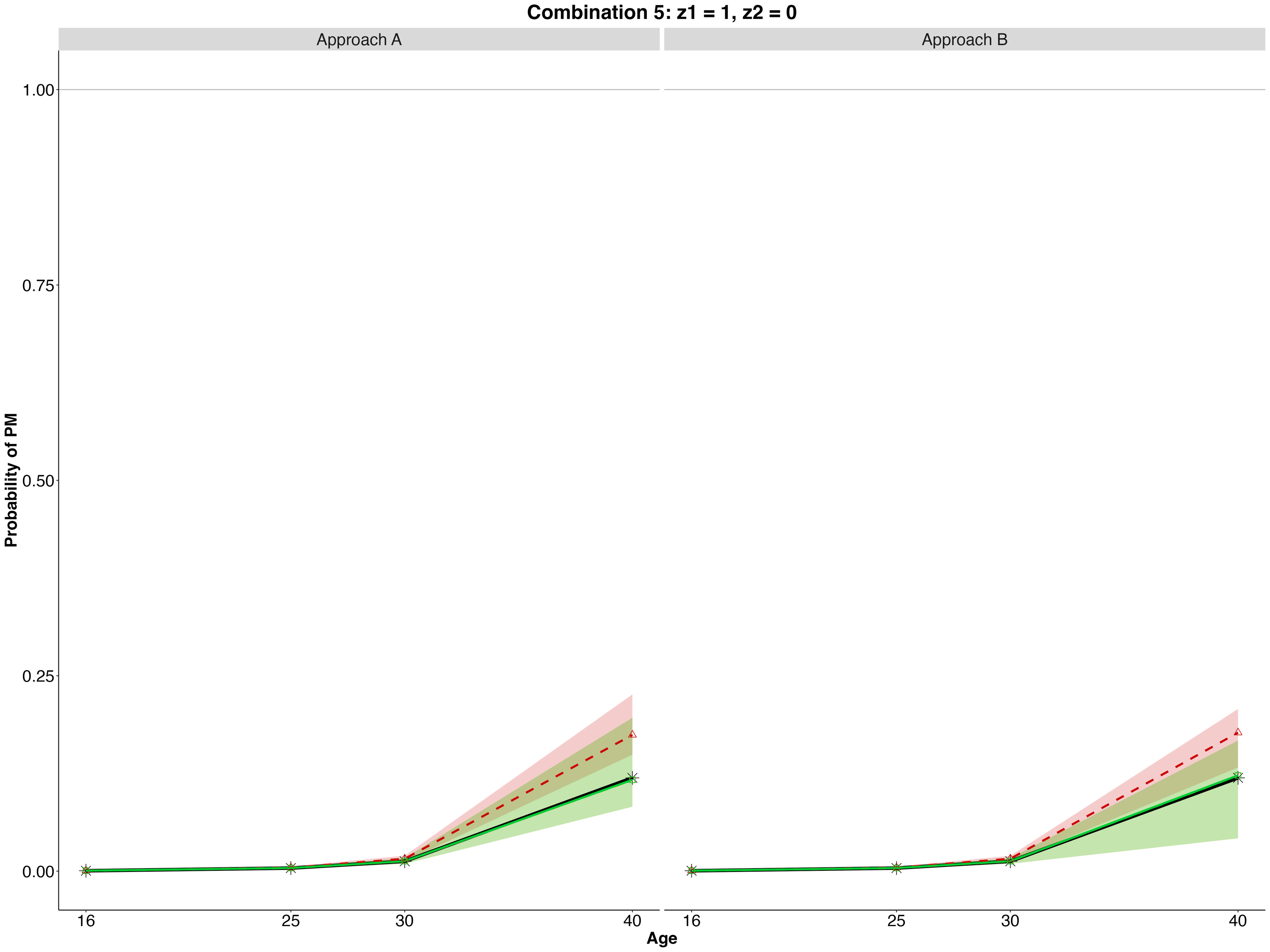}
    \caption{}
\end{subfigure}
\hfill
\begin{subfigure}[b]{0.48\textwidth}
    \centering
    \includegraphics[width=0.6\textwidth]{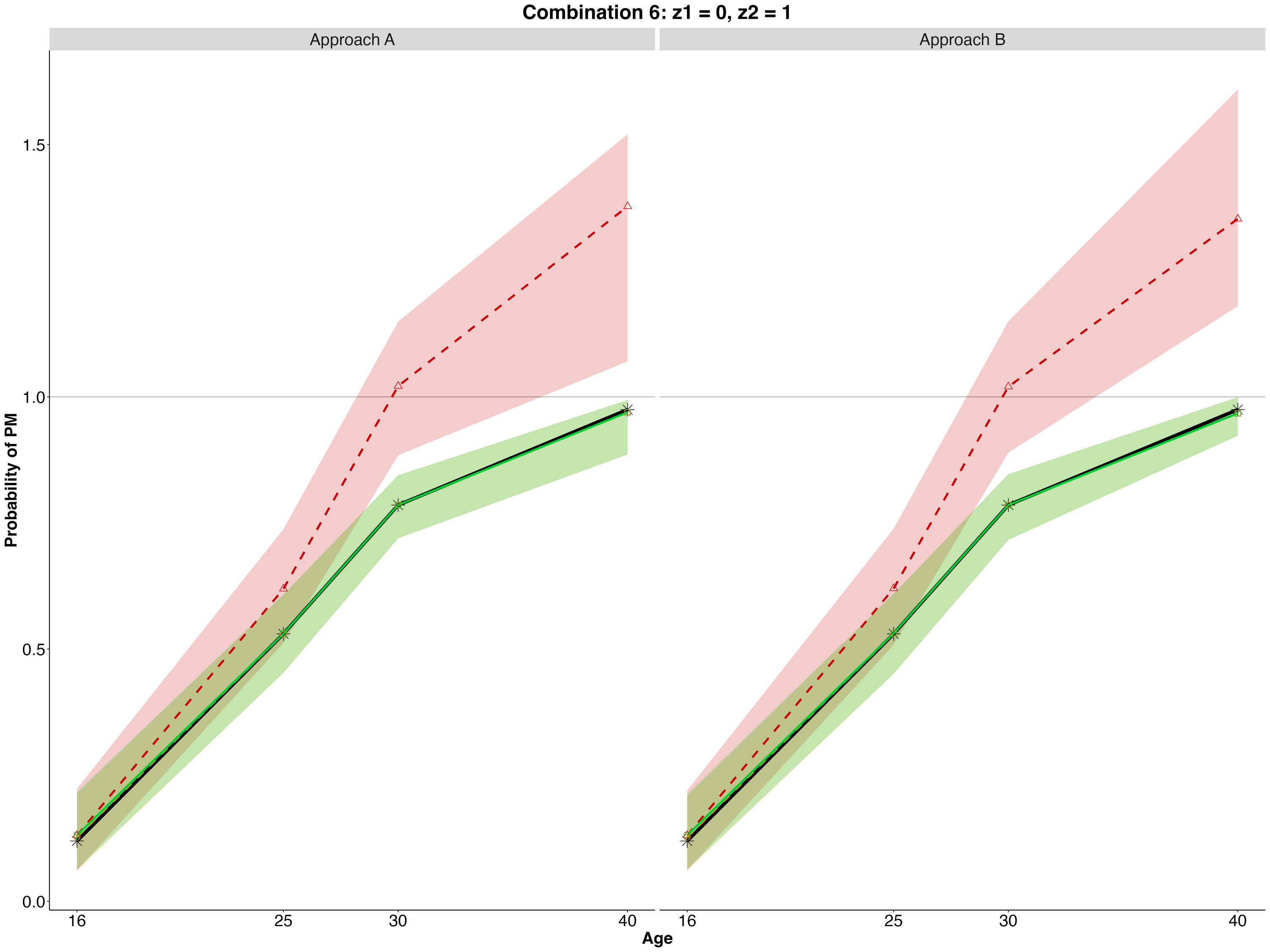}
    \caption{}
\end{subfigure}

\begin{subfigure}[b]{0.48\textwidth}
    \centering
    \includegraphics[width=0.6\textwidth]{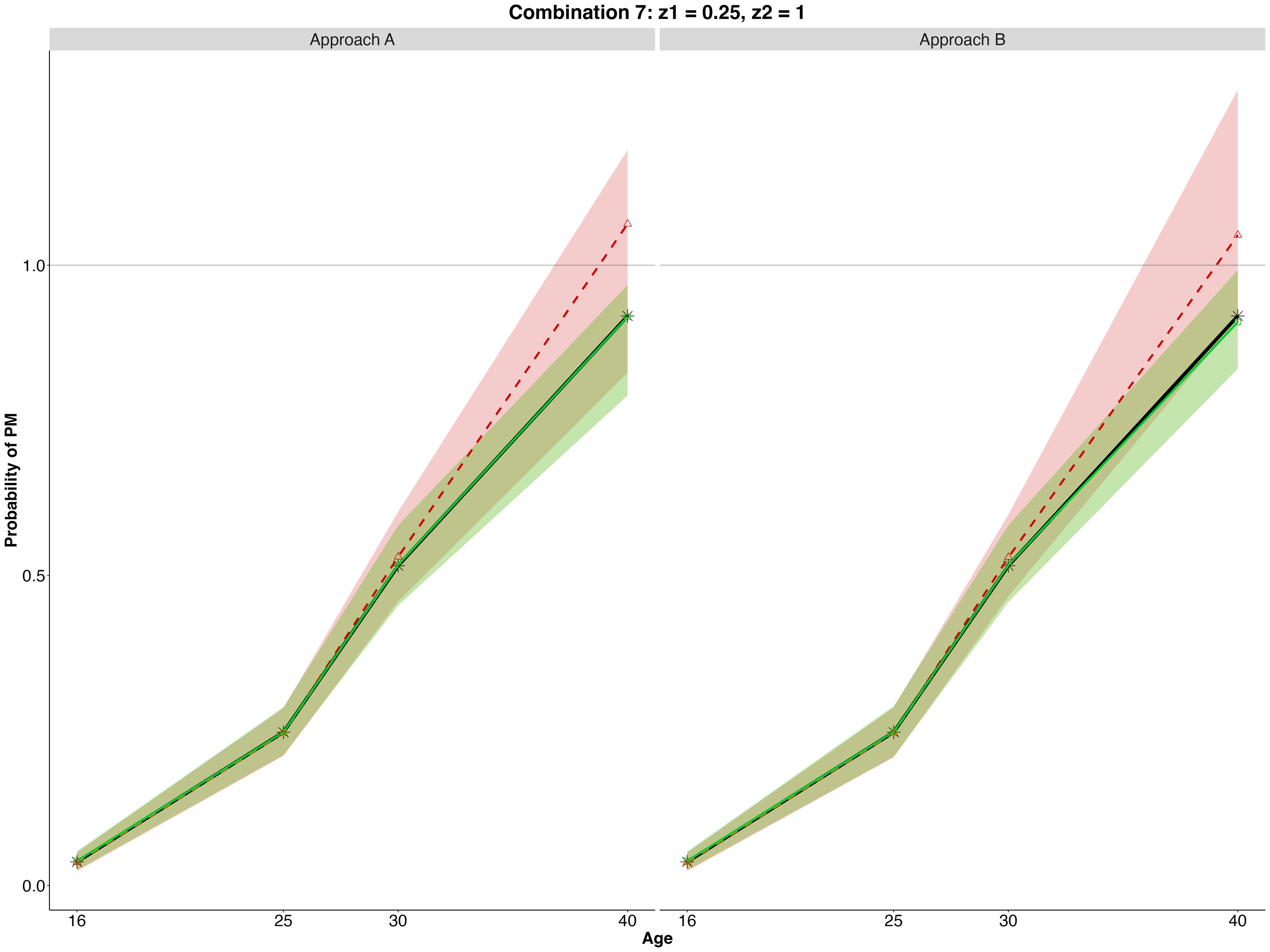}
    \caption{}
\end{subfigure}
\hfill
\begin{subfigure}[b]{0.48\textwidth}
    \centering
    \includegraphics[width=0.6\textwidth]{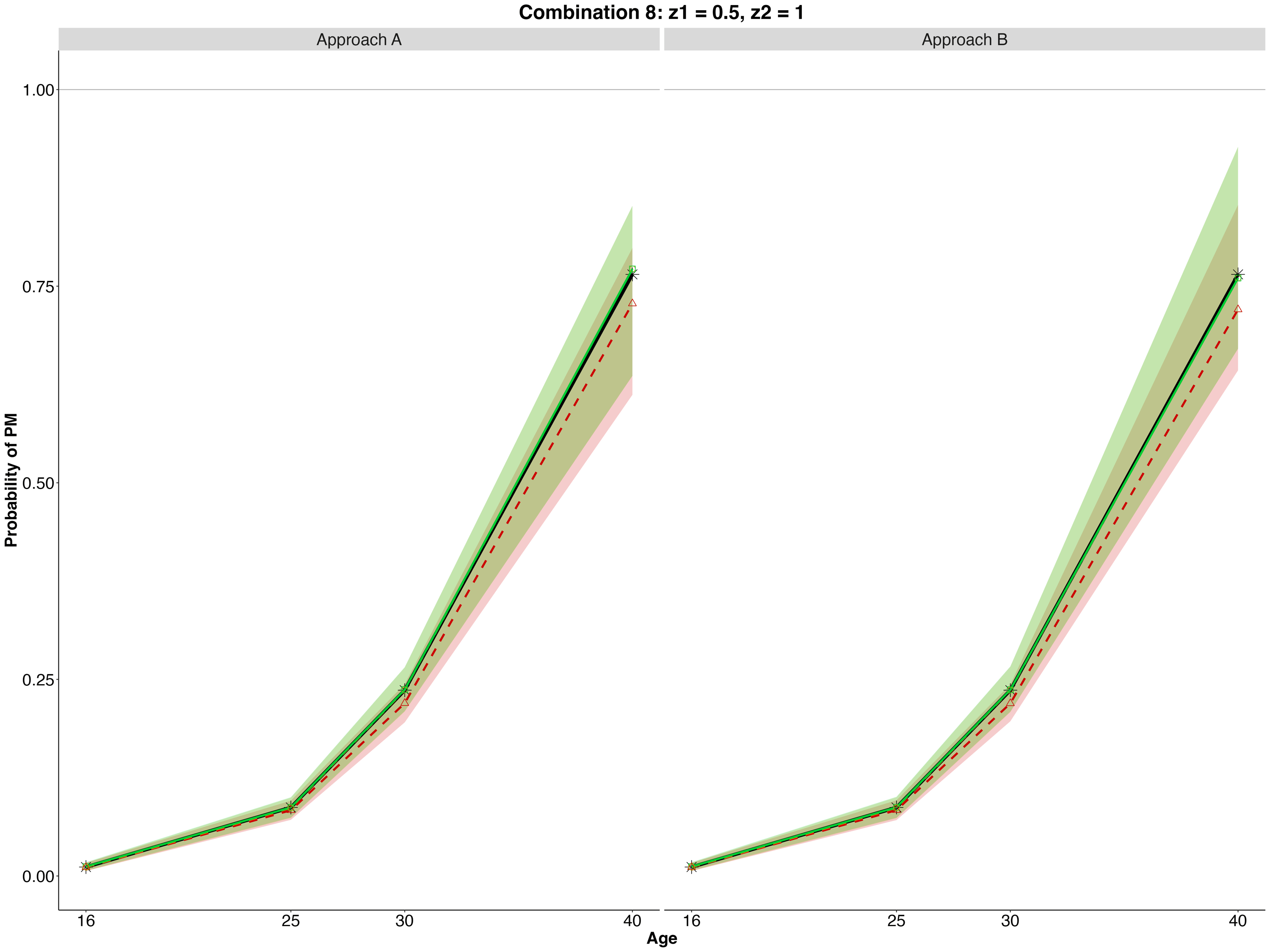}
    \caption{}
\end{subfigure}

\begin{subfigure}[b]{0.48\textwidth}
    \centering
    \includegraphics[width=0.6\textwidth]{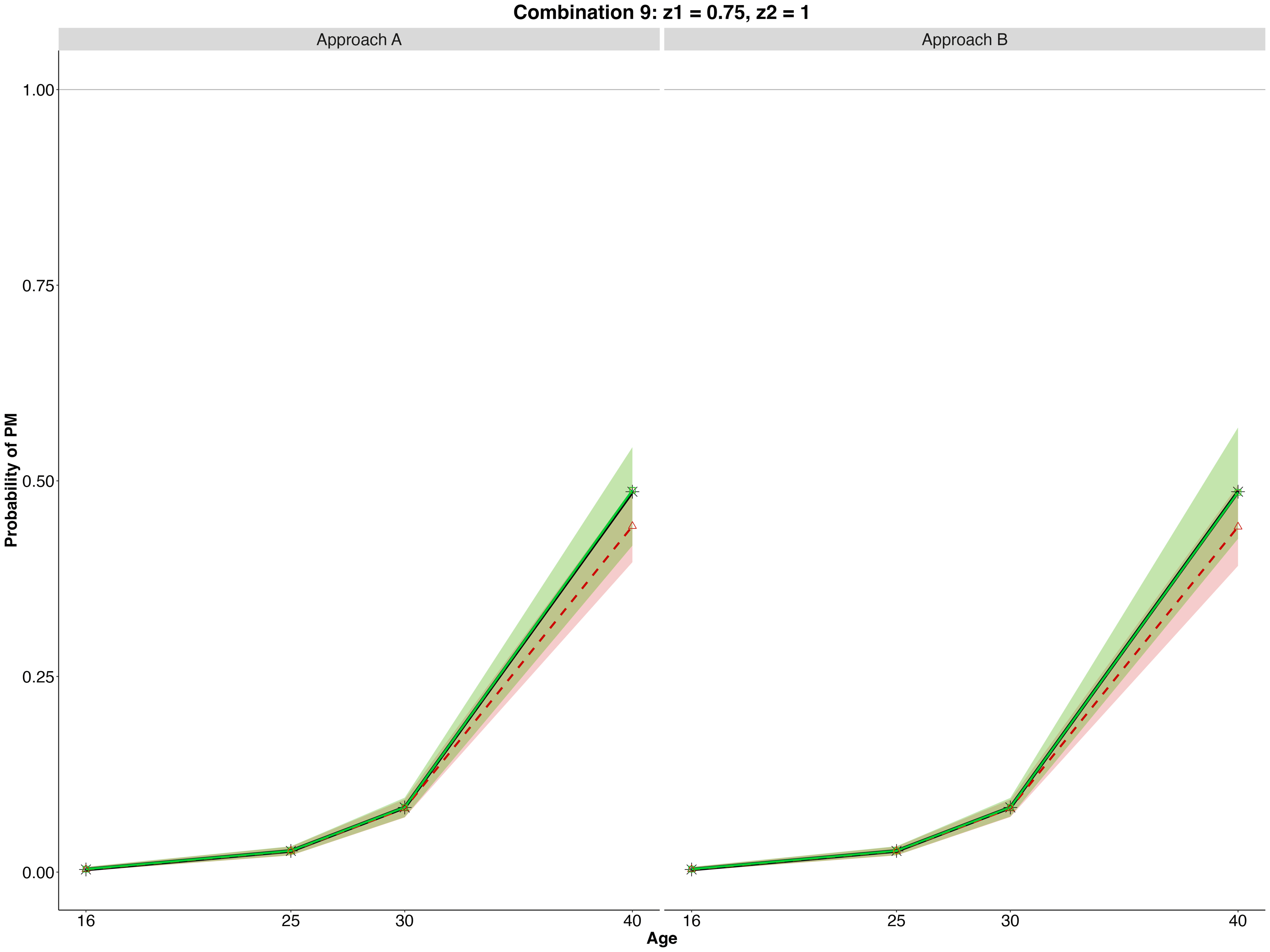}
    \caption{}
\end{subfigure}
\hfill
\begin{subfigure}[b]{0.48\textwidth}
    \centering
    \includegraphics[width=0.6\textwidth]{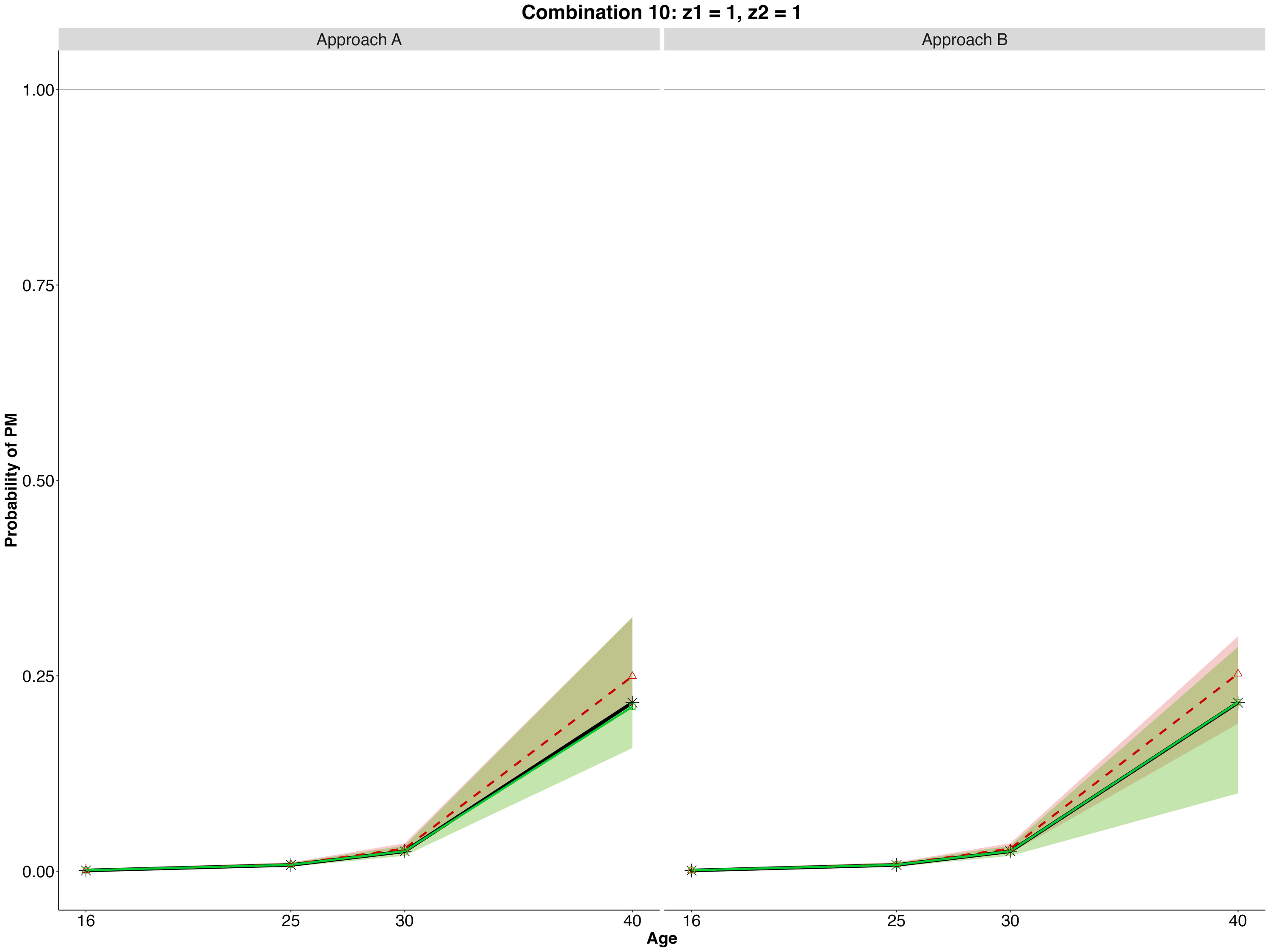}
    \caption{}
\end{subfigure}
\end{minipage}%
}
\caption{Estimated probability of PM with $95\%$ CIs across ages in Simulation~1. Red: logistic regression; yellow: multinomial logistic regression without Firth correction; green: multinomial logistic regression with Firth correction. The results from the multinomial logistic regression with and without Firth correction are nearly identical in all panels at all $t_0$. Left and right panels correspond to Approaches~A and~B, respectively. $\hat{G}(\cdot \mid \mathbf{Z})$ is estimated by SRF with the number of trees set to $100$ and the node size set to $500$.}
\label{fig: sim1_poi_all_prob}
\end{figure}

 
\medskip
\noindent{\sf \underline{Probability Estimation and the Summation Constraint}}
Figure~\ref{fig: sim1_poi_all_prob} displays the estimated probability of PM across ages under both the logistic and multinomial logistic regression models, where $\hat{G}(\cdot|\mathbf{Z})$ is estimated using SRF with the number of trees set to 100 and the node size, defined as the minimal node size to split at, set to 500. The multinomial logistic regression, with or without Firth correction, provides estimates that are close to the true probabilities at all ages, and the corresponding 95\% confidence intervals (CIs) capture the true probabilities of PM at all $t_0$. The logistic regression does not always provide accurate point estimates. Moreover, for certain covariate combinations (e.g., $Z_1 = 0$ and $Z_2 = 1$), the estimated CDF of the event exceeds one. In contrast, both approaches~A and~B under the multinomial formulation with or without Firth correction ensure that the estimated probabilities of PM remain strictly within $[0, 1]$, confirming the theoretical advantage of the multinomial model discussed in Section~\ref{sec2: estimation procedure}.
 
\centerline{ \textit{----------Insert Figure \ref{fig: sim1_poi_all_prob} Here----------}}
 
\medskip
\noindent{\sf \underline{Consistency and Standard Error Estimation}}
The coefficient estimates and the comparison of the SMESE with the SSD are reported in Supplementary Materials~\ref{supp: restul_of_simulation_set1}. For both POI and SPM outcomes, the sample means of the estimates are close to the true parameter values across all ages and the 95\% CIs cover the true values under both approaches~A and~B, confirming the consistency of the proposed estimators. The SMESE and SSD are in close agreement for POI under both approaches. For SPM, the multinomial logistic regression without Firth correction underestimates the standard error of $\hat{\beta}_{22}$, while the Firth-corrected estimator obtains SMESE that closely tracks SSD. Approach~A is more efficient than Approach~B at later ages despite the high censoring rate at age 40 (58.32\%), because a sufficient proportion of PM events are observed (on average 23.22\%). At age 40 under Approach~B, the $G(\cdot|\mathbf{Z})$ is close to zero for some subjects in certain repetitions, and consequently $\hat{G}(\cdot|\mathbf{Z})$ is as well, resulting in large IPCW that inflate the standard error estimator. This indicates that Approach~B is more sensitive to the estimation of the censoring distribution.
 
\medskip
\noindent{\sf \underline{Assessment of $\hat{G}(\cdot|\mathbf{Z})$}}
The RSMSE of the coefficient estimates under the four methods for estimating $G(\cdot|\mathbf{Z})$ is reported in Supplementary Materials~\ref{supp: restul_of_simulation_set1}. The SRF and the stratified ECDF obtain the lowest RSMSE under both approaches, while the Cox PH model on $C$ performs the worst due to the violation of the proportional hazards assumption for $C$. The Cox PH model on $C^{*}$ serves as a reasonable alternative when the proportional hazards assumption is plausible.

\subsubsection{Simulation~2: Firth Correction Under Imbalanced Data}
Simulation~2 uses the same data-generating model as Simulation~1, but with different parameter values chosen to induce data imbalance and separation for $Z_2$. For the POI outcome ($k=1$), $\gamma_{01} = -7.40$, $\gamma_{11} = 0.22$, $\beta_{11} = -4.62$, and $\beta_{21} = 2.00$. For the SPM outcome ($k=2$), $\gamma_{02} = -8.80$, $\gamma_{12} = 0.23$, $\beta_{12} = -4.83$, and $\beta_{22} = 1.00$. Descriptive statistics for Simulation~2 are summarized in Table~\ref{table: desc_sim2_data}. As shown in the table, the number of SPM events among $Z_2 = 1$ subjects is extremely small at early ages (e.g., fewer than 1 on average at $t_0 = 16$), leading to separation or quasi-complete separation. All estimates from the two approaches were evaluated at six distinct time points: 16, 18, 20, 25, 30, and 40 using the simulated data.
 
 
\medskip
\noindent{\sf \underline{Probability Estimation and the Summation Constraint}}
As in Simulation~1, the multinomial logistic regression with or without Firth correction obtains probability estimates close to the true probabilities across all ages, while the logistic regression estimates exceed one for certain covariate combinations (Figure~\ref{fig: sim2_poi_all_prob} in Supplementary Materials~\ref{supp: restul_of_simulation_set2}).
 
\begin{figure}[H]
  \centering
  \includegraphics[width=1\linewidth]{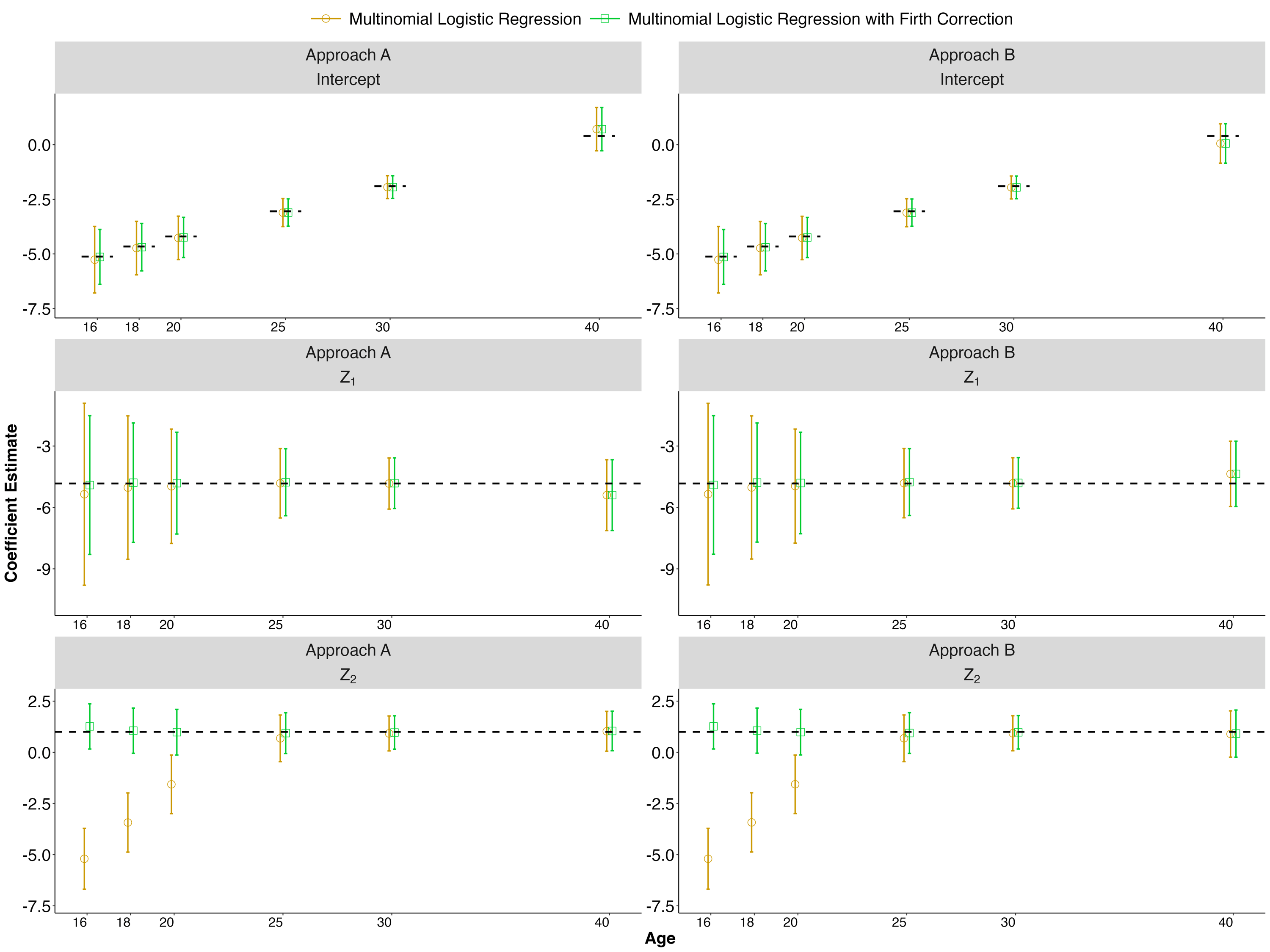}
  \caption{Estimated coefficients for the SPM outcome across ages in simulation 2. The $\hat{G}(\cdot|\mathbf{Z})$ is obtained using SRF. When using SRF, we choose number of tree equal to 100 and node size equal to 500. The dashed line represents true value of corresponding coefficient.}\label{fig: sim2_spm_all_coef}
\end{figure}

\medskip
\noindent{\sf \underline{Effect of Firth Correction on Coefficient and Standard Error Estimation}}
Figure~\ref{fig: sim2_spm_all_coef} presents the estimated coefficients for the SPM outcome. The multinomial logistic regression without Firth correction produces a severely biased $\hat{\beta}_{22}$ at early ages: at $t_0 = 16$, the sample mean is $-5.20$, far from the true value of $1.00$. The Firth-corrected model substantially reduces this bias and obtains estimates close to the true values. As $t_0$ increases beyond 25, the number of SPM events grows and the two models converge to nearly identical estimates. For the POI outcome, the sample means of the estimates from the two models are close to the true values, with 95\% CIs covering the true values across all ages, confirming that the Firth correction does not introduce additional bias when the data are balanced (Supplementary Materials~\ref{supp: restul_of_simulation_set2}).
 
\centerline{ \textit{----------Insert Figure \ref{fig: sim2_spm_all_coef} Here----------}}
 
The validity of the sandwich standard error estimator follows the same pattern as Simulation~1: the SMESE and SSD agree closely for POI under both models, while for SPM the multinomial logistic regression without Firth correction underestimates the standard error of $\hat{\beta}_{22}$ at early ages, and the Firth correction restores reliable standard error estimation. The RSMSE patterns across the four estimators of $G(\cdot|\mathbf{Z})$ are also consistent with Simulation~1, except that without Firth correction the RSMSE of $\hat{\beta}_{22}$ at early ages is extremely large across all censoring estimators including the true $G(\cdot|\mathbf{Z})$, confirming that this large RSMSE is driven by the coefficient bias from separation rather than the choice of censoring distribution estimator. Detailed results are reported in Supplementary Materials~\ref{supp: restul_of_simulation_set2}.

\section{Analysis of Premature Menopause Data} \label{sec3: real data analysis}
Having evaluated the proposed approaches in Section~\ref{sec4: simulation} under settings designed to mimic the CCSS data, we now apply them to these data.

\subsection{Descriptive Statistics}
The CCSS is a multi-institutional study that enrolled childhood cancer survivors from 31 centers across North America. Eligibility criteria for the cohort included a cancer diagnosis before the age of 21 between 1970 and 1999, and survival of at least five years post-diagnosis \citep{robison2002study,robison2009childhood}. For this analysis, the study sample is further restricted to female survivors who were at least 18 years old at their last follow-up and had available self-reported menstrual information \citep{im2023development}. The event of interest is PM, defined as the cessation of menstrual function before the age of 40. As described in Section~\ref{sec1: Introduction}, PM arises from two competing causes: POI and SPM. The right-censoring age is defined as the minimum of the age at second malignant neoplasm and the age at last follow-up; censoring ages are available for all study subjects. Under a missing at random assumption, the analysis was conducted on a study group of 6,961 subjects with complete data.

Following \citet{im2023development}, we select five exposure variables, giving six covariates $Z_1$ to $Z_6$, together with two interaction terms. The six covariates are the rescaled age at cancer diagnosis (age divided by 21 so its range is between 0 and 1, $Z_1$), race (with Caucasian as the baseline, and indicators $Z_2$ for African American and $Z_3$ for Other), and the indicators for receiving bone marrow transplant (BMT, $Z_4$), alkylating agents ($Z_5$), and radiotherapy to the abdomen, pelvis, or whole body ($Z_6$). The two interactions are the rescaled age at cancer diagnosis with the indicator for BMT ($Z_{1}Z_{4}$) and the rescaled age at cancer diagnosis with the indicator for radiotherapy ($Z_{1}Z_{6}$).

Table~\ref{table: proj2_desc_stat} presents descriptive statistics of the five exposure variables stratified by PM status, POI status, and SPM status. Among the 6,961 subjects, 15.08\% experienced premature menopause before censoring, with 11.46\% of the cohort attributed to POI and 3.62\% to SPM. The majority of subjects were Caucasian (87.67\%). More than half (50.51\%) received alkylating agents, while 22.63\% received radiotherapy and only 4.28\% received bone marrow transplantation. The median age at cancer diagnosis was 7.43 years overall, but was notably higher among those who experienced PM (median 10.73 years) compared to those who did not (median 6.96 years). Several covariate--outcome combinations exhibited sparse cell counts: only 8 African American subjects and 8 subjects of other race experienced SPM, and only 9 subjects who received BMT experienced SPM.

\centerline{\textit{----------Insert Table \ref{table: proj2_desc_stat} Here---------- }}

\subsection{Inferential Analysis}
We conduct event time analyses at prespecified ages (i.e., $t_{0}$) from 17 to 40. The $G(\cdot|\mathbf{Z})$ in the IPCW is estimated by the SRF, where both the number of trees and the node size are set to 100. All analyses are conducted in \texttt{R}. The \texttt{R} packages \texttt{ranger} and \texttt{survival} are employed to estimate $\hat{G}(\cdot\mid\mathbf{Z})$.

\begin{figure}[H]
  \centering
  \includegraphics[width=1\linewidth]{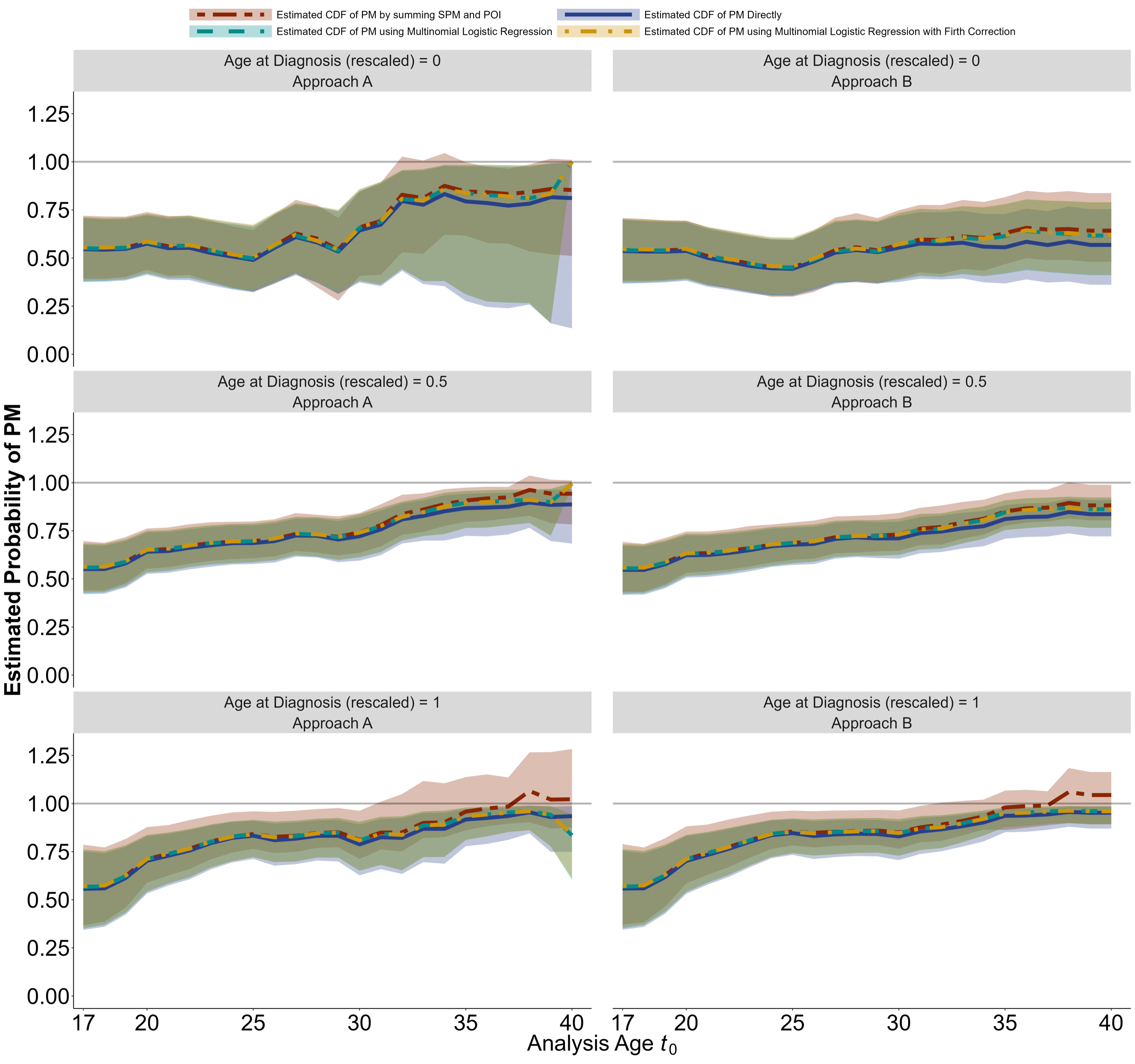}
\caption{Estimated probability of premature menopause obtained from the model directly on PM, Model 1, Model 2, and Model 2 with Firth correction. The estimates (average) represent the group of African American female who received BMT, chemotherapy, and radiotherapy (i.e., $Z_{2}=Z_{4}=Z_{5}=Z_{6}=1$ and $Z_{3}=0$), with $Z_1$ varying across subpanels: $Z_1=0$ (top), $Z_1=0.5$ (middle), and $Z_1=1$ (bottom). The $\hat{G}(\cdot|\mathbf{Z})$ was obtained using a SRF with the number of trees and node size both set to 100.}\label{fig: comp_prob_4_logs_model}
\end{figure}

A known limitation of using model~\eqref{model: project2 model 1} is that the estimated CDF of PM may exceed one. Figure~\ref{fig: comp_prob_4_logs_model} shows this occurs in the present data: the estimated probability of PM based on model~\eqref{model: project2 model 1}, computed as the sum of the estimated CIRs for POI and SPM, exceeds one at later ages. Model~\eqref{model: project2 model 2}, by construction, obtains estimated probabilities within the $[0, 1]$ interval at all ages, regardless of whether Firth correction is applied. In addition, under Approach~A, the estimated probabilities from model~\eqref{model: project2 model 2} approach 99\% at age 40 with narrow CIs, even though the corresponding coefficient standard errors are large, as summarized later in this section. For POI, the cell counts are small but no separation occurs, so the coefficient estimates remain valid and fluctuate more at later ages, because Approach~A assigns zero weight to subjects censored before $t_0$ and few subjects remain at those ages. The strong positive effect of BMT on POI risk drives the estimated CDF of PM toward one. Once a fitted probability is close to one, the curve mapping the linear predictor to the probability is nearly flat, so even a large coefficient standard error shifts the fitted probability only slightly, leaving the CIs on the CDF scale narrow. The estimated regression coefficients and sandwich standard errors from model~\eqref{model: project2 model 1} are reported in Supplementary Materials~\ref{supp: real_data_log_reg}. We now focus on the results from model~\eqref{model: project2 model 2}.

\centerline{ \textit{----------Insert Figure \ref{fig: comp_prob_4_logs_model} Here----------}}

\begin{figure}[htbp]
    \centering
    \includegraphics[width=\textwidth]{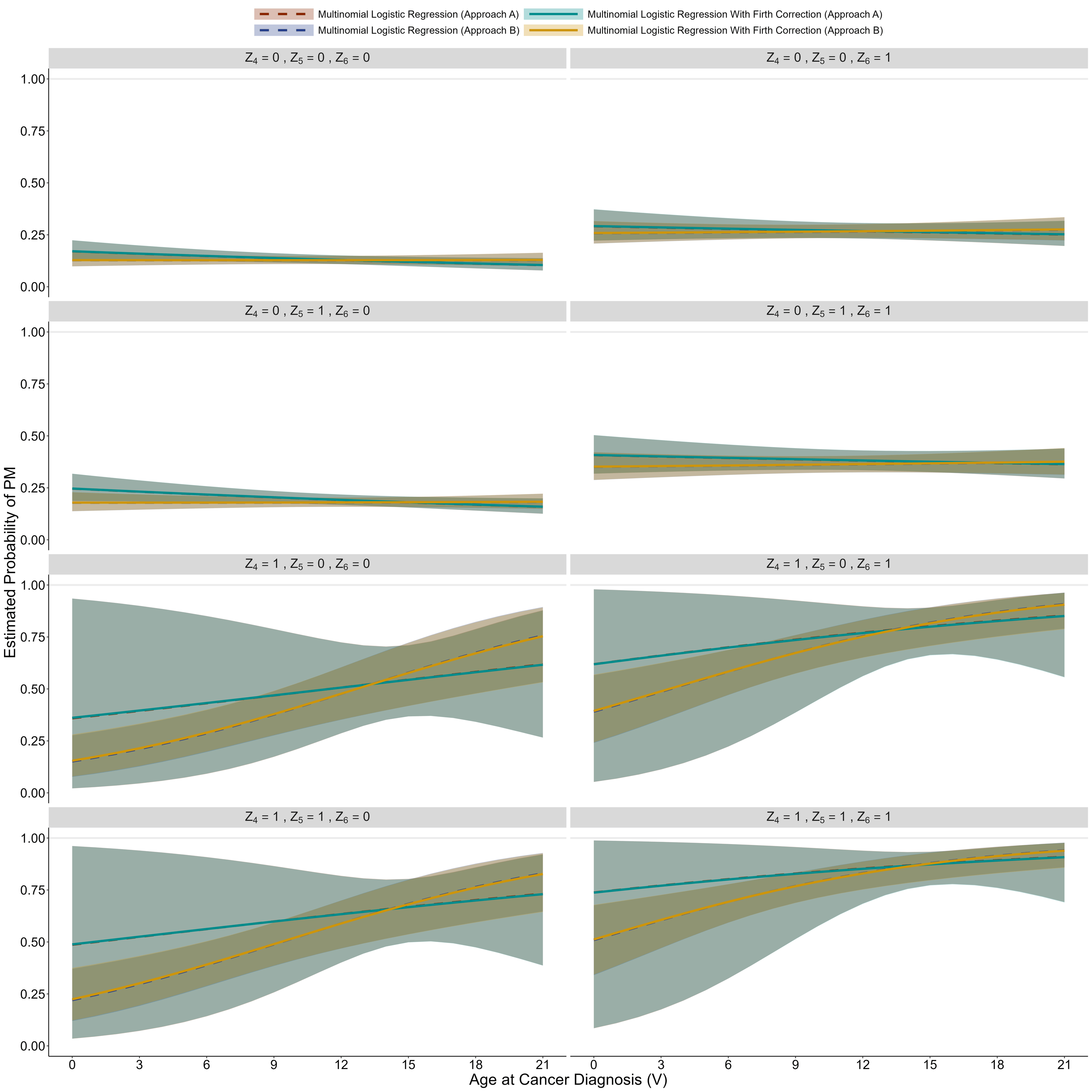}
    \caption{Estimated probability of premature menopause at $t_0=40$ versus age at cancer diagnosis ($V$) for Caucasian survivors ($Z_2=0, Z_3=0$), across eight treatment combinations of $Z_4$ (BMT), $Z_5$ (alkylating agents), and $Z_6$ (abdomen/pelvis/TBI radiation). Solid lines: Multinomial logistic regression with Firth correction; dashed lines: without correction. Red/green: Approach~A; blue/yellow: Approach~B. Shaded bands show 95\% CIs. Censoring distribution estimated via survival random forest.}
    \label{fig:allV_prob_facet6_srf_t0_40}
\end{figure}

Figure~\ref{fig:allV_prob_facet6_srf_t0_40} displays the estimated probability of PM at $t_0=40$ as a function of age at cancer diagnosis ($V$), under model~\eqref{model: project2 model 2} with and without Firth correction, for Caucasian survivors across eight treatment combinations. The Firth-corrected and uncorrected estimates are nearly identical for all covariate profiles, confirming that the correction primarily addresses coefficient instability rather than altering the estimated probabilities. The estimated probabilities under approaches~A and~B differ in magnitude but their 95\% CIs overlap at most ages, suggesting that the two approaches lead to broadly consistent inferences.

Across all treatment combinations, the estimated probability of PM increases with a later age at cancer diagnosis, and this effect is most pronounced among subjects who received BMT ($Z_4=1$). Receipt of BMT has the largest positive impact on PM risk, as panels with $Z_4=1$ show substantially increased probabilities compared to the case when $Z_4=0$. Radiation therapy ($Z_6=1$) has a moderate positive effect, shifting estimated probabilities upward across all ages at diagnosis. Alkylating agents ($Z_5=1$) also increase PM risk, though the incremental effect is smaller. When multiple treatments are received simultaneously (e.g., $Z_4=Z_5=Z_6=1$), the estimated probabilities are highest, reflecting the cumulative burden of gonadotoxic therapies.

\centerline{ \textit{----------Insert Figure \ref{fig:allV_prob_facet6_srf_t0_40} Here----------}}

Detailed results from model~\eqref{model: project2 model 2}, including coefficient estimates, sandwich standard errors, 95\% CIs, and the sensitivity analysis comparing the four estimators of $G(\cdot|\mathbf{Z})$, are reported in Figures~\ref{fig: multi_real_data_coef_srf}--\ref{fig:multi_firth_spm_se} in Supplementary Materials~\ref{supp: real_data_multi_reg}. The coefficient estimates for POI are stable across ages under both approaches, while the SPM estimates without Firth correction are highly volatile due to quasi-complete separation arising from sparse SPM events among African American subjects, BMT recipients, and radiation therapy recipients. Firth correction stabilizes the SPM estimates with a negligible effect on POI. The sandwich standard errors of $Z_4$ and $Z_{1}Z_{4}$ for POI increase sharply with age under Approach~A but remain small and stable under Approach~B, consistent with the observation of \citet{zhou202xagespecific} that Approach~B is more efficient under heavy censoring. Substantively, the effects of the treatment-related covariates on POI risk are significant and positive: the effects of BMT and alkylating agents increase with attained age, while radiation therapy increases POI risk at all ages but with an effect that diminishes over attained age. In contrast, the SPM coefficients fluctuate around zero with 95\% CIs generally covering zero, consistent with SPM arising from surgical interventions, with its risk not directly captured by the considered covariates. The choice of censoring distribution estimator (SRF, stratified ECDF, Cox PH model on $C$, or Cox PH model on $C^{*}$) has a substantial impact on coefficient estimates while its effect on standard errors is negligible (Figures~\ref{fig:multi_nofirth_poi_coef}--\ref{fig:multi_firth_spm_se}).

\subsection{Analysis with an Extended Covariate Set}\label{sec3: extended analysis}

To illustrate the variable selection procedure of Section~\ref{sec2: variable selection}, we refit model~\eqref{model: project2 model 2} on the same cohort with a larger covariate set: the radiotherapy indicator is replaced by three rescaled radiation-dose variables, and five additional chemotherapy-class indicators are added. The resulting 13 covariates $Z_{1}$ to $Z_{13}$, together with the three interactions $Z_{1}Z_{4}$, $Z_{1}Z_{5}$, and $Z_{1}Z_{6}$, give a full model of 15 terms, race counting as a single term through its two indicators $Z_{2}$ and $Z_{3}$. These labels are specific to this subsection and do not match those used elsewhere in the paper. Dropping 28 subjects with missing values on the new covariates reduces the study group from $n = 6{,}961$ to $n = 6{,}933$. Covariate definitions are given in Supplementary Materials~\ref{supp: analysis3_covariates}. Backward elimination is driven by $\mathrm{QICW}$ in four settings, namely Approaches~A and~B each with and without the Firth correction, and is run independently at each of the 24 analysis ages $t_{0} = 17, \ldots, 40$, giving 96 runs. Every run starts from the full model, uses $\varepsilon = 0$, and re-estimates $\hat{G}(\cdot \mid \mathbf{Z})$ at every step as in Algorithm~\ref{alg: backward}, with SRF held fixed at 100 trees, node size 50, $\texttt{mtry} = 2$, and the log-rank split rule.

Tables~\ref{table: vs_qicw_by_approach} and~\ref{table: vs_qicw_by_age} report the percentage of runs in which each term was retained, by setting and by analysis age respectively. Five terms are retained in every run: the rescaled age at cancer diagnosis $Z_{1}$, the BMT indicator $Z_{4}$, the abdominal and pelvic radiation-dose intensities $Z_{5}$ and $Z_{6}$, and the interaction $Z_{1}Z_{6}$. Among the remaining terms, on average the Firth correction retains more than the uncorrected fit under both approaches, and Approach~A retains more than Approach~B at each Firth setting. The mean selected model size ranges from 10.3 terms under Approach~B without the correction to 13.3 terms under Approach~A with it. The mean model size also grows with $t_{0}$, from 11.4 terms at $t_{0} = 17$--$24$ to 13.0 at $t_{0} = 36$--$40$, though not uniformly across terms: terms including the race indicators $Z_2$ and $Z_3$, antimetabolite indicator $Z_{12}$, and epipodophyllotoxin indicator $Z_{13}$ rise to 100\% at the latest ages, whereas terms such as TBI dose $Z_7$ and antibiotic indicator $Z_{11}$ fall as the analysis age increases. The selected model thus varies with the approach, the Firth setting, and the analysis age, and no single covariate set is selected throughout. The corresponding $\mathrm{QICW}_{u}$ results are reported in Supplementary Materials~\ref{supp: var_select_qicwu}.

\centerline{ \textit{----------Insert Table \ref{table: vs_qicw_by_approach} Here----------}}

\centerline{ \textit{----------Insert Table \ref{table: vs_qicw_by_age} Here----------}}

Figures~\ref{fig: vs_prob_appa_qicw} and~\ref{fig: vs_prob_appb_qicw} compare the estimated probabilities under the full model with those under the selected models, for a survivor diagnosed at age 7, Caucasian, receiving bone marrow transplant with TBI and platinum-compound chemotherapy. Each figure shows POI, SPM, and PM across the analysis ages, with and without the Firth correction. The selected models give estimated probabilities close to those of the full model under both approaches. The differences are small relative to the width of the 95\% CIs, which overlap at every analysis age, and are largest at the later ages where right censoring is heaviest.

\centerline{ \textit{----------Insert Figure \ref{fig: vs_prob_appa_qicw} Here----------}}

\centerline{ \textit{----------Insert Figure \ref{fig: vs_prob_appb_qicw} Here----------}}

\section{Final Remarks} \label{sec5: disscusion}
In this paper, we considered a multinomial logistic regression for estimating CIRs with doubly censored competing risks data. We proposed two IPCW-based estimating functions and incorporated Firth correction to handle estimation instability caused by data separation. The application to the CCSS data showed that the separate binary logistic regressions can produce an estimated CDF of the event exceeding one, while the multinomial logistic regression maintains this bound by construction. The Firth correction stabilized the SPM coefficient estimates, which exhibited extreme values due to quasi-complete separation, with a negligible effect on the POI estimates. These findings indicate that treatment-related covariates are associated with POI risk but not with SPM risk, which is consistent with SPM arising from surgical interventions, with its risk not directly captured by the considered covariates. We further adapted a weighted quasi-likelihood information criterion to the proposed estimating functions and paired it with a backward elimination, and illustrated the procedure on an extended covariate set, where the selected models gave estimated risks close to those under the full model.

Simulation studies confirmed these findings. The multinomial logistic regression yielded consistent coefficient estimates, valid sandwich standard errors, and probability estimates within $[0, 1]$. Under data imbalance, the Firth correction substantially reduced the bias induced by separation and provided reliable standard error estimation, while biased coefficients and underestimated sandwich standard errors were observed without Firth correction. We also found that the choice of censoring distribution estimator had an impact on coefficient estimation: the SRF and stratified ECDF consistently yielded the lowest RSMSE, while the Cox PH model on $C$ performed poorly due to violation of the proportional hazards assumption.

Despite its practical advantages, this study has several limitations. First, our current model treats POI and SPM as standard competing risks, assuming that the occurrence of one event precludes the other. In reality, they represent a semi-competing risks scenario, as subjects can experience SPM after experiencing POI, but not vice versa. Second, while we found that using SRF to estimate the censoring distribution performs well, the estimated censoring distribution can be sensitive to hyperparameter choices, and a comprehensive investigation into optimal tuning strategies was beyond the scope of this paper. Third, the robustness of the proposed model to substantial misspecification remains to be tested. Fourth, the variable selection procedure assumes that the censoring model and the event time model share the exact same set of covariates, and it has been derived and illustrated but not evaluated by simulation. Finally, although our simulations compared Approaches~A and~B, we have not yet fully explored the theoretical efficiency bounds or established which approach is more efficient under competing risks settings.

These limitations highlight several promising directions for future investigation. A natural extension is to explicitly account for the semi-competing risks structure of POI and SPM, potentially by incorporating copula models into the current framework to capture the dependence between the two events. In addition, comprehensive simulation studies are needed to tune SRF hyperparameters, identify scenarios where SRF is most advantageous, and provide practical guidance on tuning strategies. Methodologically, alternative penalization techniques for data separation are worth investigating, including Cauchy, normal, and $\log F$-type priors. For variable selection, the selected covariate set varies with $t_{0}$, and selecting one covariate set simultaneously for all analysis ages, rather than one set per age, would give a single interpretable model. Other routes are also available, such as a generalized score test for model adequacy \citep{wu2014} or penalizing the estimating functions with an $L_{1}$ or $L_{2}$ penalty \citep{kojima2023}. Finally, augmented inverse probability weighting could be considered to improve the efficiency of the proposed estimators.





\section*{Conflict of interest}

The authors declare no potential conflict of interests.

\bibliography{wileyNJD-AMA}

\newpage
\section*{Figures and Tables}
\begin{table}[H]
\centering
\captionof{table}{Descriptive statistics of the study population stratified by premature menopause status ($n = 6961$). Data are presented as frequency for categorical variables, and median with first and third quartiles for continuous variables. In analysis, we use rescaled age at cancer diagnosis $Z_{1}$, which is equal to $V/21$.}\label{table: proj2_desc_stat}
\scriptsize 
\setlength{\tabcolsep}{1pt}
\renewcommand{\arraystretch}{1.2} 
\resizebox{1\textwidth}{!}{
\begin{tabular}{|ccc|cccc|}
\hline
\multicolumn{3}{|c|}{}                                                                                                                              & \multicolumn{4}{c|}{\begin{tabular}[c]{@{}c@{}}Premature Menopause Status\\ (n/Median [Q1, Q3])\end{tabular}}                                                                                    \\ \cline{4-7} 
\multicolumn{3}{|c|}{}                                                                                                                              & \multicolumn{1}{c|}{\begin{tabular}[c]{@{}c@{}}Observed either \\POI or SPM\end{tabular}} &\multicolumn{1}{c|}{\begin{tabular}[c]{@{}c@{}}Observed \\POI\end{tabular}}                 & \multicolumn{1}{c|}{\begin{tabular}[c]{@{}c@{}}Observed \\SPM\end{tabular}}                 & \multicolumn{1}{c|}{\begin{tabular}[c]{@{}c@{}}No observed \\POI and SPM\end{tabular}}                  \\ \cline{4-7} 
\multicolumn{3}{|c|}{\multirow{-3}{*}{\begin{tabular}[c]{@{}c@{}}Number of Subjects \\ (total n = 6961)\end{tabular}}}                              & \multicolumn{1}{c|}{1050}                                                        & \multicolumn{1}{c|}{798}                 & \multicolumn{1}{c|}{252}                      & 5911               \\ \hline
\multicolumn{2}{|c|}{Age at Cancer Diagnosis (\textit{$V$})}                                                                                 & 7.43 (3.27, 13.86) & \multicolumn{1}{c|}{10.73 (10.21, 11.35)}                                        & \multicolumn{1}{c|}{10.21 (4.02, 14.95)} & \multicolumn{1}{c|}{11.35 (4.95, 15.34)}      & 6.96 (3.12, 13.59) \\ \hline
\multicolumn{1}{|c|}{}                                                    & \multicolumn{1}{c|}{African American(\textit{$Z_{2}$}=1)}                  & 389  & \multicolumn{1}{c|}{58}                                                          & \multicolumn{1}{c|}{50}                  & \multicolumn{1}{c|}{{ 8}} & 331                \\ \cline{2-7} 
\multicolumn{1}{|c|}{}                                                    & \multicolumn{1}{c|}{Other (\textit{$Z_{3}$}=1)}                  & 469  & \multicolumn{1}{c|}{76}                                                          & \multicolumn{1}{c|}{68}                  & \multicolumn{1}{c|}{{ 8}} & 393                \\ \cline{2-7} 
\multicolumn{1}{|c|}{\multirow{-3}{*}{Race (vs Caucasian)}}                   & \multicolumn{1}{c|}{Caucasian (\textit{$Z_{2}$}=\textit{$Z_{3}$}=0)} & 6103 & \multicolumn{1}{c|}{916}                                                         & \multicolumn{1}{c|}{680}                 & \multicolumn{1}{c|}{236}                      & 5187               \\ \hline
\multicolumn{1}{|c|}{}                                                    & \multicolumn{1}{c|}{Yes (\textit{$Z_{4}$}=1)}                    & 298  & \multicolumn{1}{c|}{178}                                                         & \multicolumn{1}{c|}{169}                 & \multicolumn{1}{c|}{{ 9}} & 120                \\ \cline{2-7} 
\multicolumn{1}{|c|}{\multirow{-2}{*}{Having Received BMT}}               & \multicolumn{1}{c|}{No (\textit{$Z_{4}$}=0)}                     & 6663 & \multicolumn{1}{c|}{872}                                                         & \multicolumn{1}{c|}{629}                 & \multicolumn{1}{c|}{243}                      & 5791               \\ \hline
\multicolumn{1}{|c|}{}                                                    & \multicolumn{1}{c|}{Yes (\textit{$Z_{5}$}=1)}                    & 3516 & \multicolumn{1}{c|}{655}                                                         & \multicolumn{1}{c|}{519}                 & \multicolumn{1}{c|}{136}                      & 2861               \\ \cline{2-7} 
\multicolumn{1}{|c|}{\multirow{-2}{*}{Having Received Alkylating agents}} & \multicolumn{1}{c|}{No (\textit{$Z_{5}$}=0)}                     & 3445 & \multicolumn{1}{c|}{395}                                                         & \multicolumn{1}{c|}{279}                 & \multicolumn{1}{c|}{116}                      & 3050               \\ \hline
\multicolumn{1}{|c|}{}                                                    & \multicolumn{1}{c|}{Yes (\textit{$Z_{6}$}=1)}                    & 1575 & \multicolumn{1}{c|}{490}                                                         & \multicolumn{1}{c|}{429}                 & \multicolumn{1}{c|}{61}                       & 1085               \\ \cline{2-7} 
\multicolumn{1}{|c|}{\multirow{-2}{*}{Having Received Radiotherapy}}      & \multicolumn{1}{c|}{No (\textit{$Z_{6}$}=0)}                     & 5386 & \multicolumn{1}{c|}{560}                                                         & \multicolumn{1}{c|}{369}                 & \multicolumn{1}{c|}{191}                      & 4826               \\ \hline
\end{tabular}
}
\end{table}

\begin{table}[H]
\centering
\footnotesize
\captionof{table}{Percentage of runs in which each term is retained by the $\mathrm{QICW}$-driven backward elimination, by approach and Firth setting.}\label{table: vs_qicw_by_approach}
\renewcommand{\arraystretch}{1.15}
\begin{tabular}{lcccc}
\toprule
Term & A, no Firth & A, Firth & B, no Firth & B, Firth \\
\midrule
$Z_{1}$, $Z_{4}$, $Z_{5}$, $Z_{6}$, $Z_{1}Z_{6}$ & 100 & 100 & 100 & 100 \\
$Z_{7}$        &  92 & 100 & 100 &  92 \\
$Z_{8}$        &  92 & 100 &  96 & 100 \\
$Z_{11}$       &  75 &  96 &  79 &  92 \\
$Z_{2}$, $Z_{3}$ & 75 & 100 &  46 & 100 \\
$Z_{12}$       &  58 & 100 &  50 & 100 \\
$Z_{1}Z_{4}$   &  42 &  38 &  88 &  88 \\
$Z_{10}$       &  54 & 100 &  25 &  75 \\
$Z_{13}$       &  58 &  96 &  33 &  75 \\
$Z_{9}$        &  25 &  67 &   8 &  67 \\
$Z_{1}Z_{5}$   &  50 &  38 &   8 &   0 \\
\midrule
Mean model size (terms) & 11.2 & 13.3 & 10.3 & 12.9 \\
\bottomrule
\end{tabular}
\par\vspace{0.2cm}
\begingroup
\raggedright \footnotesize \setlength{\parindent}{0pt}
\textit{Note:} Each column pools the 24 analysis ages $t_{0} = 17, \ldots, 40$, so an entry is the percentage of those 24 runs that retain the term. Terms are ordered by their overall retention. Race contributes the two indicators $Z_{2}$ and $Z_{3}$ but counts as a single term. The full model has 15 terms. Covariate definitions are given in Supplementary Materials~\ref{supp: analysis3_covariates}.
\par\endgroup
\end{table}

\begin{table}[H]
\centering
\footnotesize
\captionof{table}{Percentage of runs in which each term is retained by the $\mathrm{QICW}$-driven backward elimination, by analysis age.}\label{table: vs_qicw_by_age}
\renewcommand{\arraystretch}{1.15}
\begin{tabular}{lccc}
\toprule
Term & $t_{0} = 17$--$24$ & $t_{0} = 25$--$35$ & $t_{0} = 36$--$40$ \\
\midrule
$Z_{1}$, $Z_{4}$, $Z_{5}$, $Z_{6}$, $Z_{1}Z_{6}$ & 100 & 100 & 100 \\
$Z_{7}$        & 100 & 100 &  80 \\
$Z_{8}$        &  91 & 100 & 100 \\
$Z_{11}$       & 100 &  84 &  65 \\
$Z_{2}$, $Z_{3}$ & 66 &  82 & 100 \\
$Z_{12}$       &  50 &  86 & 100 \\
$Z_{1}Z_{4}$   &  59 &  73 &  50 \\
$Z_{10}$       &  69 &  50 &  85 \\
$Z_{13}$       &  47 &  64 & 100 \\
$Z_{9}$        &  50 &  25 &  65 \\
$Z_{1}Z_{5}$   &  13 &  20 &  50 \\
\midrule
Mean model size (terms) & 11.4 & 11.8 & 13.0 \\
\bottomrule
\end{tabular}
\par\vspace{0.2cm}
\begingroup
\raggedright \footnotesize \setlength{\parindent}{0pt}
\textit{Note:} Each column pools the four approach--Firth settings across the analysis ages in its range, giving 32, 44, and 20 runs respectively. Term ordering matches Table~\ref{table: vs_qicw_by_approach}.
\par\endgroup
\end{table}


\textbf{}

\begin{figure}[H]
\centering
\includegraphics[width=0.85\textwidth]{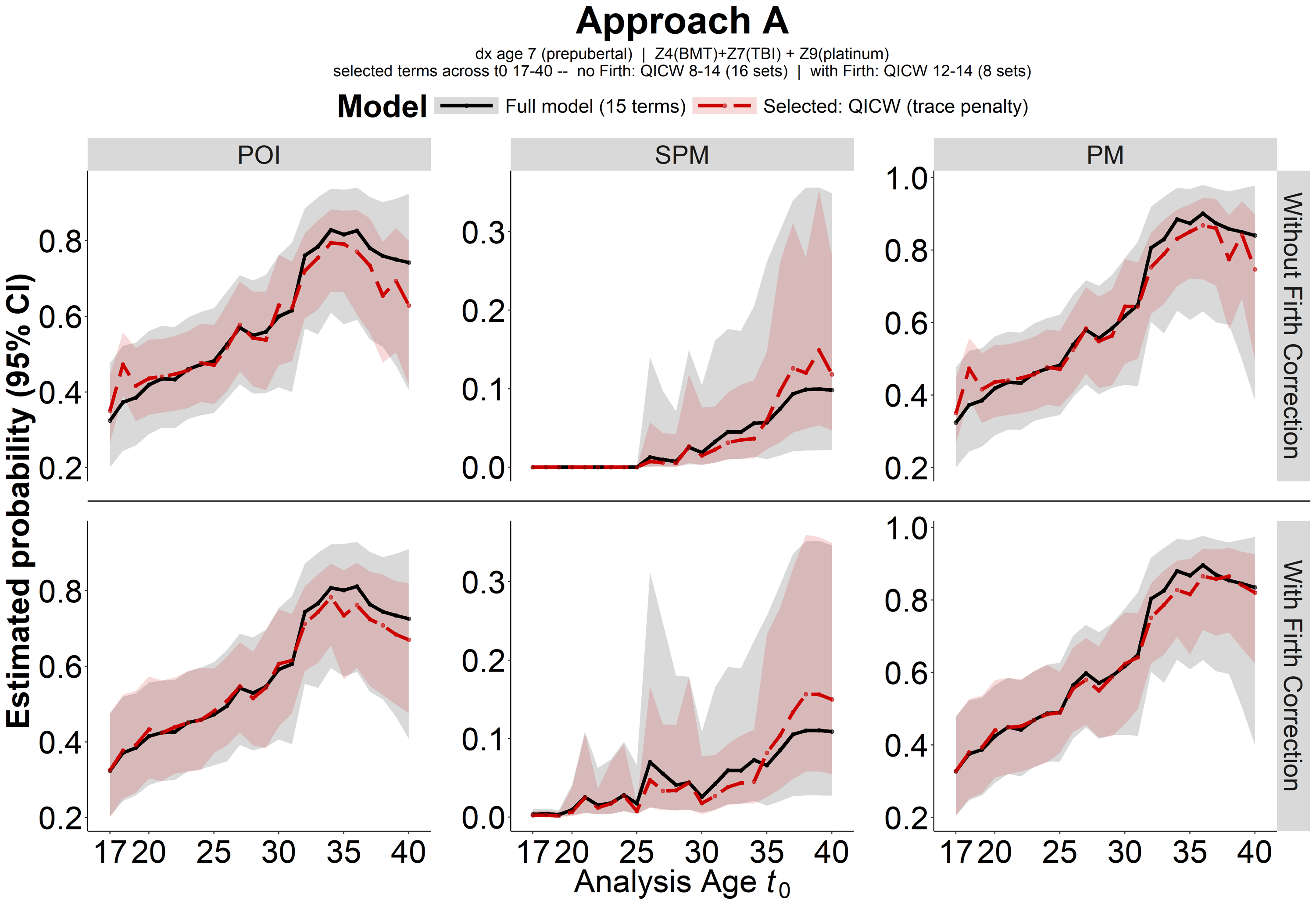}
\caption{Estimated probabilities under the full model and the $\mathrm{QICW}$-selected models, Approach~A.}\label{fig: vs_prob_appa_qicw}
\par\vspace{0.2cm}
\begingroup
\raggedright \footnotesize \setlength{\parindent}{0pt}
\textit{Note:} The covariate profile is a survivor diagnosed at age 7, Caucasian, receiving bone marrow transplant with TBI and platinum-compound chemotherapy. In the notation of Section~\ref{sec3: extended analysis}, $Z_{1} = 0.333$ ($= 7/21$), $Z_{4} = 1$, $Z_{7} = 0.762$ ($= 12/15.75$, a myeloablative TBI dose of 12~Gy rescaled by the observed maximum of 15.75~Gy), and $Z_{9} = 1$. All remaining covariates are at their reference level: $Z_{2} = Z_{3} = 0$ (Caucasian), $Z_{5} = Z_{6} = 0$ (no abdominal or pelvic radiation beyond TBI), and $Z_{8} = Z_{10} = \cdots = Z_{13} = 0$ (no other chemotherapy exposure).
\par\endgroup
\end{figure}

\begin{figure}[H]
\centering
\includegraphics[width=0.85\textwidth]{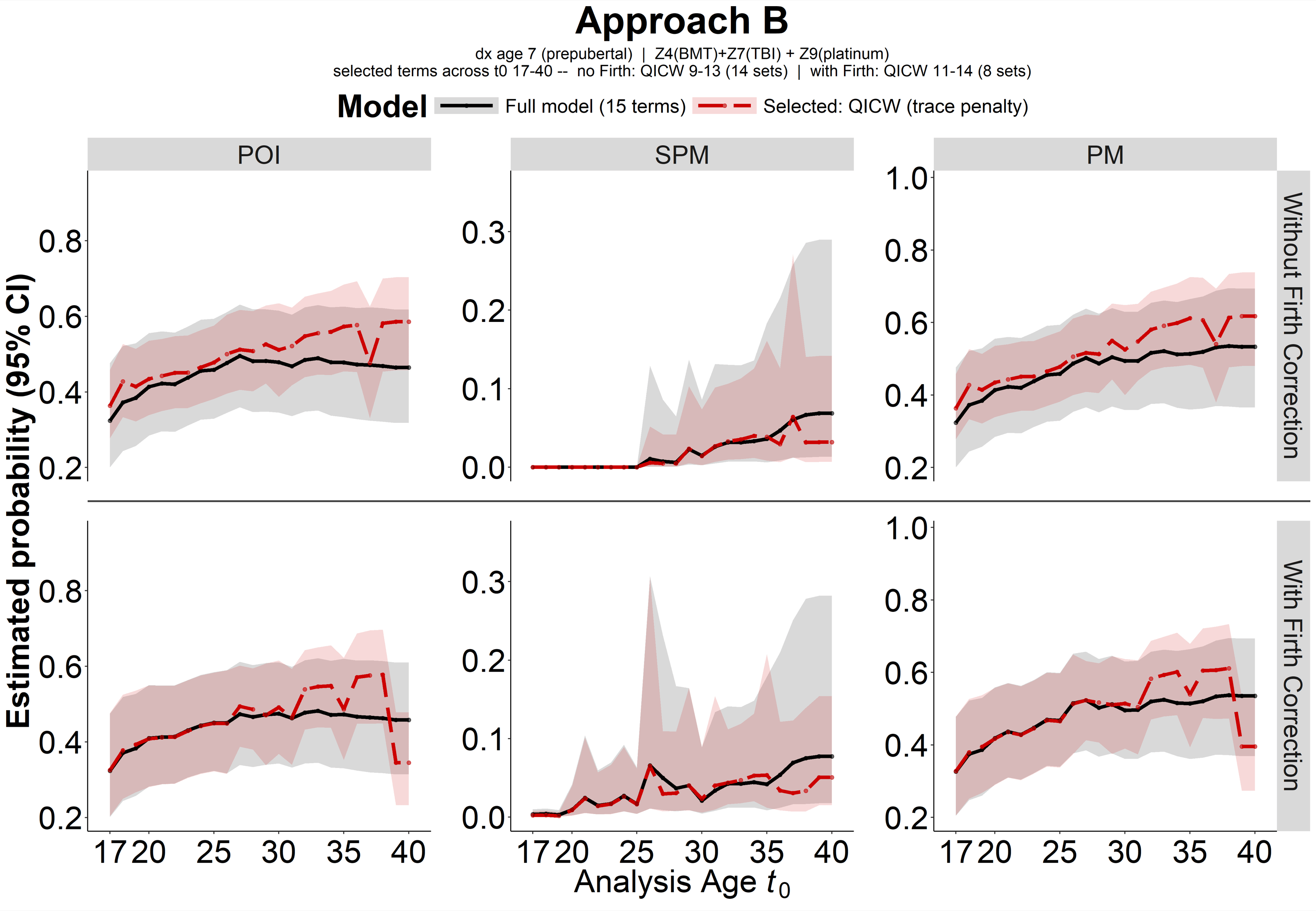}
\caption{Estimated probabilities under the full model and the $\mathrm{QICW}$-selected models, Approach~B.}\label{fig: vs_prob_appb_qicw}
\par\vspace{0.2cm}
\begingroup
\raggedright \footnotesize \setlength{\parindent}{0pt}
\textit{Note:} The covariate profile is that of Figure~\ref{fig: vs_prob_appa_qicw}.
\par\endgroup
\end{figure}




\end{document}


\maketitle
\tableofcontents
\newpage
\renewcommand{\thealgorithm}{S\arabic{algorithm}}


%
%
%
%
\section{Estimation Procedures}
%
%
%
%
\subsection{Estimation Procedure for Logistic Regression}\label{supp: logistic regression}
Recall that we consider the following logistic regression to account for competing risks:
\begin{align*} 
    \log\left[
      \frac{P(T \le t_{0},\, \eta = k \mid \mathbf{Z},\, T > V,\, V < t_{0})}
           {1 - P(T \le t_{0},\, \eta = k \mid \mathbf{Z},\, T > V,\, V < t_{0})}
    \right]
    = \kappa_{k}(t_{0}) + \bm{\tau}_{k}(t_{0})^{\top} \mathbf{Z},
\end{align*}
where $\kappa_k(t_0)$ is the intercept and $\bm{\tau}_k(t_0)$ is a $p \times 1$ vector of covariate effects at $t_0$, for cause $k = 1, 2$. 
Let $p_{ki}(t_{0},\bm{Z}_{i}) = \exp \left\{\kappa_k\left(t_0\right)+\mathbf{\tau}_k\left(t_0\right)^T \mathbf{Z}_i\right\} \big/ \left[\exp \left\{\kappa_k\left(t_0\right)+\mathbf{\tau}_k\left(t_0\right)^T \mathbf{Z}_i\right\}+1\right]$. The estimating function under Approach A is
\begin{align}\label{eq: proj2 estimating equation a}
\begin{split}
    \bm{U}_{1A}(\kappa_{k},\bm{\tau}_{k}; t_{0} \mid G) 
     =
     \displaystyle\sum_{i=1}^{n} \mathbb{I}(t_{0} \ge V_{i})
     \begin{bmatrix}
      1 \\
      \mathbf{Z}_{i} 
     \end{bmatrix}
     W_{i}(t_{0};G)
     \Bigg[
     \mathbb{I}(T_{i} \le t_{0},\eta=k)
     -p_{ki}(t_{0},\bm{Z}_{i})
    \Bigg],
\end{split}
\end{align}
The estimating function under Approach B, which weights the response variable, is
\begin{align}\label{eq: proj2 estimating equation b}
\begin{split}
    \bm{U}_{1B}(\kappa_{k},\bm{\tau}_{k}; t_{0} \mid G) 
     =
     \displaystyle\sum_{i=1}^{n} \mathbb{I}(t_{0} \ge V_{i})
     \begin{bmatrix}
      1 \\
      \mathbf{Z}_{i} 
     \end{bmatrix}
     \Bigg[
     \mathbb{I}(T_{i} \le t_{0},\eta=k) W_{i}(t_{0};G)
     -p_{ki}(t_{0},\bm{Z}_{i})
    \Bigg]
\end{split}
\end{align}
Both estimating functions extend the IPCW approaches of \citet{zhou202xagespecific} to accommodate competing risks. Under the assumption that $(T_{i},\eta_{i})$ and $C_{i}$ are conditionally independent given $\mathbf{Z}_{i}$, both estimating functions are unbiased. It is also straightforward to establish, under some regularity conditions, the consistency and asymptotic normality of the estimator derived from both estimation functions at a fixed $t_0>0$.
%
%
%
%
\subsection{Derivation of Firth-Corrected Estimating Functions}\label{supp: deriv_firth_ee}
\subsubsection{Derivation of the Firth-Corrected Estimating Functions for Logistic Regression}
\label{supp: deriv_firth_ee_model1}
We derive closed-form expressions for the Firth-corrected estimating functions $\bm{U}_{1A}^{*}$ and $\bm{U}_{1B}^{*}$. We suppress the dependence on $t_0$ and $G$ for notational convenience and write $p_{ki}$ in place of $p_{ki}(t_{0},\bm{Z}_{i})$. Let $Z_{ir}$ denote the $r$-th element of $\begin{bmatrix} 1 \\ \mathbf{Z}_{i} \end{bmatrix}$, and $\mathcal{T} = \bigl(\kappa_{k}(t_0),\, \bm{\tau}_{k}(t_0)^{\top}\bigr)^{\top}$.

\paragraph{Approach A}

The estimating function under Approach~A is
\[
  \bm{U}_{1A}(\bm{\mathcal{T}})
  = \sum_{i=1}^{n}
  \mathbb{I}(t_0 \ge V_{i})\, W_{i}\,
  \begin{bmatrix} 1 \\ \mathbf{Z}_{i} \end{bmatrix}
  \Bigg[ \mathbb{I}(T_{i} \le t_0,\, \eta_{i} = k) - p_{ki} \Bigg],
  \]
where $W_{i} \coloneqq W_{i}(t_0; G)$ is the IPCW. Taking the negative derivative with respect to $\bm{\mathcal{T}}^{\top}$ and $\partial p_{ki} / \partial \bm{\mathcal{T}}^{\top}
= p_{ki}(1 - p_{ki})
\begin{bmatrix} 1 \\ \mathbf{Z}_{i} \end{bmatrix}^{\top}$, where we obtain
\begin{align}\label{eq:I_1A_supp}
\mathcal{I}_{1A}
= -\frac{\partial \bm{U}_{1A}}{\partial \bm{\mathcal{T}}^{\top}}
= \sum_{i=1}^{n}
\mathbb{I}(t_0 \ge V_{i})\, W_{i}\,
p_{ki}\, (1 - p_{ki})\,
\begin{bmatrix} 1 \\ \mathbf{Z}_{i} \end{bmatrix}
\begin{bmatrix} 1 \\ \mathbf{Z}_{i} \end{bmatrix}^{\top}.
\end{align}

The $r$-th component of the Firth penalty requires
$\partial \mathcal{I}_{1A} / \partial \mathcal{T}_{r}$, and 
\[  \frac{\partial}{\partial \mathcal{T}_{r}}
    \bigl[p_{ki}(1 - p_{ki})\bigr]
    = (1 - 2\, p_{ki})\,
    p_{ki}\, (1 - p_{ki})\,
    Z_{ir}.
    \]
Substituting into~\eqref{eq:I_1A_supp} gives
\begin{align}\label{eq:dI_1A_supp}
\frac{\partial \mathcal{I}_{1A}}{\partial \mathcal{T}_{r}}
= \sum_{i=1}^{n}
\mathbb{I}(t_0 \ge V_{i})\, W_{i}\,
(1 - 2\, p_{ki})\,
p_{ki}\, (1 - p_{ki})\,
Z_{ir}\,
\begin{bmatrix} 1 \\ \mathbf{Z}_{i} \end{bmatrix}
\begin{bmatrix} 1 \\ \mathbf{Z}_{i} \end{bmatrix}^{\top}.
\end{align}

The $r$-th component of the Firth penalty is
\begin{align}\label{eq:A_1A_trace}
\begin{split}
A_{1A,r}
&= \frac{1}{2}\,
\mathrm{trace}\!\left[
  \mathcal{I}_{1A}^{-1}\,
  \frac{\partial \mathcal{I}_{1A}}{\partial \mathcal{T}_{r}}
  \right]\\
&= \frac{1}{2}
\sum_{i=1}^{n}
\mathbb{I}(t_0 \ge V_{i})\, W_{i}\,
(1 - 2\, p_{ki})\,
p_{ki}\, (1 - p_{ki})\,
Z_{ir}\,
\mathrm{trace}\!\left[
  \mathcal{I}_{1A}^{-1}\,
  \begin{bmatrix} 1 \\ \mathbf{Z}_{i} \end{bmatrix}
  \begin{bmatrix} 1 \\ \mathbf{Z}_{i} \end{bmatrix}^{\top}
  \right]\\
&= \frac{1}{2}
\sum_{i=1}^{n}
\mathbb{I}(t_0 \ge V_{i})\, W_{i}\,
(1 - 2\, p_{ki})\,
p_{ki}\, (1 - p_{ki})\,
Z_{ir}\,
\begin{bmatrix} 1 \\ \mathbf{Z}_{i} \end{bmatrix}^{\top}
\mathcal{I}_{1A}^{-1}
\begin{bmatrix} 1 \\ \mathbf{Z}_{i} \end{bmatrix}.
\end{split}
\end{align}
Define
\begin{align}\label{eq:h_1A_supp}
h_{1A,i}
= p_{ki}\, (1 - p_{ki})\,
\begin{bmatrix} 1 \\ \mathbf{Z}_{i} \end{bmatrix}^{\top}
\mathcal{I}_{1A}^{-1}
\begin{bmatrix} 1 \\ \mathbf{Z}_{i} \end{bmatrix}.
\end{align}
The equation~\eqref{eq:A_1A_trace} becomes
\[
  \bm{A}_{1A}(\bm{\mathcal{T}})
  = \sum_{i=1}^{n}
  \mathbb{I}(t_0 \ge V_{i})\, W_{i}\,
  h_{1A,i}\,
  \left(\tfrac{1}{2} - p_{ki}\right)\,
  \begin{bmatrix} 1 \\ \mathbf{Z}_{i} \end{bmatrix}.
  \]
Adding the penalty to the original estimating function, we obtain
\begin{align}\label{eq:A_1A_final}
\begin{split}
\bm{U}_{1A}^{*}(\bm{\mathcal{T}})
&= \bm{U}_{1A}(\bm{\mathcal{T}})
+ \bm{A}_{1A}(\bm{\mathcal{T}}) \\
&= \sum_{i=1}^{n}
\mathbb{I}(t_0 \ge V_{i})\, W_{i}\,
\begin{bmatrix} 1 \\ \mathbf{Z}_{i} \end{bmatrix}
\left[
  \mathbb{I}(T_{i} \le t_0,\, \eta_{i} = k) - p_{ki}
  + h_{1A,i}\!\left(\tfrac{1}{2} - p_{ki}\right)
  \right].
\end{split}
\end{align}

\paragraph{Approach B}

The estimating function under Approach~B is
\[
  \bm{U}_{1B}(\bm{\mathcal{T}})
  = \sum_{i=1}^{n}
  \mathbb{I}(t_0 \ge V_{i})\,
  \begin{bmatrix} 1 \\ \mathbf{Z}_{i} \end{bmatrix}
  \Bigg[ \mathbb{I}(T_{i} \le t_0,\, \eta_{i} = k)\, W_{i} - p_{ki} \Bigg],
  \]
and
\begin{align}\label{eq:I_1B_supp}
\mathcal{I}_{1B}
= -\frac{\partial \bm{U}_{1B}}{\partial \bm{\mathcal{T}}^{\top}}
= \sum_{i=1}^{n}
\mathbb{I}(t_0 \ge V_{i})\,
p_{ki}\, (1 - p_{ki})\,
\begin{bmatrix} 1 \\ \mathbf{Z}_{i} \end{bmatrix}
\begin{bmatrix} 1 \\ \mathbf{Z}_{i} \end{bmatrix}^{\top}.
\end{align}
Note that $\mathcal{I}_{1B}$ does \emph{not} involve the IPCW $W_{i}$, in contrast with $\mathcal{I}_{1A}$
  in~\eqref{eq:I_1A_supp}.

By the same calculation as in~\eqref{eq:dI_1A_supp}, the derivative of
$\mathcal{I}_{1B}$ is
\begin{align}\label{eq:dI_1B_supp}
\frac{\partial \mathcal{I}_{1B}}{\partial \mathcal{T}_{r}}
= \sum_{i=1}^{n}
\mathbb{I}(t_0 \ge V_{i})\,
(1 - 2\, p_{ki})\,
p_{ki}\, (1 - p_{ki})\,
Z_{ir}\,
\begin{bmatrix} 1 \\ \mathbf{Z}_{i} \end{bmatrix}
\begin{bmatrix} 1 \\ \mathbf{Z}_{i} \end{bmatrix}^{\top},
\end{align}
Following the same trace simplification in~\eqref{eq:A_1A_trace}, we obtain
\begin{align}\label{eq:A_1B_simplified}
A_{1B,r}
= \frac{1}{2}
\sum_{i=1}^{n}
\mathbb{I}(t_0 \ge V_{i})\,
(1 - 2\, p_{ki})\,
p_{ki}\, (1 - p_{ki})\,
Z_{ir}\,
\begin{bmatrix} 1 \\ \mathbf{Z}_{i} \end{bmatrix}^{\top}
\mathcal{I}_{1B}^{-1}
\begin{bmatrix} 1 \\ \mathbf{Z}_{i} \end{bmatrix}.
\end{align}
Define
\begin{align}\label{eq:h_1B_supp}
h_{1B,i}
= p_{ki}\, (1 - p_{ki})\,
\begin{bmatrix} 1 \\ \mathbf{Z}_{i} \end{bmatrix}^{\top}
\mathcal{I}_{1B}^{-1}
\begin{bmatrix} 1 \\ \mathbf{Z}_{i} \end{bmatrix}.
\end{align}
Then, adding the penalty to the original estimating function:
  \begin{align}\label{eq:A_1B_final}
\begin{split}
\bm{U}_{1B}^{*}(\bm{\mathcal{T}})
&= \bm{U}_{1B}(\bm{\mathcal{T}})
+ \bm{A}_{1B}(\bm{\mathcal{T}}) \\
&= \sum_{i=1}^{n}
\mathbb{I}(t_0 \ge V_{i})\,
\begin{bmatrix} 1 \\ \mathbf{Z}_{i} \end{bmatrix}
\left[
  \mathbb{I}(T_{i} \le t_0,\, \eta_{i} = k)\, W_{i} - p_{ki}
  + h_{1B,i}\!\left(\tfrac{1}{2} - p_{ki}\right)
  \right].
\end{split}
\end{align}

\subsubsection{Derivation of the Firth-Corrected Estimating Functions for Multinomial Logistic Regression}\label{supp: deriv_firth_ee_model2}
We derive the Firth-corrected estimating functions $\bm{U}_{2A}^{*}$ and $\bm{U}_{2B}^{*}$. We suppress the dependence on $t_0$ and $G$ for notational convenience and
write $\pi_{ki}$ in place of $\pi_{ki}(\bm{\theta})$. Recall the block-diagonal design matrix $\tilde{\mathbf{Z}}_{i}$, response vector $\mathbf{Y}_{i}$, probability vector $\bm{\pi}_{i} \coloneqq \bm{\pi}_{i}(t_{0},\bm{Z}_{i})$, and variance matrix $\mathbf{V}_{i}$ defined in the main text.

\paragraph{Approach A}

The estimating function under Approach~A is
\[
  \bm{U}_{2A}(\bm{\theta})
  = \sum_{i=1}^{n}
  \mathbb{I}(t_0 \ge V_{i})\, W_{i}\,
  \tilde{\mathbf{Z}}_{i}^{\top}
  \Bigg[\mathbf{Y}_{i} - \bm{\pi}_{i}\Bigg],
  \]
where $W_{i} \coloneqq W_{i}(t_0; G)$ is the IPCW. Taking the negative derivative with respect to $\bm{\theta}^{\top}$ and noting that $\partial \bm{\pi}_{i} / \partial \bm{\theta}^{\top} = \mathbf{V}_{i}(\bm{\theta})\, \tilde{\mathbf{Z}}_{i}$, where
\begin{align*}
\mathbf{V}_{i}\coloneqq\mathbf{V}_{i}(\bm{\theta})
= \mathrm{diag}\bigl(\bm{\pi}_{i}(\bm{\theta})\bigr)
- \bm{\pi}_{i}(\bm{\theta})\, \bm{\pi}_{i}(\bm{\theta})^{\top}
= \begin{pmatrix}
\pi_{1i}(1 - \pi_{1i}) & -\pi_{1i}\, \pi_{2i} \\
-\pi_{1i}\, \pi_{2i}   & \pi_{2i}(1 - \pi_{2i})
\end{pmatrix}.
\end{align*}
we obtain
\begin{align}\label{eq:I_2A_supp}
\mathcal{I}_{2A}
= -\frac{\partial \bm{U}_{2A}}{\partial \bm{\theta}^{\top}}
= \sum_{i=1}^{n}
\mathbb{I}(t_0 \ge V_{i})\, W_{i}\,
\tilde{\mathbf{Z}}_{i}^{\top}\, \mathbf{V}_{i}\,
\tilde{\mathbf{Z}}_{i}.
\end{align}
The $r$-th component of the Firth penalty requires $\partial \mathcal{I}_{2A} / \partial \theta_{r}$. Since $\mathbb{I}(t_0 \ge V_{i})$, $W_{i}$, and $\tilde{\mathbf{Z}}_{i}$ do not depend on $\bm{\theta}$, only $\mathbf{V}_{i}$ contributes:
  \begin{align}\label{eq:dI_2A_supp}
\frac{\partial \mathcal{I}_{2A}}{\partial \theta_{r}}
= \sum_{i=1}^{n}
\mathbb{I}(t_0 \ge V_{i})\, W_{i}\,
\tilde{\mathbf{Z}}_{i}^{\top}\,
\frac{\partial \mathbf{V}_{i}}{\partial \theta_{r}}\,
\tilde{\mathbf{Z}}_{i}.
\end{align}
The derivative $\partial \mathbf{V}_{i} / \partial \theta_{r}$ is a $2 \times 2$ matrix that depends on which cause the parameter $\theta_{r}$ belongs to. For the multinomial logistic model with separate parameters
$\bm{\theta}
= (\alpha_{1}, \bm{\beta}_{1}^{\top},
   \alpha_{2}, \bm{\beta}_{2}^{\top})^{\top}$,
the derivatives of $\pi_{ki}$ with respect to $\theta_{r}$ are:
  \begin{align}\label{eq:dpi_model2}
\frac{\partial \pi_{ki}}{\partial \theta_{r}}
= \begin{cases}
\pi_{ki}(1 - \pi_{ki})\, Z_{ij},
& \text{if } \theta_{r} \text{ is the $j$-th parameter of
  cause } k, \\[4pt]
-\pi_{ki}\, \pi_{ji}\, Z_{ij},
& \text{if } \theta_{r} \text{ is the $j$-th parameter of
  cause } j \ne k,
\end{cases}
\end{align}
where $Z_{ij}$ is the $j$-th element of
$\begin{bmatrix} 1 \\ \mathbf{Z}_{i} \end{bmatrix}$. The
$2 \times 2$ derivative of $\mathbf{V}_{i}$ is then
\begin{align}\label{eq:dV_general}
\frac{\partial \mathbf{V}_{i}}{\partial \theta_{r}}
= \begin{pmatrix}
(1 - 2\pi_{1i})\,
\dfrac{\partial \pi_{1i}}{\partial \theta_{r}}
&
  -\pi_{2i}\, \dfrac{\partial \pi_{1i}}{\partial \theta_{r}}
- \pi_{1i}\, \dfrac{\partial \pi_{2i}}{\partial \theta_{r}}
\\[10pt]
-\pi_{2i}\, \dfrac{\partial \pi_{1i}}{\partial \theta_{r}}
- \pi_{1i}\, \dfrac{\partial \pi_{2i}}{\partial \theta_{r}}
&
  (1 - 2\pi_{2i})\,
\dfrac{\partial \pi_{2i}}{\partial \theta_{r}}
\end{pmatrix}.
\end{align}

The $r$-th component of the Firth penalty is
\begin{align}\label{eq:A_2A_trace}
A_{2A,r}
&= \frac{1}{2}\,
\mathrm{trace}\!\left[
  \mathcal{I}_{2A}^{-1}\,
  \frac{\partial \mathcal{I}_{2A}}{\partial \theta_{r}}
  \right] \notag \\
&= \frac{1}{2}
\sum_{i=1}^{n}
\mathbb{I}(t_0 \ge V_{i})\, W_{i}\,
\mathrm{trace}\!\left[
  \mathcal{I}_{2A}^{-1}\,
  \tilde{\mathbf{Z}}_{i}^{\top}\,
  \frac{\partial \mathbf{V}_{i}}{\partial \theta_{r}}\,
  \tilde{\mathbf{Z}}_{i}
  \right].
\end{align}
Applying the cyclic property of the trace, $\mathrm{trace}[\mathbf{A}\, \mathbf{B}\, \mathbf{C}\, \mathbf{D}] = \mathrm{trace}[\mathbf{C}\, \mathbf{D}\, \mathbf{A}\, \mathbf{B}]$, and defining the $2 \times 2$ matrix
\begin{align}\label{eq:H_2A_supp}
\mathbf{H}_{2A,i}
= \tilde{\mathbf{Z}}_{i}\,
\mathcal{I}_{2A}^{-1}\,
\tilde{\mathbf{Z}}_{i}^{\top},
\end{align}
we can write
\begin{align}\label{eq:A_2A_H}
A_{2A,r}
= \frac{1}{2}
\sum_{i=1}^{n}
\mathbb{I}(t_0 \ge V_{i})\, W_{i}\,
\mathrm{trace}\!\left[
  \frac{\partial \mathbf{V}_{i}}{\partial \theta_{r}}\,
  \mathbf{H}_{2A,i}
  \right].
\end{align}

Writing the penalty in vector form and combining with the original estimating function:
  \begin{align*}
\bm{U}_{2A}^{*}(\bm{\theta})
= \sum_{i=1}^{n}
\mathbb{I}(t_0 \ge V_{i})\, W_{i}\,
\tilde{\mathbf{Z}}_{i}^{\top}
\Bigg[\mathbf{Y}_{i} - \bm{\pi}_{i}\Bigg]
+ \bm{A}_{2A}(\bm{\theta}),
\end{align*}
where the $r$-th component of $\bm{A}_{2A}$ is given by~\eqref{eq:A_2A_H}.

\bigskip
\noindent\textit{Explicit form of $\partial \mathbf{V}_{i} / \partial \theta_{r}$ for each parameter block.} Substituting~\eqref{eq:dpi_model2} into~\eqref{eq:dV_general}, the derivative takes the following forms depending on which cause $\theta_{r}$ belongs to.

If $\theta_{r}$ is the $j$-th parameter of cause~1:
  \begin{align}\label{eq:dV_cause1}
\frac{\partial \mathbf{V}_{i}}{\partial \theta_{r}}
= Z_{ij}
\begin{pmatrix}
(1 - 2\pi_{1i})\, \pi_{1i}(1 - \pi_{1i})
&
  \pi_{1i}\, \pi_{2i}(2\pi_{1i} - 1)
\\[4pt]
\pi_{1i}\, \pi_{2i}(2\pi_{1i} - 1)
&
  \pi_{1i}\, \pi_{2i}(2\pi_{2i} - 1)
\end{pmatrix}.
\end{align}

If $\theta_{r}$ is the $j$-th parameter of cause~2:
  \begin{align}\label{eq:dV_cause2}
\frac{\partial \mathbf{V}_{i}}{\partial \theta_{r}}
= Z_{ij}
\begin{pmatrix}
\pi_{1i}\, \pi_{2i}(2\pi_{1i} - 1)
&
  \pi_{1i}\, \pi_{2i}(2\pi_{2i} - 1)
\\[4pt]
\pi_{1i}\, \pi_{2i}(2\pi_{2i} - 1)
&
  (1 - 2\pi_{2i})\, \pi_{2i}(1 - \pi_{2i})
\end{pmatrix}.
\end{align}

\paragraph{Approach B}

The estimating function under Approach~B is
\[
  \bm{U}_{2B}(\bm{\theta})
  = \sum_{i=1}^{n}
  \mathbb{I}(t_0 \ge V_{i})\,
  \tilde{\mathbf{Z}}_{i}^{\top}
  \Bigg[\mathbf{Y}_{i}\, W_{i} - \bm{\pi}_{i}\Bigg],
  \]
and
\begin{align}\label{eq:I_2B_supp}
\mathcal{I}_{2B}
= -\frac{\partial \bm{U}_{2B}}{\partial \bm{\theta}^{\top}}
= \sum_{i=1}^{n}
\mathbb{I}(t_0 \ge V_{i})\,
\tilde{\mathbf{Z}}_{i}^{\top}\, \mathbf{V}_{i}\,
\tilde{\mathbf{Z}}_{i}.
\end{align}
Note that $\mathcal{I}_{2B}$ does \emph{not} involve the IPCW $W_{i}$, in contrast with $\mathcal{I}_{2A}$ in~\eqref{eq:I_2A_supp}. The derivative is
\begin{align}\label{eq:dI_2B_supp}
\frac{\partial \mathcal{I}_{2B}}{\partial \theta_{r}}
= \sum_{i=1}^{n}
\mathbb{I}(t_0 \ge V_{i})\,
\tilde{\mathbf{Z}}_{i}^{\top}\,
\frac{\partial \mathbf{V}_{i}}{\partial \theta_{r}}\,
\tilde{\mathbf{Z}}_{i},
\end{align}
which differs from~\eqref{eq:dI_2A_supp} only by the absence of $W_{i}$. Defining
\begin{align}\label{eq:H_2B_supp}
\mathbf{H}_{2B,i}
= \tilde{\mathbf{Z}}_{i}\,
\mathcal{I}_{2B}^{-1}\,
\tilde{\mathbf{Z}}_{i}^{\top},
\end{align}
the same steps as in~\eqref{eq:A_2A_trace}--\eqref{eq:A_2A_H} yield
\begin{align}\label{eq:A_2B_H}
A_{2B,r}
= \frac{1}{2}
\sum_{i=1}^{n}
\mathbb{I}(t_0 \ge V_{i})\,
\mathrm{trace}\!\left[
  \frac{\partial \mathbf{V}_{i}}{\partial \theta_{r}}\,
  \mathbf{H}_{2B,i}
  \right],
\end{align}
with $\partial \mathbf{V}_{i} / \partial \theta_{r}$ given by~\eqref{eq:dV_cause1}--\eqref{eq:dV_cause2}. The Firth-corrected estimating function is
\begin{align*}
\bm{U}_{2B}^{*}(\bm{\theta})
= \sum_{i=1}^{n}
\mathbb{I}(t_0 \ge V_{i})\,
\tilde{\mathbf{Z}}_{i}^{\top}
\Bigg[\mathbf{Y}_{i}\, W_{i} - \bm{\pi}_{i}\Bigg]
+ \bm{A}_{2B}(\bm{\theta}),
\end{align*}
where the $r$-th component of $\bm{A}_{2B}$ is given by~\eqref{eq:A_2B_H}.



%
%
%
%
\subsection{Derivation of Asymptotic Variances for Approaches~A and~B}\label{supp: derv_asy_var}
Given $G(\cdot\mid\mathbf{Z})$ is known, we derive the asymptotic variances of the estimators obtained using Approaches~A and~B. We first derive $\Sigma_{A}(\bm{\theta};G)$ and $\Sigma_{B}(\bm{\theta};G)$ 
\begin{align}
\begin{split}
\Sigma_{A}(\bm{\theta};G) 
&= Var\left[\mathbb{I}(t_0 \ge V_i)\, W_i(t_0; G)\, 
    \mathbf{\tilde{Z}}_i^{\top} \left( \mathbf{Y}_i - \bm{\pi}_{i}(t_{0},\bm{Z}_{i})\right)\right]\\
&=E\left[\mathbb{I}(t_0 \ge V_i)\mathbf{\tilde{Z}}_i^{\top}
  E\left\{ W_i(t_0; G)^{2} \left( \mathbf{Y}_i - \bm{\pi}_{i}(t_{0},\bm{Z}_{i})\right)\left( \mathbf{Y}_i - \bm{\pi}_{i}(t_{0},\bm{Z}_{i})\right)^{\top} \mid \mathbf{\tilde{Z}}_i, V_{i}\right\}
  \mathbf{\tilde{Z}}_i\right]\\
\Sigma_{B}(\bm{\theta};G) 
&= Var\left[\mathbb{I}(t_0 \ge V_i)\,
    \mathbf{\tilde{Z}}_i^{\top} \left(\mathbf{Y}_{i}W_i(t_0; G) - \bm{\pi}_{i}(t_{0},\bm{Z}_{i})\right)\right]\\
&=E\left[\mathbb{I}(t_0 \ge V_i)\mathbf{\tilde{Z}}_i^{\top}
  E\left\{\left(\mathbf{Y}_{i}W_i(t_0; G) - \bm{\pi}_{i}(t_{0},\bm{Z}_{i})\right)\left(\mathbf{Y}_{i}W_i(t_0; G) - \bm{\pi}_{i}(t_{0},\bm{Z}_{i})\right)^{\top} \mid \mathbf{\tilde{Z}}_i, V_{i}\right\}
  \mathbf{\tilde{Z}}_i\right]\\
\end{split}
\end{align}
Next, we derive $\Gamma_{A}(\bm{\theta};G)$ and $\Gamma_{B}(\bm{\theta};G)$:
\begin{align}
\begin{split}
\frac{1}{n}\frac{\partial U_{2A}(\bm{\theta}; t_{0} \mid G) }{\partial \bm{\theta}^{\top}}
= &\frac{1}{n}
\left[
- \displaystyle\sum_{i=1}^{n}
\mathbb{I}(t_0 \ge V_i)\, W_i(t_0; G)\, 
    \mathbf{\tilde{Z}}_i^{\top} \mathbf{V}_{i}
    \mathbf{\tilde{Z}}_i
\right]
\xrightarrow{a.s.}\\
&-E\left[
\mathbb{I}(t_0 \ge V_i)\, W_i(t_0; G)\, 
    \mathbf{\tilde{Z}}_i^{\top} \mathbf{V}_{i}
    \mathbf{\tilde{Z}}_i
\right]
\triangleq -\Gamma_{A}(\bm{\theta};G)\\
\frac{1}{n}\frac{\partial U_{2B}(\bm{\theta}; t_{0} \mid G) }{\partial \bm{\theta}^{\top}}
= &\frac{1}{n}
\left[
- \displaystyle\sum_{i=1}^{n}
\mathbb{I}(t_0 \ge V_i)\,
    \mathbf{\tilde{Z}}_i^{\top} \mathbf{V}_{i}
    \mathbf{\tilde{Z}}_i
\right]
\xrightarrow{a.s.}\\
&-E\left[
\mathbb{I}(t_0 \ge V_i)\,
    \mathbf{\tilde{Z}}_i^{\top} \mathbf{V}_{i}
    \mathbf{\tilde{Z}}_i
\right]
\triangleq -\Gamma_{B}(\bm{\theta};G)
\end{split}
\end{align}
The asymptotic variances of the estimators obtained using Approaches~A and~B are 
\begin{align*}
AV_{A}(\bm{\theta}; G)
&= \Gamma_{A}^{-1}(\bm{\theta}; G)\,
\Sigma_{A}(\bm{\theta}; G)\,
\bigl[\Gamma_{A}^{-1}(\bm{\theta}; G)\bigr]^{\top}, \\
AV_{B}(\bm{\theta}; G)
&= \Gamma_{B}^{-1}(\bm{\theta}; G)\,
\Sigma_{B}(\bm{\theta}; G)\,
\bigl[\Gamma_{B}^{-1}(\bm{\theta}; G)\bigr]^{\top}.
\end{align*}

%
%
%
%
\subsection{Variable Selection}\label{supp: var_select}

\subsubsection{Derivation of the QICW Criteria}\label{supp: qicw_deriv}

This section derives the weighted quasi-loglikelihoods~\eqref{eq: Qw} and the penalty terms of the criteria stated in Section~\ref{sec2: variable selection} of the main text. Arguments are suppressed throughout, so that $\pi_{ki}$ abbreviates $\pi_{ki}(t_{0}, \bm{Z}_{i})$ and $\bm{\pi}_{i}$ abbreviates $\bm{\pi}_{i}(t_{0}, \bm{Z}_{i})$, and we retain $Y_{0i} = 1 - Y_{1i} - Y_{2i}$ and $\pi_{0i} = 1 - \pi_{1i} - \pi_{2i}$.

\paragraph{The quasi-loglikelihood without censoring}

With no censoring, $\mathbf{Y}_{i}$ is multinomial. Since the multinomial law belongs to the exponential family under the logit link, its quasi-loglikelihood equals its loglikelihood, giving $Q(\bm{\theta};\, t_{0}) = \sum_{i=1}^{n} Q_{i}$ with $Q_{i} = \sum_{k=0}^{2} Y_{ki} \log \pi_{ki}$. Using $\sum_{k=0}^{2} Y_{ki} = 1$ to eliminate $Y_{0i}$,
\begin{align}\label{eq: Qi_supp}
Q_{i} = \sum_{k=1}^{2} Y_{ki} \log \frac{\pi_{ki}}{\pi_{0i}} + \log \pi_{0i}.
\end{align}
Because the logit is the canonical link, differentiating~\eqref{eq: Qi_supp} gives the per-subject score
\begin{align}\label{eq: dQi_supp}
\frac{\partial Q_{i}}{\partial \bm{\theta}} = \tilde{\mathbf{Z}}_{i}^{\top} \bigl[ \mathbf{Y}_{i} - \bm{\pi}_{i} \bigr],
\end{align}
whose sum over $i$ is the estimating function in the absence of censoring.

\paragraph{Approach A}

In~\eqref{eq: proj2 multinomial estimating equation a} the IPCW multiplies the entire per-subject score,
\[
\bm{U}_{2A}(\bm{\theta};\, t_{0} \mid \hat{G})
= \sum_{i=1}^{n} \mathbb{I}(t_{0} \ge V_{i})\, W_{i}(t_{0}; \hat{G})\,
\underbrace{\tilde{\mathbf{Z}}_{i}^{\top} \bigl[ \mathbf{Y}_{i} - \bm{\pi}_{i} \bigr]}_{\partial Q_{i} / \partial \bm{\theta}}
= \sum_{i=1}^{n} \tilde{\bm{A}}_{i},
\]
so the matching weighted quasi-loglikelihood weights $Q_{i}$ itself, which gives the expression for $Q_{w}^{A}$ in~\eqref{eq: Qw}. Writing the sandwich~\eqref{eq: varestmAapp} as $\widehat{\mathrm{Var}}\{\tilde{\bm{\theta}}_{A}\} = \hat{\mathcal{I}}_{2A}^{-1} \hat{V}\{\bm{U}_{2A}\} (\hat{\mathcal{I}}_{2A}^{-1})^{\top}$, the penalty of~\eqref{eq: qicw} reduces to
\begin{align}\label{eq: penaltyA_supp}
2\, \mathrm{trace}\Bigl( \hat{\mathcal{I}}_{2A}\, \widehat{\mathrm{Var}}\bigl\{ \tilde{\bm{\theta}}_{A} \bigr\} \Bigr)
= 2\, \mathrm{trace}\Bigl[ \hat{V}\bigl\{ \bm{U}_{2A}(\bm{\theta};\, t_{0} \mid \hat{G}) \bigr\} \bigl( \hat{\mathcal{I}}_{2A}^{-1} \bigr)^{\top} \Bigr],
\end{align}
with $\hat{V}\{\bm{U}_{2A}\} = \sum_{i=1}^{n} ( \tilde{\bm{A}}_{i} - \bar{\tilde{\bm{A}}} ) ( \tilde{\bm{A}}_{i} - \bar{\tilde{\bm{A}}} )^{\top}$ as in Section~\ref{sec2: practical implementation} of the main text. Here $\hat{\mathcal{I}}_{2A}$ plays the role of $\bm{\Omega}_{I}$ and $\widehat{\mathrm{Var}}\{\tilde{\bm{\theta}}_{A}\}$ that of $\bm{V}_{R}$ in the criterion of \citet{pan2001}.

\paragraph{Approach B}

In~\eqref{eq: proj2 multinomial estimating equation b} the IPCW sits inside the residual and multiplies $\mathbf{Y}_{i}$ only,
\[
\bm{U}_{2B}(\bm{\theta};\, t_{0} \mid \hat{G})
= \sum_{i=1}^{n} \mathbb{I}(t_{0} \ge V_{i})\, \tilde{\mathbf{Z}}_{i}^{\top}
\bigl[ W_{i}(t_{0}; \hat{G})\, \mathbf{Y}_{i} - \bm{\pi}_{i} \bigr]
= \sum_{i=1}^{n} \tilde{\bm{B}}_{i},
\]
so $Q_{w}^{B}$ is no longer a weighted version of $Q_{i}$ and must be recovered by integrating $\bm{U}_{2B}$ in $\bm{\theta}$. The integrand splits into two parts. For the first, using $\log(\pi_{ki} / \pi_{0i}) = \alpha_{k}(t_{0}) + \bm{\beta}_{k}(t_{0})^{\top} \bm{Z}_{i}$,
\begin{align}\label{eq: intB1_supp}
\int W_{i}(t_{0}; \hat{G})\, \tilde{\mathbf{Z}}_{i}^{\top} \mathbf{Y}_{i} \, d\bm{\theta}
= W_{i}(t_{0}; \hat{G})\, \mathbf{Y}_{i}^{\top} \tilde{\mathbf{Z}}_{i}\, \bm{\theta}
= W_{i}(t_{0}; \hat{G}) \sum_{k=1}^{2} Y_{ki} \log \frac{\pi_{ki}}{\pi_{0i}}.
\end{align}
For the second, since $\partial \log \pi_{0i} / \partial \bm{\theta} = -\tilde{\mathbf{Z}}_{i}^{\top} \bm{\pi}_{i}$,
\begin{align}\label{eq: intB2_supp}
\int \bigl( -\tilde{\mathbf{Z}}_{i}^{\top} \bm{\pi}_{i} \bigr) \, d\bm{\theta} = \log \pi_{0i}.
\end{align}
Adding~\eqref{eq: intB1_supp} and~\eqref{eq: intB2_supp} and summing over $i$ gives the expression for $Q_{w}^{B}$ in~\eqref{eq: Qw}. The contrast with Approach~A is that the IPCW multiplies both terms of~\eqref{eq: Qi_supp} in $Q_{w}^{A}$, but only $\sum_{k} Y_{ki} \log(\pi_{ki} / \pi_{0i})$ in $Q_{w}^{B}$, leaving $\log \pi_{0i}$ unweighted. The penalty follows the same route as~\eqref{eq: penaltyA_supp}, with the approximate meat $\hat{V}_{\mathrm{app}}\{\bm{U}_{2B}\}$ of~\eqref{eq: varestmBapp} in place of $\hat{V}\{\bm{U}_{2A}\}$:
\begin{align}\label{eq: penaltyB_supp}
2\, \mathrm{trace}\Bigl( \hat{\mathcal{I}}_{2B}\, \widehat{\mathrm{Var}}\bigl\{ \tilde{\bm{\theta}}_{B} \bigr\} \Bigr)
= 2\, \mathrm{trace}\Bigl[ \hat{V}_{\mathrm{app}}\bigl\{ \bm{U}_{2B}(\bm{\theta};\, t_{0} \mid \hat{G}) \bigr\} \bigl( \hat{\mathcal{I}}_{2B}^{-1} \bigr)^{\top} \Bigr].
\end{align}

\paragraph{With the Firth correction}

The form of the criterion does not change. Only the estimator at which $Q_{w}^{m}$, $\hat{\mathcal{I}}_{2m}$, and $\widehat{\mathrm{Var}}$ are evaluated changes: without the correction one solves $\bm{U}_{2m}(\bm{\theta};\, t_{0} \mid \hat{G}) = \bm{0}$ to obtain $\tilde{\bm{\theta}}_{m}$, and with it one solves $\bm{U}^{*}_{2m}(\bm{\theta};\, t_{0} \mid \hat{G}) = \bm{0}$ to obtain $\tilde{\bm{\theta}}_{m}^{*}$. The additional term $(1/2) \log |\mathcal{I}_{2m}|$ in $Q_{w}^{m,*}$ contributes $-\log |\hat{\mathcal{I}}_{2m}^{*}|$ to the criterion.

\subsubsection{The Criterion $\mathrm{QICW}_{u}$}\label{supp: qicwu}

When all modeling specifications are correct, $\mathrm{trace}( \hat{\mathcal{I}}_{2m}\, \widehat{\mathrm{Var}}\{\tilde{\bm{\theta}}_{m}\} ) \approx P$, so the penalty of~\eqref{eq: qicw} is approximately $2P$, where $P$ is the number of estimated coefficients of the fitted model. Replacing the trace penalty by $2P$ gives the second form
\begin{align}\label{eq: qicwu_supp}
\mathrm{QICW}_{u}^{m}(t_{0})
= -2\, Q_{w}^{m}(\tilde{\bm{\theta}}_{m};\, t_{0} \mid \hat{G}) + 2P,
\qquad m \in \{A, B\},
\end{align}
and, with the Firth correction,
\begin{align}\label{eq: qicwu_firth_supp}
\mathrm{QICW}_{u}^{m,*}(t_{0})
= -2\, Q_{w}^{m,*}(\tilde{\bm{\theta}}_{m}^{*};\, t_{0} \mid \hat{G}) + 2P.
\end{align}
Together with~\eqref{eq: qicw} and~\eqref{eq: qicw_firth} of the main text, expressions~\eqref{eq: qicwu_supp} and~\eqref{eq: qicwu_firth_supp} give eight criteria in total: for each approach $m \in \{A, B\}$, the trace penalty and the $2P$ penalty, each with and without the Firth correction.

Expanding $Q_{w}^{m,*} = Q_{w}^{m} + (1/2) \log |\mathcal{I}_{2m}|$ makes the contribution of the Firth term explicit,
\begin{align}\label{eq: qicw_firth_expanded_supp}
\mathrm{QICW}^{m,*}(t_{0})
= -2\, Q_{w}^{m}(\tilde{\bm{\theta}}_{m}^{*};\, t_{0} \mid \hat{G})
- \log \bigl| \hat{\mathcal{I}}_{2m}^{*} \bigr|
+ 2\, \mathrm{trace}\Bigl( \hat{\mathcal{I}}_{2m}^{*}\, \widehat{\mathrm{Var}}\bigl\{ \tilde{\bm{\theta}}_{m}^{*} \bigr\} \Bigr),
\end{align}
and similarly for $\mathrm{QICW}_{u}^{m,*}$ with $2P$ in place of the trace penalty.

The two penalty forms differ only in the penalty term. The approximation leading to~\eqref{eq: qicwu_supp} relies on correct specification of the censoring model as well as of the outcome model, and $\hat{G}(\cdot \mid \mathbf{Z})$ is not always well specified in practice. The main text therefore reports $\mathrm{QICW}$ throughout; the corresponding $\mathrm{QICW}_{u}$ results are given in Section~\ref{supp: var_select_qicwu}.

\subsubsection{Choosing the Tolerance $\varepsilon$}\label{supp: tolerance}

The analyses of the main text use $\varepsilon = 0$ throughout. This section records how a positive tolerance would be calibrated, since the acceptance rule of Algorithm~\ref{alg: backward} admits any $\varepsilon \ge 0$.

Write $P_{\mathcal{M}}$ for the number of coefficients in model $\mathcal{M}$, and $P_{z} = P_{\mathcal{M}} - P_{\mathcal{M} \setminus \{z\}}$ for the number carried by term $z$ across both causes. In the extended covariate set of the main text, $P_{z} = 2$ for every term except race, which has three levels and so carries $P_{z} = 4$. Splitting $d^{\star}$ into its fit and penalty parts under $\mathrm{QICW}_{u}^{m}$,
\[
d^{\star}
= \mathrm{QICW}_{u}^{m}(\mathcal{M} \setminus \{z\}) - \mathrm{QICW}_{u}^{m}(\mathcal{M})
= e_{z} - 2 P_{z},
\qquad
e_{z} = 2 \bigl\{ Q_{w}^{m}(\tilde{\bm{\theta}}_{\mathcal{M}}) - Q_{w}^{m}(\tilde{\bm{\theta}}_{\mathcal{M} \setminus \{z\}}) \bigr\},
\]
so the acceptance rule $d^{\star} < \varepsilon$ is exactly $e_{z} < 2 P_{z} + \varepsilon$. The tolerance therefore shifts the critical value of a likelihood-ratio-type test of $H_{0}: \bm{\beta}_{1z}(t_{0}) = \bm{\beta}_{2z}(t_{0}) = \bm{0}$, and three choices follow.

Setting $\varepsilon = 0$ removes $z$ whenever $e_{z} < 2 P_{z}$, which is the AIC-type rule and the one used in the main text. Since a constant $\varepsilon$ enters only the acceptance test and not the ranking of the $q_{z}$, the elimination path is then the same for every such $\varepsilon$, and only the stopping point moves along it.

Alternatively, $\varepsilon$ may be calibrated to a level $\xi$ by setting $\varepsilon(P_{z}) = \chi^{2}_{P_{z},\, 1 - \xi} - 2 P_{z}$, so that each step reproduces a level-$\xi$ test. At $\xi = 0.05$ this gives $\varepsilon = 1.991$ for a term with $P_{z} = 2$, since $\chi^{2}_{2,\, 0.95} = 5.991$, and $\varepsilon = 1.488$ for race, since $\chi^{2}_{4,\, 0.95} = 9.488$. Because $\varepsilon$ is now term-specific, the candidates are ranked by $q_{z} - \varepsilon(P_{z})$ and a removal is accepted if the minimum falls below $c$; unlike the constant-$\varepsilon$ case, the elimination path itself may change.

A third option guards against drift from the full model, since steps that are individually acceptable can accumulate. Writing $d^{\dagger} = \mathrm{QICW}_{u}^{m}(\mathcal{M} \setminus \{z\}) - \mathrm{QICW}_{u}^{m}(\mathcal{M}_{\mathrm{full}})$, a removal is accepted only if both $d^{\star} < \varepsilon(P_{z})$ and $d^{\dagger} < \varepsilon(P_{\mathrm{full}} - P_{\mathcal{M} \setminus \{z\}})$. The second condition tightens as elimination proceeds: since $\chi^{2}_{P,\, 1 - \xi}$ grows more slowly than $2P$, the quantity $\varepsilon(P)$ falls as more coefficients are removed, taking the values $1.991$, $1.488$, and $0.592$ at $P = 2$, $4$, and $6$, and $-0.493$ at $P = 8$. Once it turns negative the guard permits no increase at all, and a candidate must score strictly below the full model to be accepted.

The derivation above uses $\mathrm{QICW}_{u}^{m}$ only to obtain a workable value for $\varepsilon$. The same $\varepsilon$ would then serve as the stopping rule for all eight criteria, including the Firth-corrected $\mathrm{QICW}^{m,*}$ and $\mathrm{QICW}_{u}^{m,*}$.

Finally, when $\varepsilon > 0$ an accepted removal may raise the criterion, so $\mathrm{QICW}$ need not decrease monotonically along the elimination path. The selected model $\mathcal{M}^{\star}$ is then the model at the stopping point rather than the model with the smallest criterion value encountered along it. Under the strict-decrease rule $\varepsilon = 0$ used in the main text, the criterion decreases at every accepted step and the two coincide.

%
%
%
%
\section{Additional Results of the PM Analysis}\label{supp: real_data}
%
%
\subsection{Logistic Regression}\label{supp: real_data_log_reg}
\label{supp: logistic_real_data}

\begin{figure}[H]
    \centering
    \begin{subfigure}[b]{0.48\textwidth}
        \centering
        \includegraphics[width=\textwidth]{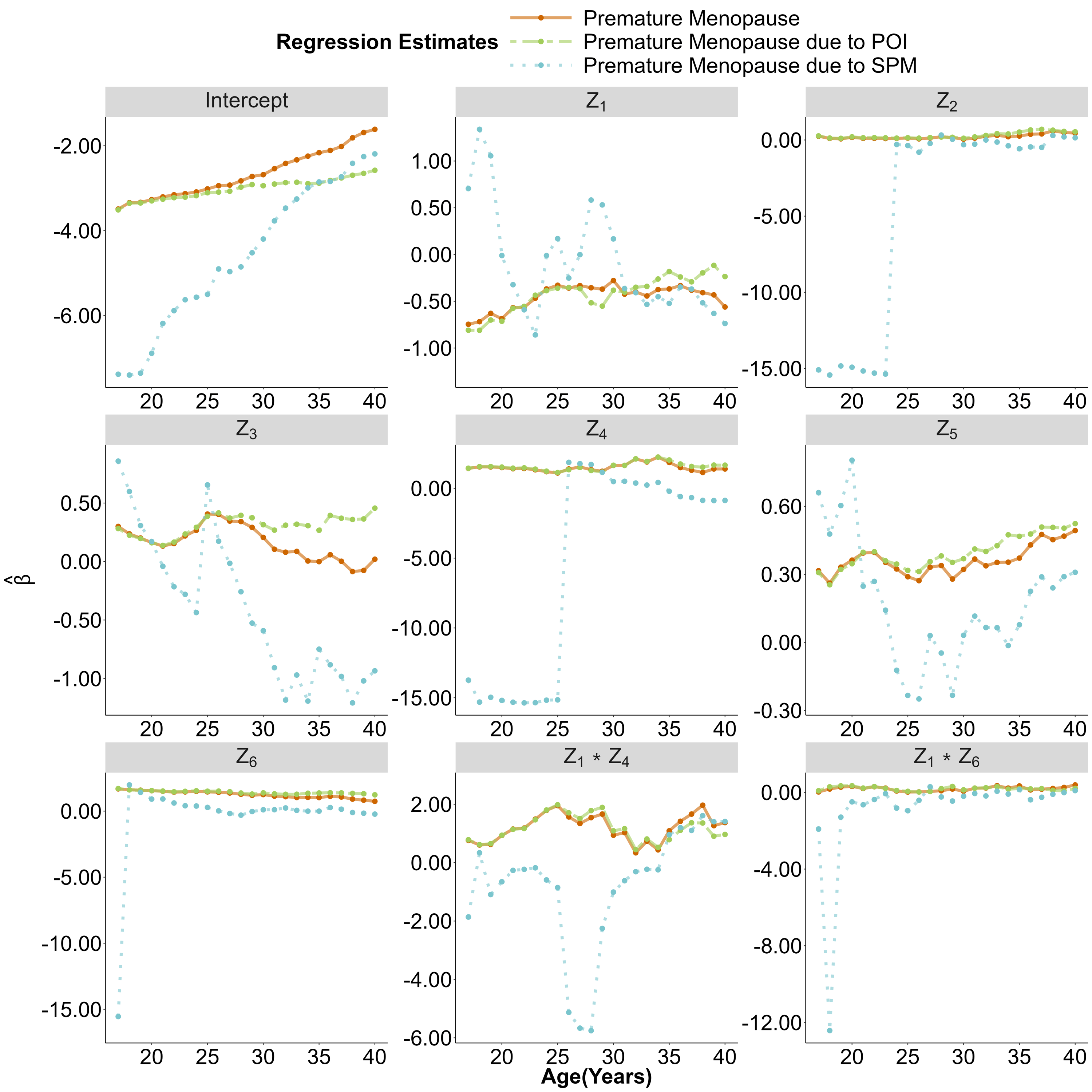}
        \caption{Approach A (Without Firth Correction)}
        \label{fig:log_coef_appa_mlr}
    \end{subfigure}
    \hfill
    \begin{subfigure}[b]{0.48\textwidth}
        \centering
        \includegraphics[width=\textwidth]{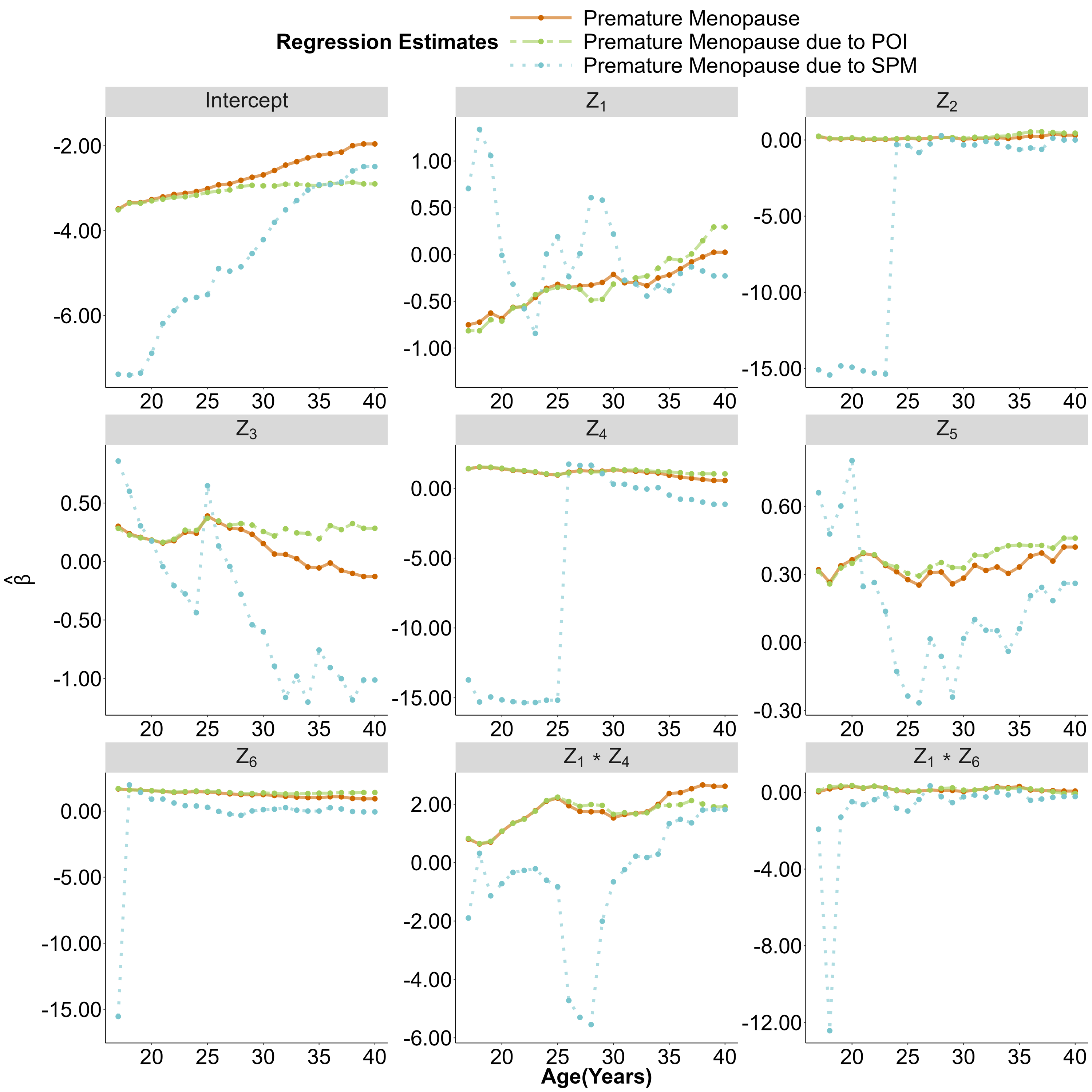}
        \caption{Approach B (Without Firth Correction)}
        \label{fig:log_coef_appb_mlr}
    \end{subfigure}

    \vspace{0.5cm}

    \begin{subfigure}[b]{0.48\textwidth}
        \centering
        \includegraphics[width=\textwidth]{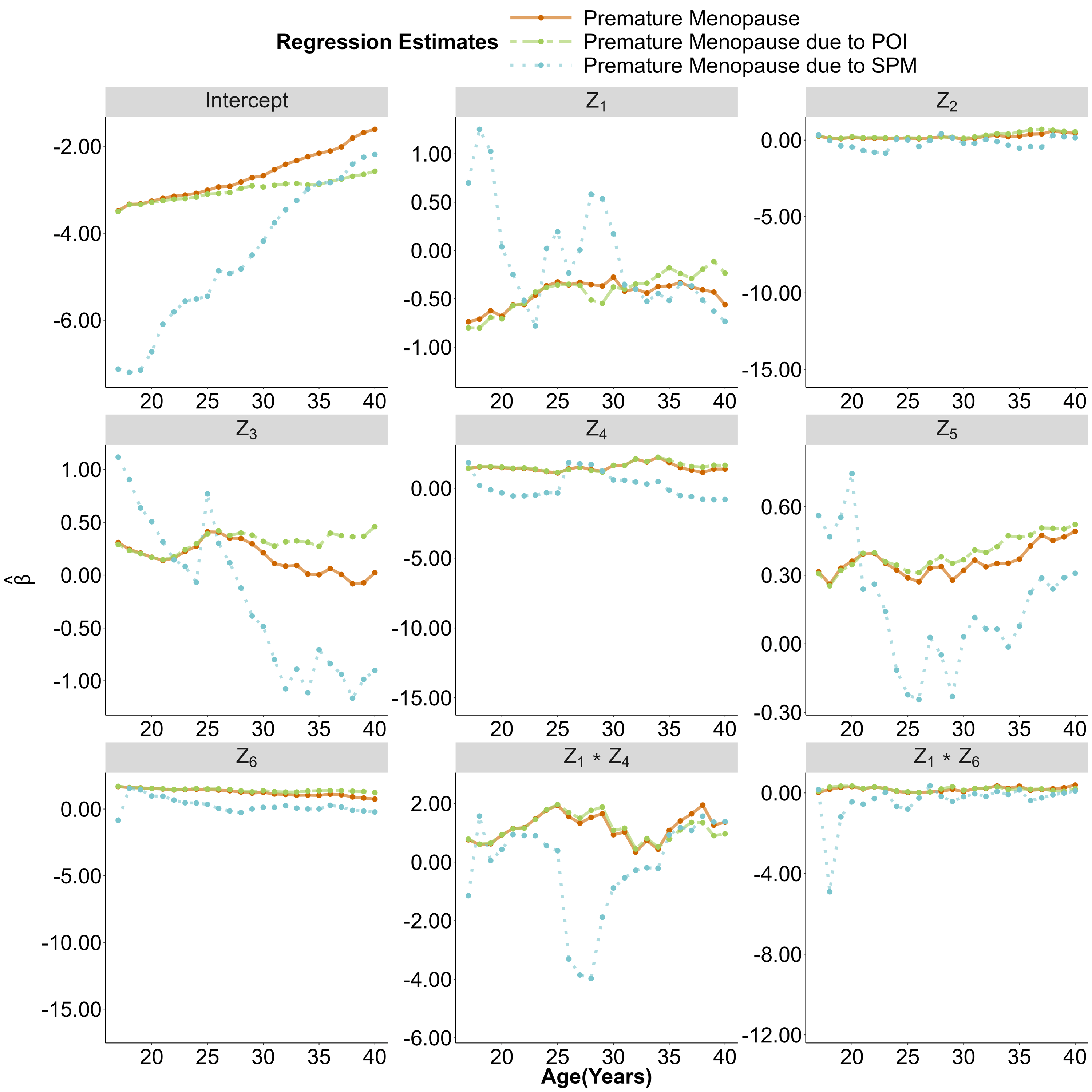}
        \caption{Approach A (With Firth Correction)}
        \label{fig:log_coef_appa_firth}
    \end{subfigure}
    \hfill
    \begin{subfigure}[b]{0.48\textwidth}
        \centering
        \includegraphics[width=\textwidth]{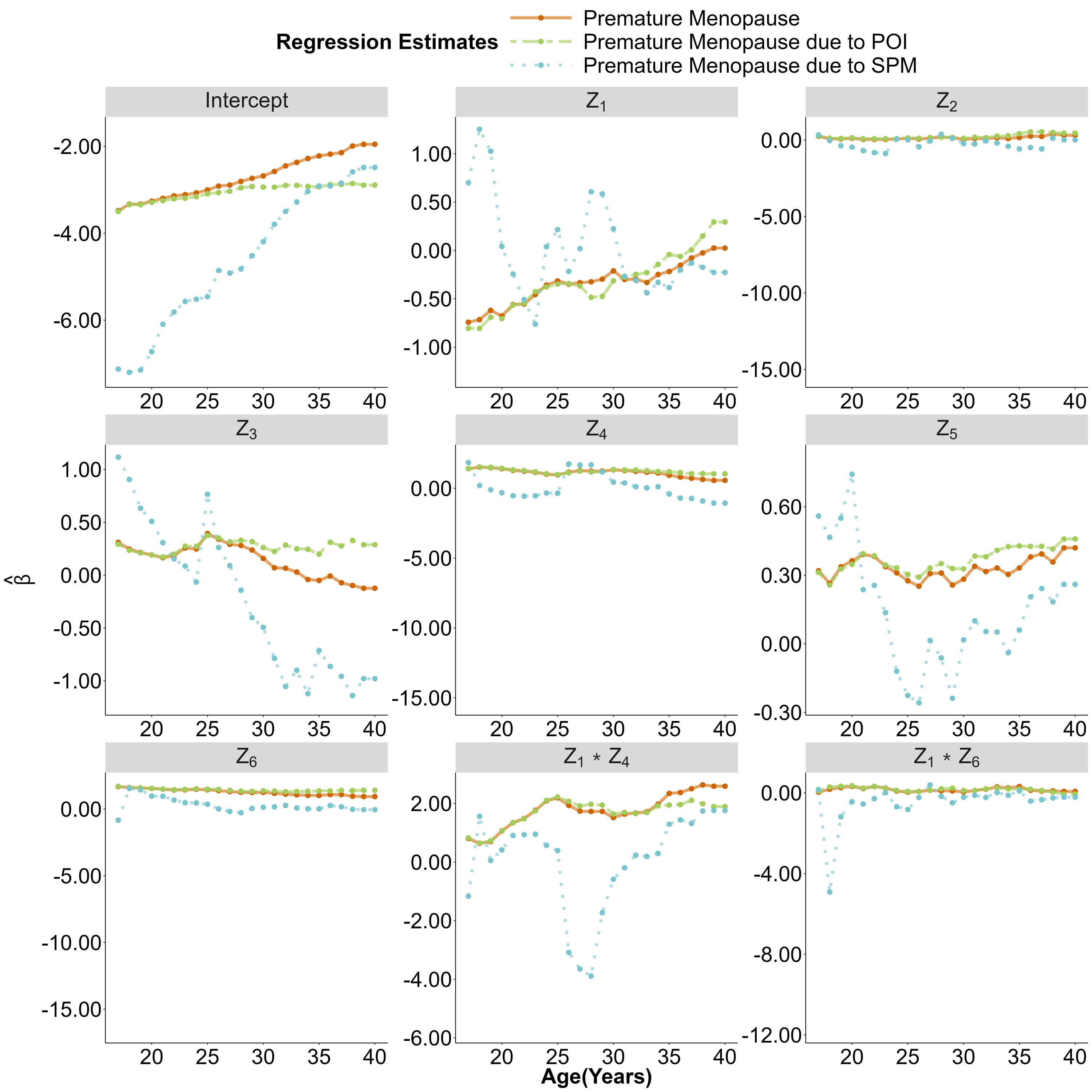}
        \caption{Approach B (With Firth Correction)}
        \label{fig:log_coef_appb_firth}
    \end{subfigure}

    \caption{The estimated coefficients when using logistic regression with and without Firth correction among Approaches~A and~B. The $\hat{G}(\cdot \mid \mathbf{Z})$ is obtained using SRF. The number of trees and node size are set to 100.  $Z_{1}\text{: Rescaled age at cancer diagnosis}$,
             $Z_{2}\text{: Race-African American}$,
             $Z_{3}\text{: Race-Other}$,
             $Z_{4}\text{: Receipt of BMT}$,
             $Z_{5}\text{: Receipt of Alkylating agents}$
             $Z_{6}\text{: Receipt of radiation to the abdomen/pelvis/total body}$}
    \label{fig: log_real_data_coef_srf}
\end{figure}

\begin{figure}[H]
    \centering
    \begin{subfigure}[b]{0.48\textwidth}
        \centering
        \includegraphics[width=\textwidth]{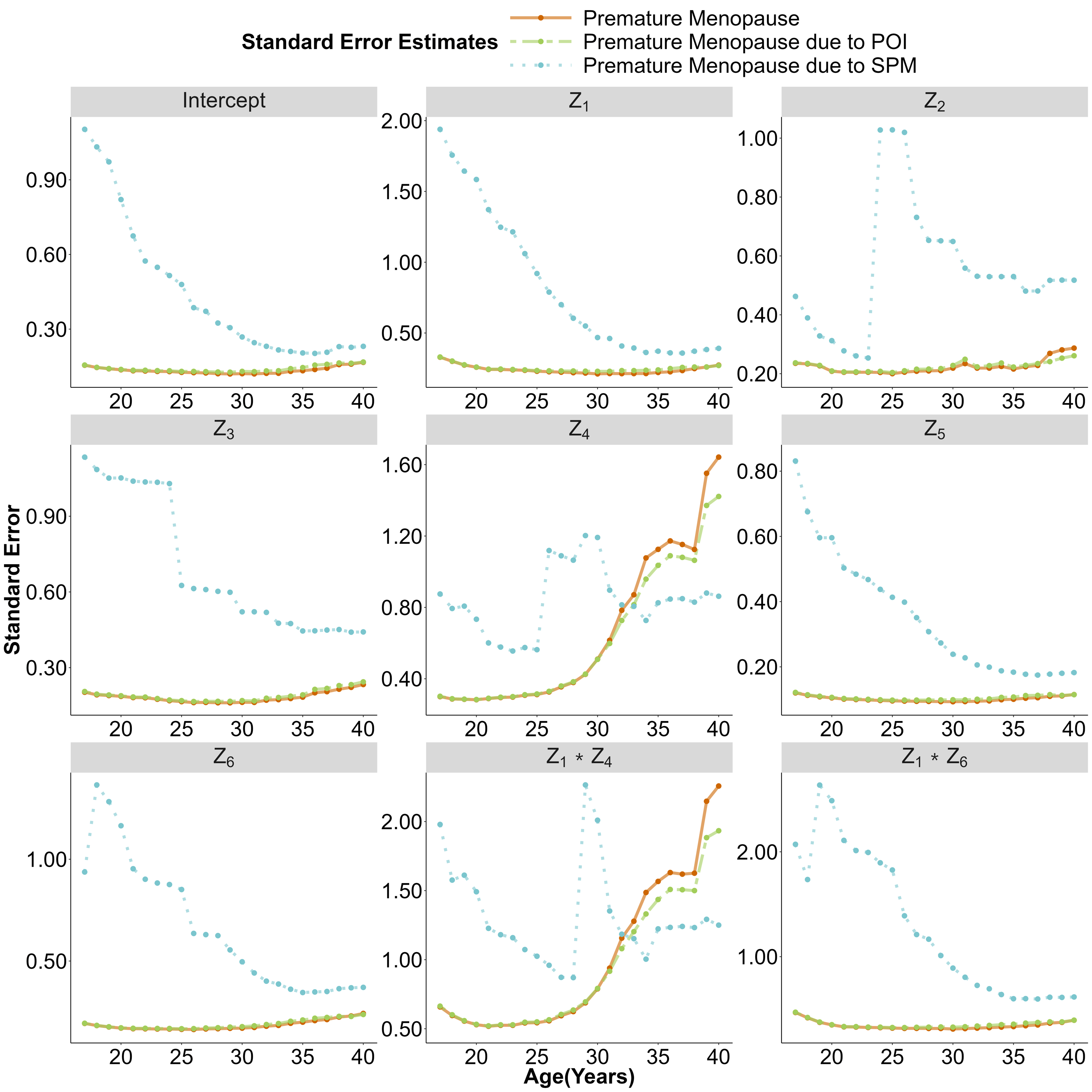}
        \caption{Approach A (Without Firth Correction)}
        \label{fig:log_se_appa_mlr}
    \end{subfigure}
    \hfill
    \begin{subfigure}[b]{0.48\textwidth}
        \centering
        \includegraphics[width=\textwidth]{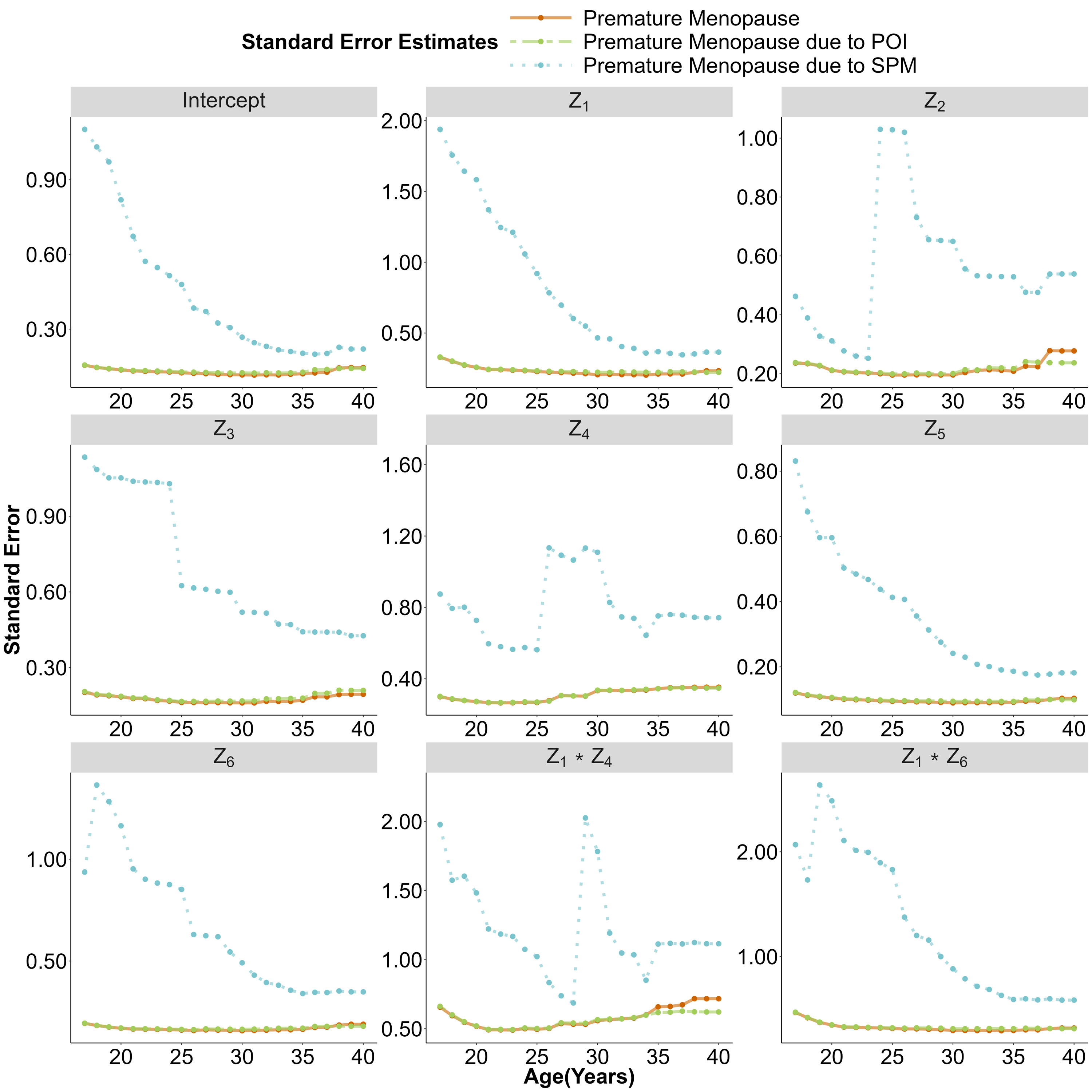}
        \caption{Approach B (Without Firth Correction)}
        \label{fig:log_se_appb_mlr}
    \end{subfigure}

    \vspace{0.5cm}

    \begin{subfigure}[b]{0.48\textwidth}
        \centering
        \includegraphics[width=\textwidth]{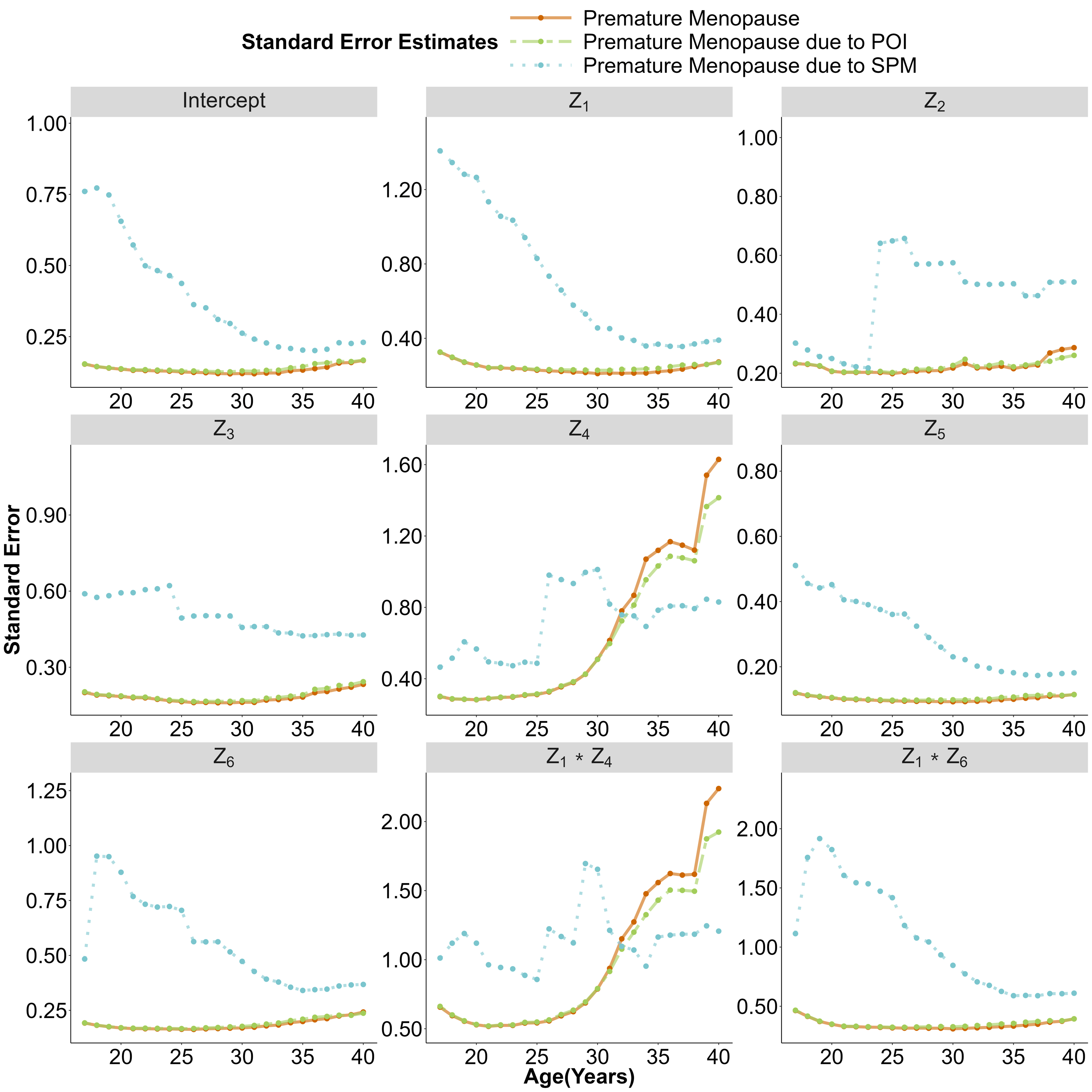}
        \caption{Approach A (With Firth Correction)}
        \label{fig:log_se_appa_firth}
    \end{subfigure}
    \hfill
    \begin{subfigure}[b]{0.48\textwidth}
        \centering
        \includegraphics[width=\textwidth]{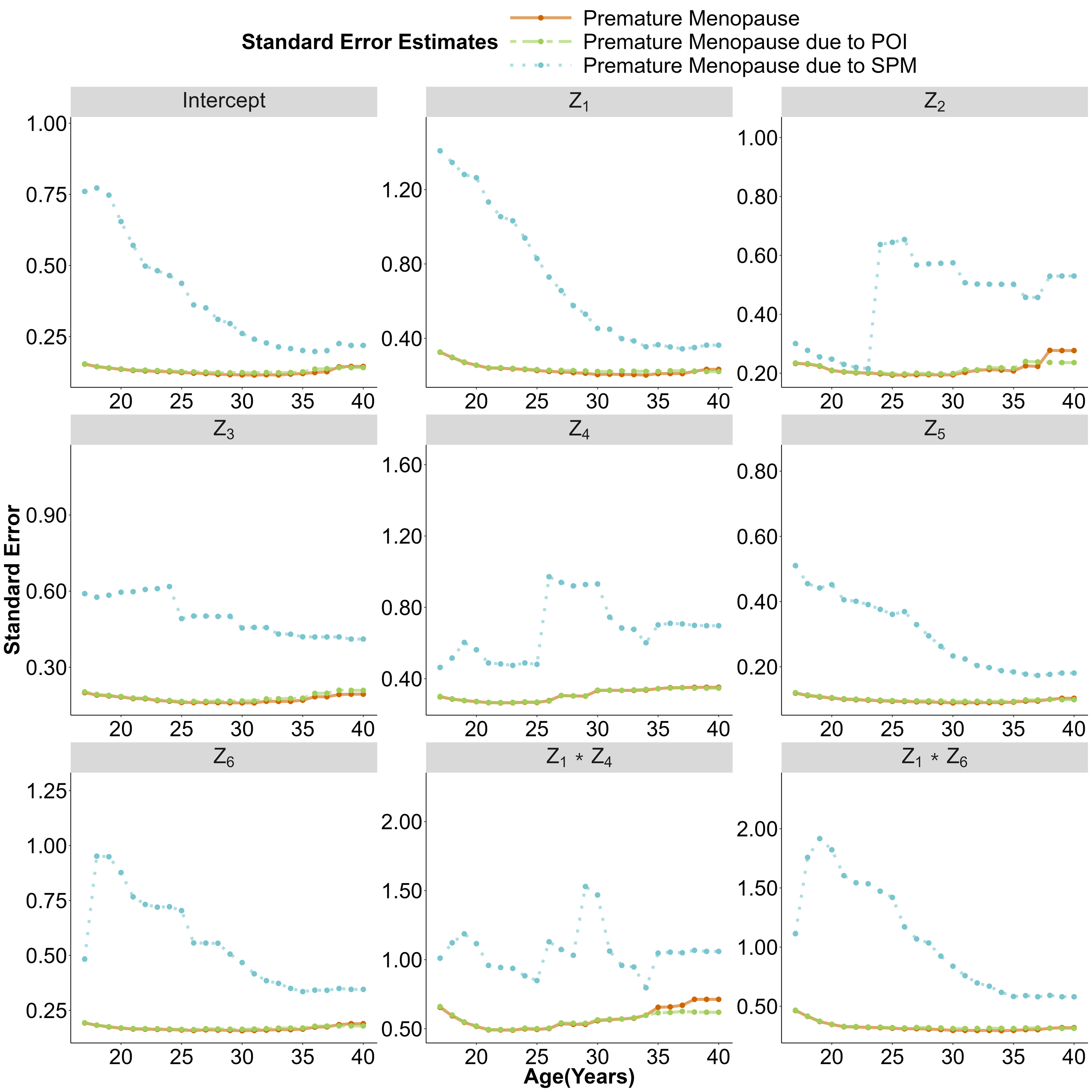}
        \caption{Approach B (With Firth Correction)}
        \label{fig:log_se_appb_firth}
    \end{subfigure}

    \caption{The sandwich standard error of estimated coefficients when using logistic regression with and without Firth correction among Approaches~A and~B. The $\hat{G}(\cdot \mid \mathbf{Z})$ is obtained using SRF. The number of trees and node size are set to 100.  $Z_{1}\text{: Rescaled age at cancer diagnosis}$,
             $Z_{2}\text{: Race-African American}$,
             $Z_{3}\text{: Race-Other}$,
             $Z_{4}\text{: Receipt of BMT}$,
             $Z_{5}\text{: Receipt of Alkylating agents}$
             $Z_{6}\text{: Receipt of radiation to the abdomen/pelvis/total body}$}
    \label{fig: log_real_data_se_srf}
\end{figure}

Figures~\ref{fig: log_real_data_coef_srf} and~\ref{fig: log_real_data_se_srf} present the estimated regression coefficients and sandwich standard errors from model~\eqref{model: project2 model 1}, where $\hat{G}(\cdot \mid \mathbf{Z})$ is estimated via SRF. As shown in Figure~\ref{fig: log_real_data_coef_srf}, the coefficient estimates for PM and POI are stable across ages under both Approaches~A and~B. In contrast, the SPM estimates are highly volatile, and show extreme values for $Z_2$, $Z_4$, $Z_{1}Z_{4}$, $Z_{6}$, and $Z_{1}Z_{6}$ at early ages. This instability is consistent with the sparse distribution of African American subjects, BMT recipients, and radiation therapy recipients among SPM cases, which leads to quasi-complete separation. Applying Firth correction (Figures~\ref{fig:log_coef_appa_firth} and~\ref{fig:log_coef_appb_firth}) substantially attenuates these extreme excursions in the SPM estimates while leaving the PM and POI estimates unchanged. Differences between Approaches~A and~B are modest, as both yield similar coefficient trajectories across ages.

The sandwich standard errors (Figure~\ref{fig: log_real_data_se_srf}) reinforce these findings. The standard errors for PM and POI are small and stable, whereas those for SPM are substantially larger and exhibit sharp increases at ages corresponding to coefficient instability. In the presence of quasi-complete separation, sandwich standard errors obtained without Firth correction are unreliable because the sandwich variance is evaluated at unstable coefficient estimates. With Firth correction, the estimated coefficients remain finite and well-behaved, yielding more reliable standard errors; we return to this point in the simulation studies.


\begin{figure}[H]
    \centering
    \begin{subfigure}[b]{0.48\textwidth}
        \centering
        \includegraphics[width=\textwidth]{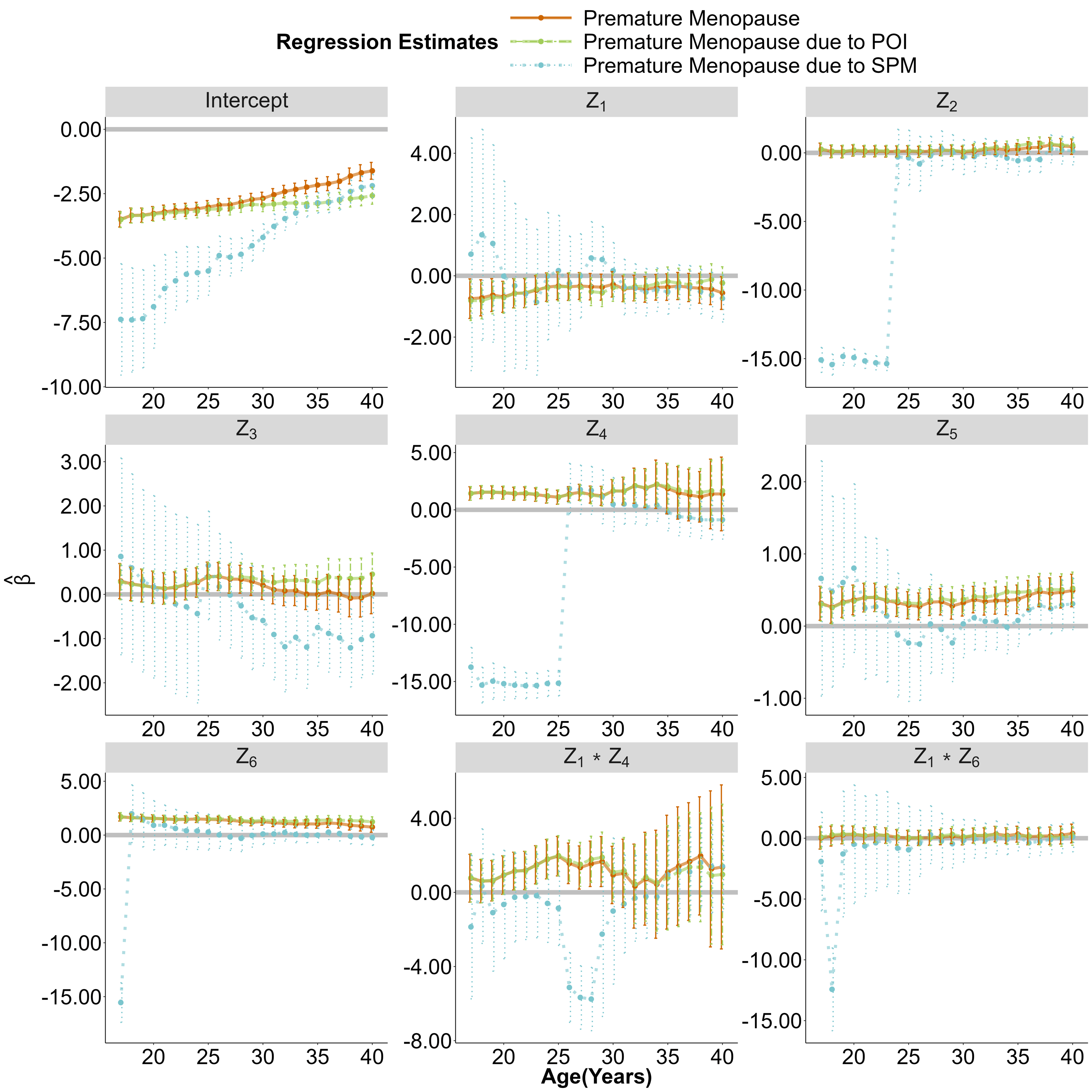}
        \caption{Approach A (Without Firth Correction)}
        \label{fig:log_ci_appa_mlr}
    \end{subfigure}
    \hfill
    \begin{subfigure}[b]{0.48\textwidth}
        \centering
        \includegraphics[width=\textwidth]{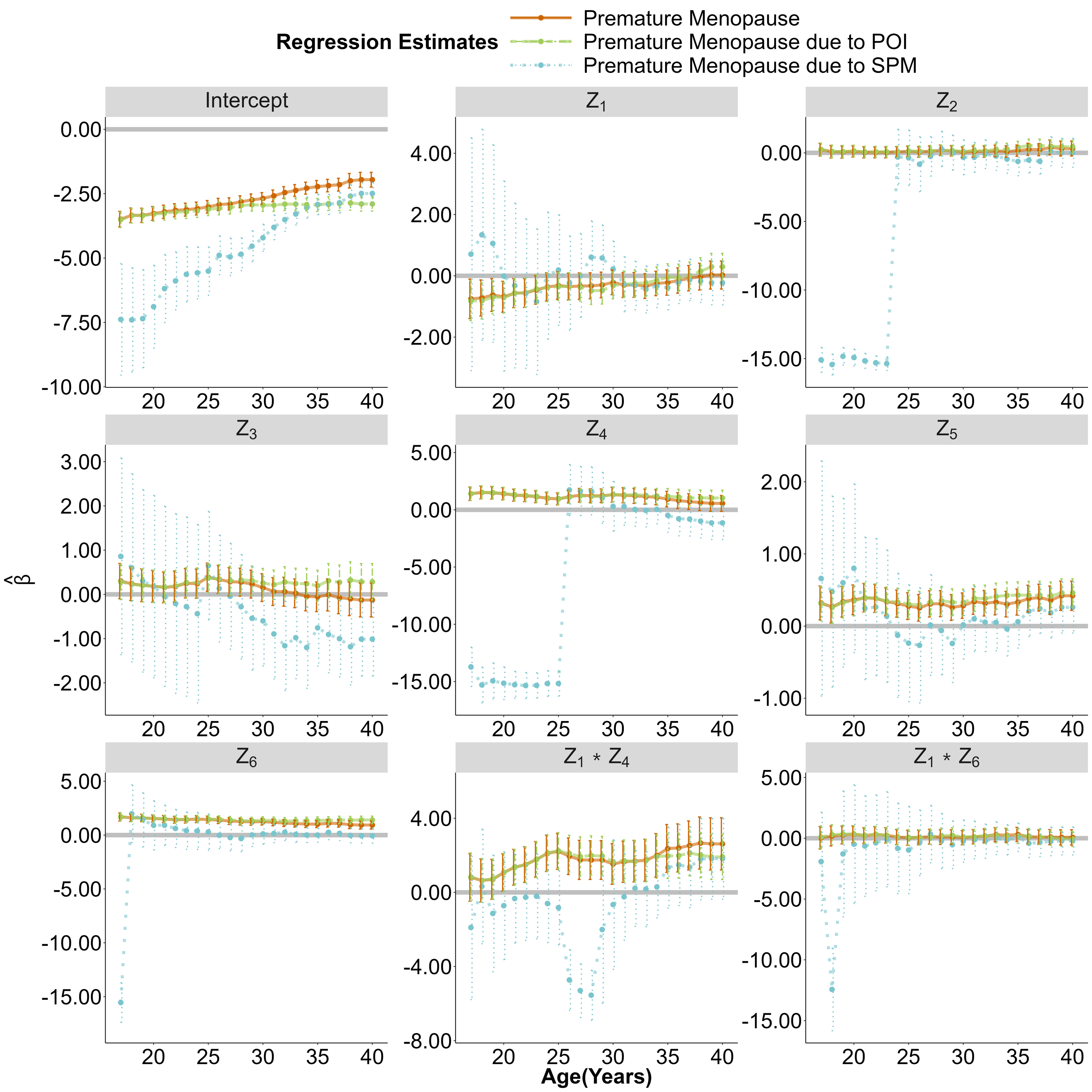}
        \caption{Approach B (Without Firth Correction)}
        \label{fig:log_ci_appb_mlr}
    \end{subfigure}

    \vspace{0.5cm}

    \begin{subfigure}[b]{0.48\textwidth}
        \centering
        \includegraphics[width=\textwidth]{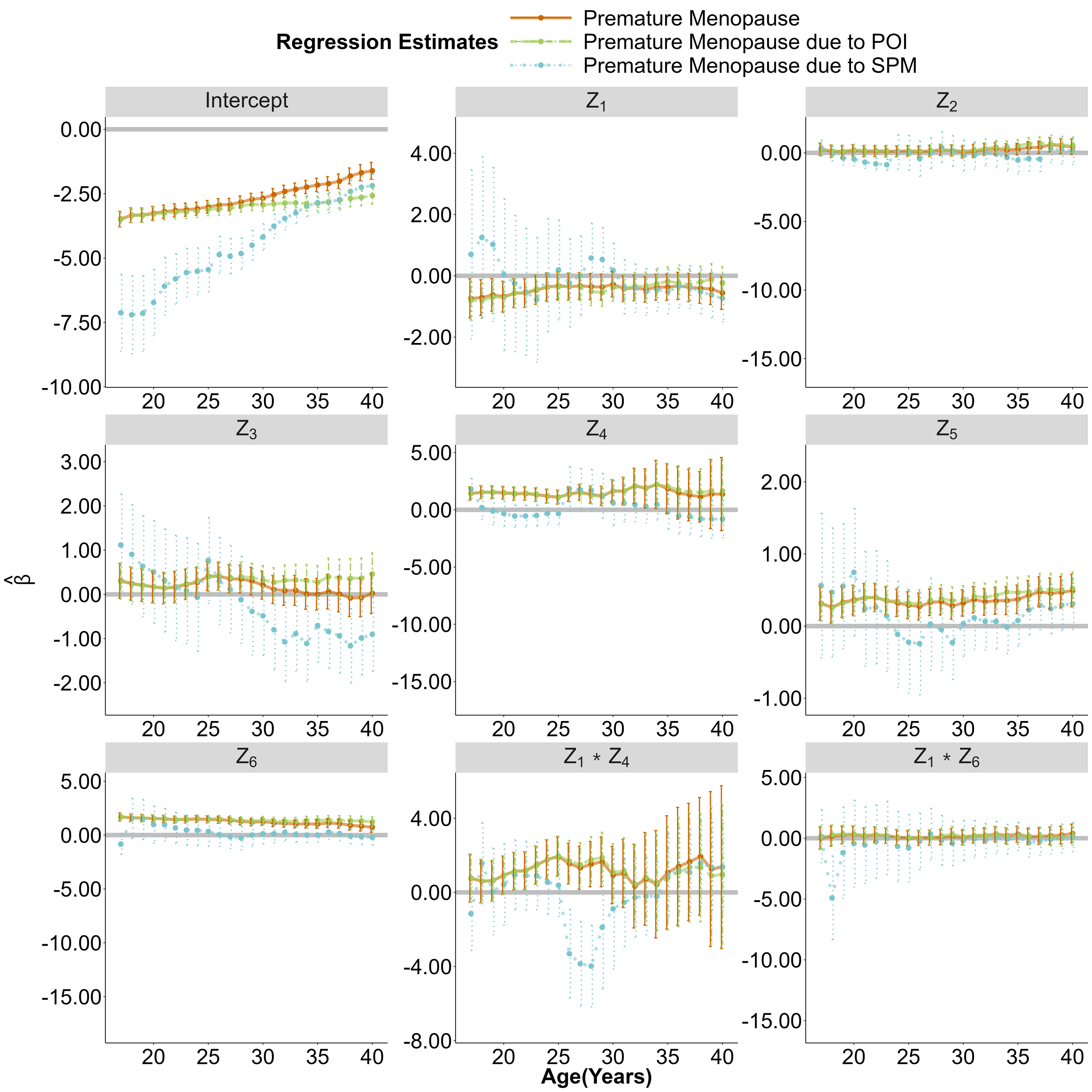}
        \caption{Approach A (With Firth Correction)}
        \label{fig:log_ci_appa_firth}
    \end{subfigure}
    \hfill
    \begin{subfigure}[b]{0.48\textwidth}
        \centering
        \includegraphics[width=\textwidth]{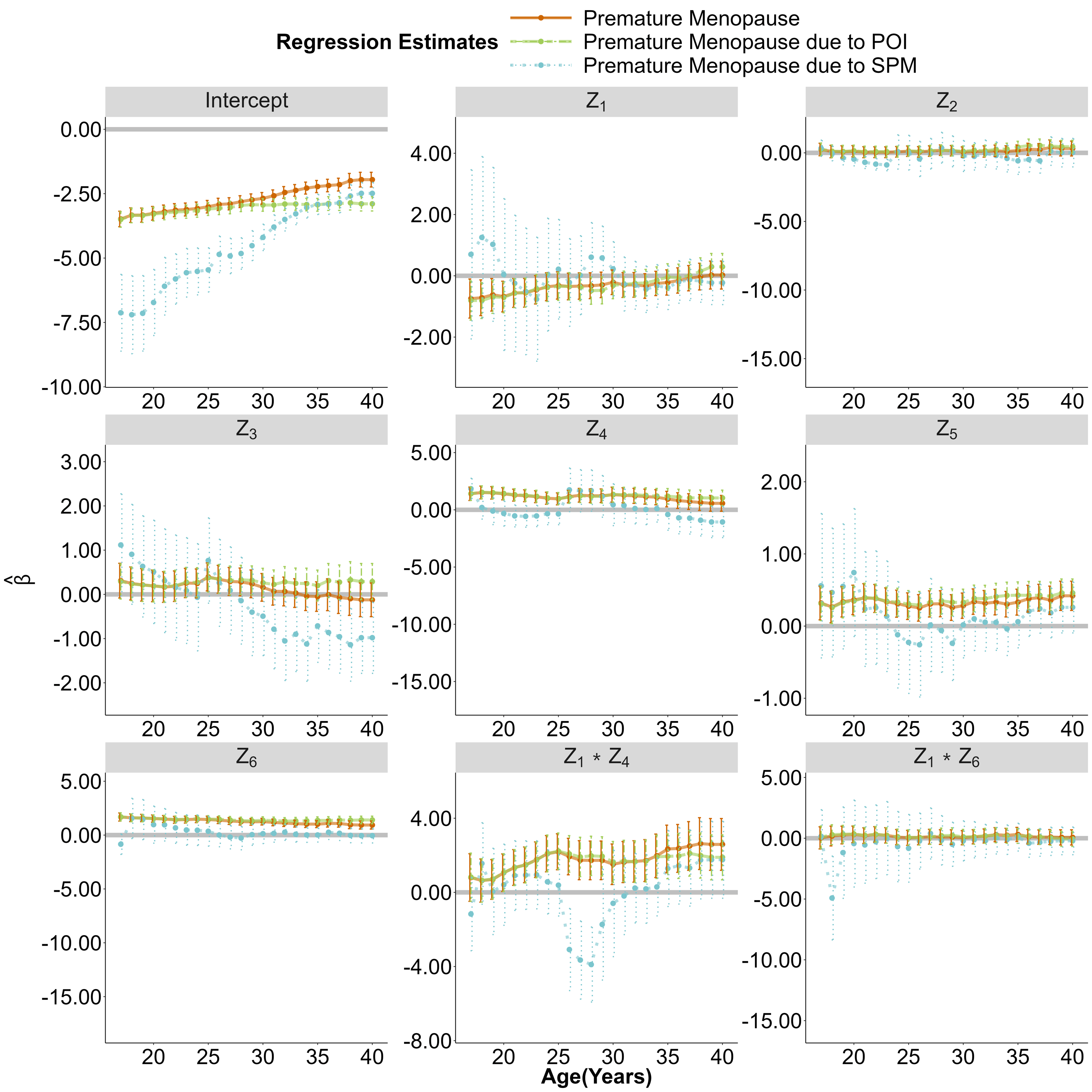}
        \caption{Approach B (With Firth Correction)}
        \label{fig:log_ci_appb_firth}
    \end{subfigure}

    \caption{The estimated coefficient with 95\% CI when using logistic regression with and without Firth correction among Approaches~A and~B. The $\hat{G}(\cdot \mid \mathbf{Z})$ is obtained using SRF. The number of trees and node size are set to 100.  $Z_{1}\text{: Rescaled age at cancer diagnosis}$,
             $Z_{2}\text{: Race-African American}$,
             $Z_{3}\text{: Race-Other}$,
             $Z_{4}\text{: Receipt of BMT}$,
             $Z_{5}\text{: Receipt of Alkylating agents}$
             $Z_{6}\text{: Receipt of radiation to the abdomen/pelvis/total body}$}
    \label{fig: log_real_data_ci_srf}
\end{figure}
\subsubsection{Without Firth Correction While Using Different $\hat{G}(\cdot \mid \mathbf{Z})$}
\begin{figure}[H]
    \centering
    \begin{subfigure}[t]{0.6\textwidth}
        \centering
        \includegraphics[width=1\textwidth]{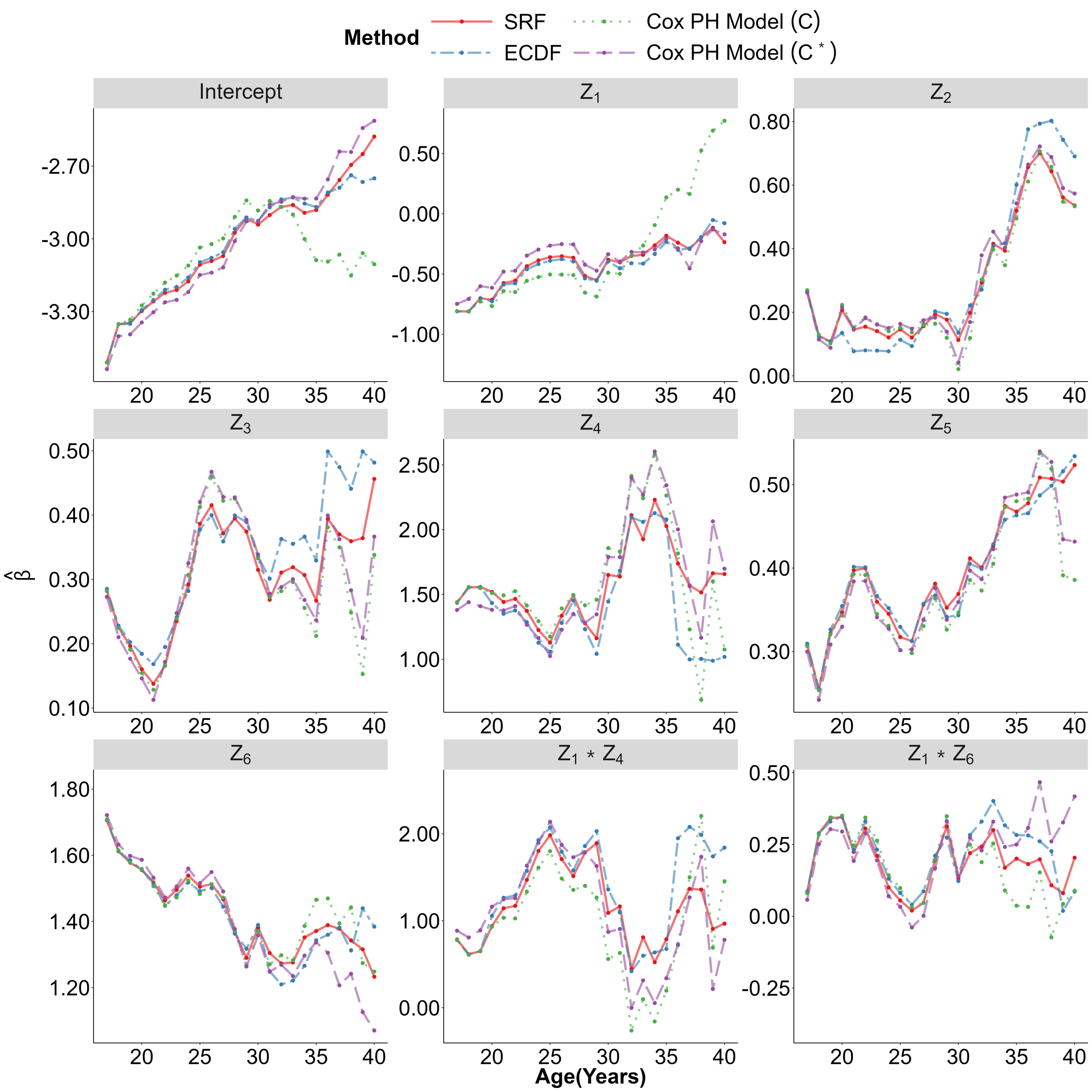}
        \caption{Estimated Coefficients using Approach A}
        \label{fig:log_nofirth_poi_coef_appa}
    \end{subfigure}%
    \vfill
    \begin{subfigure}[t]{0.6\textwidth}
        \centering
        \includegraphics[width=1\textwidth]{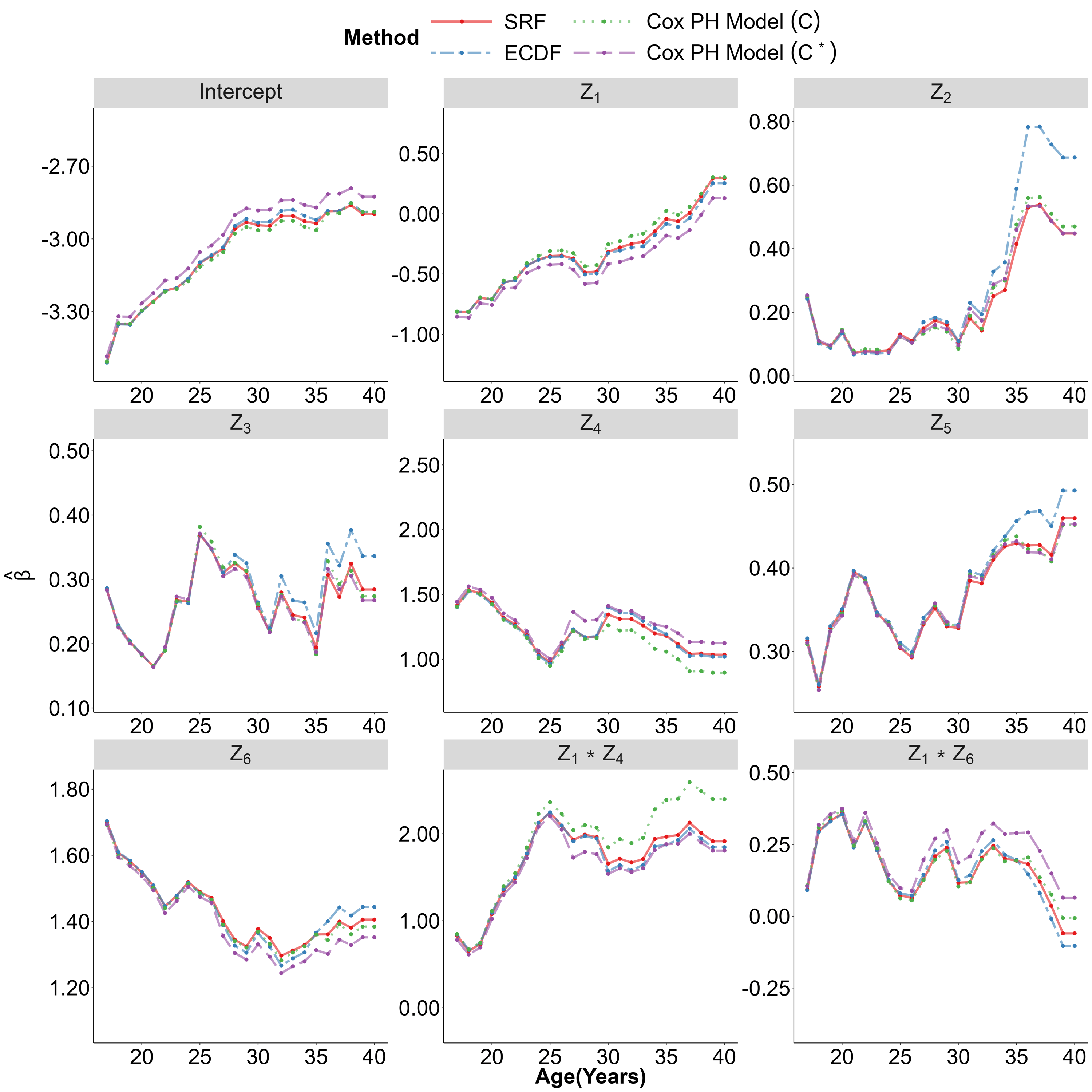}
        \caption{Estimated Coefficients using Approach B}
        \label{fig:log_nofirth_poi_coef_appb}
    \end{subfigure}%
    \caption{Logistic regression without Firth correction (outcome: POI). The estimated coefficients using Approaches~A and~B with different methods to obtain $\hat{G}(\cdot \mid \mathbf{Z})$. The number of trees and the node size in SRF are set to 100.
             $Z_{1}\text{: Rescaled age at cancer diagnosis}$,
             $Z_{2}\text{: Race-African American}$,
             $Z_{3}\text{: Race-Other}$,
             $Z_{4}\text{: Receipt of BMT}$,
             $Z_{5}\text{: Receipt of Alkylating agents}$,
             $Z_{6}\text{: Receipt of radiation to the abdomen/pelvis/total body}$.}
    \label{fig:log_nofirth_poi_coef}
\end{figure}

\begin{figure}[H]
    \centering
    \begin{subfigure}[b]{0.6\textwidth}
        \centering
        \includegraphics[width=1\textwidth]{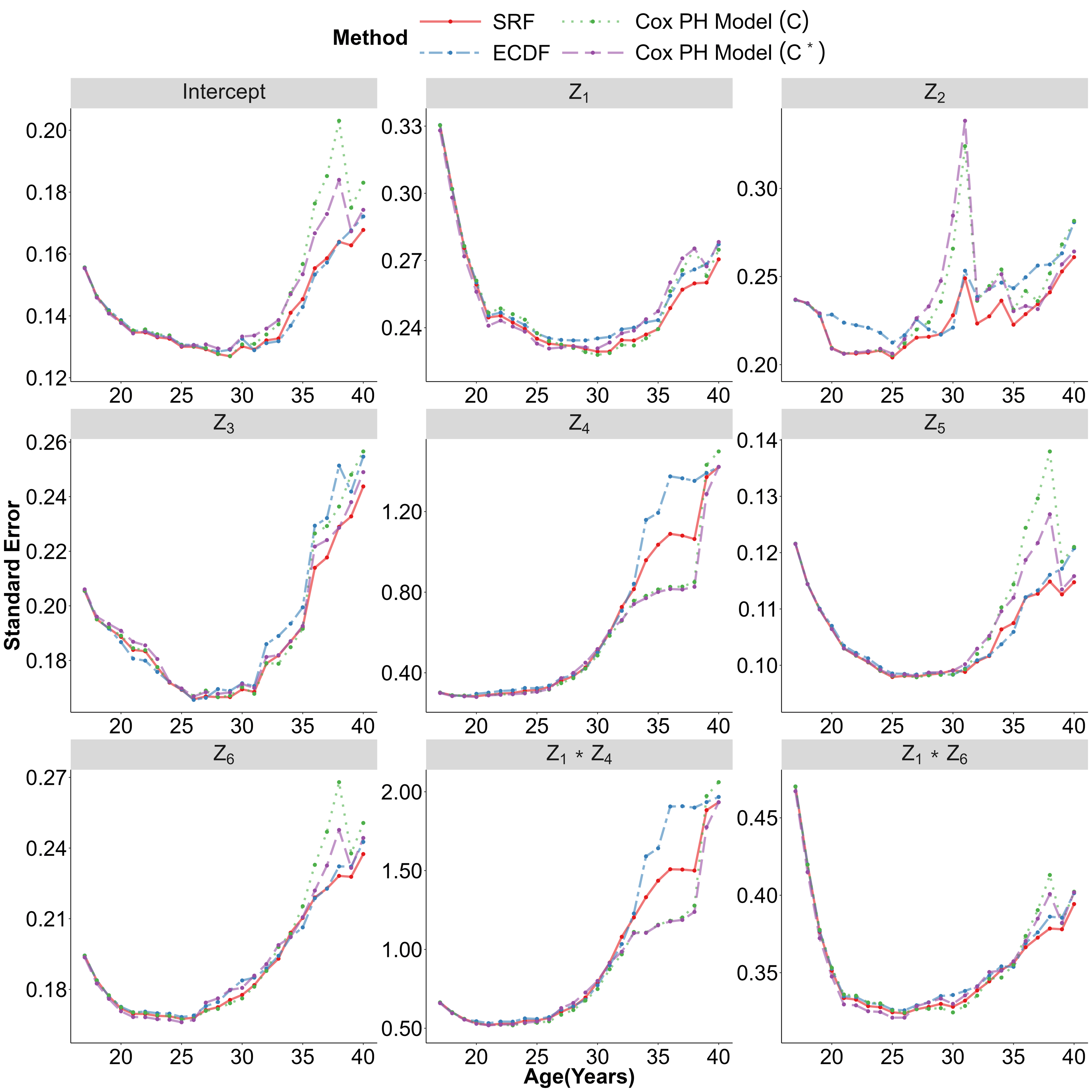}
        \caption{Standard Error Estimates using Approach A}
        \label{fig:log_nofirth_poi_se_appa}
    \end{subfigure}
    \vfill
    \begin{subfigure}[b]{0.6\textwidth}
        \centering
        \includegraphics[width=1\textwidth]{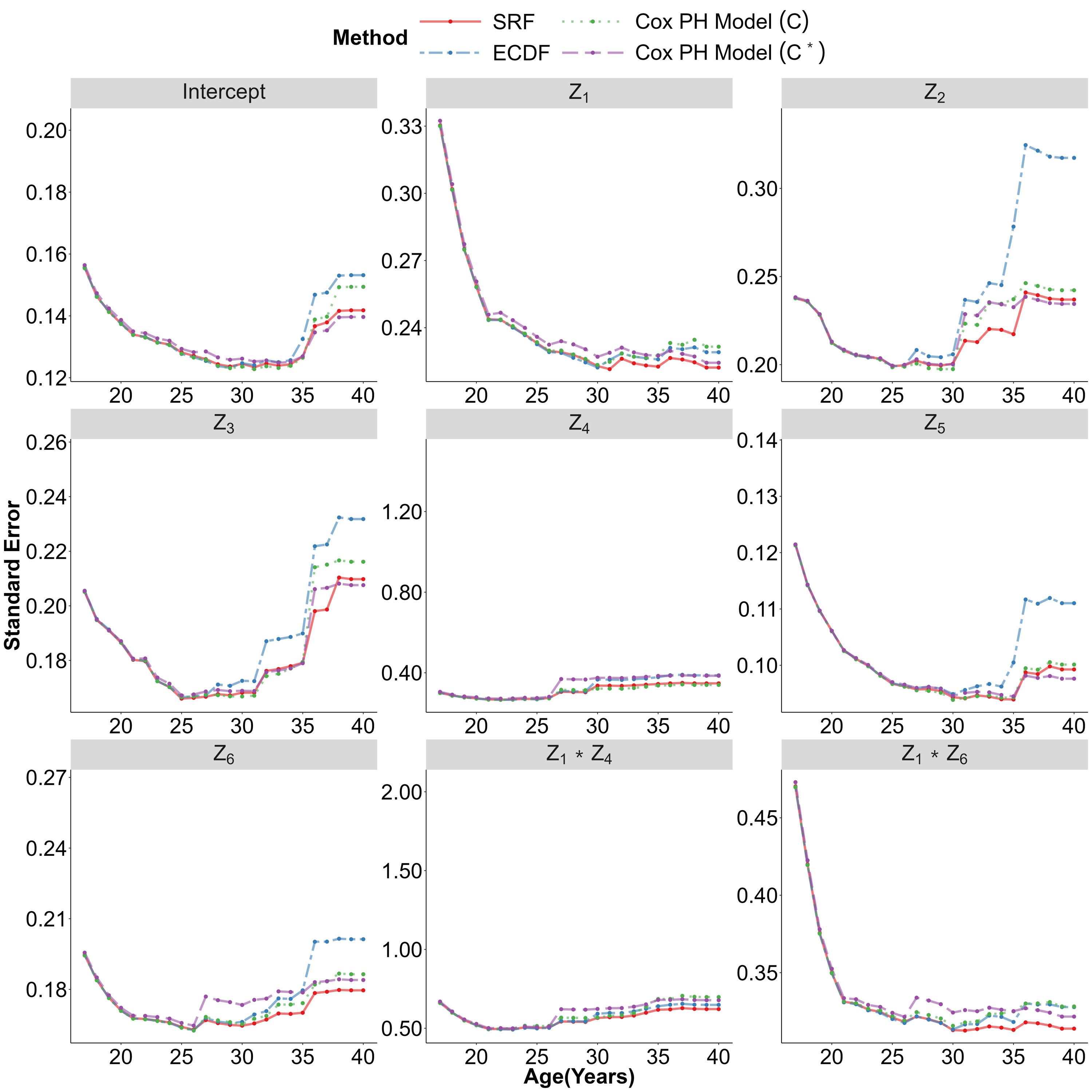}
        \caption{Standard Error Estimates using Approach B}
        \label{fig:log_nofirth_poi_se_appb}
    \end{subfigure}
    \caption{Logistic regression without Firth correction (outcome: POI). The standard error of estimated coefficients using Approaches~A and~B with different methods to obtain $\hat{G}(\cdot \mid \mathbf{Z})$. The number of trees and the node size in SRF are set to 100.
             $Z_{1}\text{: Rescaled age at cancer diagnosis}$,
             $Z_{2}\text{: Race-African American}$,
             $Z_{3}\text{: Race-Other}$,
             $Z_{4}\text{: Receipt of BMT}$,
             $Z_{5}\text{: Receipt of Alkylating agents}$,
             $Z_{6}\text{: Receipt of radiation to the abdomen/pelvis/total body}$.}
    \label{fig:log_nofirth_poi_se}
\end{figure}


\begin{figure}[H]
    \centering
    \begin{subfigure}[t]{0.6\textwidth}
        \centering
        \includegraphics[width=1\textwidth]{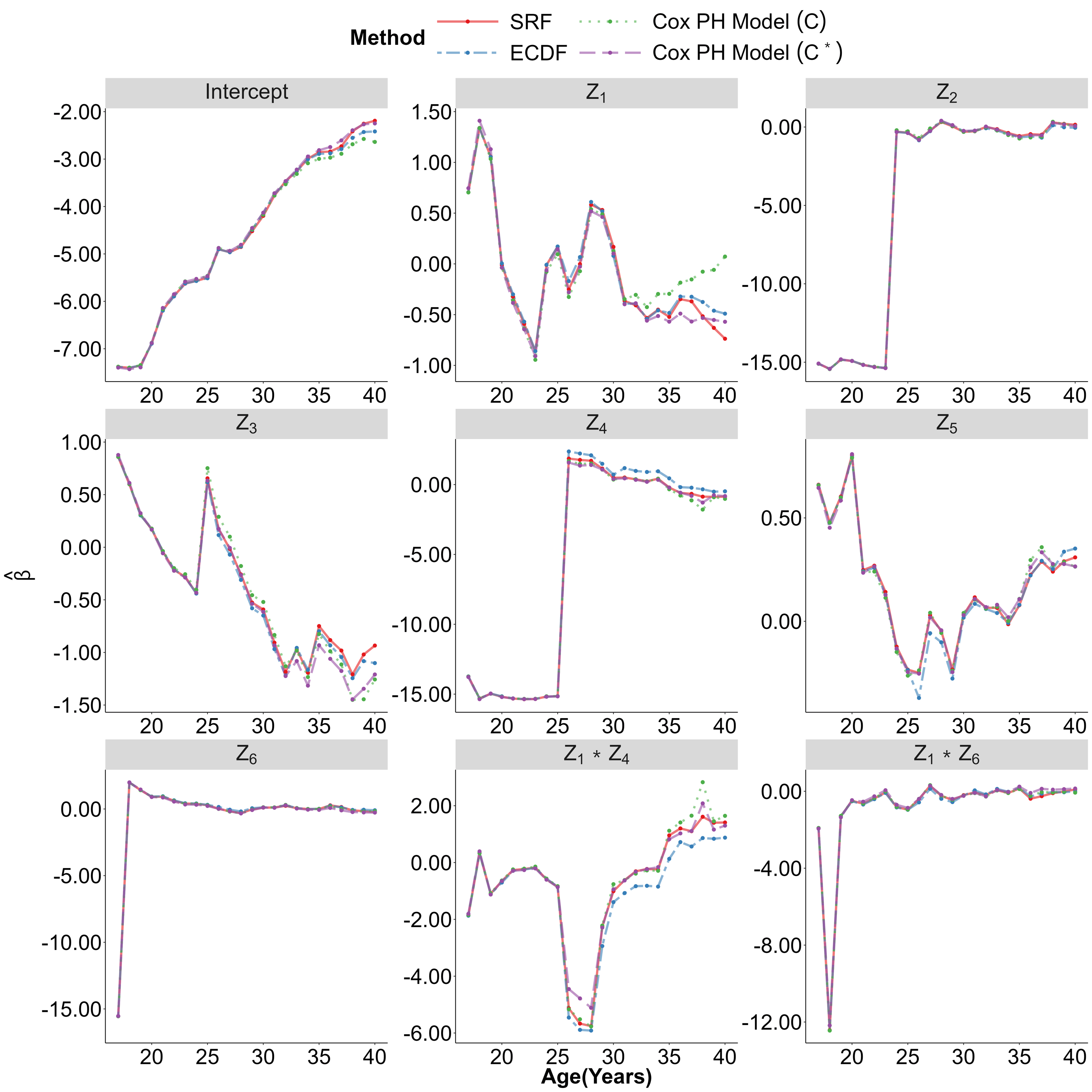}
        \caption{Estimated Coefficients using Approach A}
        \label{fig:log_nofirth_spm_coef_appa}
    \end{subfigure}%
    \vfill
    \begin{subfigure}[t]{0.6\textwidth}
        \centering
        \includegraphics[width=1\textwidth]{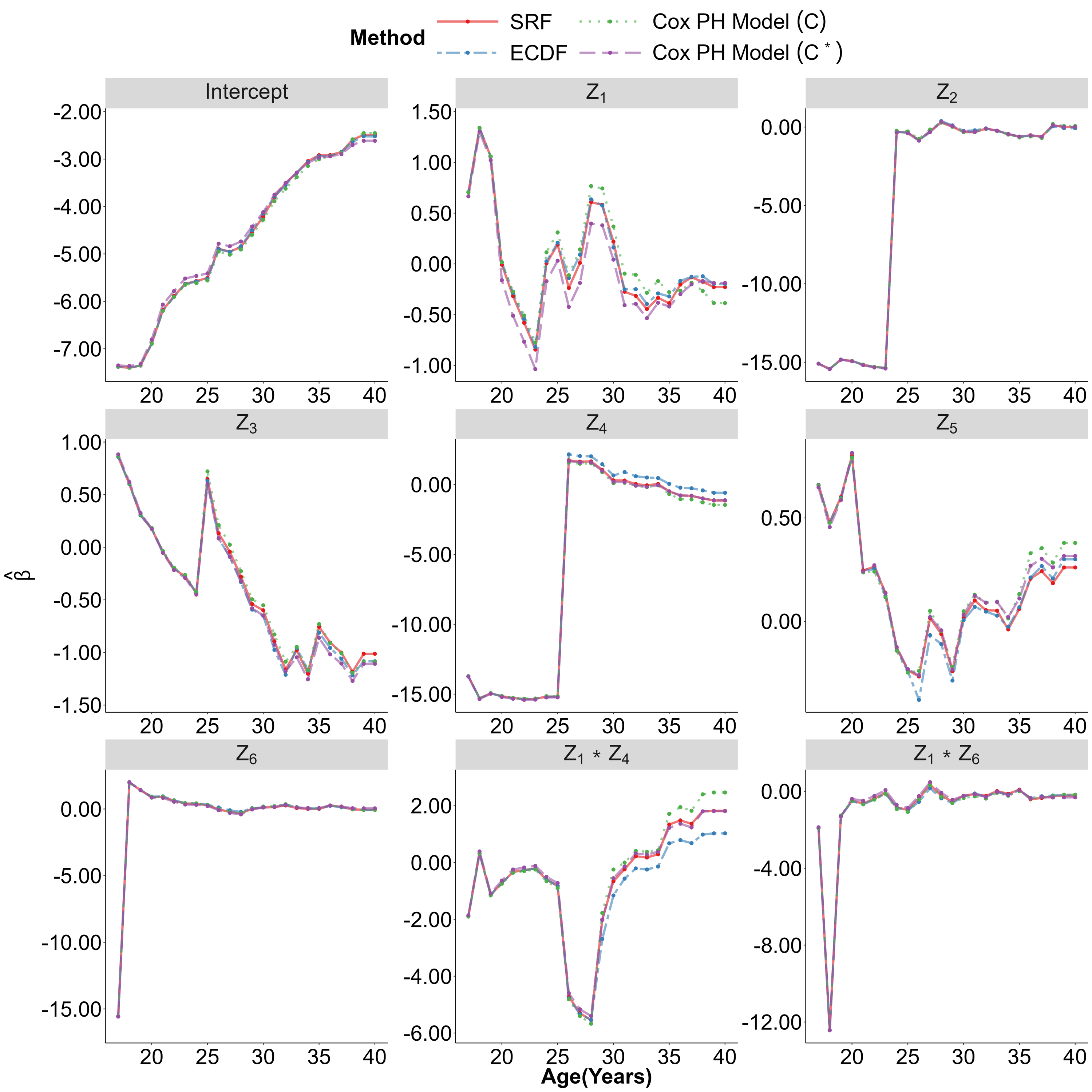}
        \caption{Estimated Coefficients using Approach B}
        \label{fig:log_nofirth_spm_coef_appb}
    \end{subfigure}%
    \caption{Logistic regression without Firth correction (outcome: SPM). The estimated coefficients using Approaches~A and~B with different methods to obtain $\hat{G}(\cdot \mid \mathbf{Z})$. The number of trees and the node size in SRF are set to 100.
             $Z_{1}\text{: Rescaled age at cancer diagnosis}$,
             $Z_{2}\text{: Race-African American}$,
             $Z_{3}\text{: Race-Other}$,
             $Z_{4}\text{: Receipt of BMT}$,
             $Z_{5}\text{: Receipt of Alkylating agents}$,
             $Z_{6}\text{: Receipt of radiation to the abdomen/pelvis/total body}$.}
    \label{fig:log_nofirth_spm_coef}
\end{figure}

\begin{figure}[H]
    \centering
    \begin{subfigure}[b]{0.6\textwidth}
        \centering
        \includegraphics[width=1\textwidth]{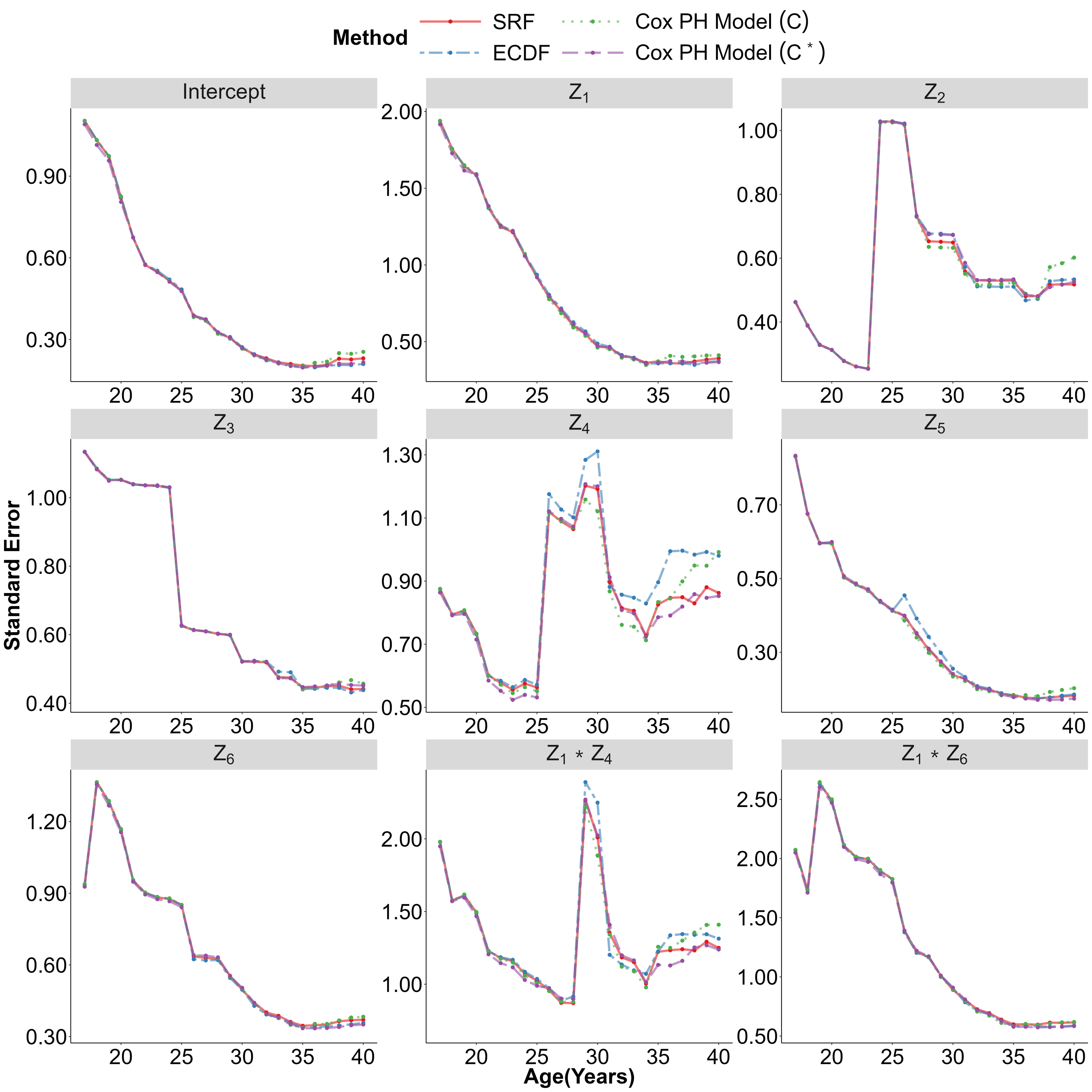}
        \caption{Standard Error Estimates using Approach A}
        \label{fig:log_nofirth_spm_se_appa}
    \end{subfigure}
    \vfill
    \begin{subfigure}[b]{0.6\textwidth}
        \centering
        \includegraphics[width=1\textwidth]{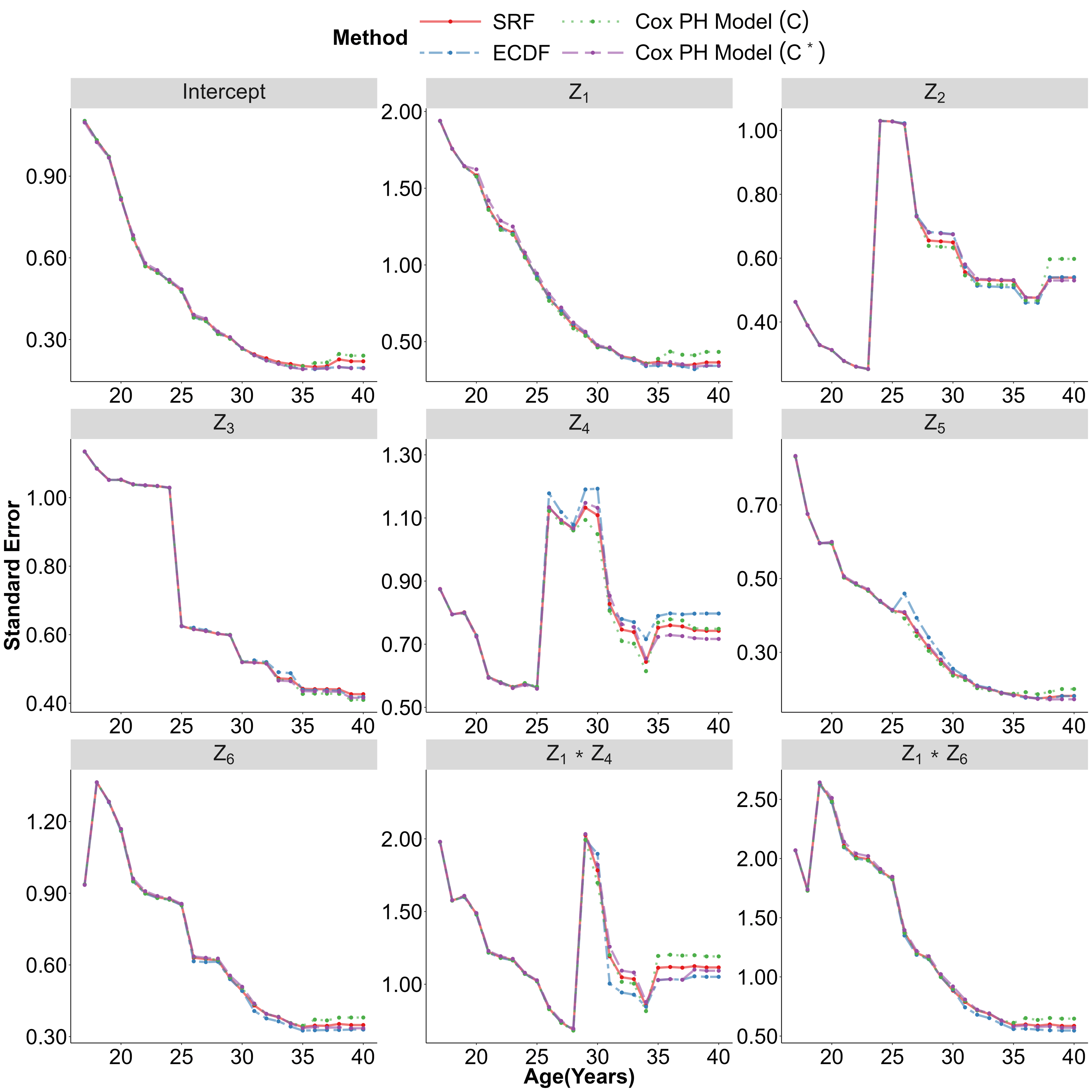}
        \caption{Standard Error Estimates using Approach B}
        \label{fig:log_nofirth_spm_se_appb}
    \end{subfigure}
    \caption{Logistic regression without Firth correction (outcome: SPM). The standard error of estimated coefficients using Approaches~A and~B with different methods to obtain $\hat{G}(\cdot \mid \mathbf{Z})$. The number of trees and the node size in SRF are set to 100.
             $Z_{1}\text{: Rescaled age at cancer diagnosis}$,
             $Z_{2}\text{: Race-African American}$,
             $Z_{3}\text{: Race-Other}$,
             $Z_{4}\text{: Receipt of BMT}$,
             $Z_{5}\text{: Receipt of Alkylating agents}$,
             $Z_{6}\text{: Receipt of radiation to the abdomen/pelvis/total body}$.}
    \label{fig:log_nofirth_spm_se}
\end{figure}

\subsubsection{With Firth Correction While Using Different $\hat{G}(\cdot \mid \mathbf{Z})$}
\begin{figure}[H]
    \centering
    \begin{subfigure}[t]{0.6\textwidth}
        \centering
        \includegraphics[width=1\textwidth]{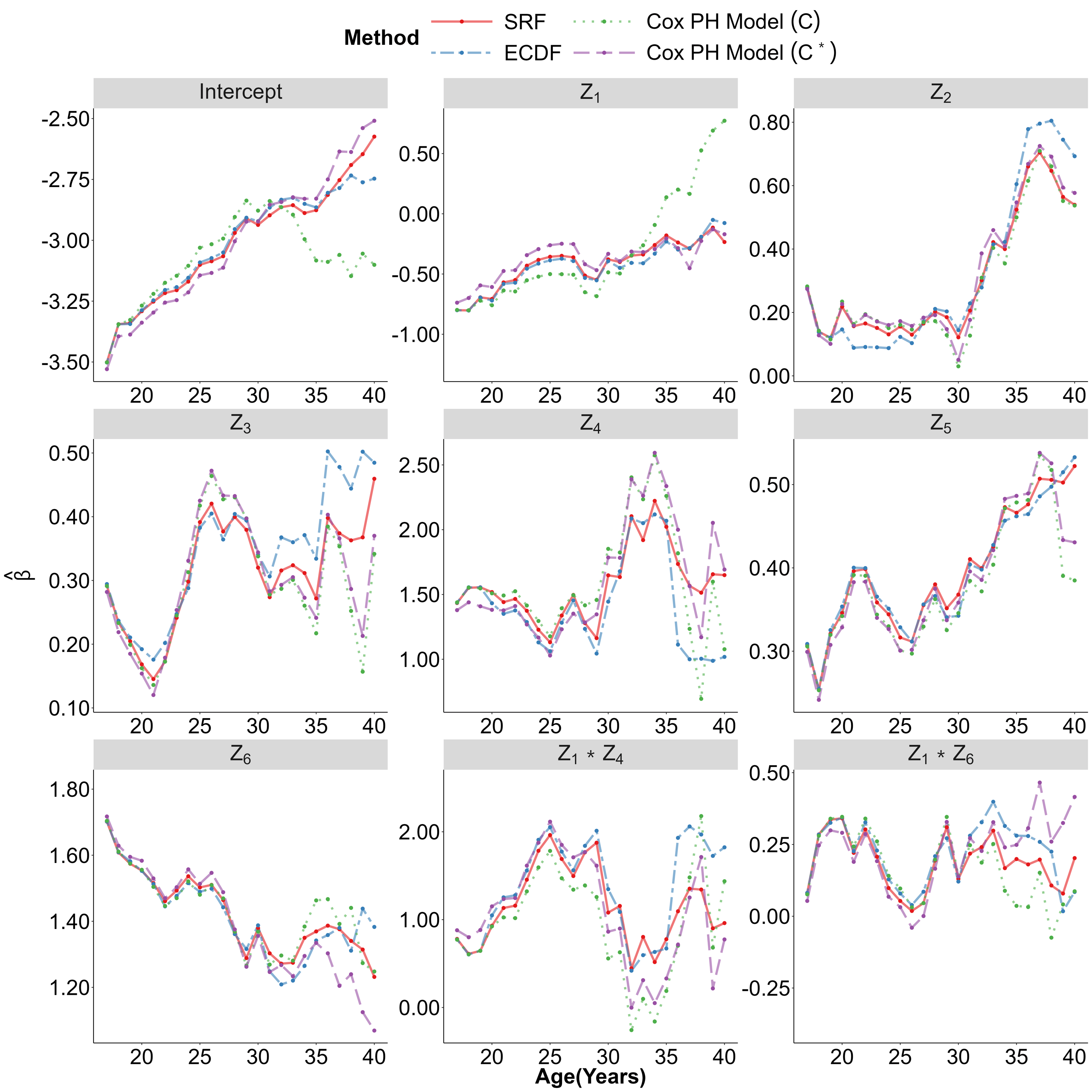}
        \caption{Estimated Coefficients using Approach A}
        \label{fig:log_firth_poi_coef_appa}
    \end{subfigure}%
    \vfill
    \begin{subfigure}[t]{0.6\textwidth}
        \centering
        \includegraphics[width=1\textwidth]{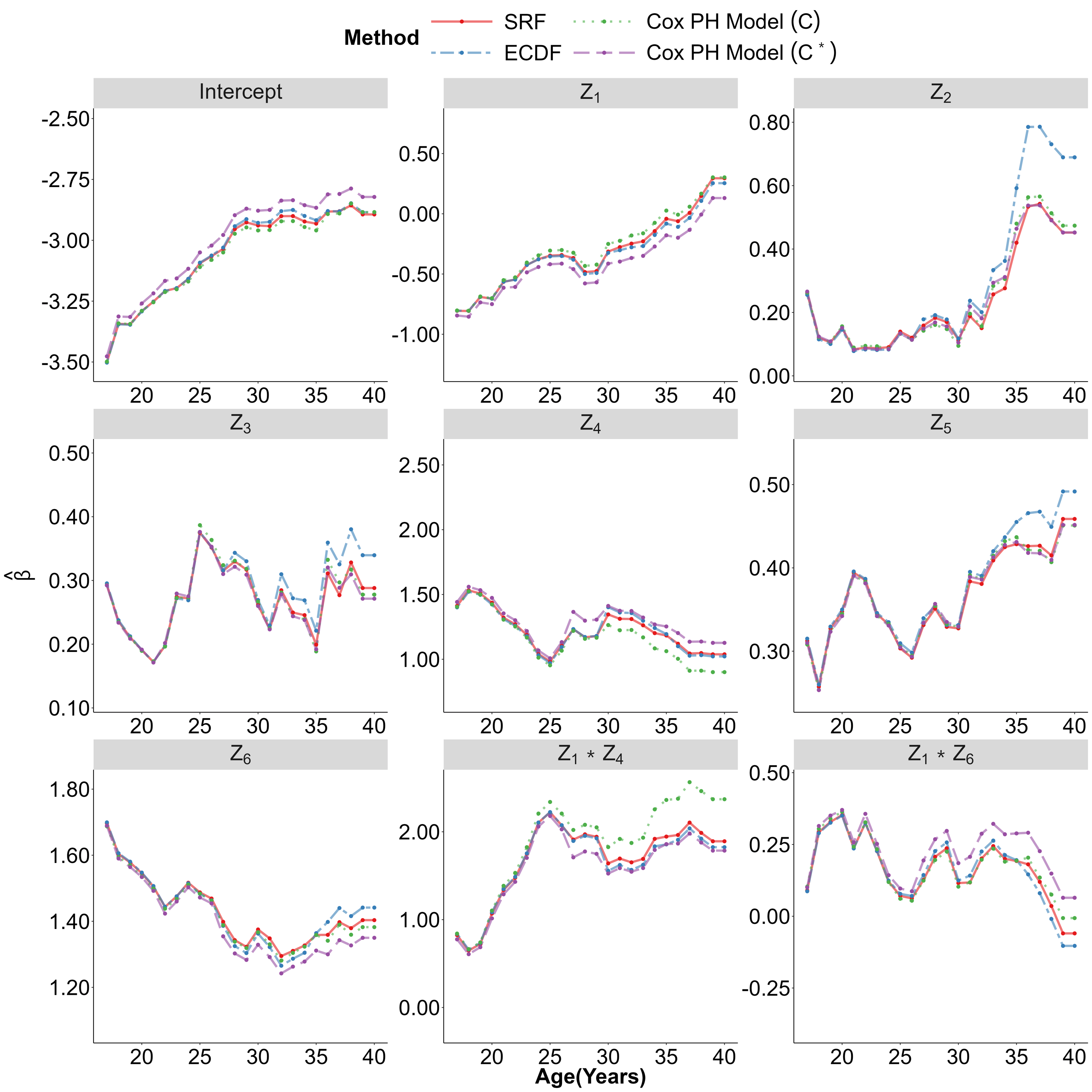}
        \caption{Estimated Coefficients using Approach B}
        \label{fig:log_firth_poi_coef_appb}
    \end{subfigure}%
    \caption{Logistic regression with Firth correction (outcome: POI). The estimated coefficients using Approaches~A and~B with different methods to obtain $\hat{G}(\cdot \mid \mathbf{Z})$. The number of trees and the node size in SRF are set to 100.
             $Z_{1}\text{: Rescaled age at cancer diagnosis}$,
             $Z_{2}\text{: Race-African American}$,
             $Z_{3}\text{: Race-Other}$,
             $Z_{4}\text{: Receipt of BMT}$,
             $Z_{5}\text{: Receipt of Alkylating agents}$,
             $Z_{6}\text{: Receipt of radiation to the abdomen/pelvis/total body}$.}
    \label{fig:log_firth_poi_coef}
\end{figure}

\begin{figure}[H]
    \centering
    \begin{subfigure}[b]{0.6\textwidth}
        \centering
        \includegraphics[width=1\textwidth]{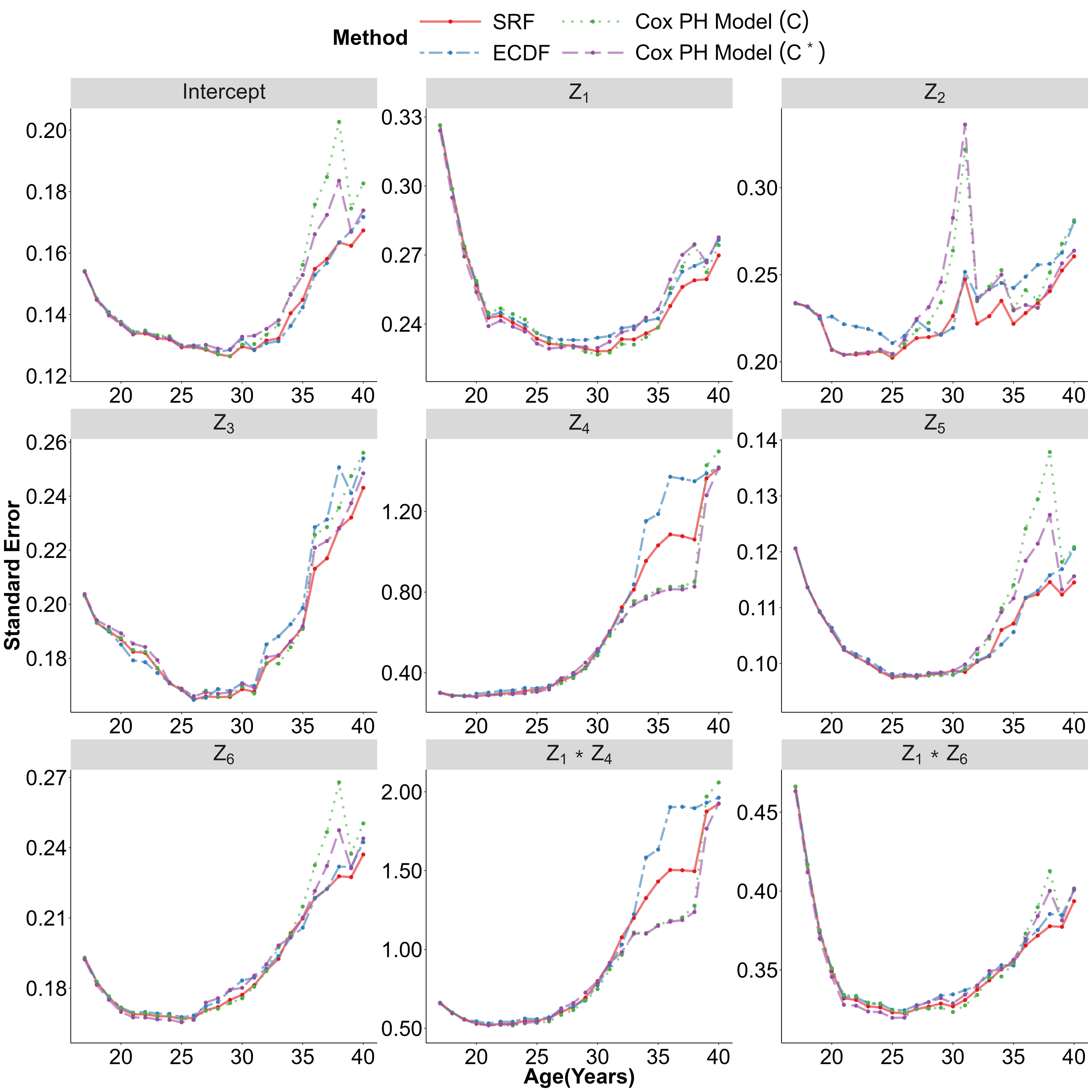}
        \caption{Standard Error Estimates using Approach A}
        \label{fig:log_firth_poi_se_appa}
    \end{subfigure}
    \vfill
    \begin{subfigure}[b]{0.6\textwidth}
        \centering
        \includegraphics[width=1\textwidth]{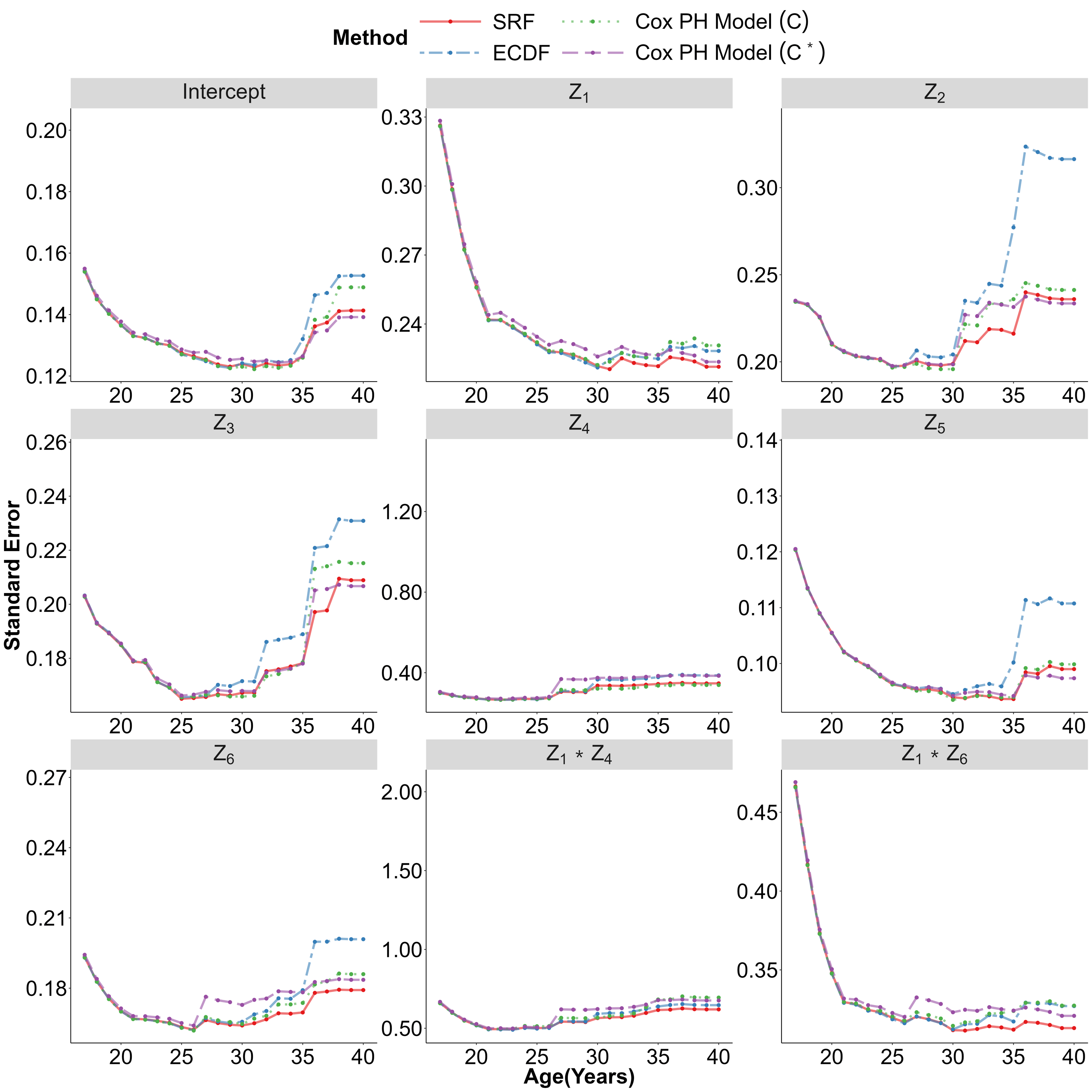}
        \caption{Standard Error Estimates using Approach B}
        \label{fig:log_firth_poi_se_appb}
    \end{subfigure}
    \caption{Logistic regression with Firth correction (outcome: POI). The standard error of estimated coefficients using Approaches~A and~B with different methods to obtain $\hat{G}(\cdot \mid \mathbf{Z})$. The number of trees and the node size in SRF are set to 100.
             $Z_{1}\text{: Rescaled age at cancer diagnosis}$,
             $Z_{2}\text{: Race-African American}$,
             $Z_{3}\text{: Race-Other}$,
             $Z_{4}\text{: Receipt of BMT}$,
             $Z_{5}\text{: Receipt of Alkylating agents}$,
             $Z_{6}\text{: Receipt of radiation to the abdomen/pelvis/total body}$.}
    \label{fig:log_firth_poi_se}
\end{figure}


\begin{figure}[H]
    \centering
    \begin{subfigure}[t]{0.6\textwidth}
        \centering
        \includegraphics[width=1\textwidth]{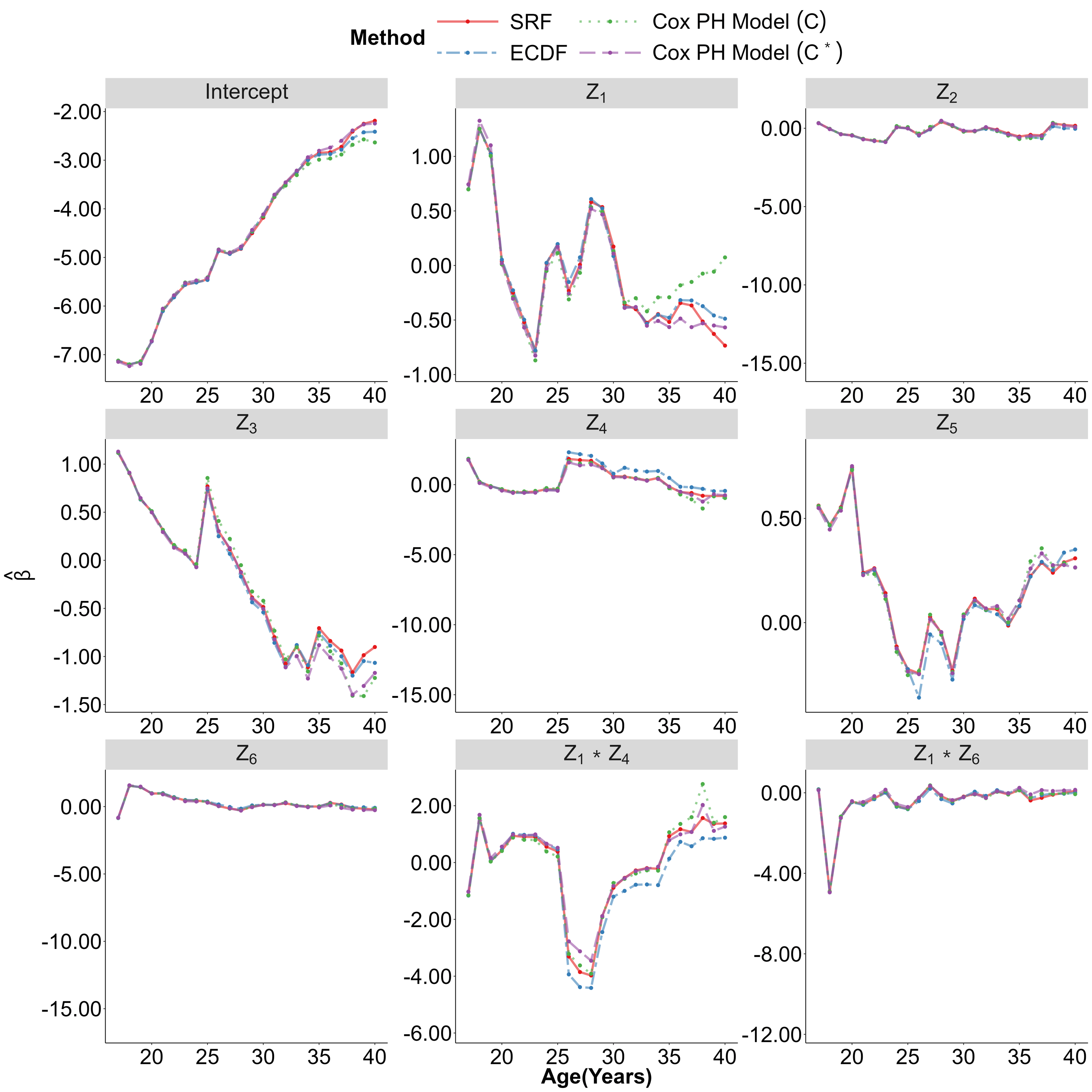}
        \caption{Estimated Coefficients using Approach A}
        \label{fig:log_firth_spm_coef_appa}
    \end{subfigure}%
    \vfill
    \begin{subfigure}[t]{0.6\textwidth}
        \centering
        \includegraphics[width=1\textwidth]{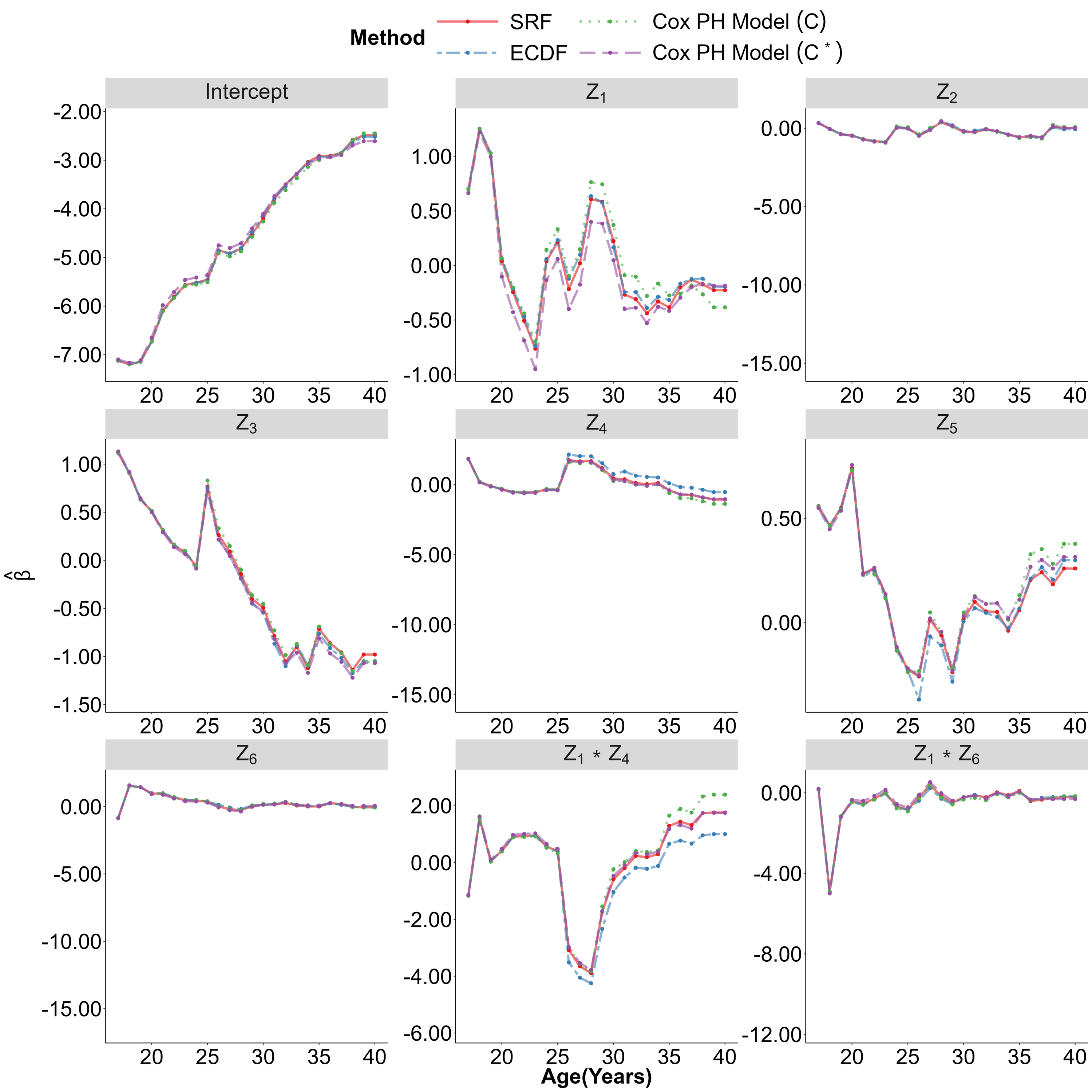}
        \caption{Estimated Coefficients using Approach B}
        \label{fig:log_firth_spm_coef_appb}
    \end{subfigure}%
    \caption{Logistic regression with Firth correction (outcome: SPM). The estimated coefficients using Approaches~A and~B with different methods to obtain $\hat{G}(\cdot \mid \mathbf{Z})$. The number of trees and the node size in SRF are set to 100.
             $Z_{1}\text{: Rescaled age at cancer diagnosis}$,
             $Z_{2}\text{: Race-African American}$,
             $Z_{3}\text{: Race-Other}$,
             $Z_{4}\text{: Receipt of BMT}$,
             $Z_{5}\text{: Receipt of Alkylating agents}$,
             $Z_{6}\text{: Receipt of radiation to the abdomen/pelvis/total body}$.}
    \label{fig:log_firth_spm_coef}
\end{figure}

\begin{figure}[H]
    \centering
    \begin{subfigure}[b]{0.6\textwidth}
        \centering
        \includegraphics[width=1\textwidth]{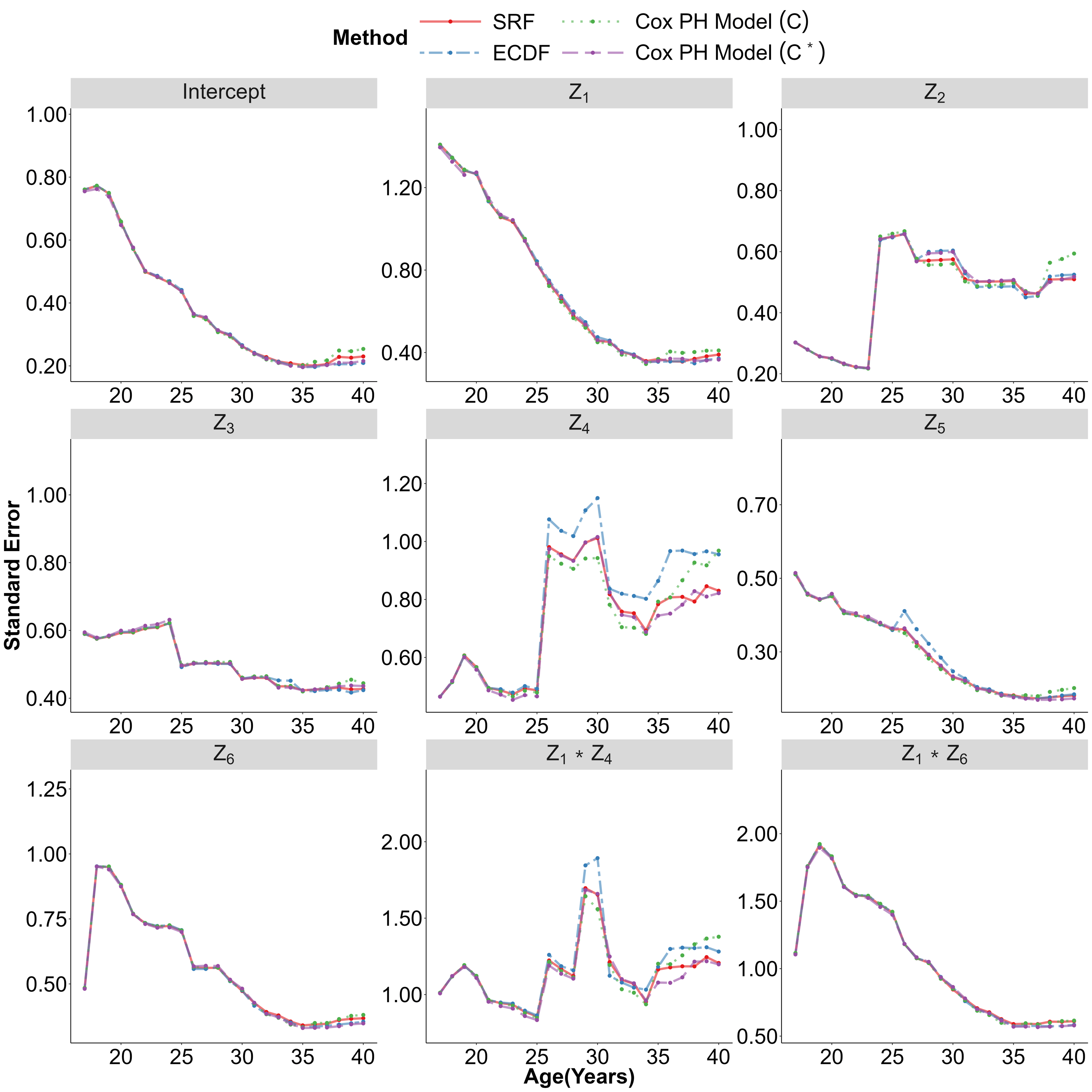}
        \caption{Standard Error Estimates using Approach A}
        \label{fig:log_firth_spm_se_appa}
    \end{subfigure}
    \vfill
    \begin{subfigure}[b]{0.6\textwidth}
        \centering
        \includegraphics[width=1\textwidth]{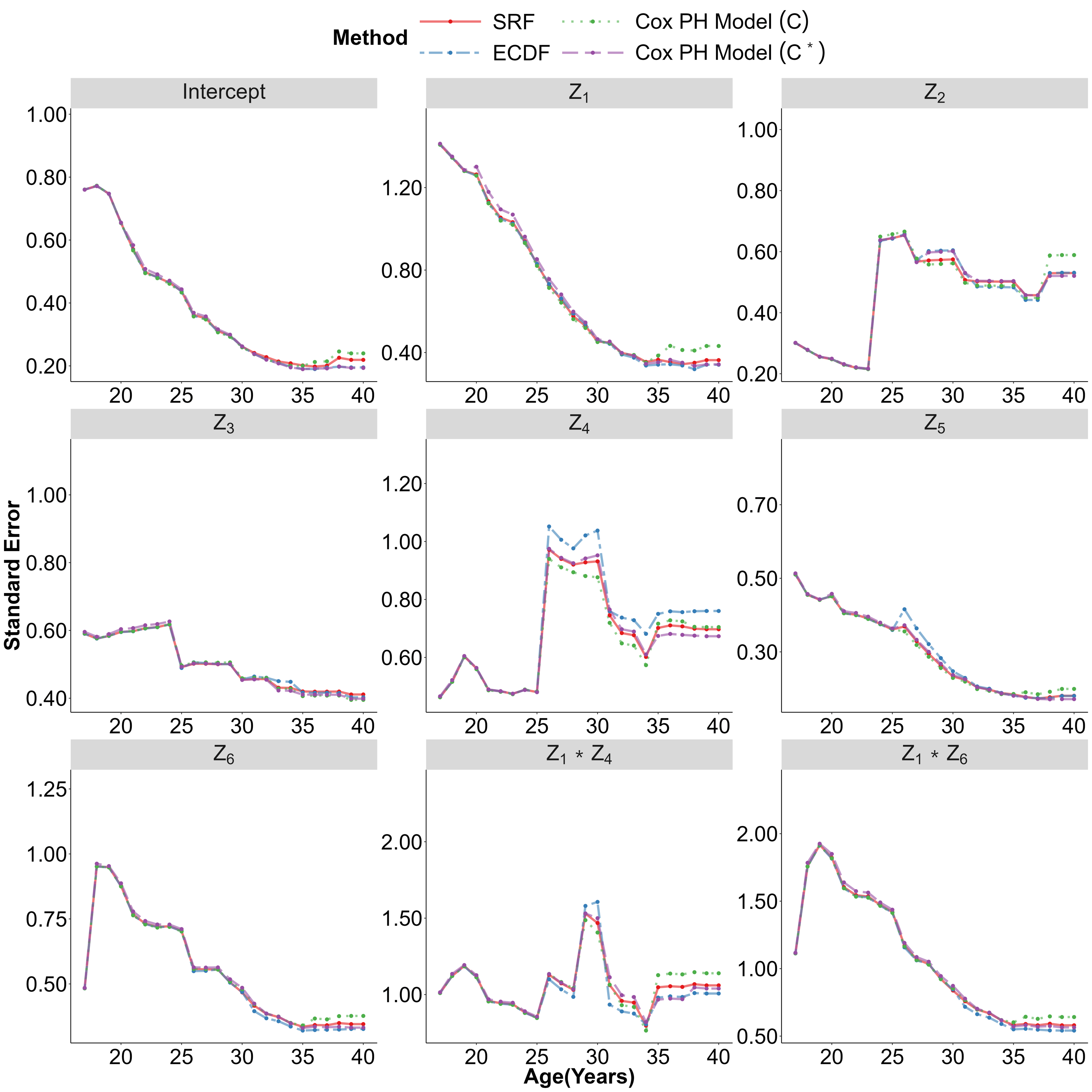}
        \caption{Standard Error Estimates using Approach B}
        \label{fig:log_firth_spm_se_appb}
    \end{subfigure}
    \caption{Logistic regression with Firth correction (outcome: SPM). The standard error of estimated coefficients using Approaches~A and~B with different methods to obtain $\hat{G}(\cdot \mid \mathbf{Z})$. The number of trees and the node size in SRF are set to 100.
             $Z_{1}\text{: Rescaled age at cancer diagnosis}$,
             $Z_{2}\text{: Race-African American}$,
             $Z_{3}\text{: Race-Other}$,
             $Z_{4}\text{: Receipt of BMT}$,
             $Z_{5}\text{: Receipt of Alkylating agents}$,
             $Z_{6}\text{: Receipt of radiation to the abdomen/pelvis/total body}$.}
    \label{fig:log_firth_spm_se}
\end{figure}

%
%
\subsection{Multinomial Logistic Regression}\label{supp: real_data_multi_reg}

\begin{figure}[H]
    \centering
    \begin{subfigure}[b]{0.48\textwidth}
        \centering
        \includegraphics[width=\textwidth]{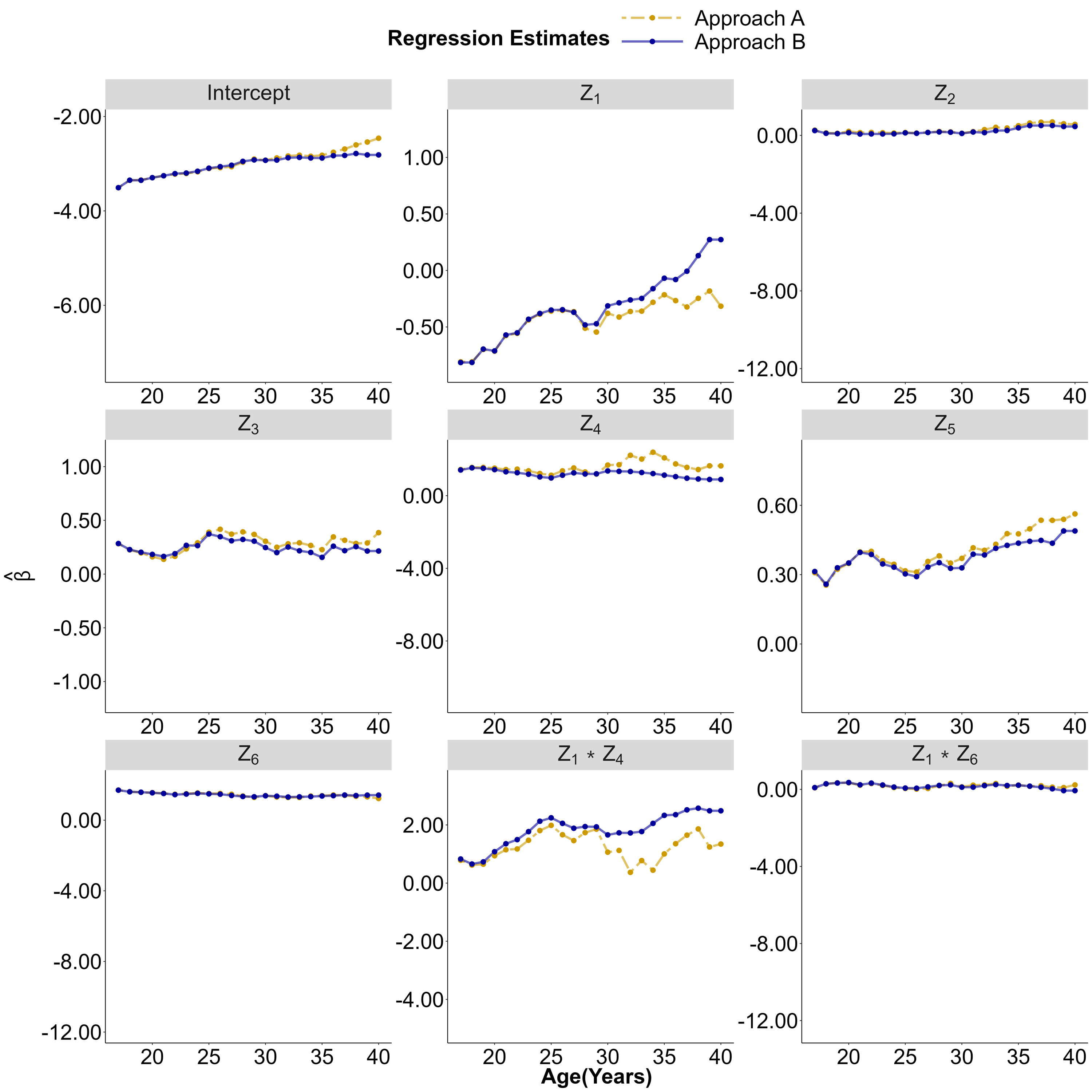}
        \caption{POI (Without Firth Correction)}
        \label{fig:multi_coef_appa_mlr}
    \end{subfigure}
    \hfill
    \begin{subfigure}[b]{0.48\textwidth}
        \centering
        \includegraphics[width=\textwidth]{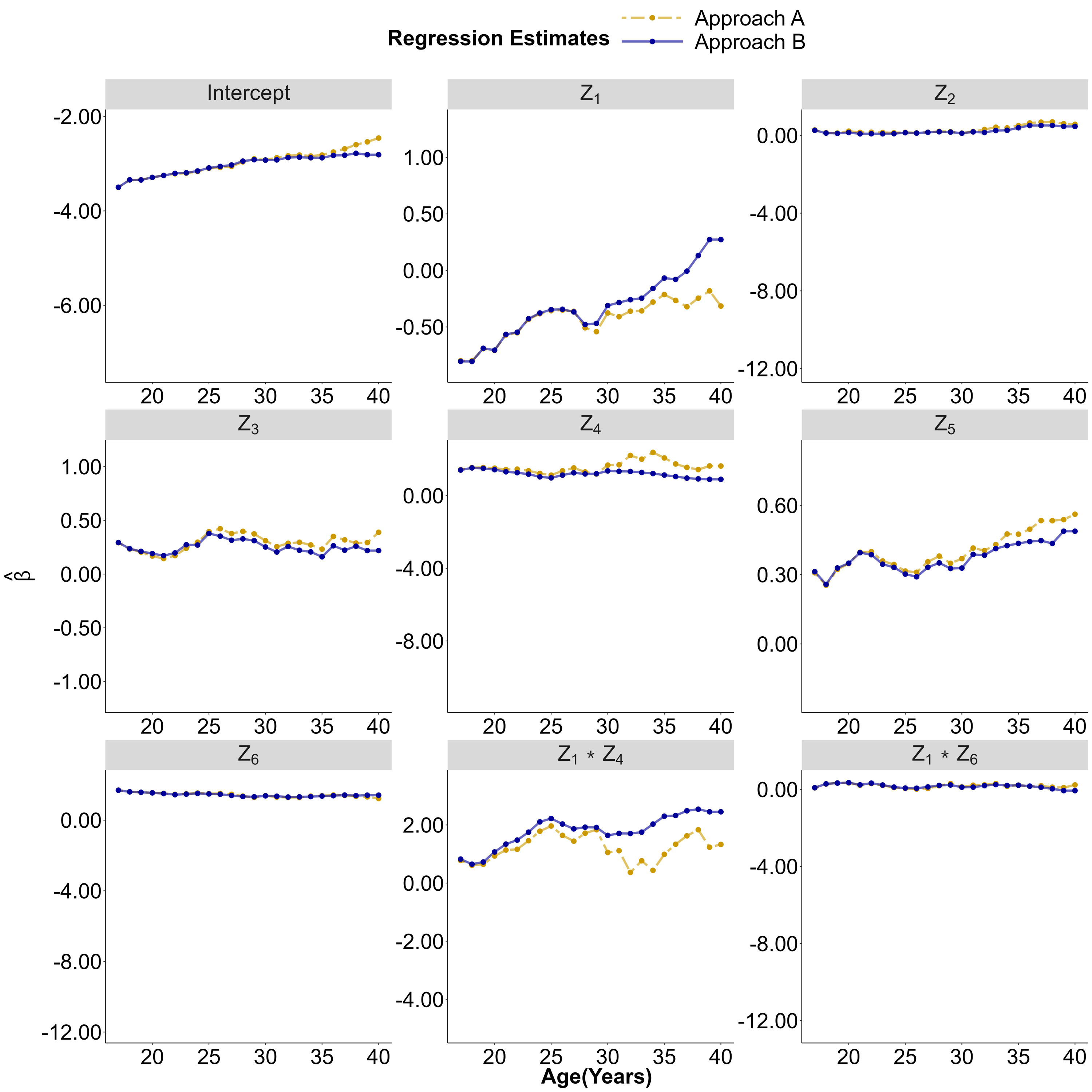}
        \caption{POI (With Firth Correction)}
        \label{fig:multi_coef_appa_firth}
    \end{subfigure}

    \vspace{0.5cm}

    \begin{subfigure}[b]{0.48\textwidth}
        \centering
        \includegraphics[width=\textwidth]{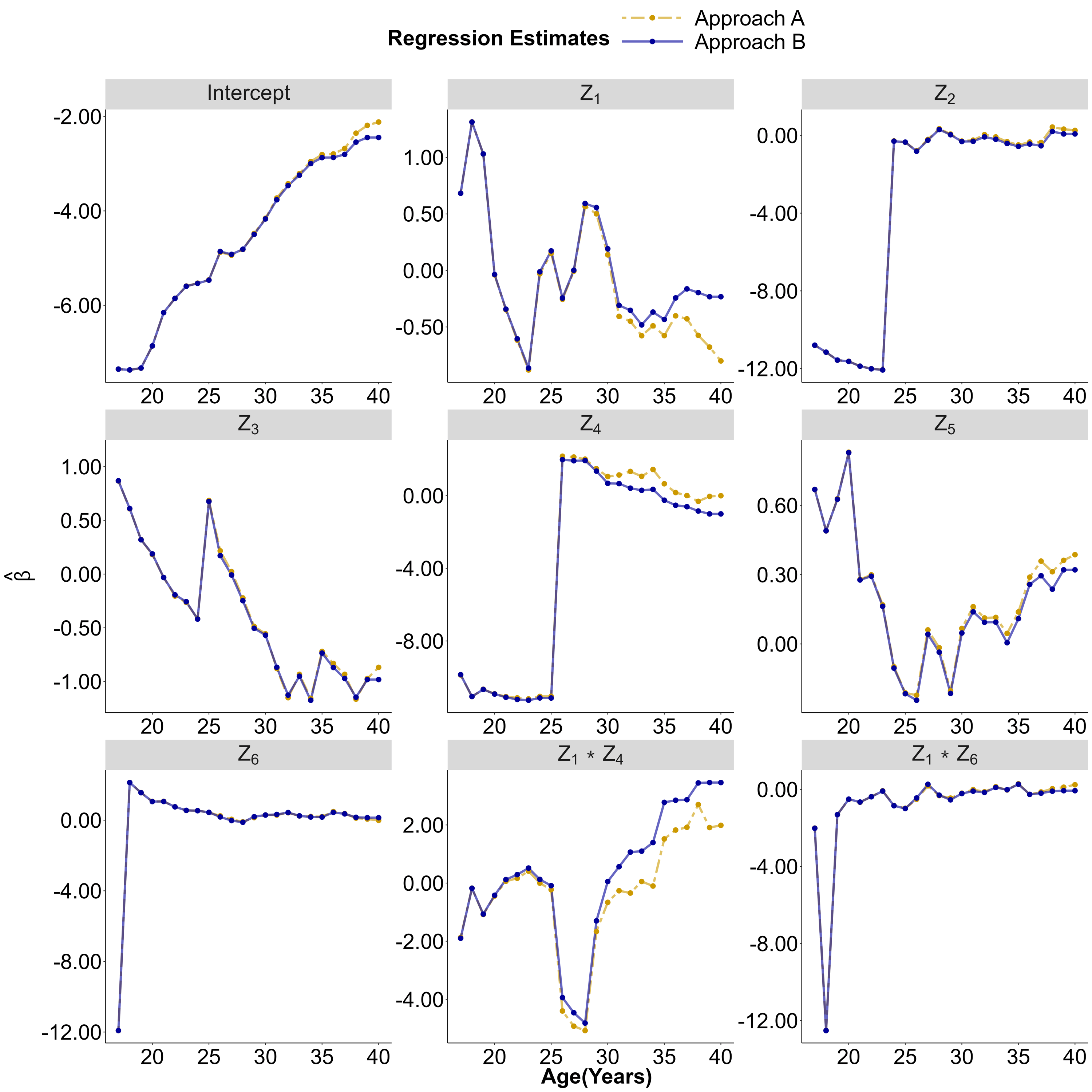}
        \caption{SPM (Without Firth Correction)}
        \label{fig:multi_coef_appb_mlr}
    \end{subfigure}
    \hfill
    \begin{subfigure}[b]{0.48\textwidth}
        \centering
        \includegraphics[width=\textwidth]{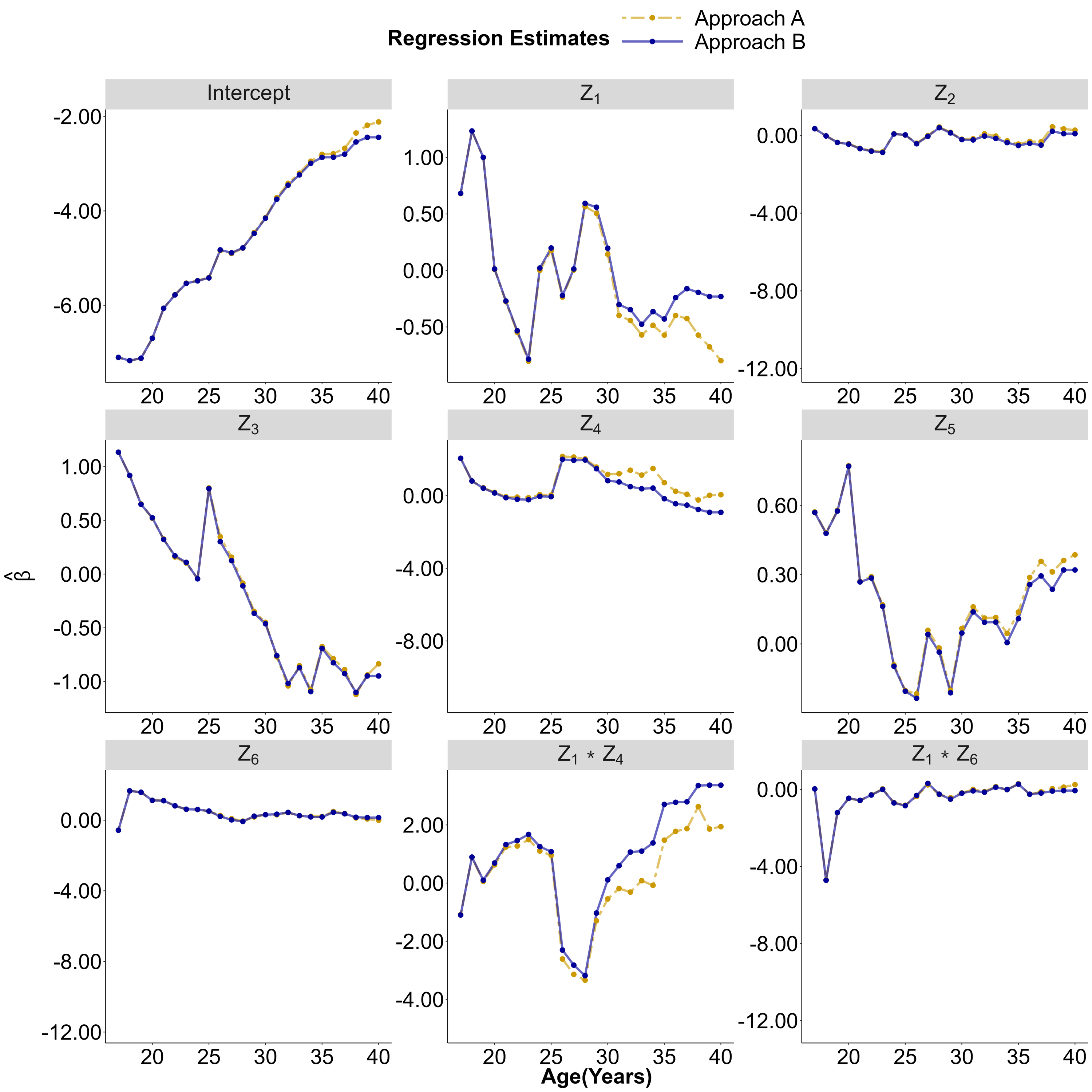}
        \caption{SPM (With Firth Correction)}
        \label{fig:multi_coef_appb_firth}
    \end{subfigure}

    \caption{The estimated coefficients when using multinomial logistic regression with and without Firth correction when outcome is POI or SPM. The $\hat{G}(\cdot \mid \mathbf{Z})$ is obtained using SRF. The number of trees and node size are set to 100.  $Z_{1}\text{: Rescaled age at cancer diagnosis}$,
             $Z_{2}\text{: Race-African American}$,
             $Z_{3}\text{: Race-Other}$,
             $Z_{4}\text{: Receipt of BMT}$,
             $Z_{5}\text{: Receipt of Alkylating agents}$,
             $Z_{6}\text{: Receipt of radiation to the abdomen/pelvis/total body}$}
    \label{fig: multi_real_data_coef_srf}
\end{figure}

\begin{figure}[H]
    \centering
    \begin{subfigure}[b]{0.48\textwidth}
        \centering
        \includegraphics[width=\textwidth]{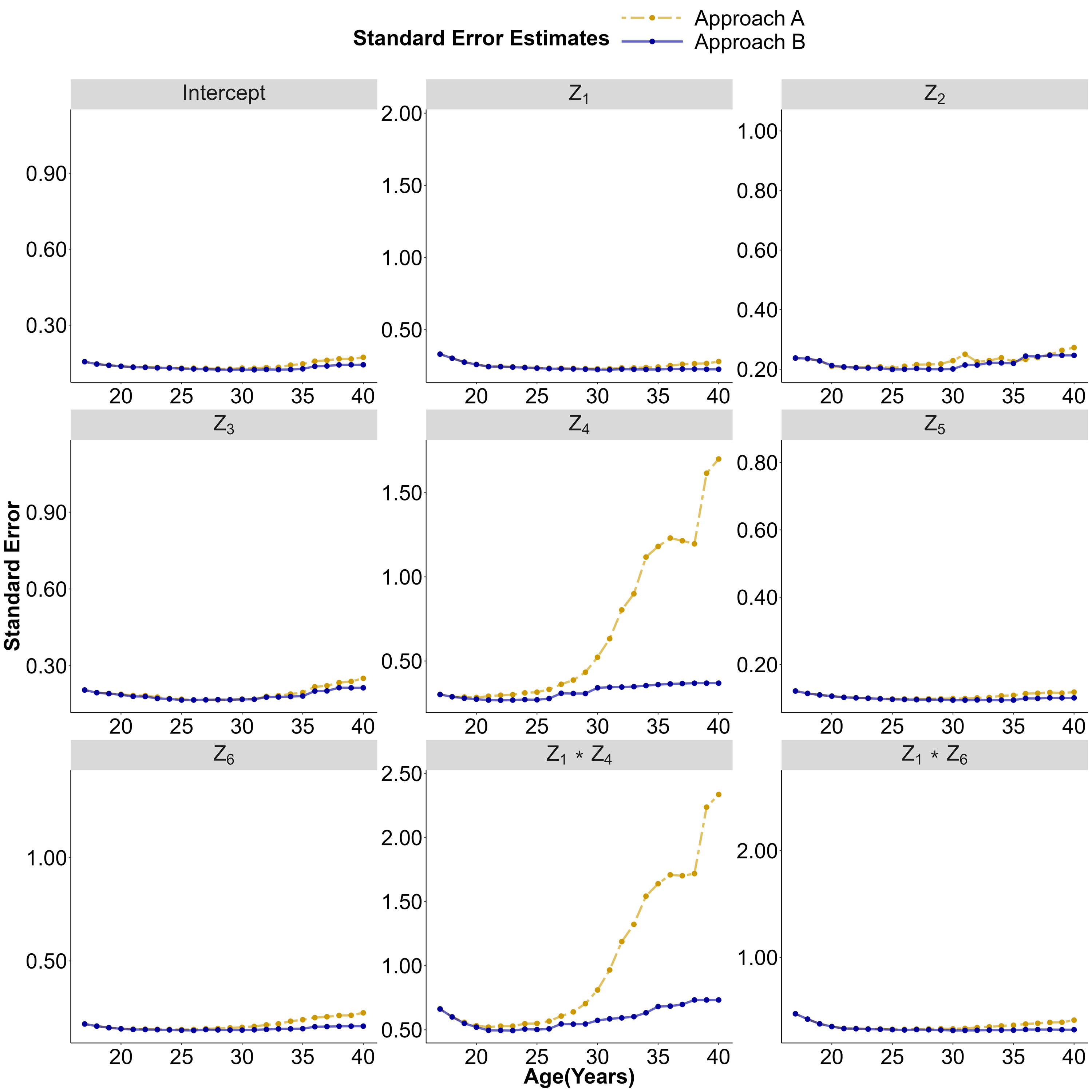}
        \caption{POI (Without Firth Correction)}
        \label{fig:multi_se_appa_mlr}
    \end{subfigure}
    \hfill
    \begin{subfigure}[b]{0.48\textwidth}
        \centering
        \includegraphics[width=\textwidth]{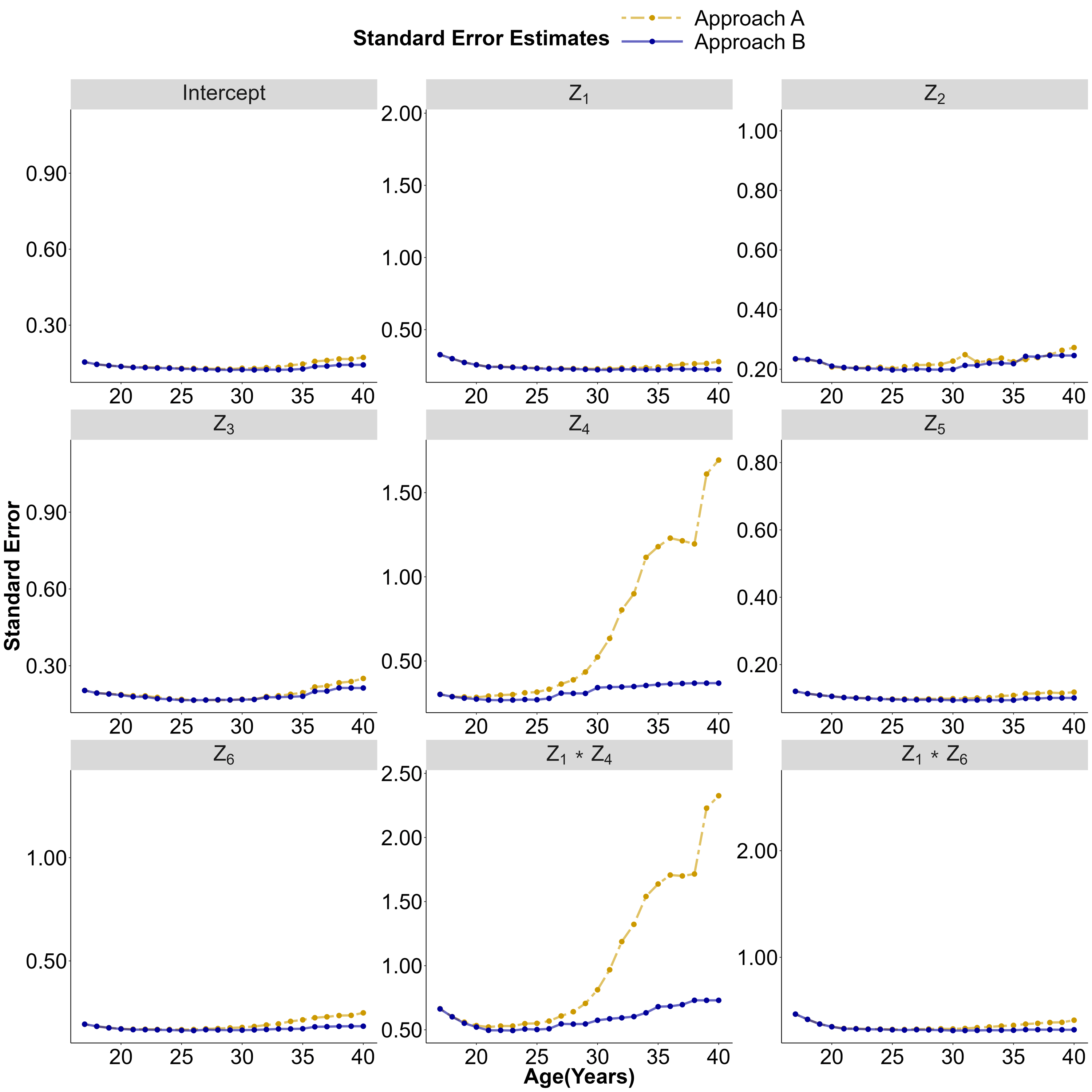}
        \caption{POI (With Firth Correction)}
        \label{fig:multi_se_appa_firth}
    \end{subfigure}

    \vspace{0.5cm}

    \begin{subfigure}[b]{0.48\textwidth}
        \centering
        \includegraphics[width=\textwidth]{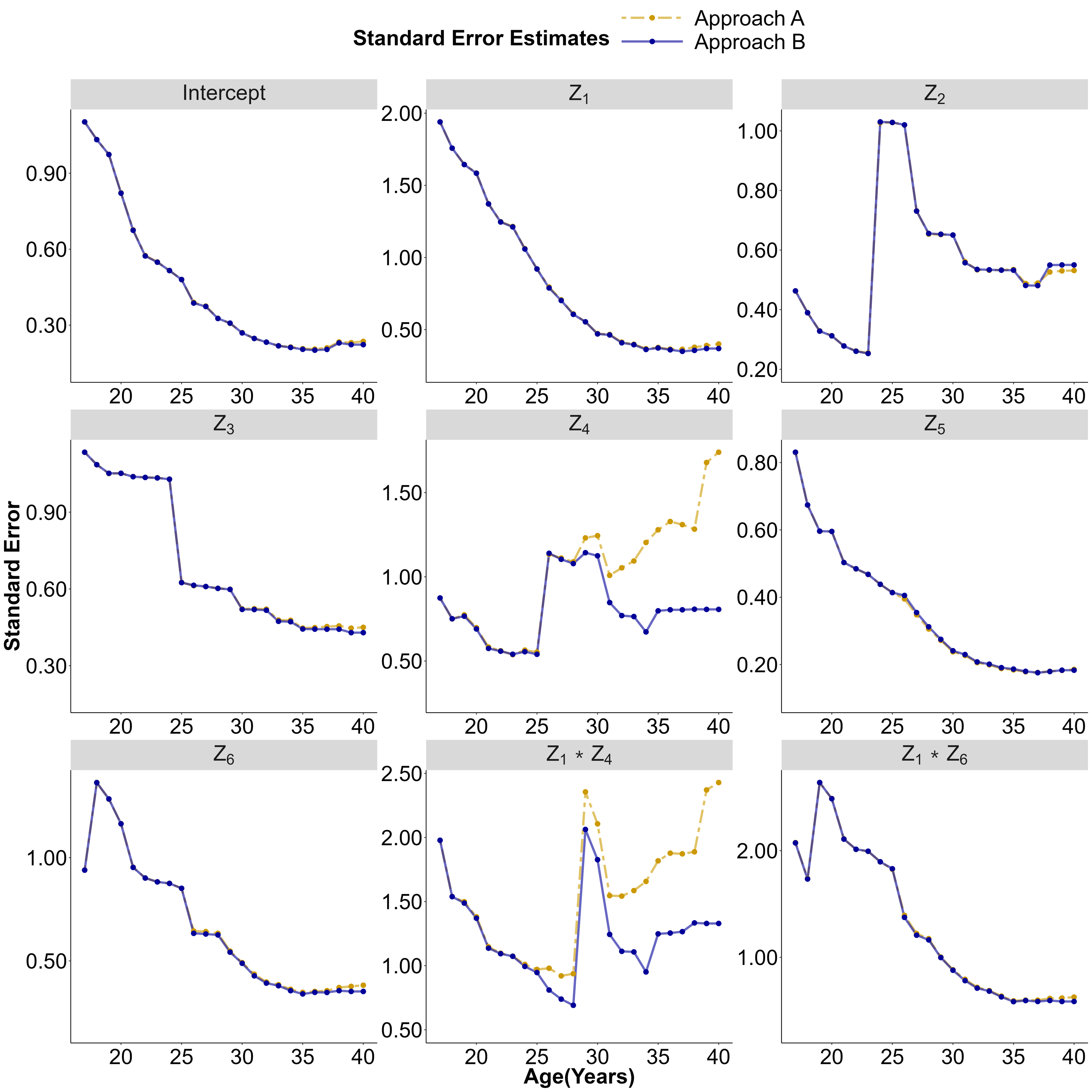}
        \caption{SPM (Without Firth Correction)}
        \label{fig:multi_se_appb_mlr}
    \end{subfigure}
    \hfill
    \begin{subfigure}[b]{0.48\textwidth}
        \centering
        \includegraphics[width=\textwidth]{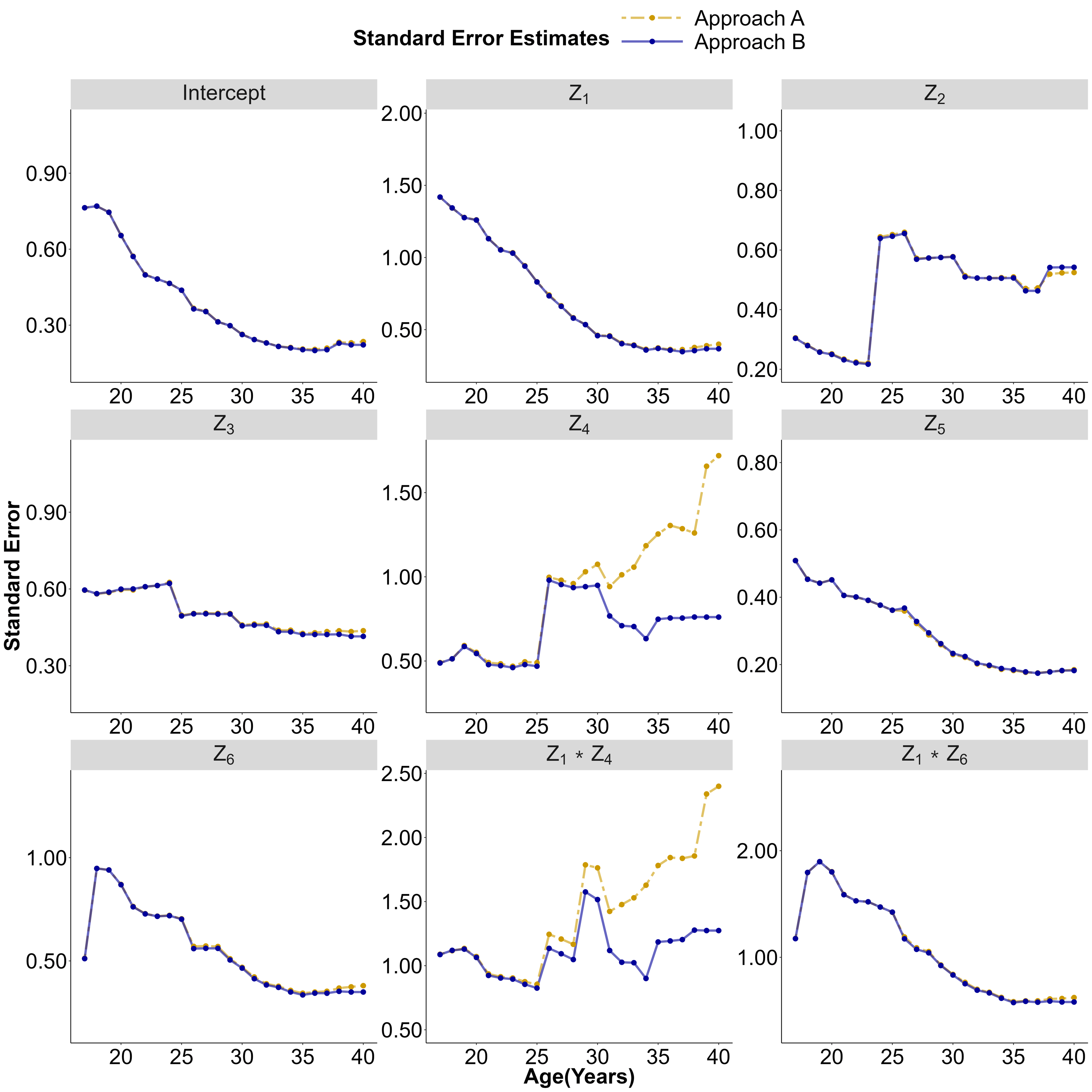}
        \caption{SPM (With Firth Correction)}
        \label{fig:multi_se_appb_firth}
    \end{subfigure}

    \caption{The sandwich standard error of the estimated coefficients when using multinomial logistic regression with and without Firth correction when outcome is POI or SPM. The $\hat{G}(\cdot \mid \mathbf{Z})$ is obtained using SRF. The number of trees and node size are set to 100.  $Z_{1}\text{: Rescaled age at cancer diagnosis}$,
             $Z_{2}\text{: Race-African American}$,
             $Z_{3}\text{: Race-Other}$,
             $Z_{4}\text{: Receipt of BMT}$,
             $Z_{5}\text{: Receipt of Alkylating agents}$,
             $Z_{6}\text{: Receipt of radiation to the abdomen/pelvis/total body}$}
    \label{fig: multi_real_data_se_srf}
\end{figure}

Figures~\ref{fig: multi_real_data_coef_srf} and~\ref{fig: multi_real_data_se_srf} present the estimated regression coefficients and sandwich standard errors from model~\eqref{model: project2 model 2}. The coefficient estimates for POI are stable across ages under both Approaches~A and~B. In contrast, the SPM estimates without Firth correction are highly volatile, and show extreme values at early ages for $Z_2$, $Z_4$, $Z_6$, $Z_{1}Z_{4}$, and $Z_{1}Z_{6}$. This instability is consistent with the sparse distribution of African American subjects, BMT recipients, and radiation therapy recipients among SPM cases, which leads to quasi-complete separation. The overall patterns are consistent with those from model~\eqref{model: project2 model 1}: Firth correction stabilizes the SPM estimates with a negligible effect on POI.

The sandwich standard errors from model~\eqref{model: project2 model 2} are shown in Figure~\ref{fig: multi_real_data_se_srf}. For the POI outcome under Approach~A, the standard errors of $Z_4$ and $Z_{1}Z_{4}$ increase sharply at older ages, whereas those under Approach~B remain small and stable. This is consistent with \citet{zhou202xagespecific}, who noted that Approach~B is more efficient under heavy censoring. Furthermore, Firth correction has a negligible impact on POI standard errors but reduces those for SPM. For the SPM outcome under Approach~B, the standard errors are uniformly larger than those for POI and decrease with age, consistent with the increasing cumulative number of SPM events at older ages.

\begin{figure}[H]
    \centering
    \begin{subfigure}[b]{0.48\textwidth}
        \centering
        \includegraphics[width=\textwidth]{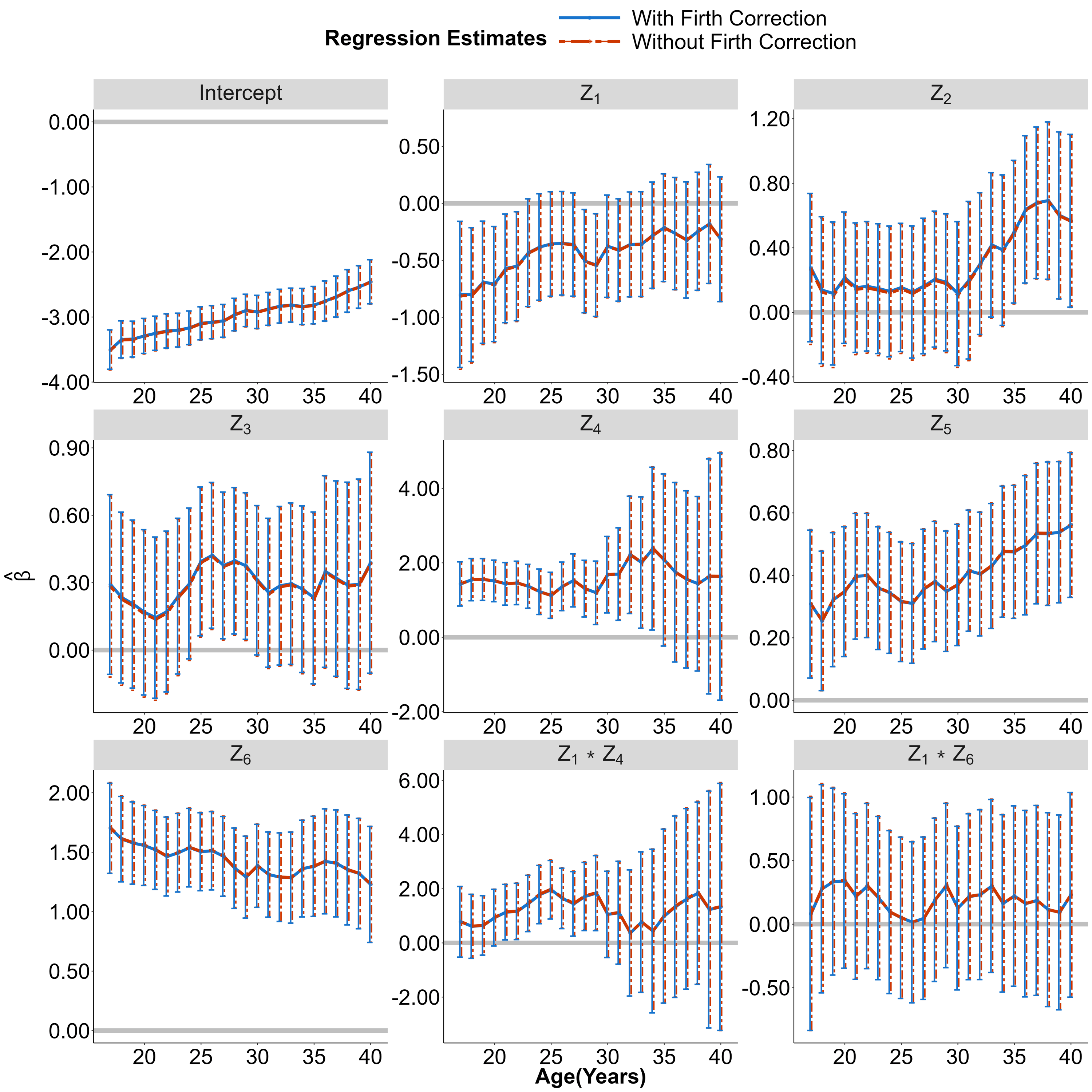}
        \caption{Approach A (POI)}
        \label{fig:mlr_poi_ci_appa}
    \end{subfigure}
    \hfill
    \begin{subfigure}[b]{0.48\textwidth}
        \centering
        \includegraphics[width=\textwidth]{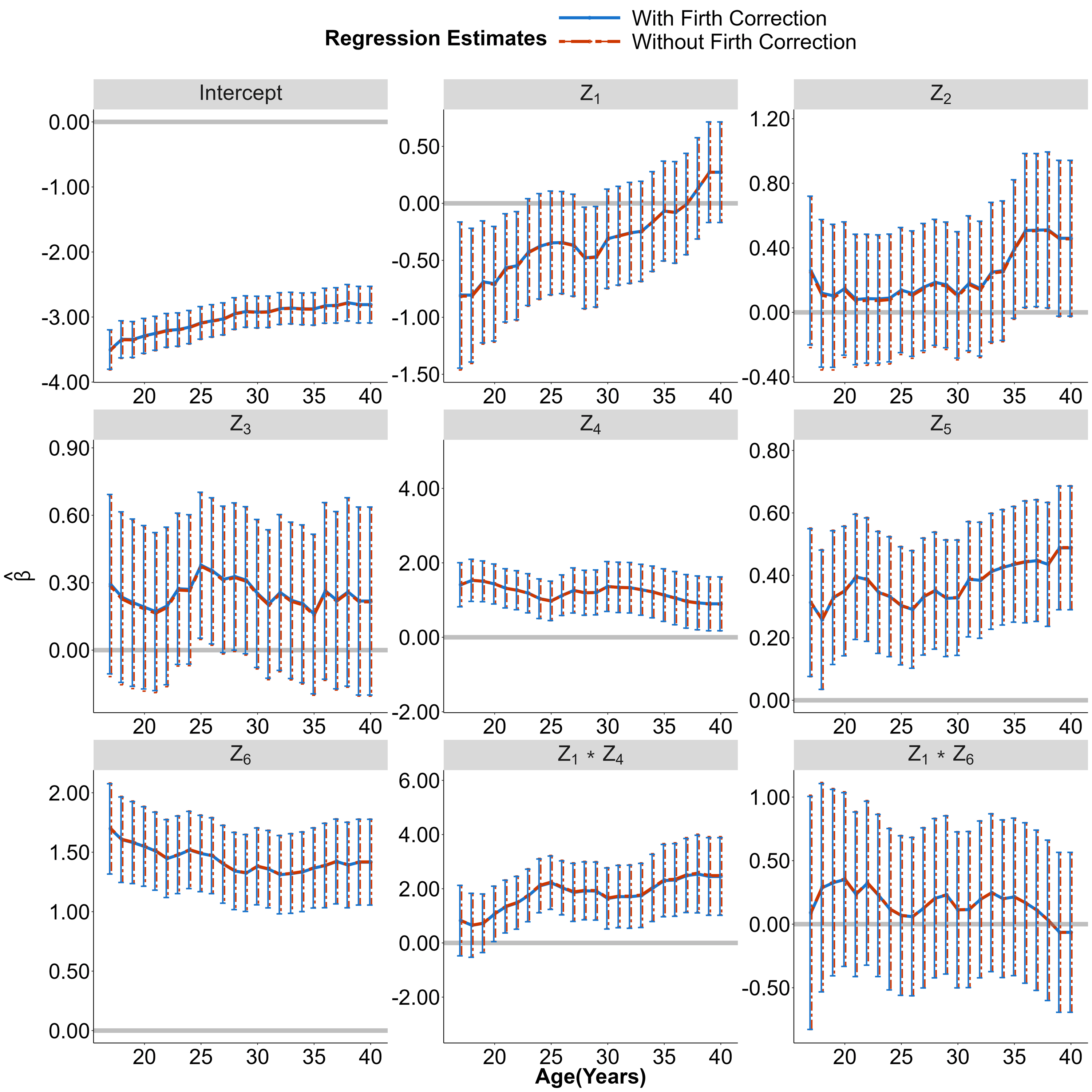}
        \caption{Approach B (POI)}
        \label{fig:mlr_poi_ci_appb}
    \end{subfigure}

    \vspace{0.5cm}

    \begin{subfigure}[b]{0.48\textwidth}
        \centering
        \includegraphics[width=\textwidth]{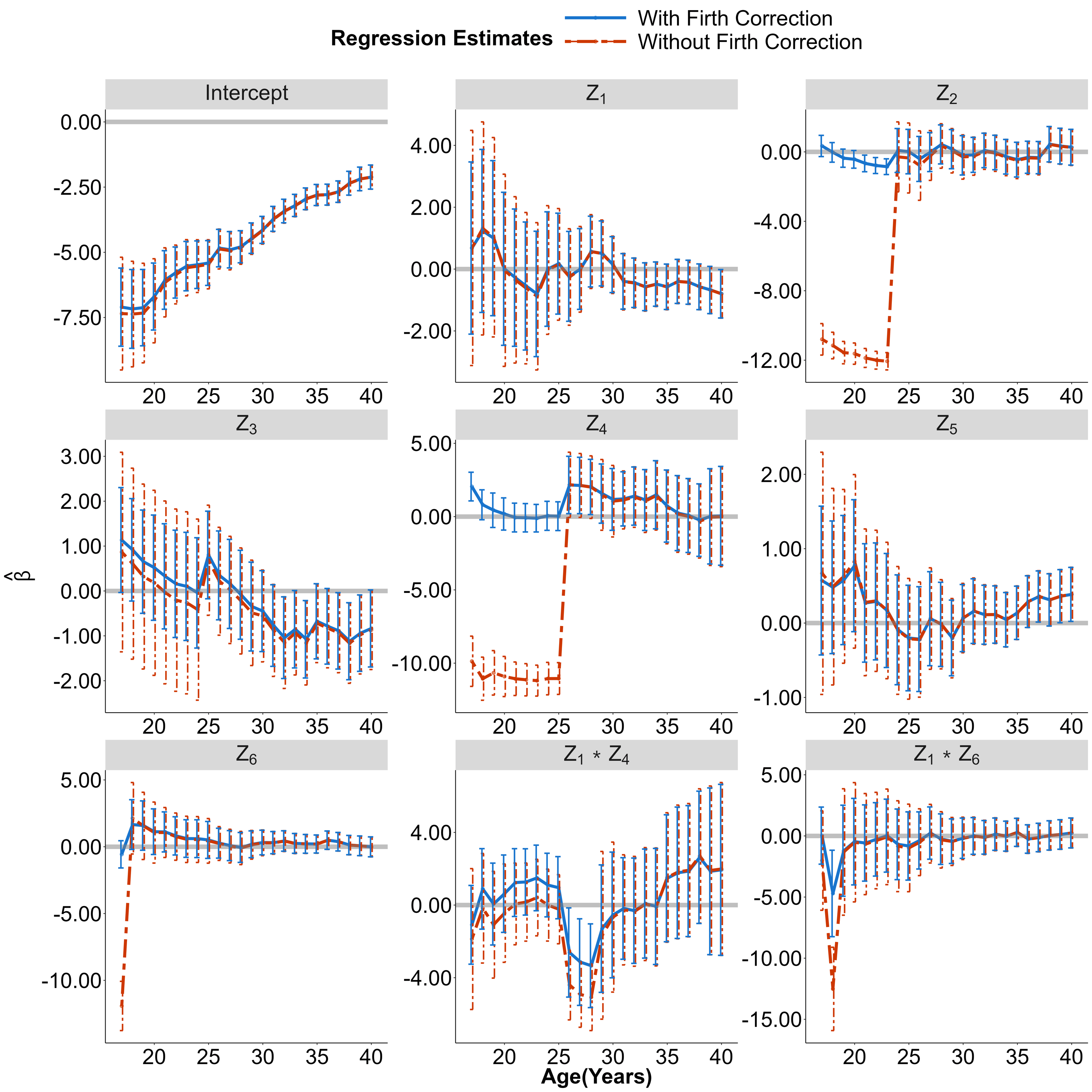}
        \caption{Approach A (SPM)}
        \label{fig:mlr_spm_ci_appa}
    \end{subfigure}
    \hfill
    \begin{subfigure}[b]{0.48\textwidth}
        \centering
        \includegraphics[width=\textwidth]{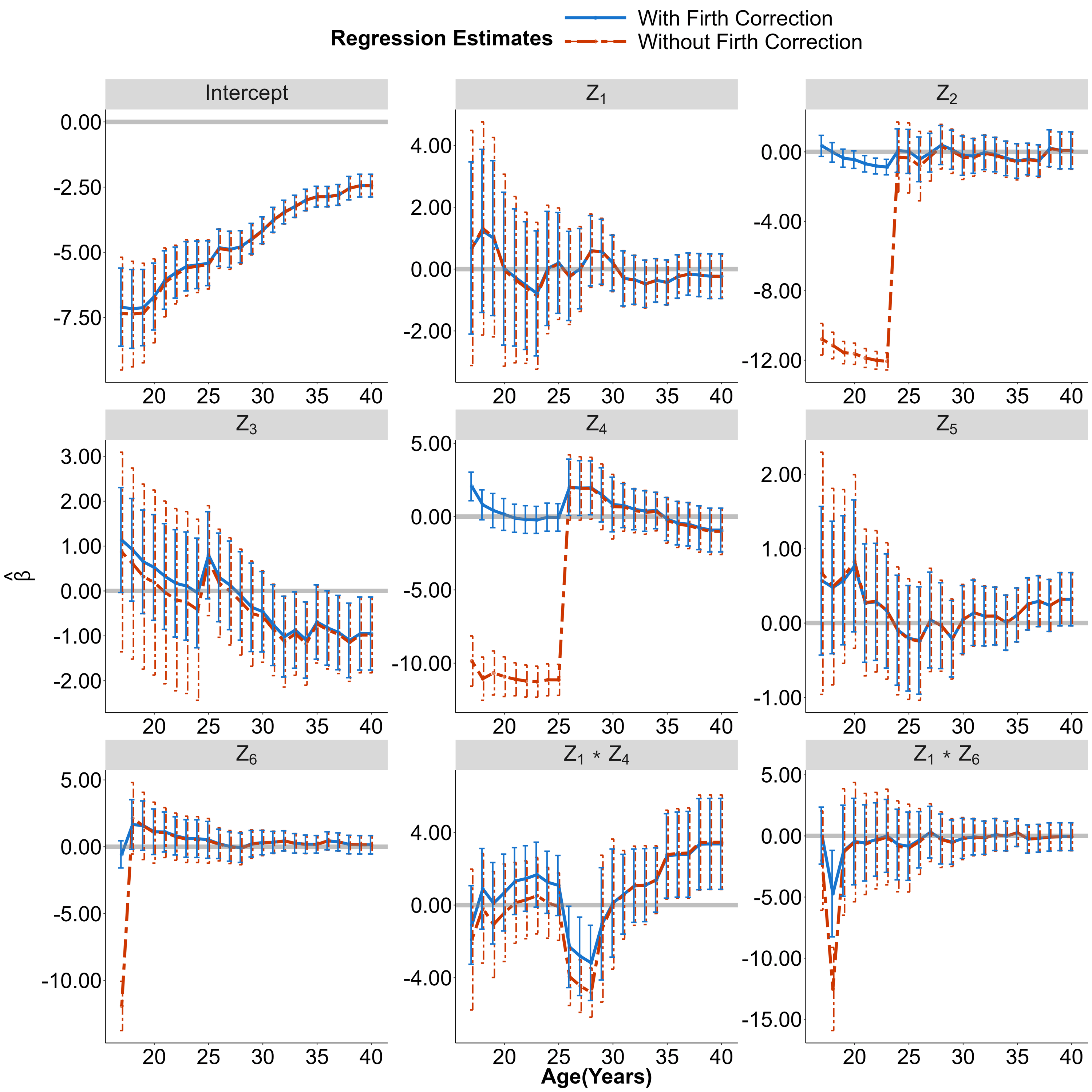}
        \caption{Approach B (SPM)}
        \label{fig:mlr_spm_ci_appb}
    \end{subfigure}

    \caption{The estimated coefficient with 95\% CI when using multinomial logistic regression with and without Firth correction among Approaches~A and~B. The $\hat{G}(\cdot \mid \mathbf{Z})$ is obtained using SRF. The number of trees and node size are set to 100.  $Z_{1}\text{: Rescaled age at cancer diagnosis}$,
             $Z_{2}\text{: Race-African American}$,
             $Z_{3}\text{: Race-Other}$,
             $Z_{4}\text{: Receipt of BMT}$,
             $Z_{5}\text{: Receipt of Alkylating agents}$
             $Z_{6}\text{: Receipt of radiation to the abdomen/pelvis/total body}$}
    \label{fig: multi_real_data_ci_srf}
\end{figure}

We constructed 95\% CIs for all model parameters under Approaches~A and~B. Since the model includes the interactions $Z_{1}Z_{4}$ and $Z_{1}Z_{6}$, the coefficients of $Z_{1}$, $Z_{4}$, and $Z_{6}$ are the effects at the reference level of the covariate they interact with. For the POI outcome (Figures~\ref{fig:mlr_poi_ci_appa} and~\ref{fig:mlr_poi_ci_appb}), among subjects who received neither BMT nor radiotherapy, the risk is lower for subjects with a younger age at cancer diagnosis ($Z_1$), though this effect weakens as $t_0$ increases. Non-Caucasian subjects tend to have higher risk than Caucasian subjects, but these differences are not statistically significant at most ages. The estimated effect of BMT is influenced by the approach used. Under Approach~B, subjects exposed to BMT ($Z_4$) have significantly higher risk than those not exposed at the cancer diagnosis age of zero, and this effect is modified by the age at cancer diagnosis: the BMT effect increases with older diagnosis age. The BMT effect on POI risk increases as attained age ($t_0$) increases to around 25 and then stabilizes. The effect of alkylating agents ($Z_5$) on POI risk is significant and positive, and grows over $t_0$. Radiation therapy ($Z_6$) increases POI risk at all ages, although its effect diminishes over $t_0$. The age at cancer diagnosis does not modify the radiation effect ($Z_{1}Z_{6}$). These findings are consistent with those reported in \citet{zhou202xagespecific}. For the SPM outcome (Figures~\ref{fig:mlr_spm_ci_appa} and~\ref{fig:mlr_spm_ci_appb}), all coefficients fluctuate around zero after Firth correction is used to handle quasi-complete separation, and the 95\% CIs generally cover zero. This is expected, as SPM is linked to surgical interventions that are not directly related to cancer-diagnosis age, cancer treatments, or race.

\begin{figure}[H]
    \centering
    \begin{subfigure}[t]{0.6\textwidth}
        \centering
        \includegraphics[width=1\textwidth]{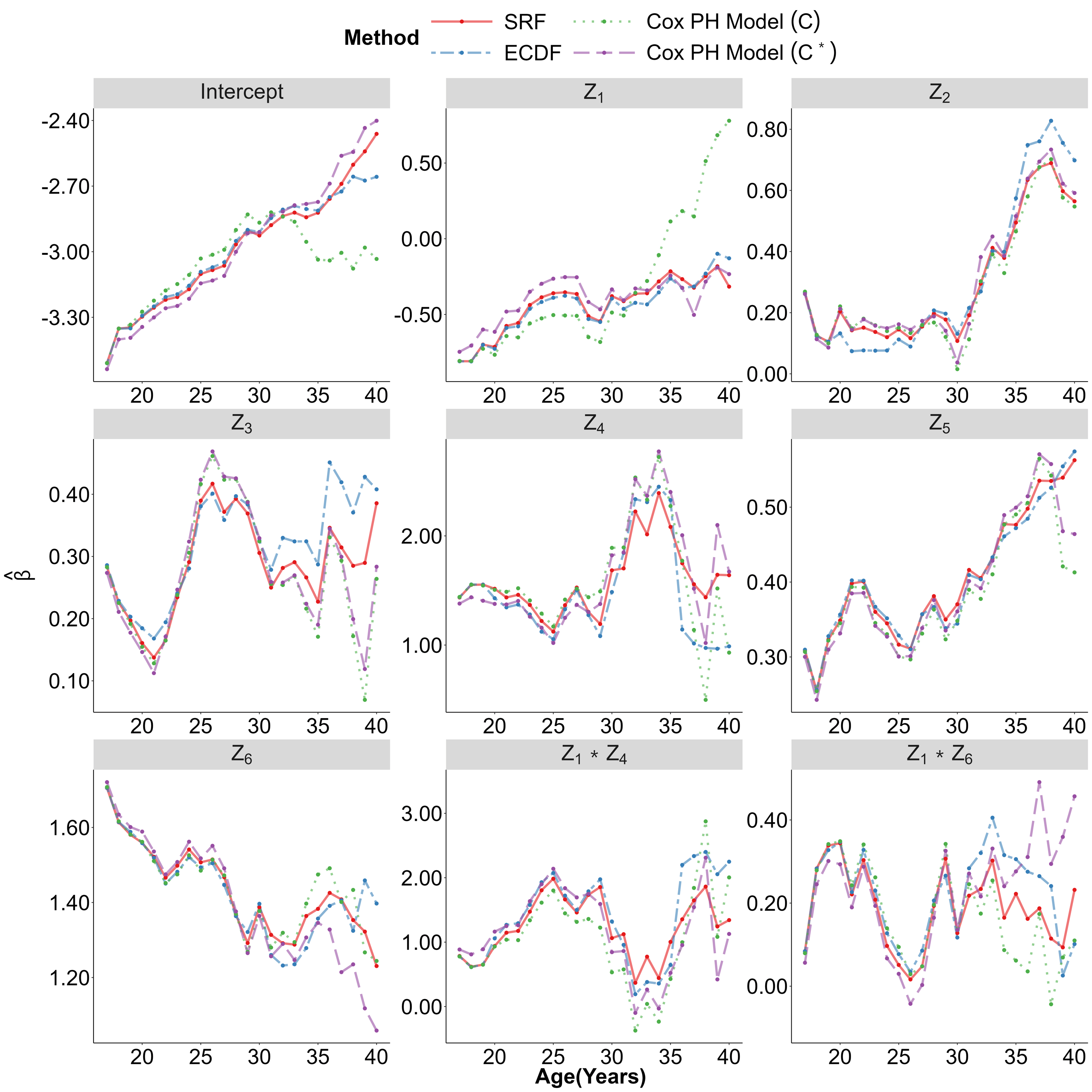}
        \caption{Estimated Coefficients using Approach A}
        \label{fig:multi_nofirth_poi_coef_appa}
    \end{subfigure}%
    \vfill
    \begin{subfigure}[t]{0.6\textwidth}
        \centering
        \includegraphics[width=1\textwidth]{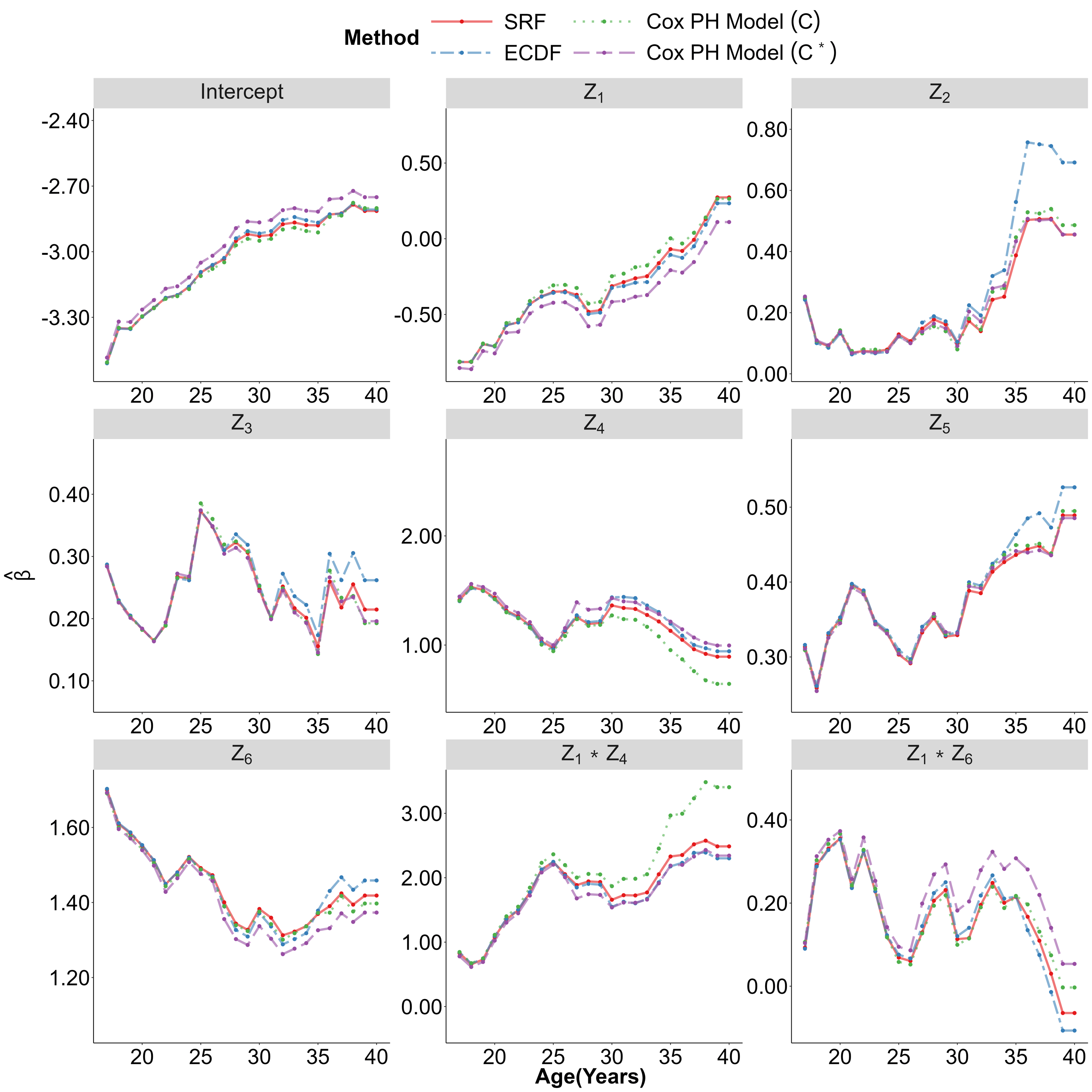}
        \caption{Estimated Coefficients using Approach B}
        \label{fig:multi_nofirth_poi_coef_appb}
    \end{subfigure}%
    \caption{Multinomial logistic regression without Firth correction (outcome: POI). The estimated coefficients using Approaches~A and~B with different methods to obtain $\hat{G}(\cdot \mid \mathbf{Z})$. The number of trees and the node size in SRF are set to 100.
             $Z_{1}\text{: Rescaled age at cancer diagnosis}$,
             $Z_{2}\text{: Race-African American}$,
             $Z_{3}\text{: Race-Other}$,
             $Z_{4}\text{: Receipt of BMT}$,
             $Z_{5}\text{: Receipt of Alkylating agents}$,
             $Z_{6}\text{: Receipt of radiation to the abdomen/pelvis/total body}$.}
    \label{fig:multi_nofirth_poi_coef}
\end{figure}

\begin{figure}[H]
    \centering
    \begin{subfigure}[b]{0.6\textwidth}
        \centering
        \includegraphics[width=1\textwidth]{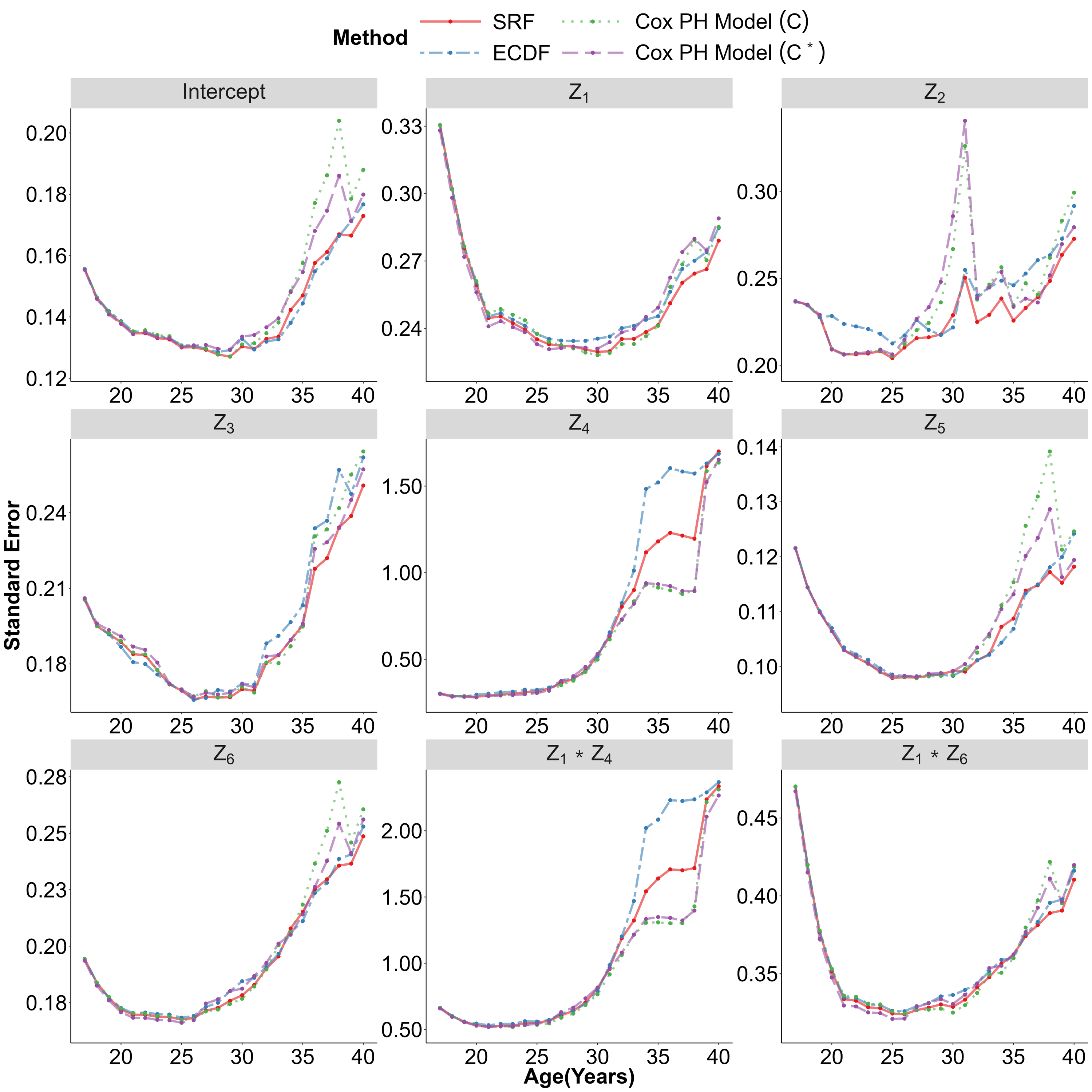}
        \caption{Standard Error Estimates using Approach A}
        \label{fig:multi_nofirth_poi_se_appa}
    \end{subfigure}
    \vfill
    \begin{subfigure}[b]{0.6\textwidth}
        \centering
        \includegraphics[width=1\textwidth]{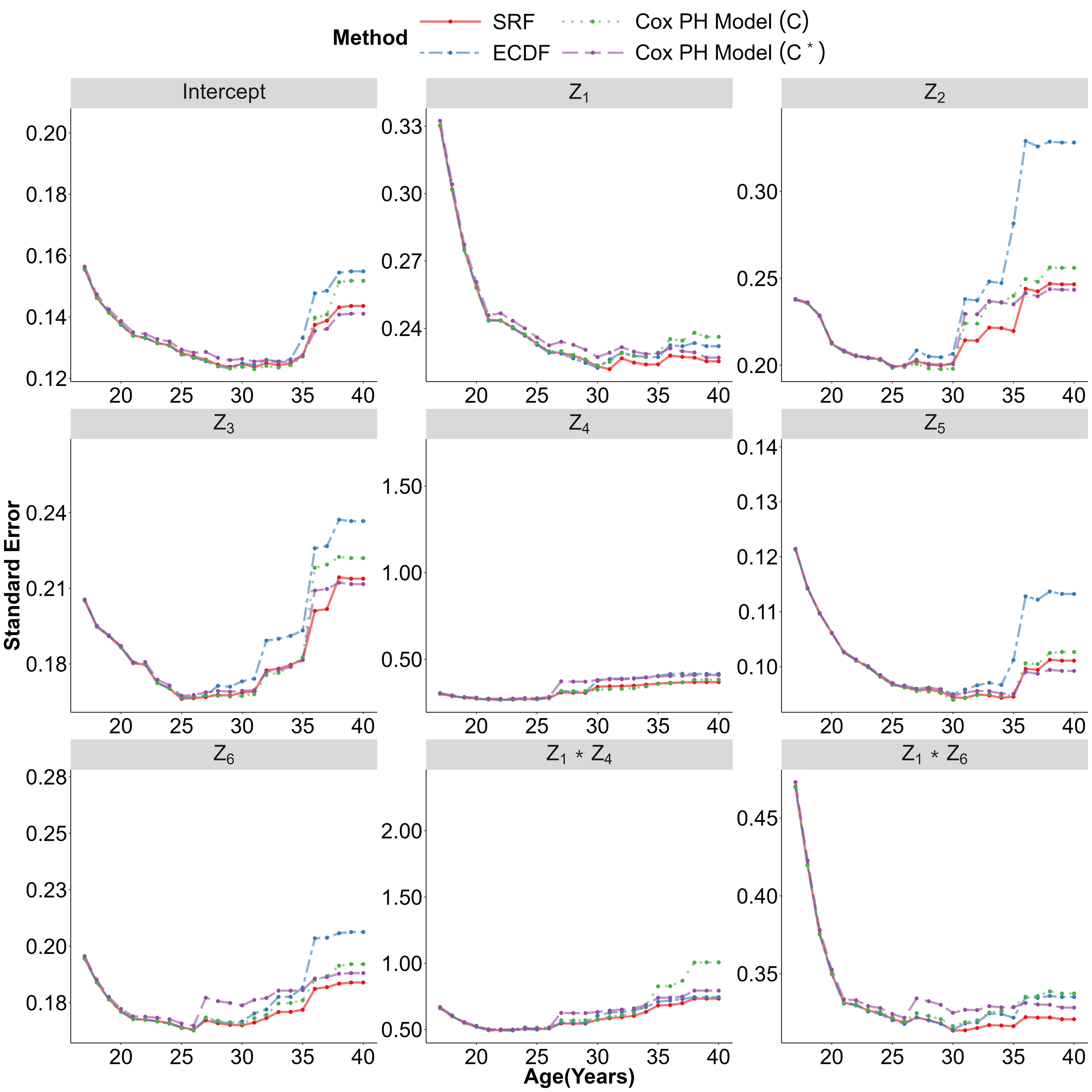}
        \caption{Standard Error Estimates using Approach B}
        \label{fig:multi_nofirth_poi_se_appb}
    \end{subfigure}
    \caption{Multinomial logistic regression without Firth correction (outcome: POI). The standard error of estimated coefficients using Approaches~A and~B with different methods to obtain $\hat{G}(\cdot \mid \mathbf{Z})$. The number of trees and the node size in SRF are set to 100.
             $Z_{1}\text{: Rescaled age at cancer diagnosis}$,
             $Z_{2}\text{: Race-African American}$,
             $Z_{3}\text{: Race-Other}$,
             $Z_{4}\text{: Receipt of BMT}$,
             $Z_{5}\text{: Receipt of Alkylating agents}$,
             $Z_{6}\text{: Receipt of radiation to the abdomen/pelvis/total body}$.}
    \label{fig:multi_nofirth_poi_se}
\end{figure}

\begin{figure}[H]
    \centering
    \begin{subfigure}[t]{0.6\textwidth}
        \centering
        \includegraphics[width=1\textwidth]{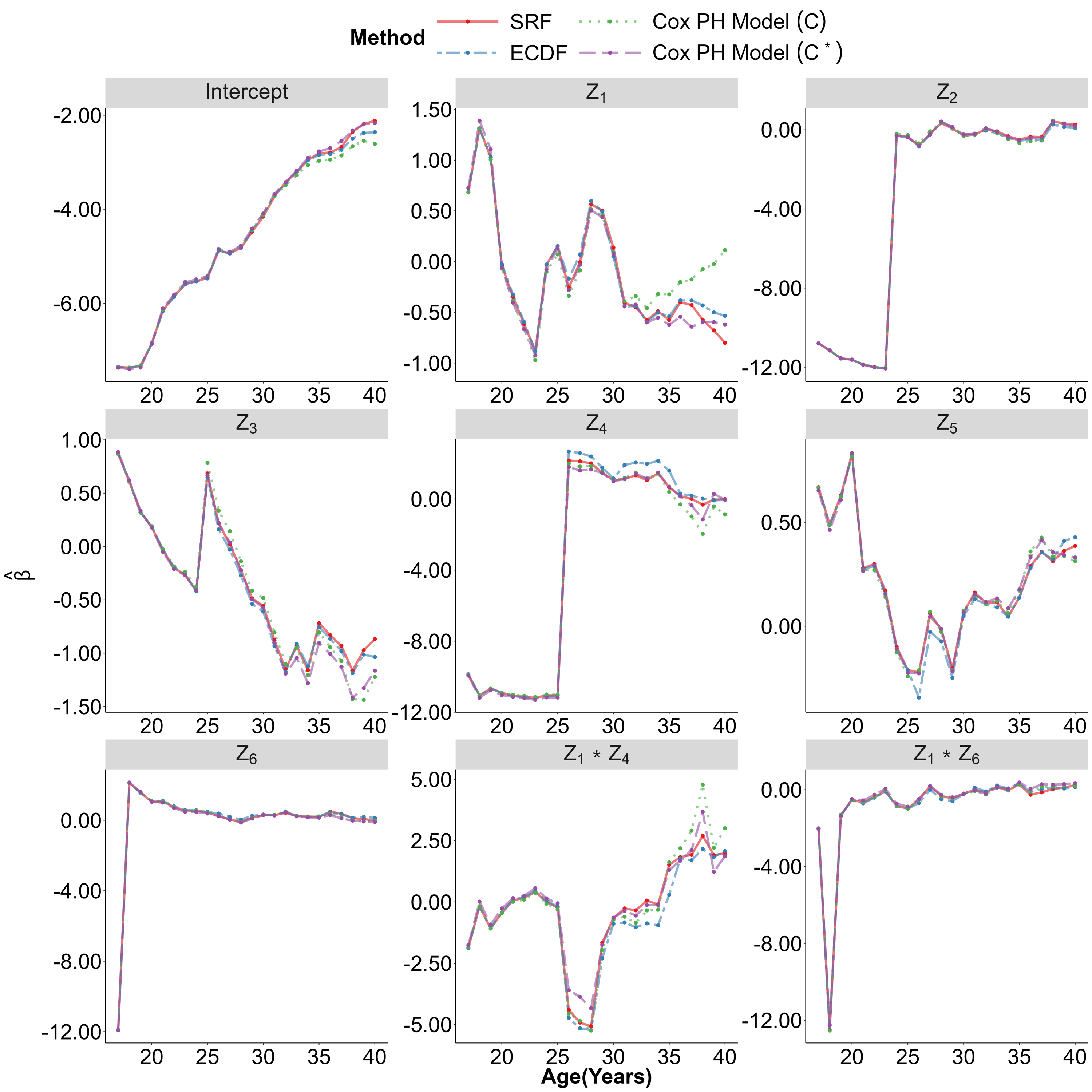}
        \caption{Estimated Coefficients using Approach A}
        \label{fig:multi_nofirth_spm_coef_appa}
    \end{subfigure}%
    \vfill
    \begin{subfigure}[t]{0.6\textwidth}
        \centering
        \includegraphics[width=1\textwidth]{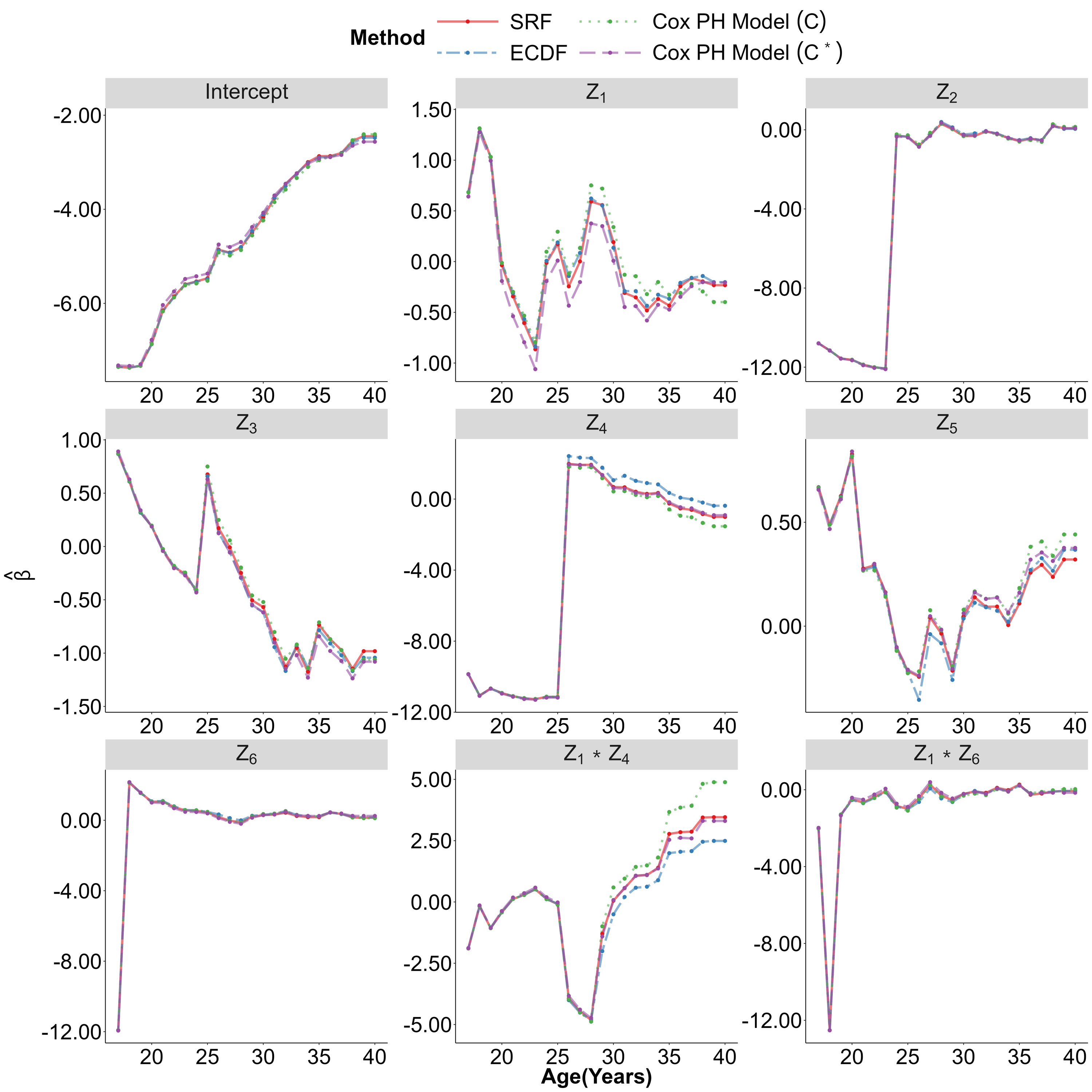}
        \caption{Estimated Coefficients using Approach B}
        \label{fig:multi_nofirth_spm_coef_appb}
    \end{subfigure}%
    \caption{Multinomial logistic regression without Firth correction (outcome: SPM). The estimated coefficients using Approaches~A and~B with different methods to obtain $\hat{G}(\cdot \mid \mathbf{Z})$. The number of trees and the node size in SRF are set to 100.
             $Z_{1}\text{: Rescaled age at cancer diagnosis}$,
             $Z_{2}\text{: Race-African American}$,
             $Z_{3}\text{: Race-Other}$,
             $Z_{4}\text{: Receipt of BMT}$,
             $Z_{5}\text{: Receipt of Alkylating agents}$,
             $Z_{6}\text{: Receipt of radiation to the abdomen/pelvis/total body}$.}
    \label{fig:multi_nofirth_spm_coef}
\end{figure}

\begin{figure}[H]
    \centering
    \begin{subfigure}[b]{0.6\textwidth}
        \centering
        \includegraphics[width=1\textwidth]{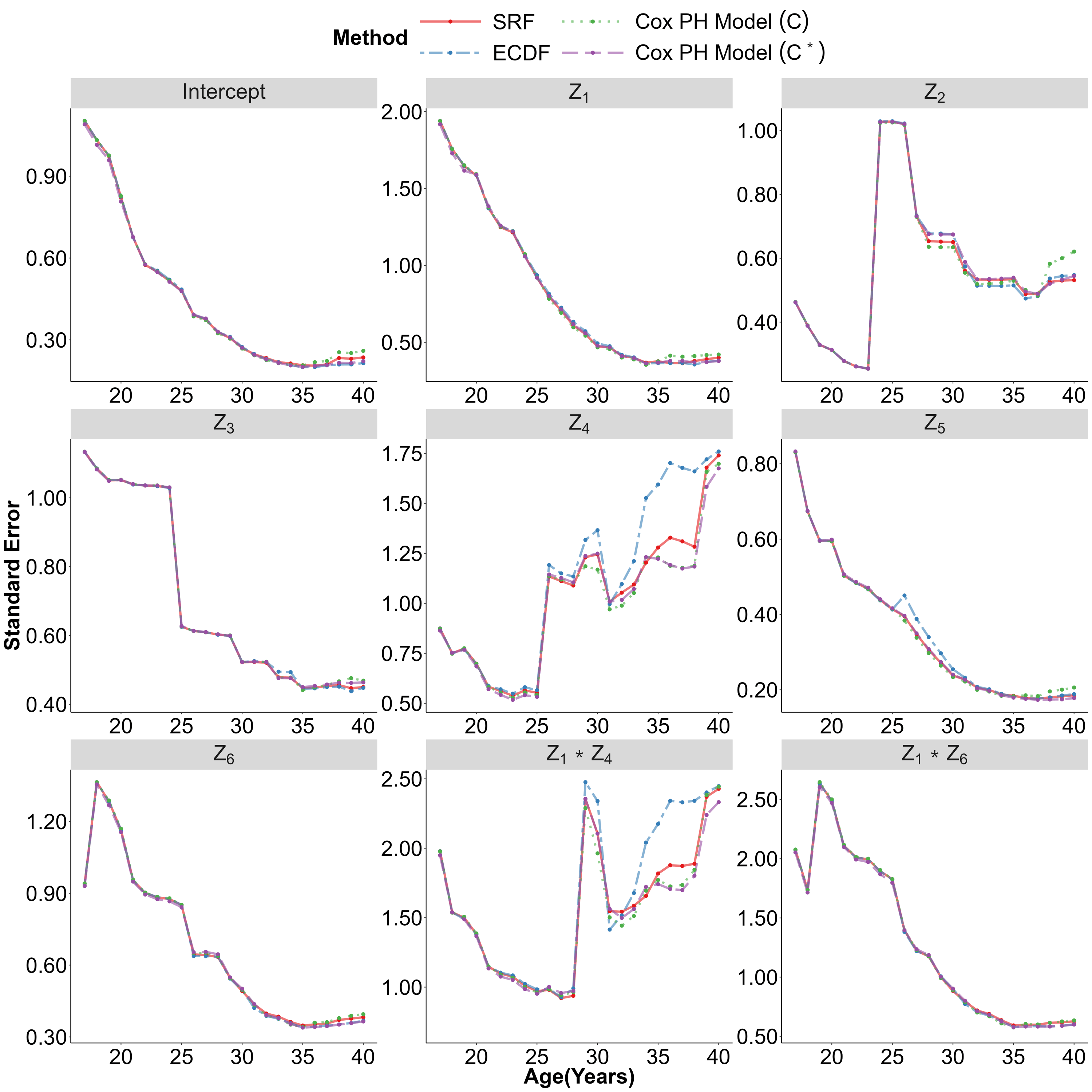}
        \caption{Standard Error Estimates using Approach A}
        \label{fig:multi_nofirth_spm_se_appa}
    \end{subfigure}
    \vfill
    \begin{subfigure}[b]{0.6\textwidth}
        \centering
        \includegraphics[width=1\textwidth]{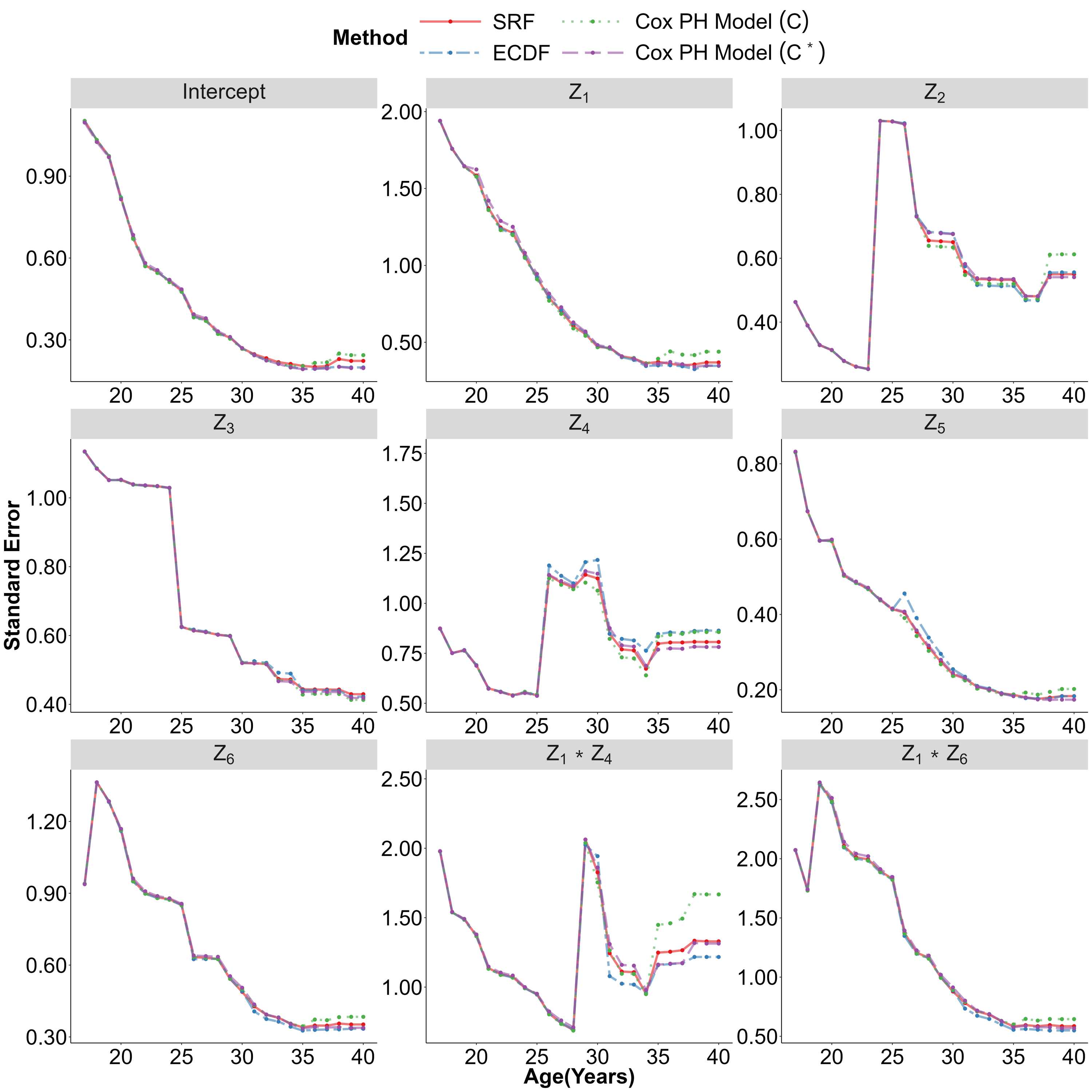}
        \caption{Standard Error Estimates using Approach B}
        \label{fig:multi_nofirth_spm_se_appb}
    \end{subfigure}
    \caption{Multinomial logistic regression without Firth correction (outcome: SPM). The standard error of estimated coefficients using Approaches~A and~B with different methods to obtain $\hat{G}(\cdot \mid \mathbf{Z})$. The number of trees and the node size in SRF are set to 100.
             $Z_{1}\text{: Rescaled age at cancer diagnosis}$,
             $Z_{2}\text{: Race-African American}$,
             $Z_{3}\text{: Race-Other}$,
             $Z_{4}\text{: Receipt of BMT}$,
             $Z_{5}\text{: Receipt of Alkylating agents}$,
             $Z_{6}\text{: Receipt of radiation to the abdomen/pelvis/total body}$.}
    \label{fig:multi_nofirth_spm_se}
\end{figure}

\begin{figure}[H]
    \centering
    \begin{subfigure}[t]{0.6\textwidth}
        \centering
        \includegraphics[width=1\textwidth]{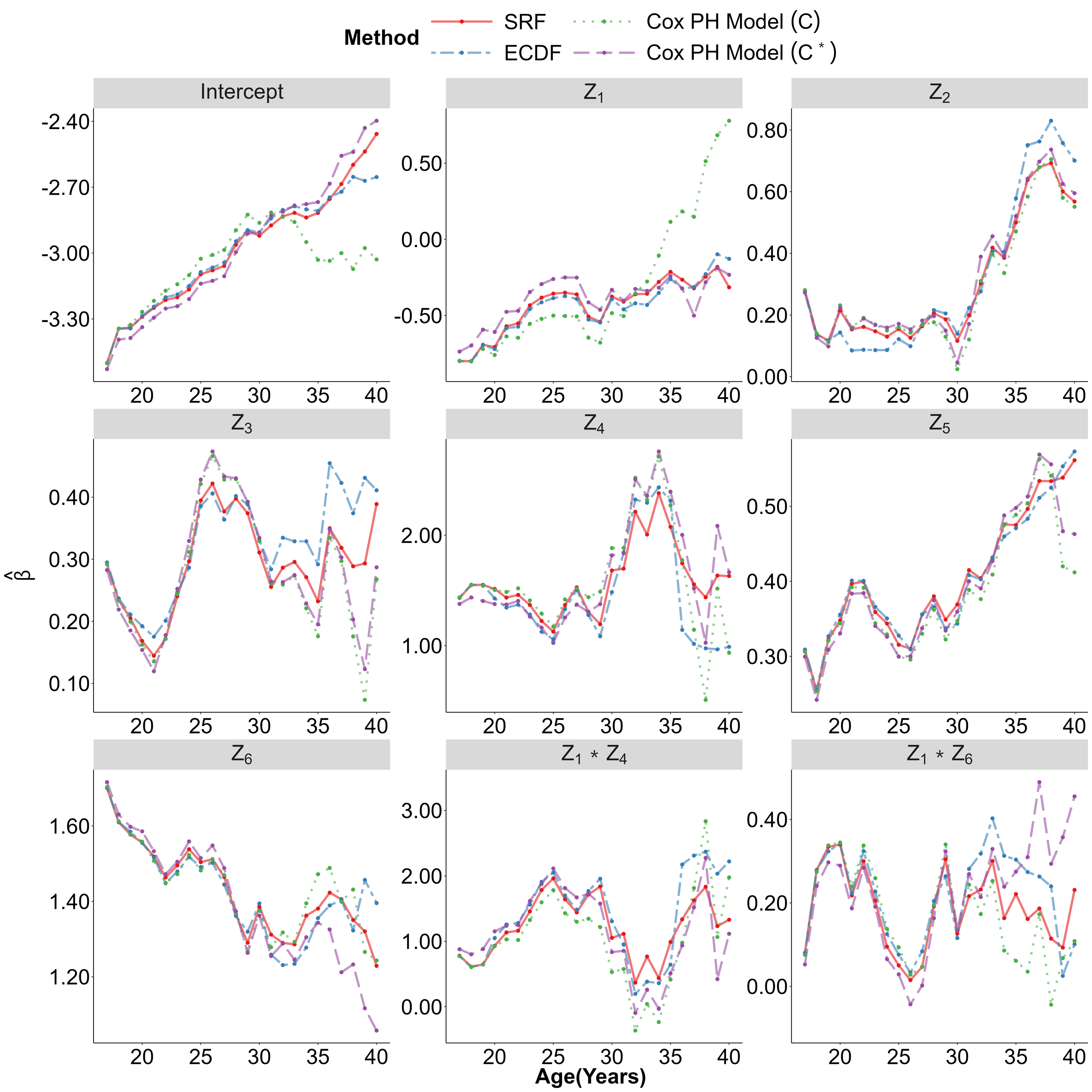}
        \caption{Estimated Coefficients using Approach A}
        \label{fig:multi_firth_poi_coef_appa}
    \end{subfigure}%
    \vfill
    \begin{subfigure}[t]{0.6\textwidth}
        \centering
        \includegraphics[width=1\textwidth]{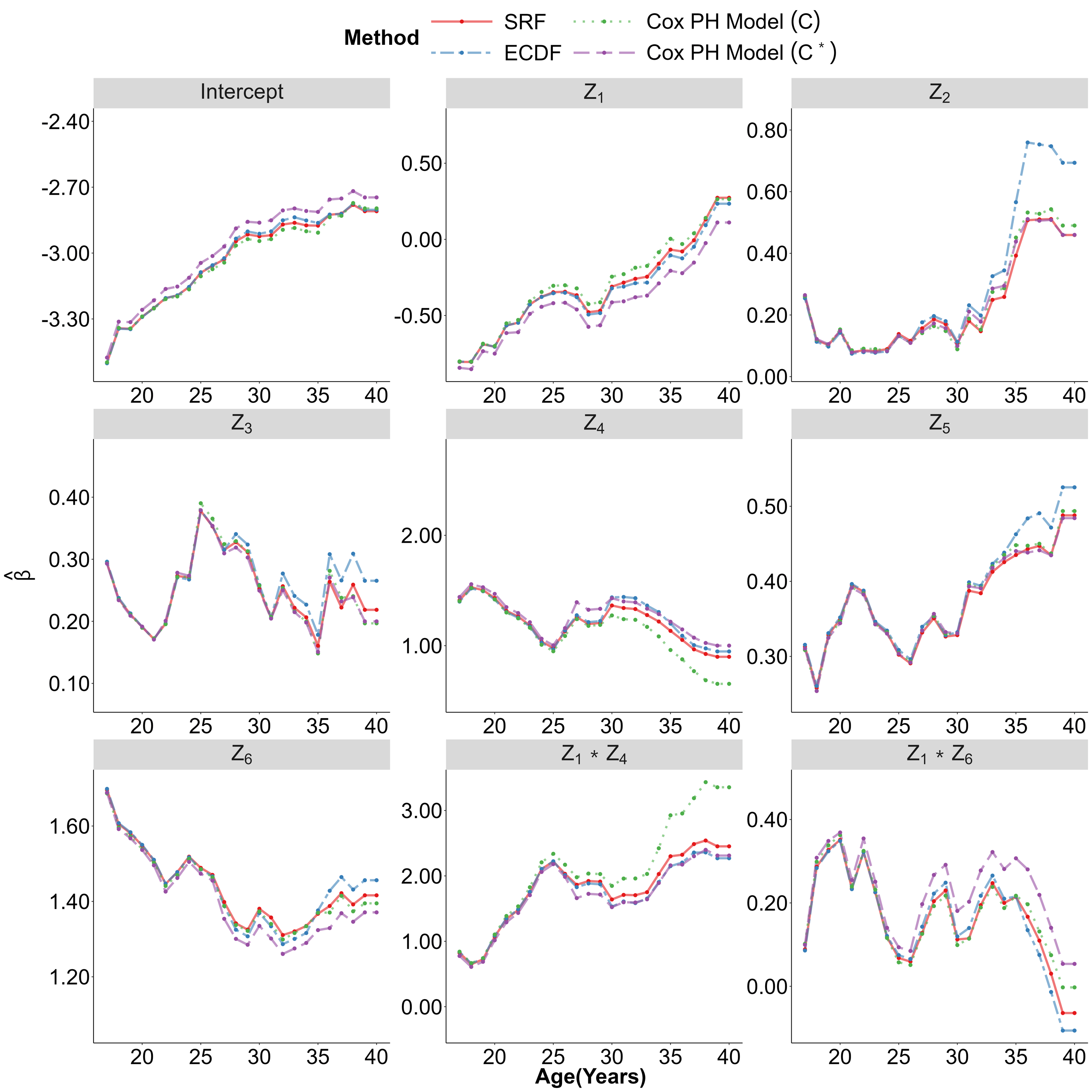}
        \caption{Estimated Coefficients using Approach B}
        \label{fig:multi_firth_poi_coef_appb}
    \end{subfigure}%
    \caption{Multinomial logistic regression with Firth correction (outcome: POI). The estimated coefficients using Approaches~A and~B with different methods to obtain $\hat{G}(\cdot \mid \mathbf{Z})$. The number of trees and the node size in SRF are set to 100.
             $Z_{1}\text{: Rescaled age at cancer diagnosis}$,
             $Z_{2}\text{: Race-African American}$,
             $Z_{3}\text{: Race-Other}$,
             $Z_{4}\text{: Receipt of BMT}$,
             $Z_{5}\text{: Receipt of Alkylating agents}$,
             $Z_{6}\text{: Receipt of radiation to the abdomen/pelvis/total body}$.}
    \label{fig:multi_firth_poi_coef}
\end{figure}

\begin{figure}[H]
    \centering
    \begin{subfigure}[b]{0.6\textwidth}
        \centering
        \includegraphics[width=1\textwidth]{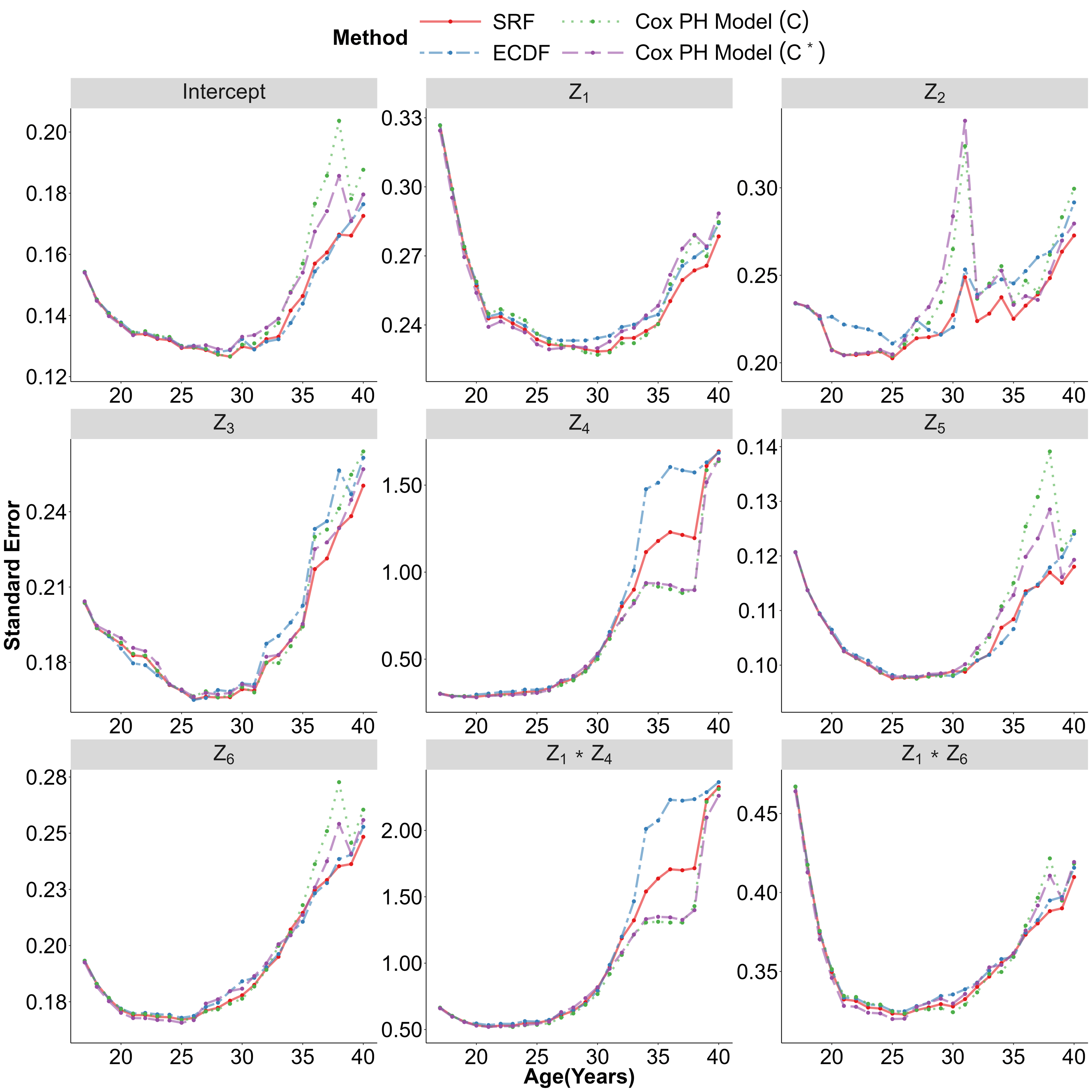}
        \caption{Standard Error Estimates using Approach A}
        \label{fig:multi_firth_poi_se_appa}
    \end{subfigure}
    \vfill
    \begin{subfigure}[b]{0.6\textwidth}
        \centering
        \includegraphics[width=1\textwidth]{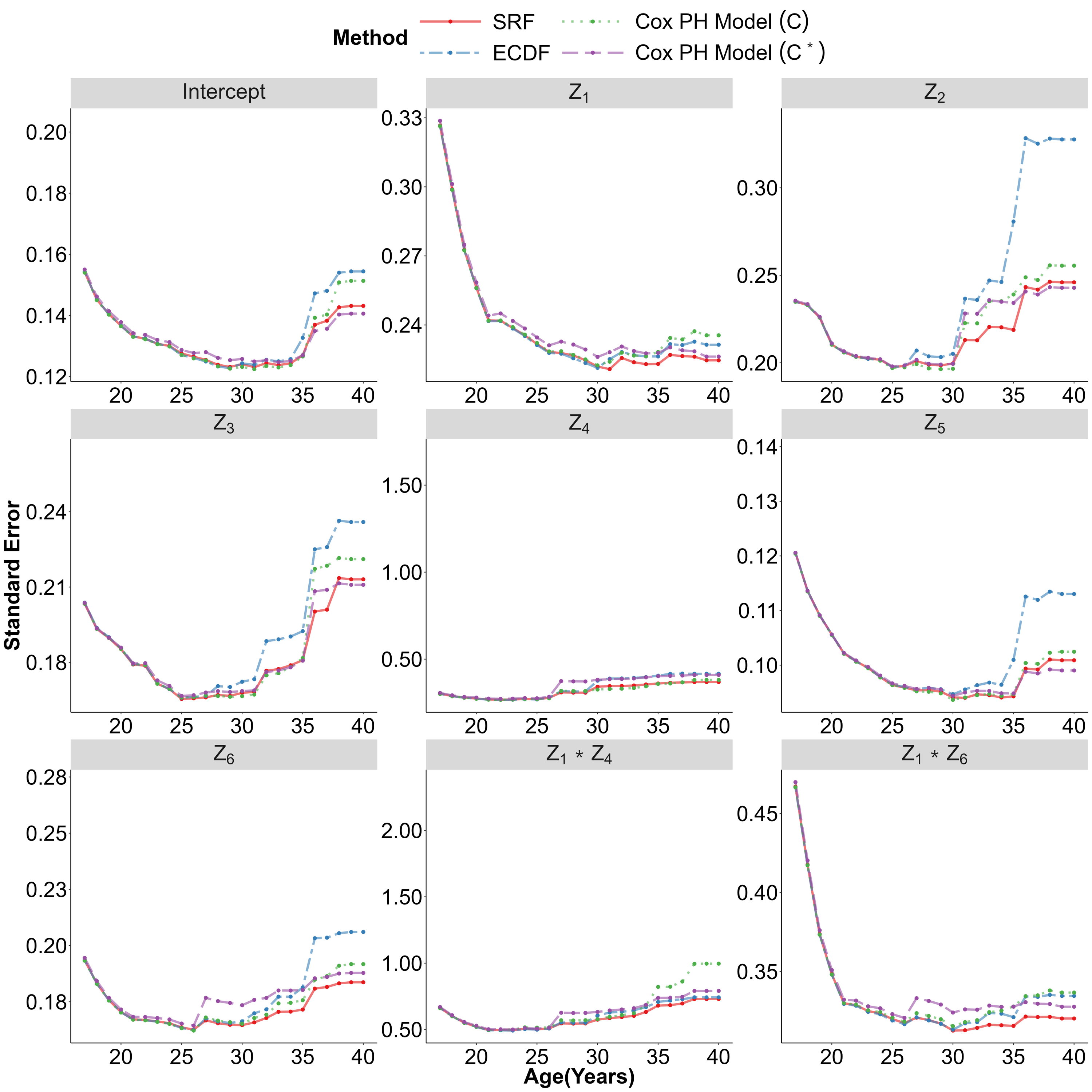}
        \caption{Standard Error Estimates using Approach B}
        \label{fig:multi_firth_poi_se_appb}
    \end{subfigure}
    \caption{Multinomial logistic regression with Firth correction (outcome: POI). The standard error of estimated coefficients using Approaches~A and~B with different methods to obtain $\hat{G}(\cdot \mid \mathbf{Z})$. The number of trees and the node size in SRF are set to 100.
             $Z_{1}\text{: Rescaled age at cancer diagnosis}$,
             $Z_{2}\text{: Race-African American}$,
             $Z_{3}\text{: Race-Other}$,
             $Z_{4}\text{: Receipt of BMT}$,
             $Z_{5}\text{: Receipt of Alkylating agents}$,
             $Z_{6}\text{: Receipt of radiation to the abdomen/pelvis/total body}$.}
    \label{fig:multi_firth_poi_se}
\end{figure}

\begin{figure}[H]
    \centering
    \begin{subfigure}[t]{0.6\textwidth}
        \centering
        \includegraphics[width=1\textwidth]{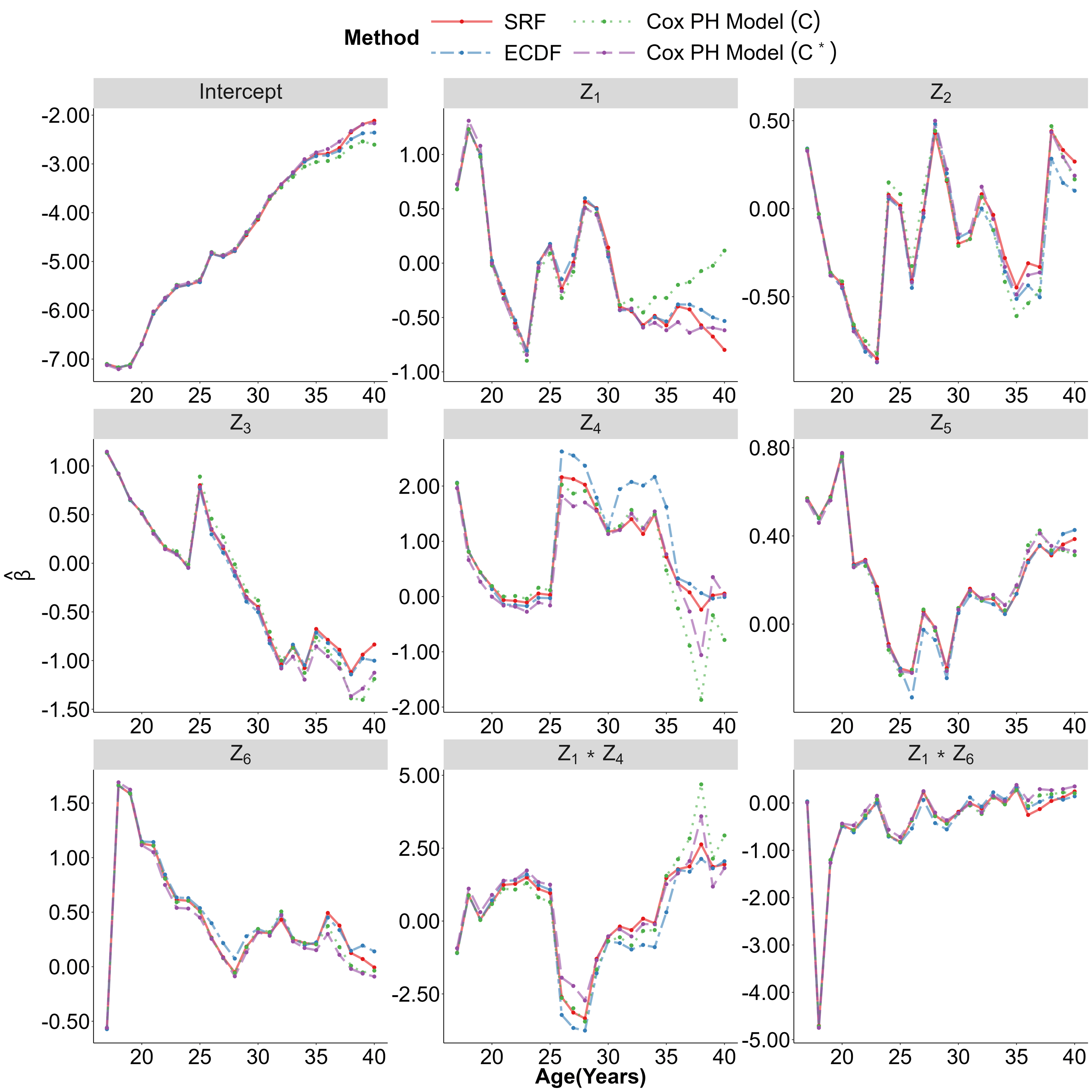}
        \caption{Estimated Coefficients using Approach A}
        \label{fig:multi_firth_spm_coef_appa}
    \end{subfigure}%
    \vfill
    \begin{subfigure}[t]{0.6\textwidth}
        \centering
        \includegraphics[width=1\textwidth]{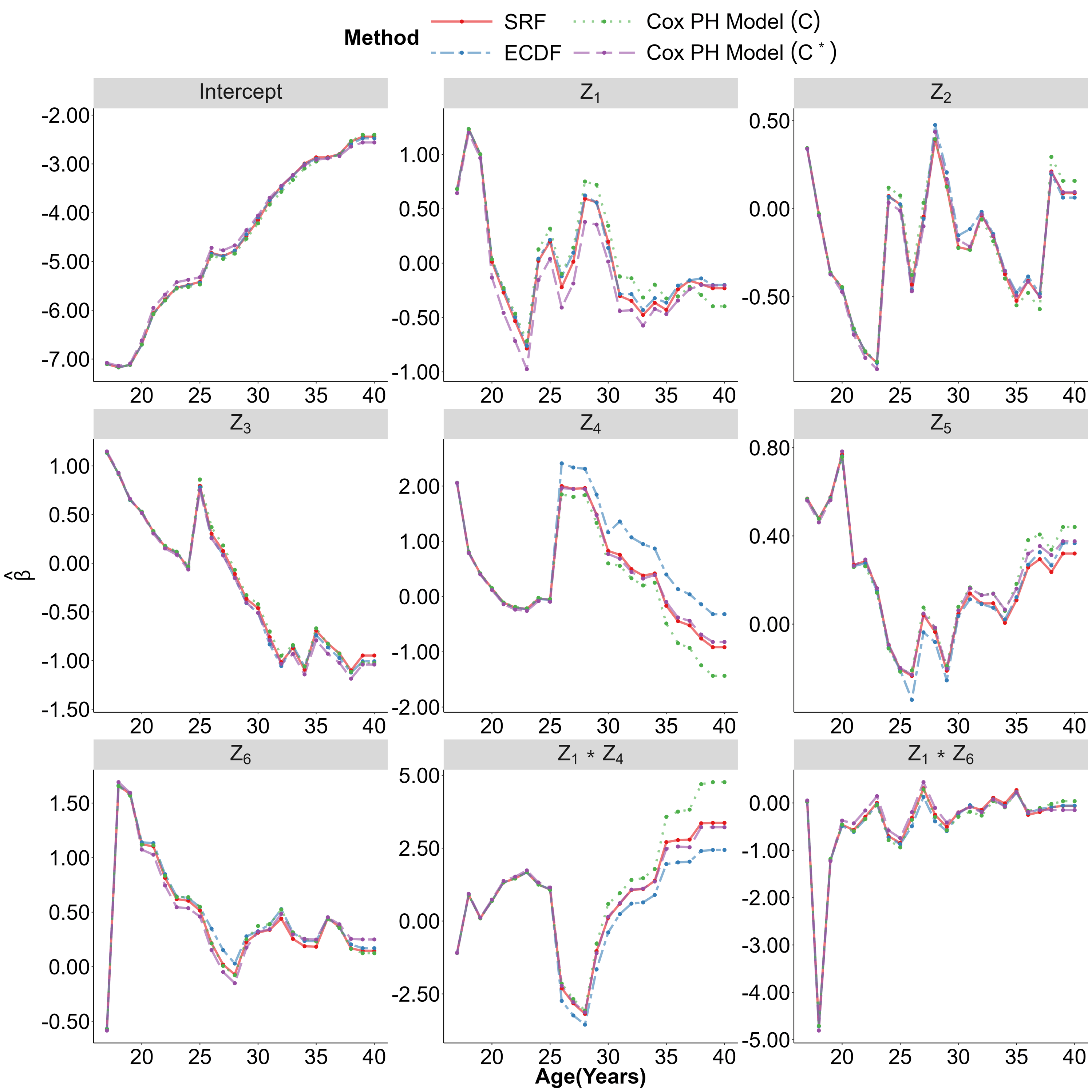}
        \caption{Estimated Coefficients using Approach B}
        \label{fig:multi_firth_spm_coef_appb}
    \end{subfigure}%
    \caption{Multinomial logistic regression with Firth correction (outcome: SPM). The estimated coefficients using Approaches~A and~B with different methods to obtain $\hat{G}(\cdot \mid \mathbf{Z})$. The number of trees and the node size in SRF are set to 100.
             $Z_{1}\text{: Rescaled age at cancer diagnosis}$,
             $Z_{2}\text{: Race-African American}$,
             $Z_{3}\text{: Race-Other}$,
             $Z_{4}\text{: Receipt of BMT}$,
             $Z_{5}\text{: Receipt of Alkylating agents}$,
             $Z_{6}\text{: Receipt of radiation to the abdomen/pelvis/total body}$.}
    \label{fig:multi_firth_spm_coef}
\end{figure}

\begin{figure}[H]
    \centering
    \begin{subfigure}[b]{0.6\textwidth}
        \centering
        \includegraphics[width=1\textwidth]{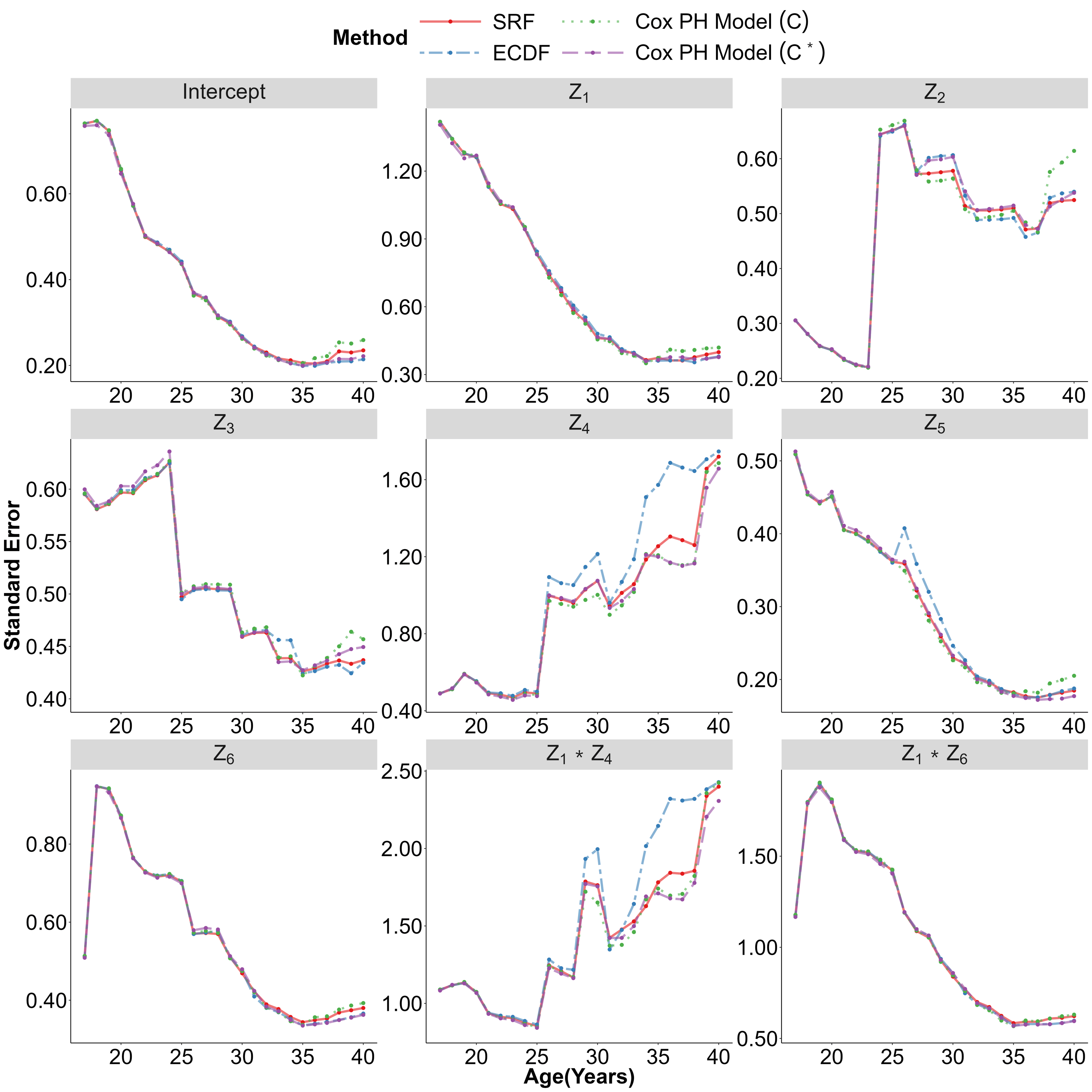}
        \caption{Standard Error Estimates using Approach A}
        \label{fig:multi_firth_spm_se_appa}
    \end{subfigure}
    \vfill
    \begin{subfigure}[b]{0.6\textwidth}
        \centering
        \includegraphics[width=1\textwidth]{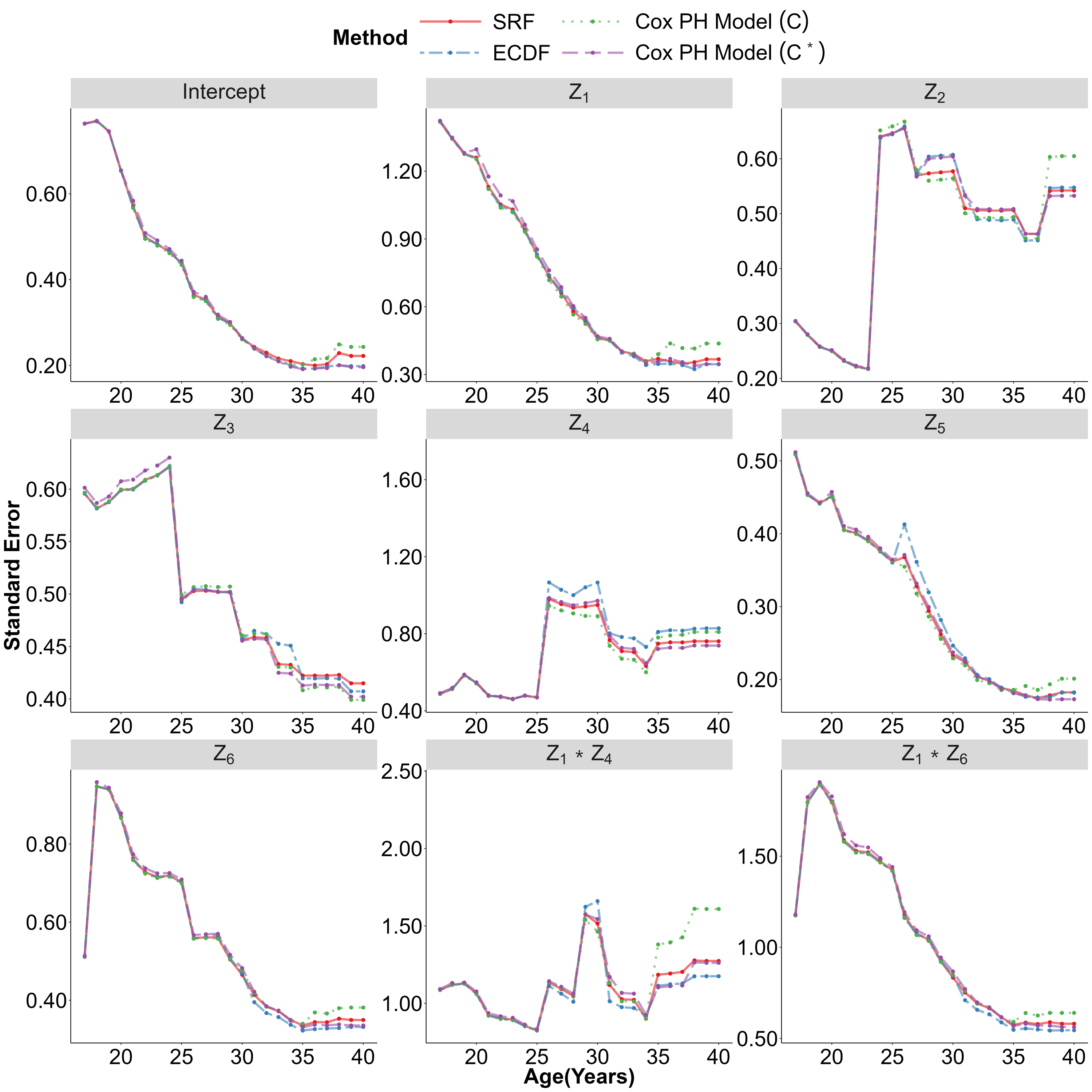}
        \caption{Standard Error Estimates using Approach B}
        \label{fig:multi_firth_spm_se_appb}
    \end{subfigure}
    \caption{Multinomial logistic regression with Firth correction (outcome: SPM). The standard error of estimated coefficients using Approaches~A and~B with different methods to obtain $\hat{G}(\cdot \mid \mathbf{Z})$. The number of trees and the node size in SRF are set to 100.
             $Z_{1}\text{: Rescaled age at cancer diagnosis}$,
             $Z_{2}\text{: Race-African American}$,
             $Z_{3}\text{: Race-Other}$,
             $Z_{4}\text{: Receipt of BMT}$,
             $Z_{5}\text{: Receipt of Alkylating agents}$,
             $Z_{6}\text{: Receipt of radiation to the abdomen/pelvis/total body}$.}
    \label{fig:multi_firth_spm_se}
\end{figure}

The results above use SRF to estimate $G(\cdot \mid \mathbf{Z})$. We also evaluated the censoring distribution using a stratified ECDF, a Cox PH model on $C$, and a Cox PH model on $C^{*}$. The choice of censoring estimator has a substantial impact on coefficient estimates, while its effect on standard errors is negligible (Figures~\ref{fig:multi_nofirth_poi_coef} to~\ref{fig:multi_firth_spm_se}).


\subsubsection{Without Firth Correction While Using Different $\hat{G}(\cdot \mid \mathbf{Z})$}

\subsubsection{With Firth Correction While Using Different $\hat{G}(\cdot \mid \mathbf{Z})$}

%
%
\subsection{Extended Covariate Set}\label{supp: analysis3_covariates}

This section gives the covariate definitions for the analysis of Section~\ref{sec3: extended analysis} of the main text, which refits model~\eqref{model: project2 model 2} with a larger covariate set.

Thirteen covariates are used. The continuous covariate $Z_1$ is the rescaled age at primary cancer diagnosis. Two race indicators $Z_2$ and $Z_3$ correspond to African American and Other, with Caucasian as the reference category. The BMT indicator $Z_4$ equals one when a bone marrow transplant was received. Three rescaled radiation-dose variables $Z_5$, $Z_6$, and $Z_7$ measure abdominal, pelvic, and total body irradiation (TBI) doses. Because TBI irradiates the abdomen and the pelvis, the recorded abdominal and pelvic field maxima include the TBI contribution. We therefore subtract the TBI dose from each field maximum before rescaling, so that $Z_5$ and $Z_6$ carry the non-TBI component only and remain distinct from $Z_7$. Each of the three doses is then divided by its own observed maximum, so that all three lie in $[0, 1]$. After the subtraction, the observed maxima are 69~Gy for the abdominal field and 78~Gy for the pelvic field, which reverses the ordering of the two before the subtraction, and 15.75~Gy for TBI. Six chemotherapy-class indicators $Z_8$ to $Z_{13}$ record exposure to alkylating, platinum-compound, anthracycline--anthraquinone, antibiotic, antimetabolite, and epipodophyllotoxin agents. Detailed CCSS variable definitions are listed in Table~\ref{supp_table: analysis3_covariates}.

The full model comprises 15 terms: the thirteen main effects $Z_1$ to $Z_{13}$, with race contributing the two indicators $Z_2$ and $Z_3$ but counting as a single term, together with the three interactions $Z_1 Z_4$, $Z_1 Z_5$, and $Z_1 Z_6$. The three interactions enter the outcome model and the two Cox PH censoring models; SRF and the stratified ECDF are fitted on the main-effects covariate list only. The stratified ECDF is formed by crossing the quartiles of $Z_1$, the mean splits of the first two principal components from a PCA on $(Z_5, Z_6, Z_7)$, the three race categories, and the BMT indicator.

\begin{table}[H]
\centering
\footnotesize
\caption{Covariate definitions for the extended covariate set of Section~\ref{sec3: extended analysis}.}\label{supp_table: analysis3_covariates}
\renewcommand{\arraystretch}{1.1}
\resizebox{\textwidth}{!}{%
\begin{tabular}{cll}
\toprule
Label    & CCSS variable                       & Description \\
\midrule
$Z_1$    & \texttt{age\_dx\_rescale}           & Rescaled age at primary cancer diagnosis \\
$Z_2$    & \texttt{race\_3 = ``Black''}        & Race: African American (vs.\ Caucasian reference) \\
$Z_3$    & \texttt{race\_3 = ``Other''}        & Race: Other (vs.\ Caucasian reference) \\
$Z_4$    & \texttt{bmt\_tbi = ``Yes''}         & Bone marrow transplant \\
$Z_5$    & \texttt{abdmaxrtdose\_rescale}      & Rescaled max abdominal radiation dose, net of TBI \\
$Z_6$    & \texttt{pelvismaxrtdose\_rescale}   & Rescaled max pelvic radiation dose, net of TBI \\
$Z_7$    & \texttt{tbidose\_rescale}           & Rescaled TBI dose \\
$Z_8$    & \texttt{alky\_ever = ``Yes''}       & Any alkylating-agent exposure \\
$Z_9$    & \texttt{pc\_ever = ``Yes''}         & Any platinum-compound exposure \\
$Z_{10}$ & \texttt{aa\_ever = ``Yes''}         & Any anthracycline--anthraquinone exposure \\
$Z_{11}$ & \texttt{ab\_ever = ``Yes''}         & Any antibiotic-chemotherapy exposure \\
$Z_{12}$ & \texttt{am\_ever = ``Yes''}         & Any antimetabolite exposure \\
$Z_{13}$ & \texttt{ep\_ever = ``Yes''}         & Any epipodophyllotoxin exposure \\
\midrule
\multicolumn{3}{l}{\textit{Three interactions, added to the outcome model and to the two Cox PH censoring models:}} \\
$Z_1 Z_4$ & \texttt{age\_dx\_rescale} $\times$ \texttt{bmt\_tbi}                 & Age $\times$ BMT \\
$Z_1 Z_5$ & \texttt{age\_dx\_rescale} $\times$ \texttt{abdmaxrtdose\_rescale}    & Age $\times$ abdominal dose \\
$Z_1 Z_6$ & \texttt{age\_dx\_rescale} $\times$ \texttt{pelvismaxrtdose\_rescale} & Age $\times$ pelvic dose \\
\bottomrule
\end{tabular}
}
\par\vspace{0.2cm}
\begingroup
\raggedright \footnotesize \setlength{\parindent}{0pt}
\textit{Note:} The three interactions $Z_1 Z_4$, $Z_1 Z_5$, and $Z_1 Z_6$ are added to the outcome model and to the two Cox PH censoring models. The abdominal and pelvic field maxima recorded in the CCSS include the TBI dose; $Z_5$ and $Z_6$ are those maxima net of the TBI dose, so that they carry the non-TBI component only. Each of $Z_5$, $Z_6$, and $Z_7$ is then divided by its own observed maximum and lies in $[0, 1]$; the maxima are 69~Gy, 78~Gy, and 15.75~Gy respectively. Race contributes two indicators but counts as a single term, so the full model comprises 15 terms. The $Z_j$ labels in this section are specific to the extended covariate set and do not match those used elsewhere in the paper.
\par\endgroup
\end{table}

%
%
\subsection{Variable Selection Results under $\mathrm{QICW}_{u}$}\label{supp: var_select_qicwu}

This section reports the $2P$-penalty counterpart of the results in Section~\ref{sec3: extended analysis} of the main text. Backward elimination is run under $\mathrm{QICW}_{u}$ in the same four settings and at the same 24 analysis ages, giving a further 96 runs, and is otherwise carried out exactly as described there. Tables~\ref{supp_table: vs_qicwu_by_approach} and~\ref{supp_table: vs_qicwu_by_age} correspond to Tables~\ref{table: vs_qicw_by_approach} and~\ref{table: vs_qicw_by_age} of the main text and are constructed in the same way.

The same five terms are retained in every run, and the two criteria give broadly similar orderings among the remaining terms. The TBI dose $Z_{7}$ is retained in all runs under $\mathrm{QICW}_{u}$, whereas it drops at the latest analysis ages under $\mathrm{QICW}$. The mean selected model size is larger under $\mathrm{QICW}_{u}$ in all four settings, though only slightly, and it grows with $t_{0}$ under both criteria.

\begin{table}[H]
\centering
\footnotesize
\caption{Percentage of analysis ages at which each term is retained by the $\mathrm{QICW}_{u}$-driven backward elimination, by setting.}\label{supp_table: vs_qicwu_by_approach}
\renewcommand{\arraystretch}{1.15}
\begin{tabular}{lcccc}
\toprule
Term & A, no Firth & A, Firth & B, no Firth & B, Firth \\
\midrule
$Z_{1}$, $Z_{4}$, $Z_{5}$, $Z_{6}$, $Z_{1}Z_{6}$ & 100 & 100 & 100 & 100 \\
$Z_{7}$        & 100 & 100 & 100 & 100 \\
$Z_{8}$        &  92 & 100 &  96 & 100 \\
$Z_{11}$       &  83 &  88 &  83 & 100 \\
$Z_{2}$, $Z_{3}$ & 71 & 100 &  46 & 100 \\
$Z_{12}$       &  63 & 100 &  46 & 100 \\
$Z_{1}Z_{4}$   &  67 &  63 &  83 &  83 \\
$Z_{10}$       &  63 & 100 &  25 &  83 \\
$Z_{13}$       &  58 &  92 &  33 &  83 \\
$Z_{9}$        &  42 &  58 &  25 &  63 \\
$Z_{1}Z_{5}$   &  46 &  38 &   4 &   0 \\
\midrule
Mean model size (terms) & 11.8 & 13.4 & 10.4 & 13.1 \\
\bottomrule
\end{tabular}
\par\vspace{0.2cm}
\begingroup
\raggedright \footnotesize \setlength{\parindent}{0pt}
\textit{Note:} The $2P$-penalty counterpart of Table~\ref{table: vs_qicw_by_approach} of the main text, with the same term ordering and the same pooling over the 24 analysis ages. Covariate definitions are given in Table~\ref{supp_table: analysis3_covariates}.
\par\endgroup
\end{table}

\begin{table}[H]
\centering
\footnotesize
\caption{Percentage of runs at which each term is retained by the $\mathrm{QICW}_{u}$-driven backward elimination, by analysis age.}\label{supp_table: vs_qicwu_by_age}
\renewcommand{\arraystretch}{1.15}
\begin{tabular}{lccc}
\toprule
Term & $t_{0} = 17$--$24$ & $t_{0} = 25$--$35$ & $t_{0} = 36$--$40$ \\
\midrule
$Z_{1}$, $Z_{4}$, $Z_{5}$, $Z_{6}$, $Z_{1}Z_{6}$ & 100 & 100 & 100 \\
$Z_{7}$        & 100 & 100 & 100 \\
$Z_{8}$        &  91 & 100 & 100 \\
$Z_{11}$       & 100 &  82 &  85 \\
$Z_{2}$, $Z_{3}$ & 53 &  89 & 100 \\
$Z_{12}$       &  50 &  86 & 100 \\
$Z_{1}Z_{4}$   &  44 &  91 &  85 \\
$Z_{10}$       &  66 &  61 &  85 \\
$Z_{13}$       &  44 &  68 & 100 \\
$Z_{9}$        &  28 &  41 &  90 \\
$Z_{1}Z_{5}$   &   6 &  20 &  50 \\
\midrule
Mean model size (terms) & 10.8 & 12.4 & 14.0 \\
\bottomrule
\end{tabular}
\par\vspace{0.2cm}
\begingroup
\raggedright \footnotesize \setlength{\parindent}{0pt}
\textit{Note:} The $2P$-penalty counterpart of Table~\ref{table: vs_qicw_by_age} of the main text, with the same pooling into 32, 44, and 20 runs. Term ordering matches Table~\ref{supp_table: vs_qicwu_by_approach}.
\par\endgroup
\end{table}

\section{Additional Results of Simulation Studies}
%
%
%
%
\subsection{Numerical Results of Simulation 1}\label{supp: restul_of_simulation_set1}
\begin{table}[H]
\centering
\caption{Descriptive statistics of Simulation~1 datasets stratified by event type. The $n$ row reports the average number of subjects across simulation iterations.}\label{table: desc_sim1_data}
\renewcommand{\arraystretch}{1.2}
\setlength{\tabcolsep}{1pt}
\resizebox{1\textwidth}{!}{
\begin{tabular}{|cc|cccc|}
\hline
\multicolumn{2}{|c|}{\multirow{3}{*}{\textbf{Simulation Scenarios}}}                              & \multicolumn{4}{c|}{\textbf{\begin{tabular}[c]{@{}c@{}}Premature Menopause Status\\ (Average count/ Average Median (Q1, Q3))\end{tabular}}}                                                                                                          \\ \cline{3-6} 
\multicolumn{2}{|c|}{}                                                                            & \multicolumn{3}{c|}{\textbf{Observed Event}}                                                                                                                                                              & \multirow{2}{*}{\textbf{No Observed Event}} \\ \cline{3-5}
\multicolumn{2}{|c|}{}                                                                            & \multicolumn{1}{c|}{\textbf{Any (POI/SPM)}} & \multicolumn{1}{c|}{\textbf{POI}}         & \multicolumn{1}{c|}{\textbf{SPM}}         &                                             \\ \hline
\multicolumn{1}{|c|}{\multirow{2}{*}{\textbf{Overall\textsuperscript{$\dagger$}}}} & $V$\textsuperscript{$\S$}                      & \multicolumn{1}{c|}{11.63 (7.79, 15.48)}    & \multicolumn{1}{c|}{11.50 (7.70, 15.36)}  & \multicolumn{1}{c|}{11.81 (7.94, 15.64)}  & 17.09 (14.04, 19.24)                        \\ \cline{2-6} 
\multicolumn{1}{|c|}{}                                  & $C$\textsuperscript{$\S$}                      & \multicolumn{1}{c|}{32.73 (27.66, 37.80)}   & \multicolumn{1}{c|}{32.63 (27.57, 37.70)} & \multicolumn{1}{c|}{32.89 (27.80, 37.94)} & 37.55 (33.02, 41.92)                        \\ \hline
\multicolumn{1}{|c|}{\multirow{2}{*}{\begin{tabular}[c]{@{}c@{}}$t_0 = 16$\textsuperscript{$\ddagger$}\\ (0.43\% censored)\end{tabular}}}       & $n$    & \multicolumn{1}{c|}{46.24}                  & \multicolumn{1}{c|}{25.08}                & \multicolumn{1}{c|}{21.16}                & \textit{6953.76}                            \\ \cline{2-6} 
\multicolumn{1}{|c|}{}                                  & Observed $T$\textsuperscript{$\S$}             & \multicolumn{1}{c|}{13.98 (11.90, 15.36)}   & \multicolumn{1}{c|}{14.00 (12.03, 15.34)} & \multicolumn{1}{c|}{13.83 (11.82, 15.23)} & -                                           \\ \hline
\multicolumn{1}{|c|}{\multirow{2}{*}{\begin{tabular}[c]{@{}c@{}}$t_0 = 25$\textsuperscript{$\ddagger$}\\ (8.17\% censored)\end{tabular}}}       & $n$    & \multicolumn{1}{c|}{310.69}                 & \multicolumn{1}{c|}{173.10}               & \multicolumn{1}{c|}{137.58}               & \textit{6689.31}                            \\ \cline{2-6} 
\multicolumn{1}{|c|}{}                                  & Observed $T$\textsuperscript{$\S$}             & \multicolumn{1}{c|}{21.54 (18.54, 23.49)}   & \multicolumn{1}{c|}{21.60 (18.67, 23.51)} & \multicolumn{1}{c|}{21.43 (18.40, 23.44)} & -                                           \\ \hline
\multicolumn{1}{|c|}{\multirow{2}{*}{\begin{tabular}[c]{@{}c@{}}$t_0 = 30$\textsuperscript{$\ddagger$}\\ (21.05\% censored)\end{tabular}}}       & $n$    & \multicolumn{1}{c|}{664.76}                 & \multicolumn{1}{c|}{376.49}               & \multicolumn{1}{c|}{288.28}               & \textit{6335.24}                            \\ \cline{2-6} 
\multicolumn{1}{|c|}{}                                  & Observed $T$\textsuperscript{$\S$}             & \multicolumn{1}{c|}{25.44 (21.85, 27.96)}   & \multicolumn{1}{c|}{25.51 (21.97, 27.99)} & \multicolumn{1}{c|}{25.35 (21.70, 27.91)} & -                                           \\ \hline
\multicolumn{1}{|c|}{\multirow{2}{*}{\begin{tabular}[c]{@{}c@{}}$t_0 = 40$\textsuperscript{$\ddagger$}\\ (58.32\% censored)\end{tabular}}}       & $n$    & \multicolumn{1}{c|}{1625.93}                & \multicolumn{1}{c|}{940.60}               & \multicolumn{1}{c|}{685.33}               & \textit{5374.07}                            \\ \cline{2-6} 
\multicolumn{1}{|c|}{}                                  & Observed $T$\textsuperscript{$\S$}             & \multicolumn{1}{c|}{31.67 (26.65, 35.72)}   & \multicolumn{1}{c|}{31.81 (26.84, 35.81)} & \multicolumn{1}{c|}{31.48 (26.41, 35.59)} & -                                           \\ \hline
\end{tabular}
}
\vspace{0.2cm}
\par
\begingroup
\raggedright \footnotesize \setlength{\parindent}{0pt}
\textsuperscript{$\S$} For continuous variables ($V$, $C$, and Observed $T$), values are reported as the average median (1st quartile, 3rd quartile).\\[0.1cm]
\textsuperscript{$\dagger$} Overall descriptive statistics for total subjects: cancer diagnosis age ($V$) is 14.85 (10.51, 18.18); censoring age ($C$) is 35.40 (30.24, 40.29). \\[0.1cm]
\textsuperscript{$\ddagger$} The censoring rate (percentage censored before $t_0$) is shown beneath each $t_0$ label. 
\par\endgroup
\end{table}
\newpage

\begin{figure}[H]
  \centering
  \includegraphics[width=1\linewidth]{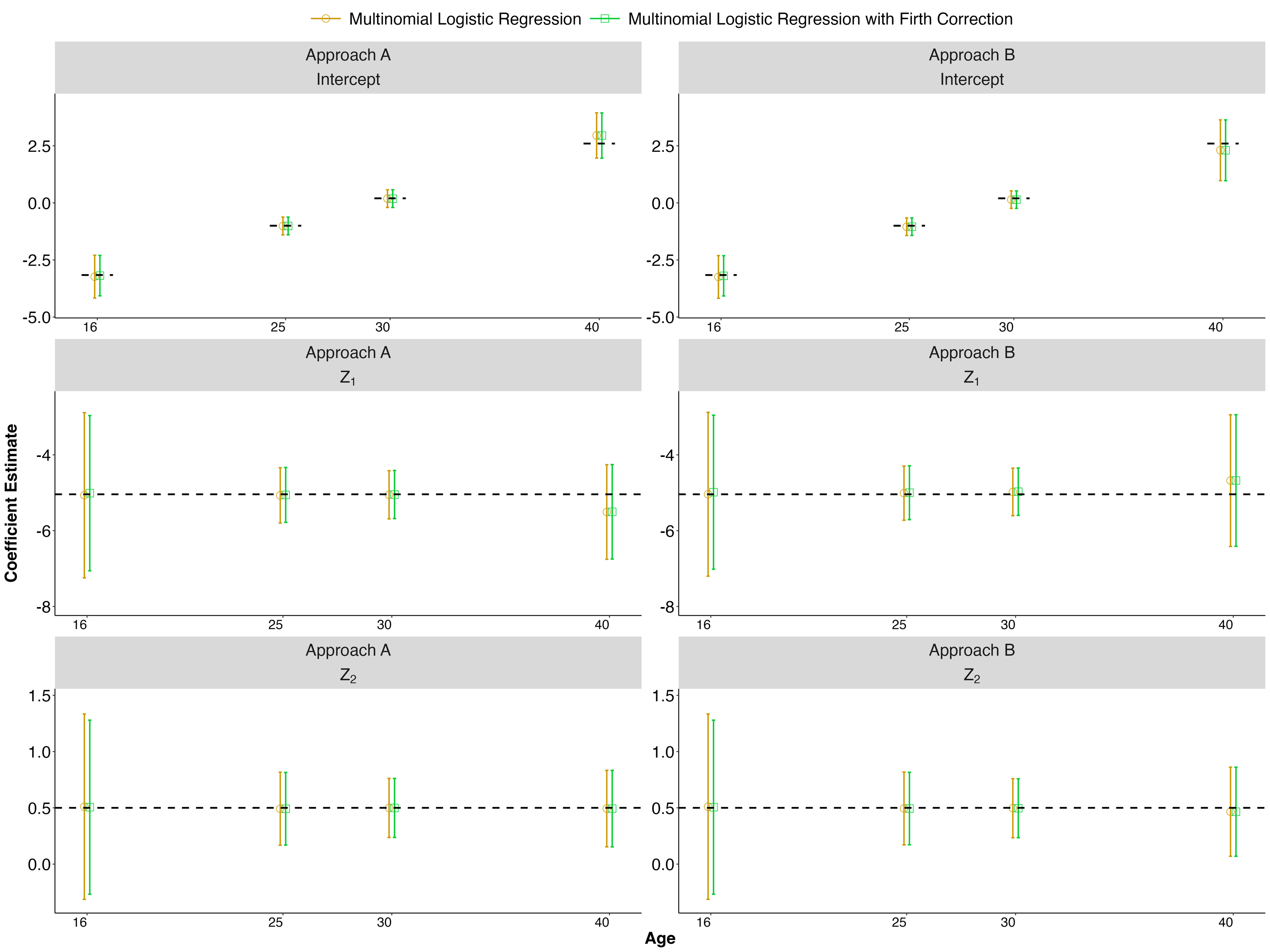}
  \caption{Estimated coefficients for the POI outcome across ages in simulation 1. The $\hat{G}(\cdot \mid \mathbf{Z})$ is obtained using SRF. When using SRF, we choose number of tree equal to 100 and node size equal to 500. The dashed line represents true value of corresponding coefficient.}\label{fig: sim1_poi_all_coef}
\end{figure}

\begin{figure}[H]
  \centering
  \includegraphics[width=1\linewidth]{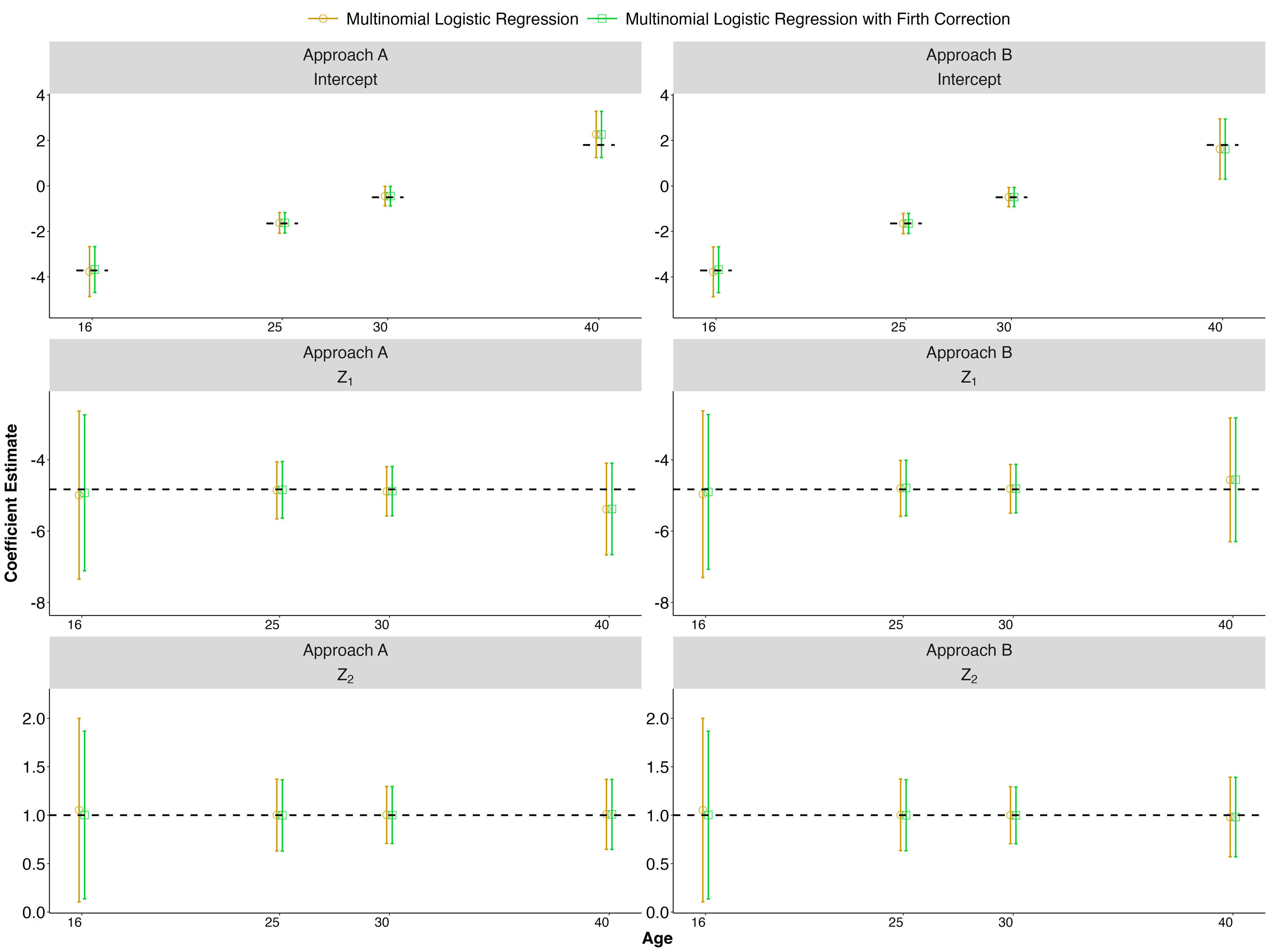}
  \caption{Estimated coefficients for the SPM outcome across ages in simulation 1. The $\hat{G}(\cdot \mid \mathbf{Z})$ is obtained using SRF. When using SRF, we choose number of tree equal to 100 and node size equal to 500. The dashed line represents true value of corresponding coefficient.}\label{fig: sim1_spm_all_coef}
\end{figure}

Figures~\ref{fig: sim1_poi_all_coef} and~\ref{fig: sim1_spm_all_coef} present the estimated coefficients for the POI and SPM outcomes, respectively. For both outcomes, the sample means of the estimates are close to the true parameter values across all ages, and the 95\% CIs consistently cover the true values under both Approaches~A and~B, confirming the consistency of the proposed estimators.

\begin{figure}[H]
  \centering
  \includegraphics[width=1\linewidth]{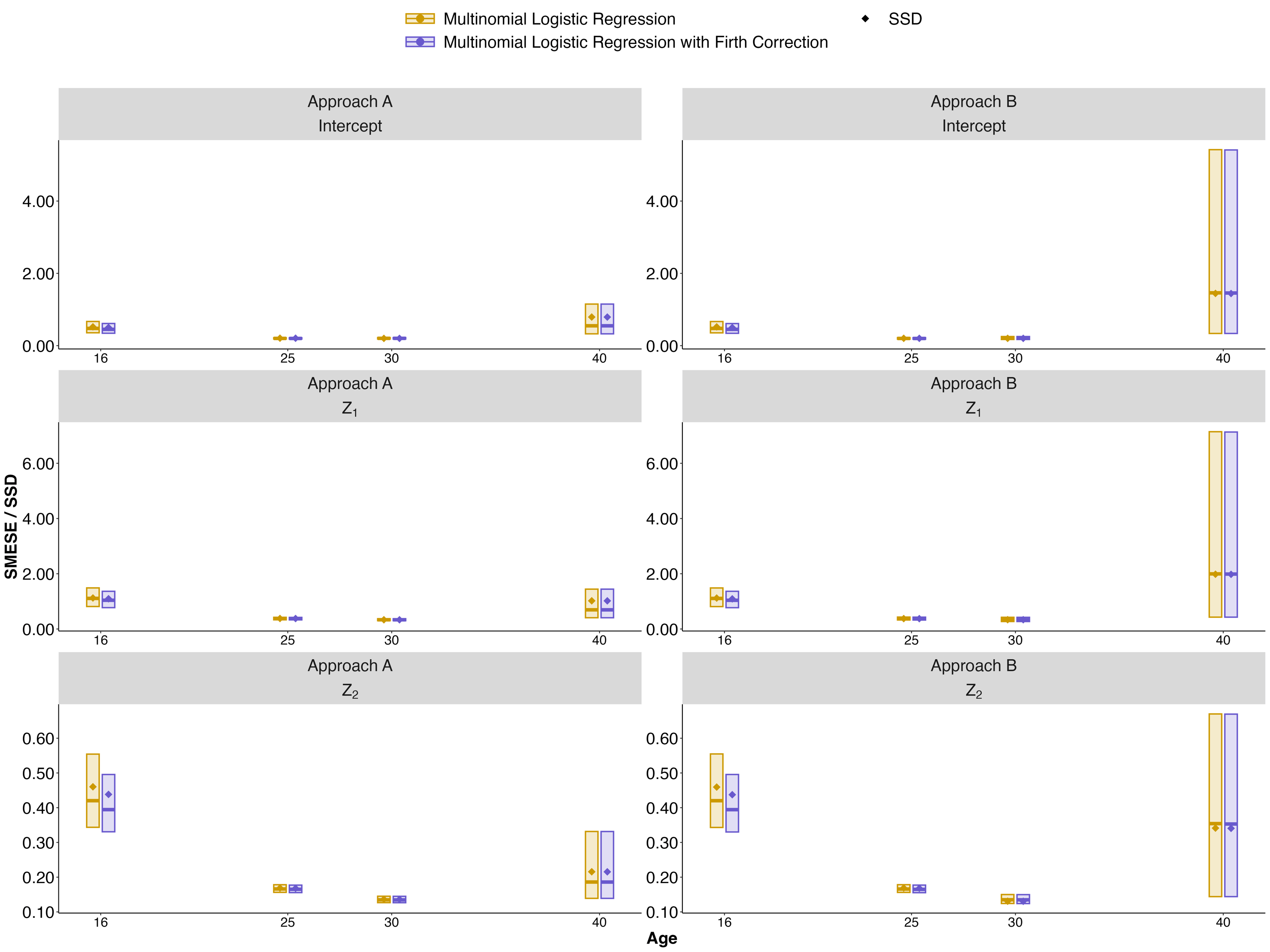}
  \caption{Sample mean of estimated standard errors (SMESE) with 2.5\% - 97.5\% quantile compared to sample standard deviation of estimates (SSD) while outcome is POI in simulation 1. The $\hat{G}(\cdot \mid \mathbf{Z})$ is obtained using SRF. When using SRF, we choose number of tree equal to 100 and node size equal to 500.}\label{fig: sim1_poi_all_sse_smese}
\end{figure}

\begin{figure}[H]
  \centering
  \includegraphics[width=1\linewidth]{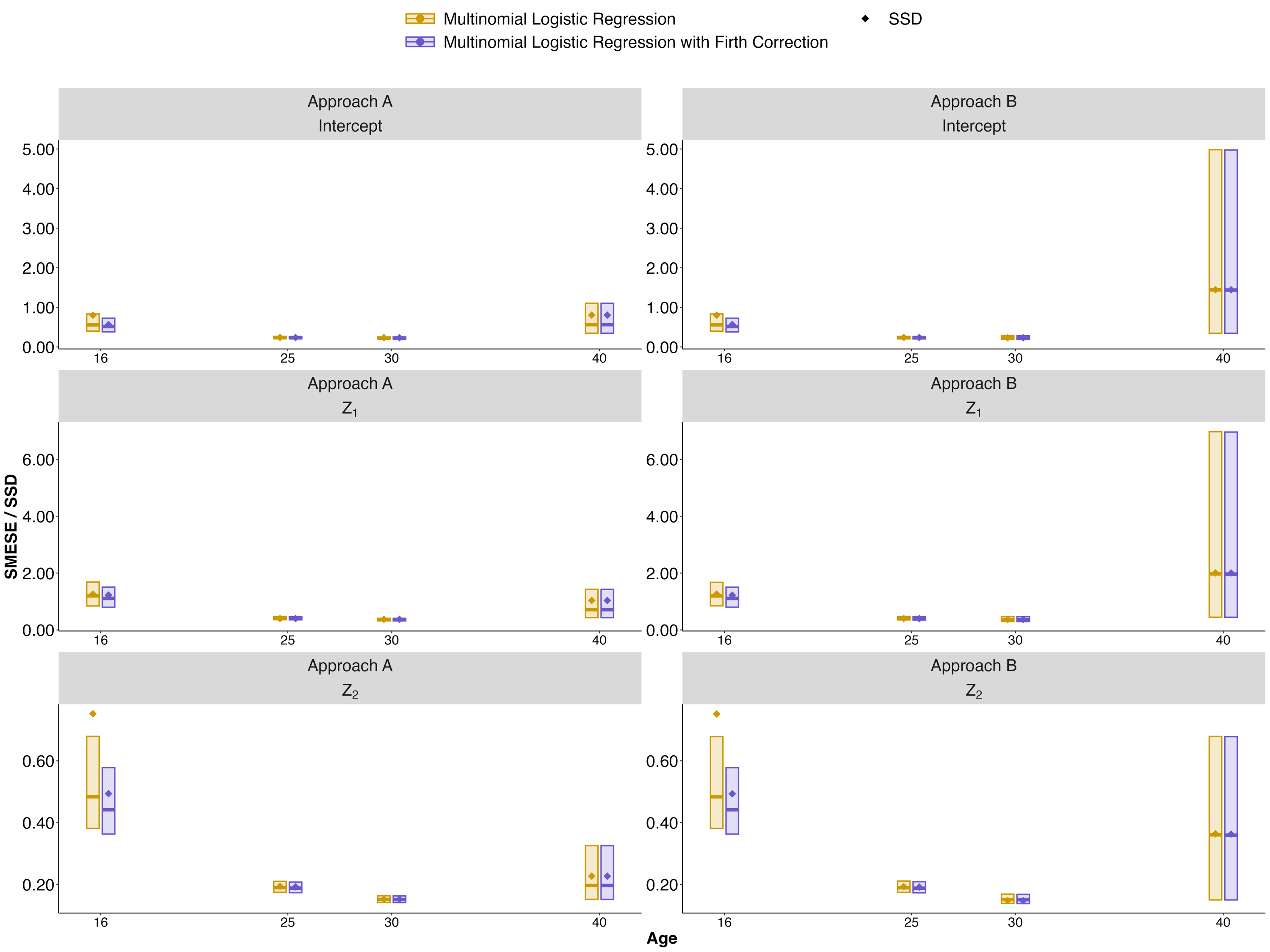}
  \caption{Sample mean of estimated standard errors (SMESE) with 2.5\% - 97.5\% quantile compared to sample standard deviation of estimates (SSD) while outcome is SPM in simulation 1. The $\hat{G}(\cdot \mid \mathbf{Z})$ is obtained using SRF. When using SRF, we choose number of tree equal to 100 and node size equal to 500.}\label{fig: sim1_spm_all_sse_smese}
\end{figure}

We assess the validity of the sandwich standard error estimator by comparing the sample mean of the estimated standard errors (SMESE), along with the 2.5\%--97.5\% quantile range, with the sample standard deviation of the estimates (SSD). For the POI outcome (Figure~\ref{fig: sim1_poi_all_sse_smese}), the SMESE and SSD are in close agreement across all coefficients and ages, indicating that the sandwich standard error estimator performs well. For the SPM outcome (Figure~\ref{fig: sim1_spm_all_sse_smese}), the multinomial logistic regression without Firth correction underestimates the standard error of $\hat{\beta}_{22}$, as the SSD exceeds the SMESE. With Firth correction, the SMESE and SSD are in close agreement for all coefficients, demonstrating that the Firth-corrected estimator obtains reliable standard error estimation even for the rarer SPM outcome.

Under Approach~A, the SSD is generally smaller than under Approach~B, particularly at later ages, indicating that Approach~A is more efficient in this setting despite the relatively high censoring rate at age 40 (58.32\%). This is because a sufficient proportion of PM events are observed (on average 23.22\%), providing Approach~A with enough information. However, the variation of the sandwich standard error estimator is large at age 16 under both approaches and at age 40 under Approach~B. At age 16, this is due to the small number of observed events. At age 40 under Approach~B, the $G(\cdot \mid \mathbf{Z})$ is close to zero for some subjects in certain repetitions, and consequently $\hat{G}(\cdot \mid \mathbf{Z})$ is as well, resulting in large IPCW that inflate the standard error estimator. This indicates that Approach~B is more sensitive to the estimation of the censoring distribution.

\begin{figure}[H]
    \centering
    \begin{subfigure}[t]{1\textwidth}
        \centering
        \includegraphics[width=0.8\textwidth]{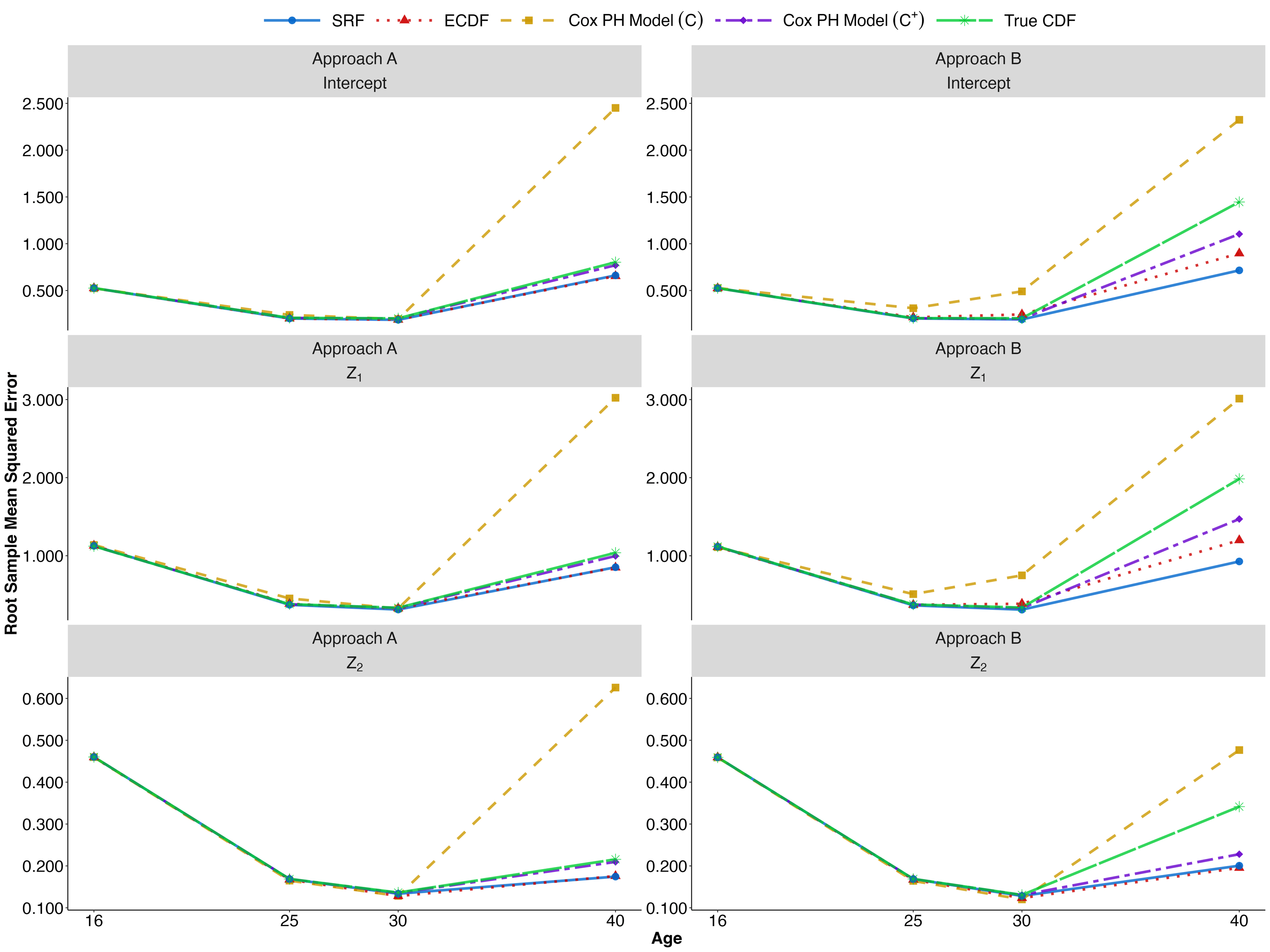}
        \caption{Without Firth Correction}
        \label{fig: proj2_sim1_poi_rmse_nf}
    \end{subfigure}%
    \vfill
    \begin{subfigure}[t]{0.8\textwidth}
        \centering
        \includegraphics[width=1\textwidth]{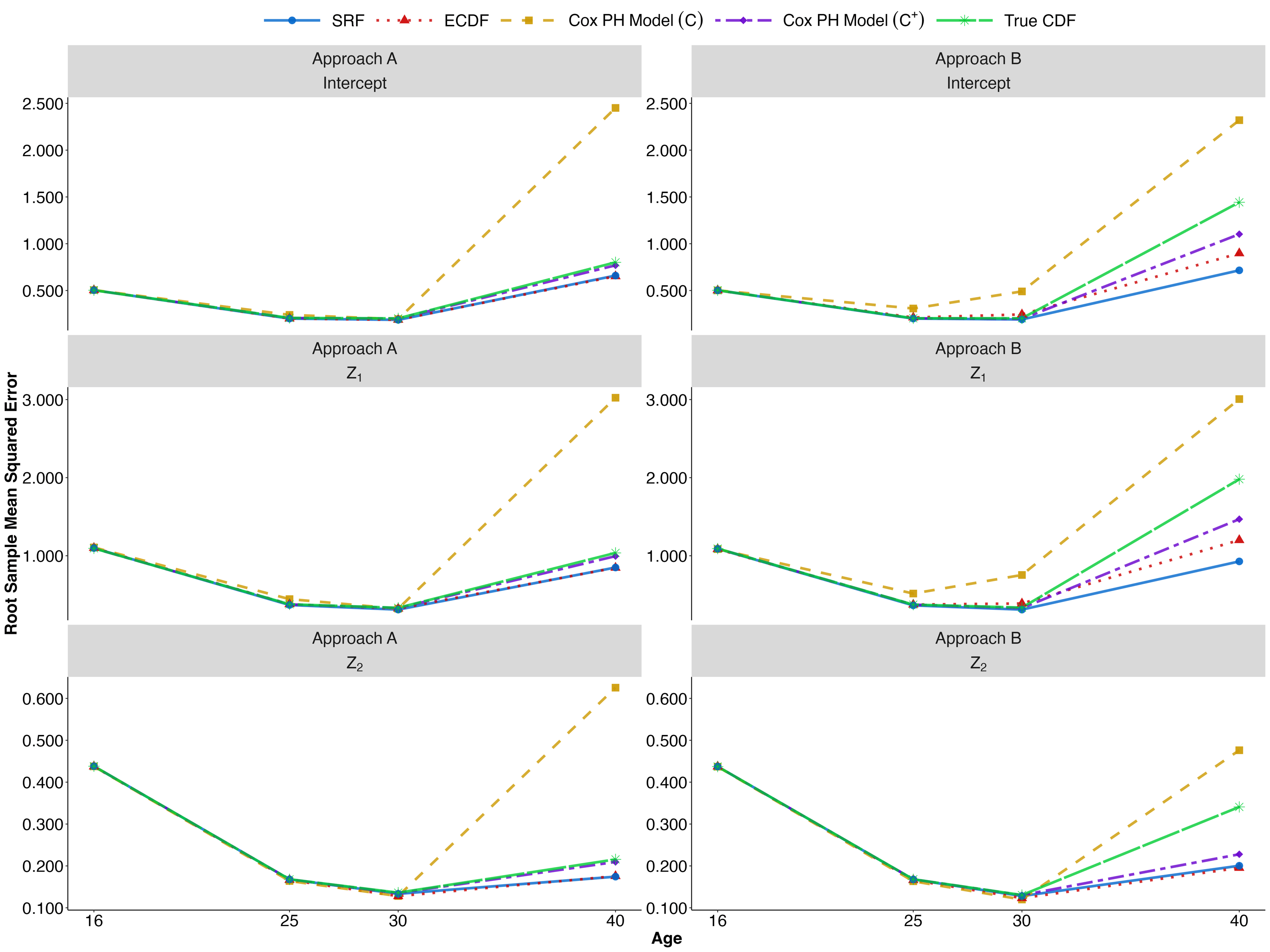}
        \caption{With Firth Correction}
        \label{fig: proj2_sim1_poi_rmse_f}
    \end{subfigure}%
  \caption{The comparison of the root sample mean squared error of the estimates (RSMSE) while outcome is POI in simulation 1. The $\hat{G}(\cdot \mid \mathbf{Z})$ is obtained using stratified ECDF, Cox PH model on $C$, Cox PH model on $C^{*}$, True CDF, and SRF. When using SRF, we choose number of tree equal to 100 and node size equal to 500.}\label{fig: proj2_sim1_poi_rmse_all}
\end{figure}

\begin{figure}[H]
    \centering
    \begin{subfigure}[t]{0.8\textwidth}
        \centering
        \includegraphics[width=1\textwidth]{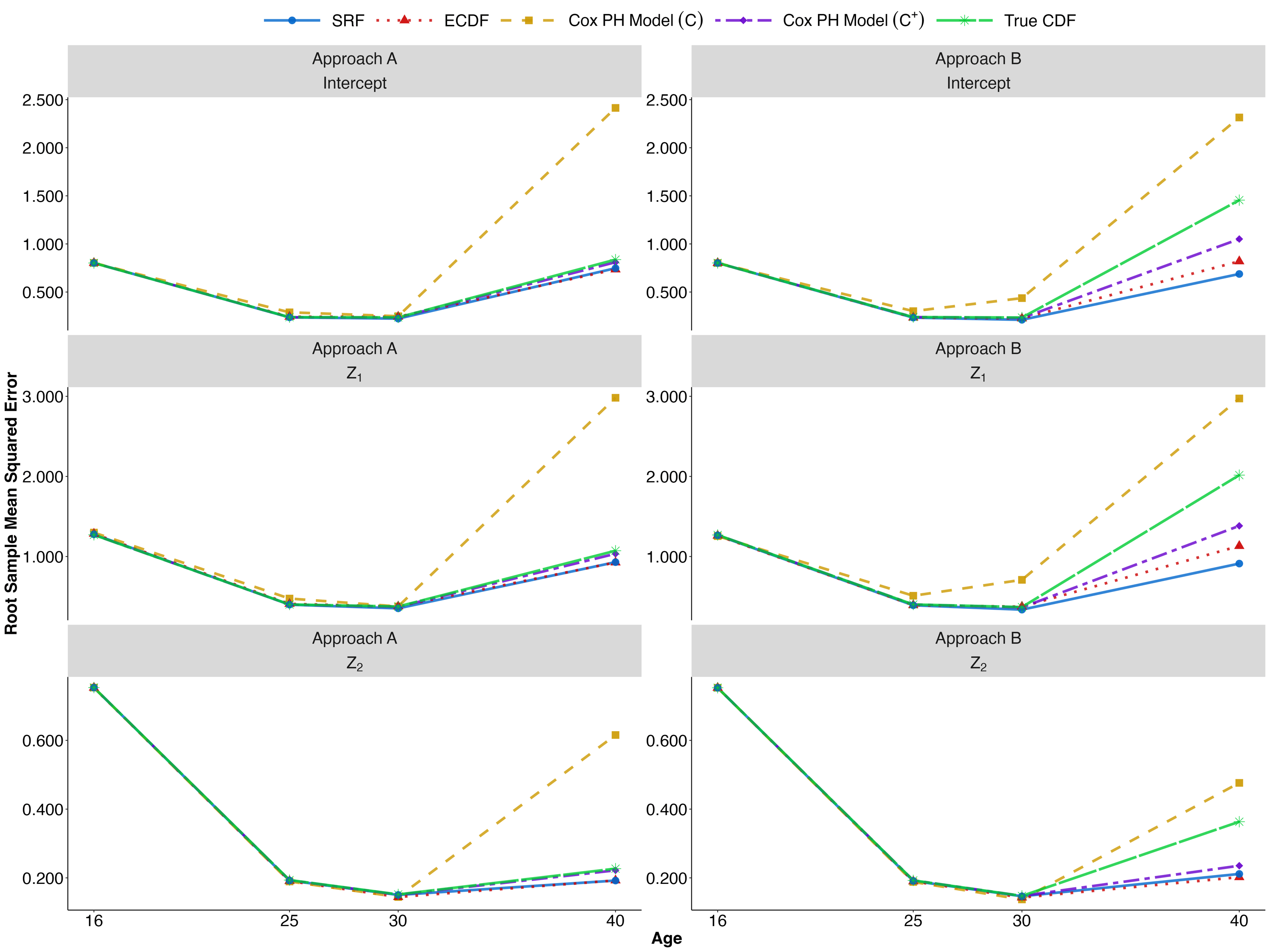}
        \caption{Without Firth Correction}
        \label{fig: proj2_sim1_spm_rmse_nf}
    \end{subfigure}%
    \hfill
    \begin{subfigure}[t]{0.8\textwidth}
        \centering
        \includegraphics[width=1\textwidth]{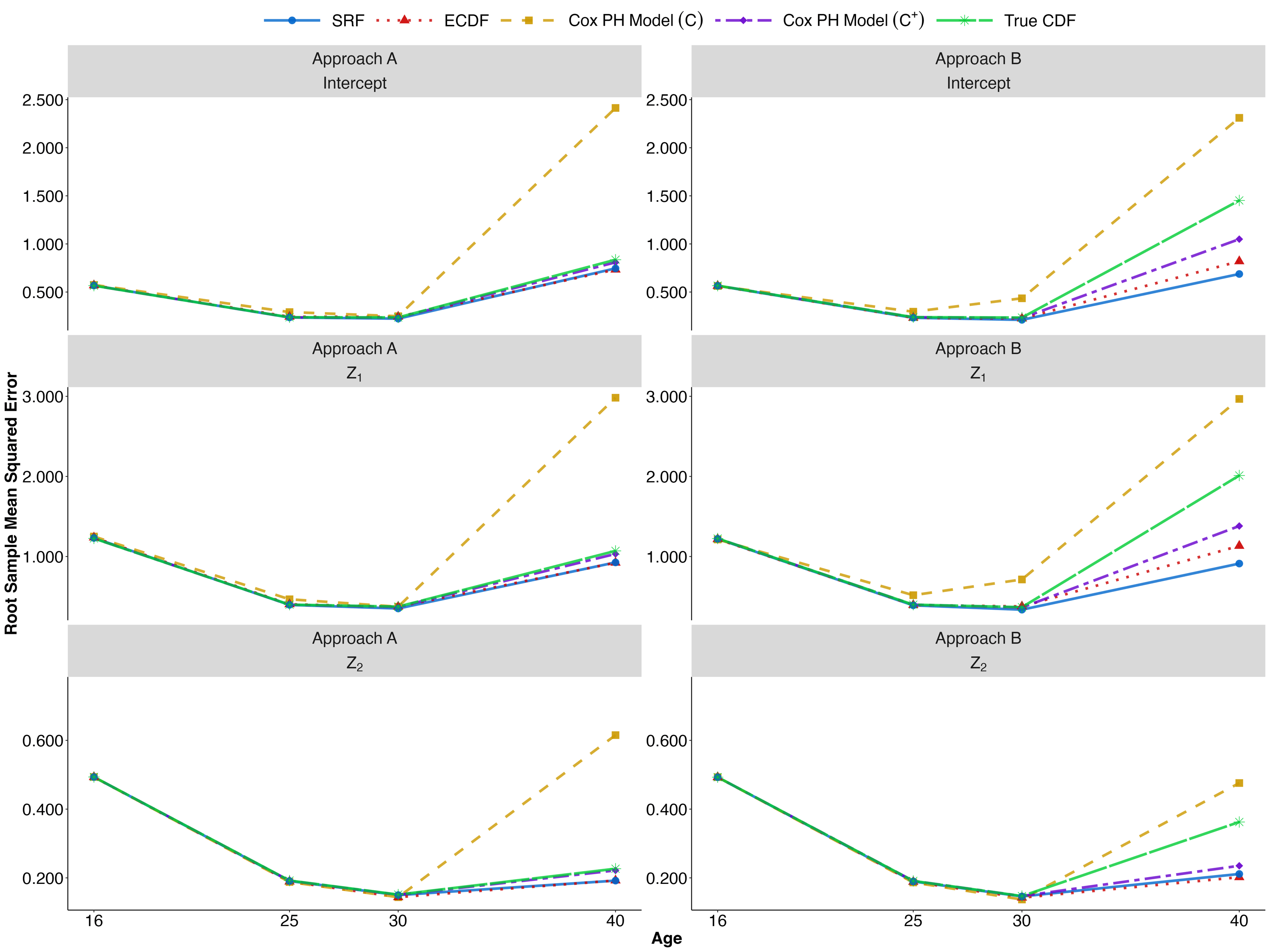}
        \caption{With Firth Correction}
        \label{fig: proj2_sim1_spm_rmse_f}
    \end{subfigure}%
  \caption{The comparison of the root sample mean squared error of the estimates (RSMSE) while outcome is SPM in simulation 1. The $\hat{G}(\cdot \mid \mathbf{Z})$ is obtained using stratified ECDF, Cox PH model on $C$, Cox PH model on $C^{*}$, True CDF, and SRF. When using SRF, we choose number of tree equal to 100 and node size equal to 500.}\label{fig: proj2_sim1_spm_rmse_all}
\end{figure}

We now compare the impact of the censoring distribution estimator on coefficient estimation by evaluating the RSMSE across the four methods described in Section~\ref{sec2: estimation procedure}. Figures~\ref{fig: proj2_sim1_poi_rmse_all} and~\ref{fig: proj2_sim1_spm_rmse_all} present the RSMSE for the POI and SPM outcomes, respectively. For both outcomes, the results are similar with and without Firth correction. Under Approach~A, the SRF, stratified ECDF, and Cox PH model on $C^{*}$ all obtain comparable RSMSEs that are close to the benchmark obtained using the true censoring distribution, while the Cox PH model on $C$ obtains substantially higher RSMSE, particularly at later ages. Under Approach~B, the differences among methods are more pronounced at later ages: the SRF obtains the lowest RSMSE, followed by the stratified ECDF and the Cox PH model on $C^{*}$, while the Cox PH model on $C$ again performs the worst. The poor performance of the Cox PH model on $C$ is due to the violation of the proportional hazards assumption for the censoring time $C$. Overall, the SRF and the stratified ECDF provide the most accurate coefficient estimation across both approaches, while the Cox PH model on $C^{*}$ serves as a reasonable alternative when the proportional hazards assumption for the censoring time is plausible.


%
%
%
%
\subsection{Numerical Results of Simulation 2}\label{supp: restul_of_simulation_set2}
\begin{table}[H]
\centering
\caption{Descriptive statistics of Simulation~2 datasets stratified by event type. The $n$ row reports the average number of subjects across simulation iterations; the $n$ ($Z_2=1$) row reports the corresponding counts among subjects with $Z_2=1$. The sparse SPM counts among $Z_2=1$ subjects (e.g., 0.83 at $t_0=16$) illustrate the data imbalance that induces separation.}\label{table: desc_sim2_data}
\renewcommand{\arraystretch}{1.2}
\setlength{\tabcolsep}{1pt}
\resizebox{1\textwidth}{!}{
\begin{tabular}{|cc|cccc|}
\hline
\multicolumn{2}{|c|}{\multirow{3}{*}{\textbf{Simulation Scenarios}}}                              & \multicolumn{4}{c|}{\textbf{\begin{tabular}[c]{@{}c@{}}Premature Menopause Status\\ (Average count/ Average Median (Q1, Q3))\end{tabular}}}                                                                                                          \\ \cline{3-6} 
\multicolumn{2}{|c|}{}                                                                            & \multicolumn{3}{c|}{\textbf{Observed Event}}                                                                                                                                                              & \multirow{2}{*}{\textbf{No Observed Event}} \\ \cline{3-5}
\multicolumn{2}{|c|}{}                                                                            & \multicolumn{1}{c|}{\textbf{Any (POI/SPM)}} & \multicolumn{1}{c|}{\textbf{POI}}         & \multicolumn{1}{c|}{\textbf{SPM}}         &                                             \\ \hline
\multicolumn{1}{|c|}{\multirow{2}{*}{\textbf{Overall\textsuperscript{$\dagger$}}}} & $V$\textsuperscript{$\S$}                      & \multicolumn{1}{c|}{5.82 (3.46, 8.86)}      & \multicolumn{1}{c|}{5.86 (3.50, 8.91)}    & \multicolumn{1}{c|}{5.68 (3.43, 8.62)}    & 10.67 (7.07, 14.26)                         \\ \cline{2-6} 
\multicolumn{1}{|c|}{}                                  & $C$\textsuperscript{$\S$}                      & \multicolumn{1}{c|}{28.77 (24.18, 33.43)}   & \multicolumn{1}{c|}{28.78 (24.20, 33.47)} & \multicolumn{1}{c|}{28.69 (24.19, 33.18)} & 32.74 (27.82, 37.59)                        \\ \hline
\multicolumn{1}{|c|}{\multirow{3}{*}{\begin{tabular}[c]{@{}c@{}}$t_0 = 16$\textsuperscript{$\ddagger$}\\ (0.82\% censored)\end{tabular}}}       & $n$              & \multicolumn{1}{c|}{36.54}                  & \multicolumn{1}{c|}{29.88}                & \multicolumn{1}{c|}{6.66}                 & \textit{6963.46}                            \\ \cline{2-6} 
\multicolumn{1}{|c|}{}                                  & $n$ ($Z_2=1$)    & \multicolumn{1}{c|}{8.99}                   & \multicolumn{1}{c|}{8.16}                 & \multicolumn{1}{c|}{0.83}                 & \textit{341.22}                             \\ \cline{2-6} 
\multicolumn{1}{|c|}{}                                  & Observed $T$\textsuperscript{$\S$}             & \multicolumn{1}{c|}{14.64 (12.81, 15.81)}   & \multicolumn{1}{c|}{14.58 (12.74, 15.77)} & \multicolumn{1}{c|}{14.65 (13.38, 15.47)} & -                                           \\ \hline
\multicolumn{1}{|c|}{\multirow{3}{*}{\begin{tabular}[c]{@{}c@{}}$t_0 = 18$\textsuperscript{$\ddagger$}\\ (1.90\% censored)\end{tabular}}}       & $n$              & \multicolumn{1}{c|}{55.96}                  & \multicolumn{1}{c|}{45.67}                & \multicolumn{1}{c|}{10.28}                & \textit{6944.04}                            \\ \cline{2-6} 
\multicolumn{1}{|c|}{}                                  & $n$ ($Z_2=1$)    & \multicolumn{1}{c|}{13.22}                  & \multicolumn{1}{c|}{12.09}                & \multicolumn{1}{c|}{1.13}                 & \textit{337.00}                             \\ \cline{2-6} 
\multicolumn{1}{|c|}{}                                  & Observed $T$\textsuperscript{$\S$}             & \multicolumn{1}{c|}{16.07 (13.96, 17.50)}   & \multicolumn{1}{c|}{16.02 (13.90, 17.46)} & \multicolumn{1}{c|}{16.14 (14.49, 17.32)} & -                                           \\ \hline
\multicolumn{1}{|c|}{\multirow{3}{*}{\begin{tabular}[c]{@{}c@{}}$t_0 = 20$\textsuperscript{$\ddagger$}\\ (3.83\% censored)\end{tabular}}}       & $n$              & \multicolumn{1}{c|}{85.17}                  & \multicolumn{1}{c|}{69.52}                & \multicolumn{1}{c|}{15.65}                & \textit{6914.83}                            \\ \cline{2-6} 
\multicolumn{1}{|c|}{}                                  & $n$ ($Z_2=1$)    & \multicolumn{1}{c|}{19.83}                  & \multicolumn{1}{c|}{18.10}                & \multicolumn{1}{c|}{1.73}                 & \textit{330.40}                             \\ \cline{2-6} 
\multicolumn{1}{|c|}{}                                  & Observed $T$\textsuperscript{$\S$}             & \multicolumn{1}{c|}{17.58 (15.21, 19.15)}   & \multicolumn{1}{c|}{17.53 (15.16, 19.12)} & \multicolumn{1}{c|}{17.67 (15.61, 19.09)} & -                                           \\ \hline
\multicolumn{1}{|c|}{\multirow{3}{*}{\begin{tabular}[c]{@{}c@{}}$t_0 = 25$\textsuperscript{$\ddagger$}\\ (14.30\% censored)\end{tabular}}}      & $n$              & \multicolumn{1}{c|}{214.06}                 & \multicolumn{1}{c|}{172.62}               & \multicolumn{1}{c|}{41.44}                & \textit{6785.94}                            \\ \cline{2-6} 
\multicolumn{1}{|c|}{}                                  & $n$ ($Z_2=1$)    & \multicolumn{1}{c|}{45.25}                  & \multicolumn{1}{c|}{41.15}                & \multicolumn{1}{c|}{4.10}                 & \textit{304.96}                             \\ \cline{2-6} 
\multicolumn{1}{|c|}{}                                  & Observed $T$\textsuperscript{$\S$}             & \multicolumn{1}{c|}{21.49 (18.50, 23.51)}   & \multicolumn{1}{c|}{21.46 (18.45, 23.49)} & \multicolumn{1}{c|}{21.60 (18.83, 23.50)} & -                                           \\ \hline
\multicolumn{1}{|c|}{\multirow{3}{*}{\begin{tabular}[c]{@{}c@{}}$t_0 = 30$\textsuperscript{$\ddagger$}\\ (33.95\% censored)\end{tabular}}}      & $n$              & \multicolumn{1}{c|}{443.23}                 & \multicolumn{1}{c|}{353.75}               & \multicolumn{1}{c|}{89.48}                & \textit{6556.77}                            \\ \cline{2-6} 
\multicolumn{1}{|c|}{}                                  & $n$ ($Z_2=1$)    & \multicolumn{1}{c|}{82.12}                  & \multicolumn{1}{c|}{74.30}                & \multicolumn{1}{c|}{7.82}                 & \textit{268.11}                             \\ \cline{2-6} 
\multicolumn{1}{|c|}{}                                  & Observed $T$\textsuperscript{$\S$}             & \multicolumn{1}{c|}{25.32 (21.69, 27.87)}   & \multicolumn{1}{c|}{25.27 (21.60, 27.84)} & \multicolumn{1}{c|}{25.50 (22.01, 27.94)} & -                                           \\ \hline
\multicolumn{1}{|c|}{\multirow{3}{*}{\begin{tabular}[c]{@{}c@{}}$t_0 = 40$\textsuperscript{$\ddagger$}\\ (74.60\% censored)\end{tabular}}}      & $n$              & \multicolumn{1}{c|}{946.49}                 & \multicolumn{1}{c|}{744.10}               & \multicolumn{1}{c|}{202.40}               & \textit{6053.51}                            \\ \cline{2-6} 
\multicolumn{1}{|c|}{}                                  & $n$ ($Z_2=1$)    & \multicolumn{1}{c|}{136.60}                 & \multicolumn{1}{c|}{123.21}               & \multicolumn{1}{c|}{13.39}                & \textit{213.62}                             \\ \cline{2-6} 
\multicolumn{1}{|c|}{}                                  & Observed $T$\textsuperscript{$\S$}             & \multicolumn{1}{c|}{30.62 (25.73, 34.80)}   & \multicolumn{1}{c|}{30.50 (25.59, 34.71)} & \multicolumn{1}{c|}{31.02 (26.28, 35.08)} & -                                           \\ \hline
\end{tabular}
}
\vspace{0.2cm}
\par
\begingroup
\raggedright \footnotesize \setlength{\parindent}{0pt}
\textsuperscript{$\S$} For continuous variables ($V$, $C$, and Observed $T$), values are reported as the average median (1st quartile, 3rd quartile).\\[0.1cm]
\textsuperscript{$\dagger$} Overall descriptive statistics for total subjects: cancer diagnosis age ($V$) is 10.50 (6.86, 14.15); censoring age ($C$) is 32.60 (27.66, 37.48). \\[0.1cm]
\textsuperscript{$\ddagger$} The censoring rate (percentage censored before $t_0$) is shown beneath each $t_0$ label. 
\par\endgroup
\end{table}
\newpage

\begin{figure}[H]
  \centering
  \includegraphics[width=1\linewidth]{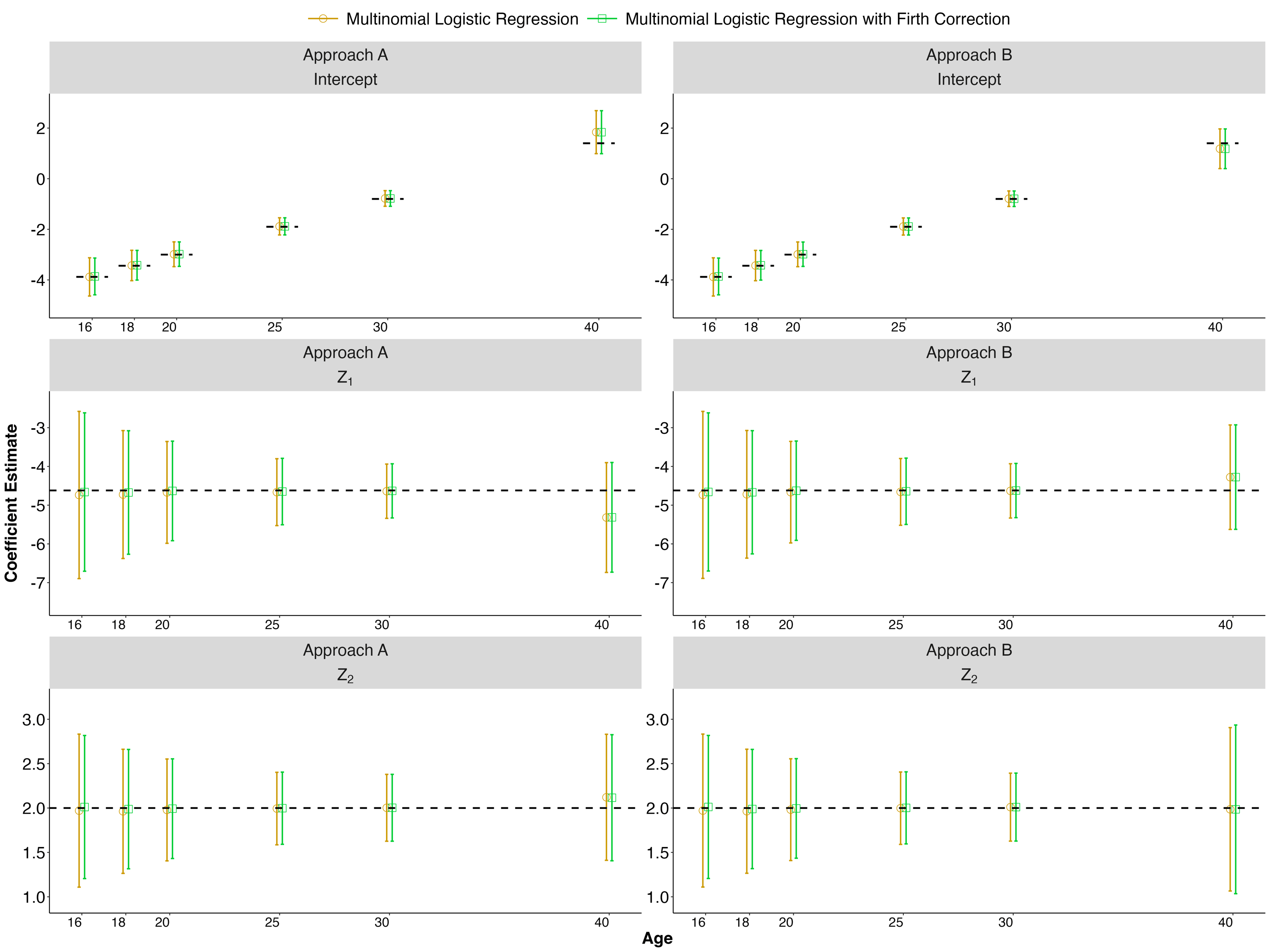}
  \caption{Estimated coefficients for the POI outcome across ages in simulation 2. The $\hat{G}(\cdot \mid \mathbf{Z})$ is obtained using SRF. When using SRF, we choose number of tree equal to 100 and node size equal to 500. The dashed line represents true value of corresponding coefficient.}\label{fig: sim2_poi_all_coef}
\end{figure}

Figure~\ref{fig: sim2_poi_all_coef} presents the estimated coefficients for the POI outcome. The multinomial logistic regression obtains consistent estimates both with and without Firth correction, with the 95\% CIs covering the true values across all ages. The two models produce nearly identical coefficient estimates, confirming that the Firth correction does not introduce additional bias when the data are balanced. The corresponding result for the SPM outcome is presented in the main text.

\begin{figure}[H]
  \centering
  \includegraphics[width=1\linewidth]{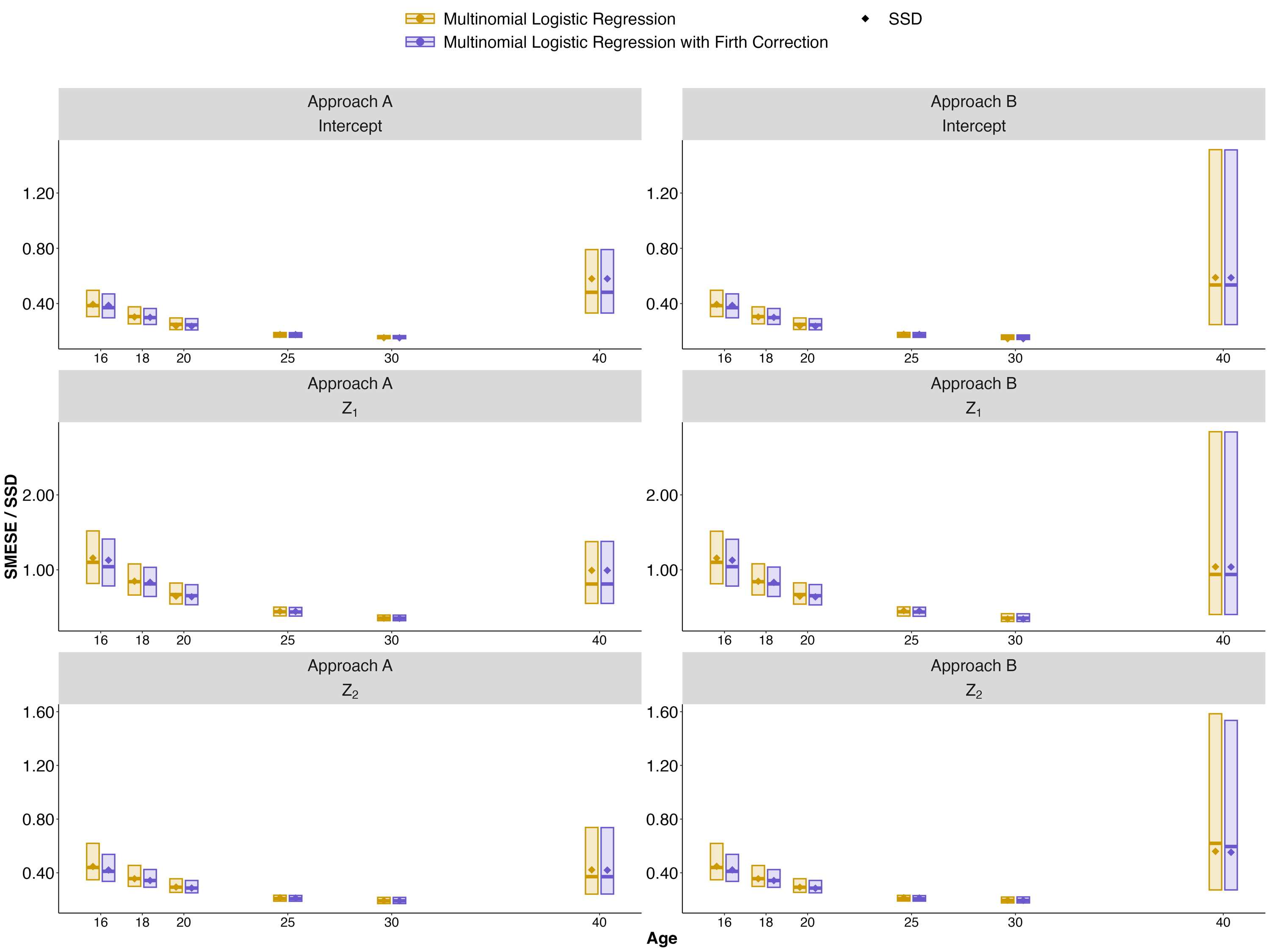}
  \caption{Sample mean of estimated standard errors (SMESE) with 2.5\% - 97.5\% quantile compared to sample standard deviation of estimates (SSD) while outcome is POI in simulation 2. The $\hat{G}(\cdot \mid \mathbf{Z})$ is obtained using SRF. When using SRF, we choose number of tree equal to 100 and node size equal to 500.}\label{fig: sim2_poi_all_sse_smese}
\end{figure}

\begin{figure}[H]
  \centering
  \includegraphics[width=1\linewidth]{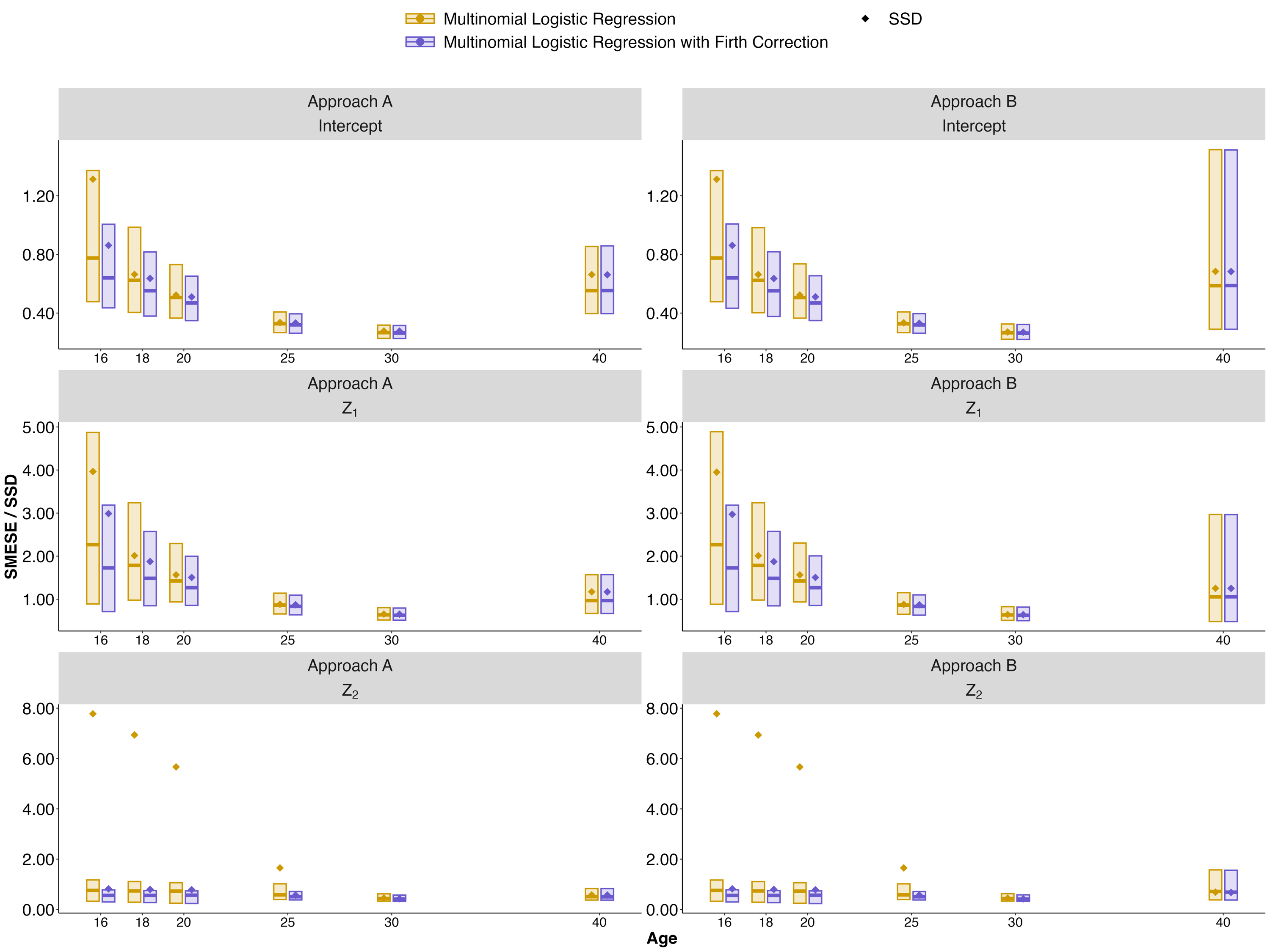}
  \caption{Sample mean of estimated standard errors (SMESE) with 2.5\% - 97.5\% quantile compared to sample standard deviation of estimates (SSD) while outcome is SPM in simulation 2. The $\hat{G}(\cdot \mid \mathbf{Z})$ is obtained using SRF. When using SRF, we choose number of tree equal to 100 and node size equal to 500.}\label{fig: sim2_spm_all_sse_smese}
\end{figure}

We also assess the validity of the sandwich standard error estimator by comparing the SMESE with the SSD. For the POI outcome (Figure~\ref{fig: sim2_poi_all_sse_smese}), the SMESE and SSD are in close agreement for both models, confirming valid standard error estimation when the data are relatively balanced. For the SPM outcome (Figure~\ref{fig: sim2_spm_all_sse_smese}), the multinomial logistic regression without Firth correction underestimates the standard error of $\hat{\beta}_{22}$ at early ages, as the sandwich standard error is evaluated at unstable coefficient estimates caused by quasi-complete separation. With Firth correction, the coefficient estimates are stable, the SSD is substantially reduced, and the SMESE closely tracks the SSD.

\begin{figure}[H]
    \centering
    \begin{subfigure}[t]{1\textwidth}
        \centering
        \includegraphics[width=0.8\textwidth]{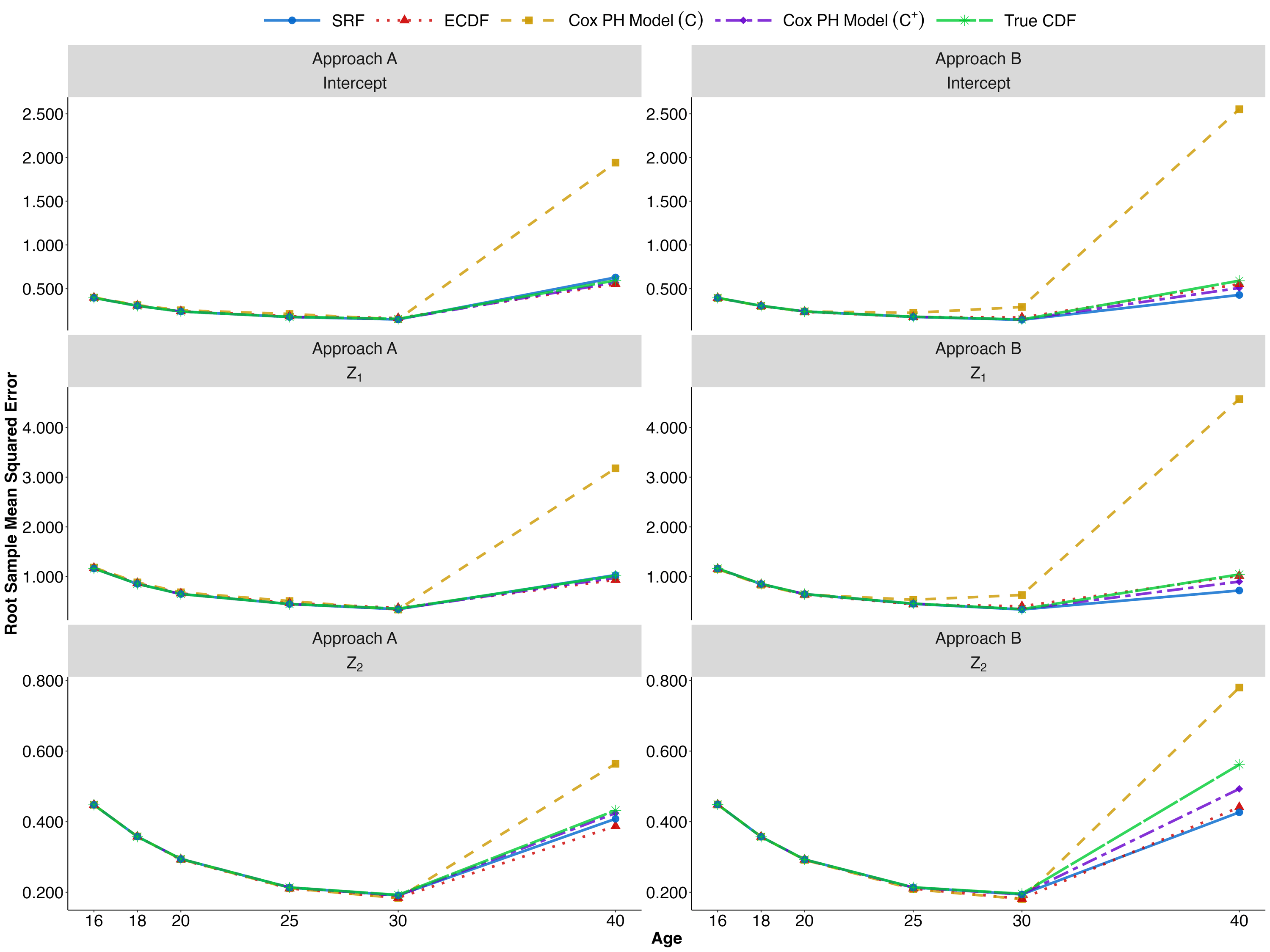}
        \caption{Without Firth Correction}
        \label{fig: proj2_sim2_poi_rmse_nf}
    \end{subfigure}%
    \vfill
    \begin{subfigure}[t]{0.8\textwidth}
        \centering
        \includegraphics[width=1\textwidth]{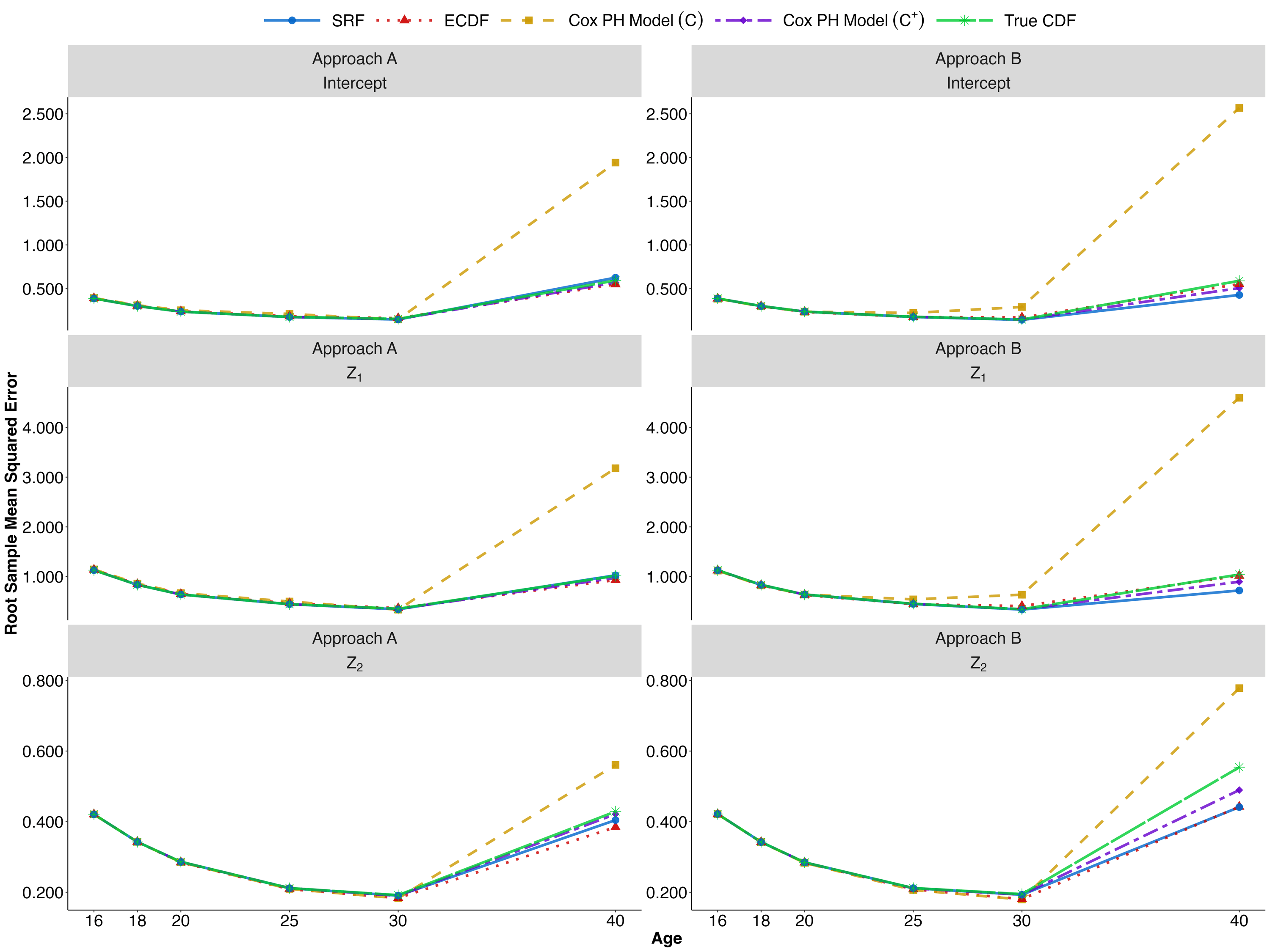}
        \caption{With Firth Correction}
        \label{fig: proj2_sim2_poi_rmse_f}
    \end{subfigure}%
  \caption{The comparison of the root sample mean squared error of the estimates (RSMSE) while outcome is POI in simulation 2. The $\hat{G}(\cdot \mid \mathbf{Z})$ is obtained using stratified ECDF, Cox PH model, Cox PH model on $C^{*}$, True CDF, and SRF. When using SRF, we choose number of tree equal to 100 and node size equal to 500.}\label{fig: proj2_sim2_poi_rmse_all}
\end{figure}

\begin{figure}[H]
    \centering
    \begin{subfigure}[t]{0.8\textwidth}
        \centering
        \includegraphics[width=1\textwidth]{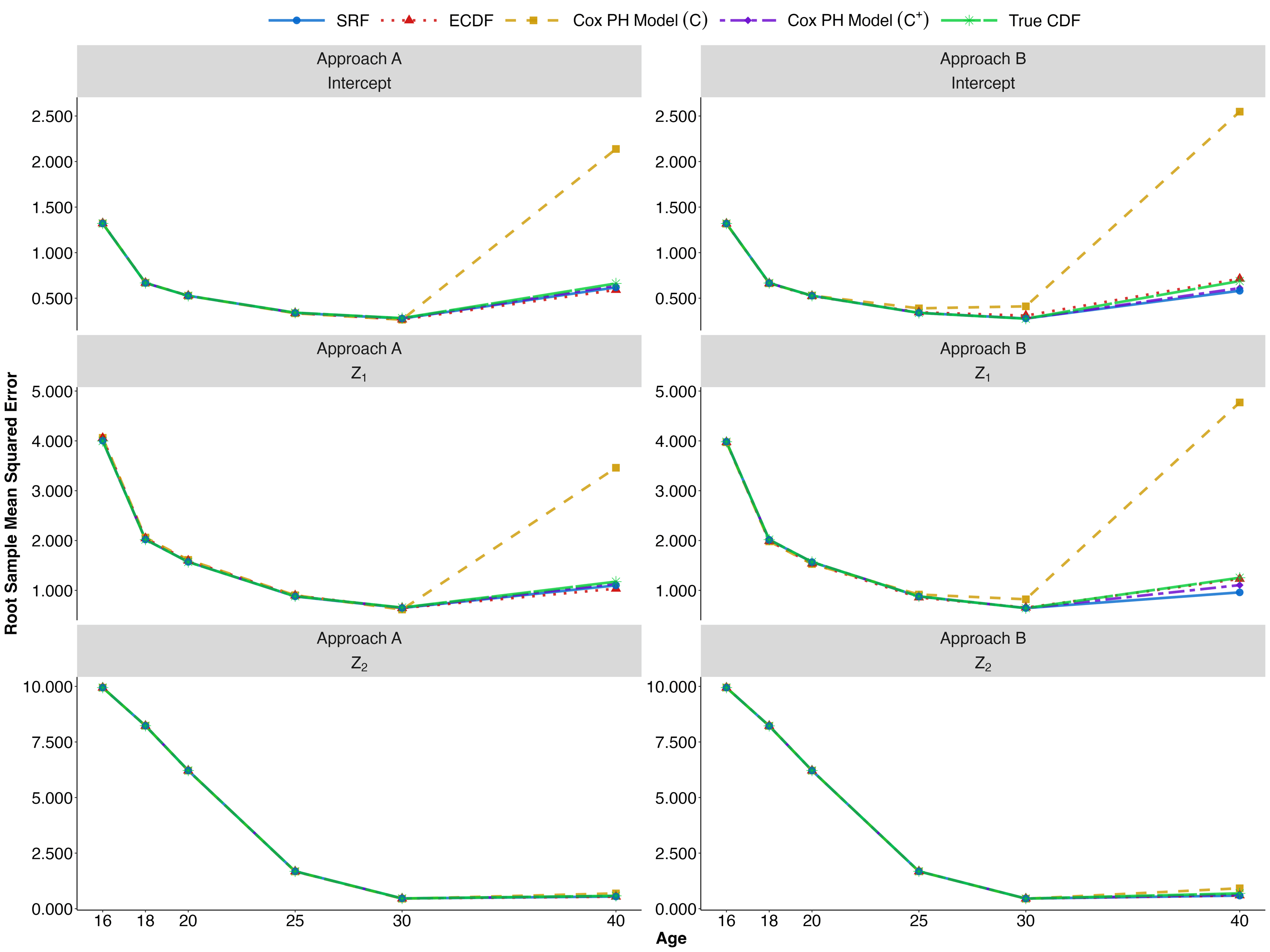}
        \caption{Without Firth Correction}
        \label{fig: proj2_sim2_spm_rmse_nf}
    \end{subfigure}%
    \hfill
    \begin{subfigure}[t]{0.8\textwidth}
        \centering
        \includegraphics[width=1\textwidth]{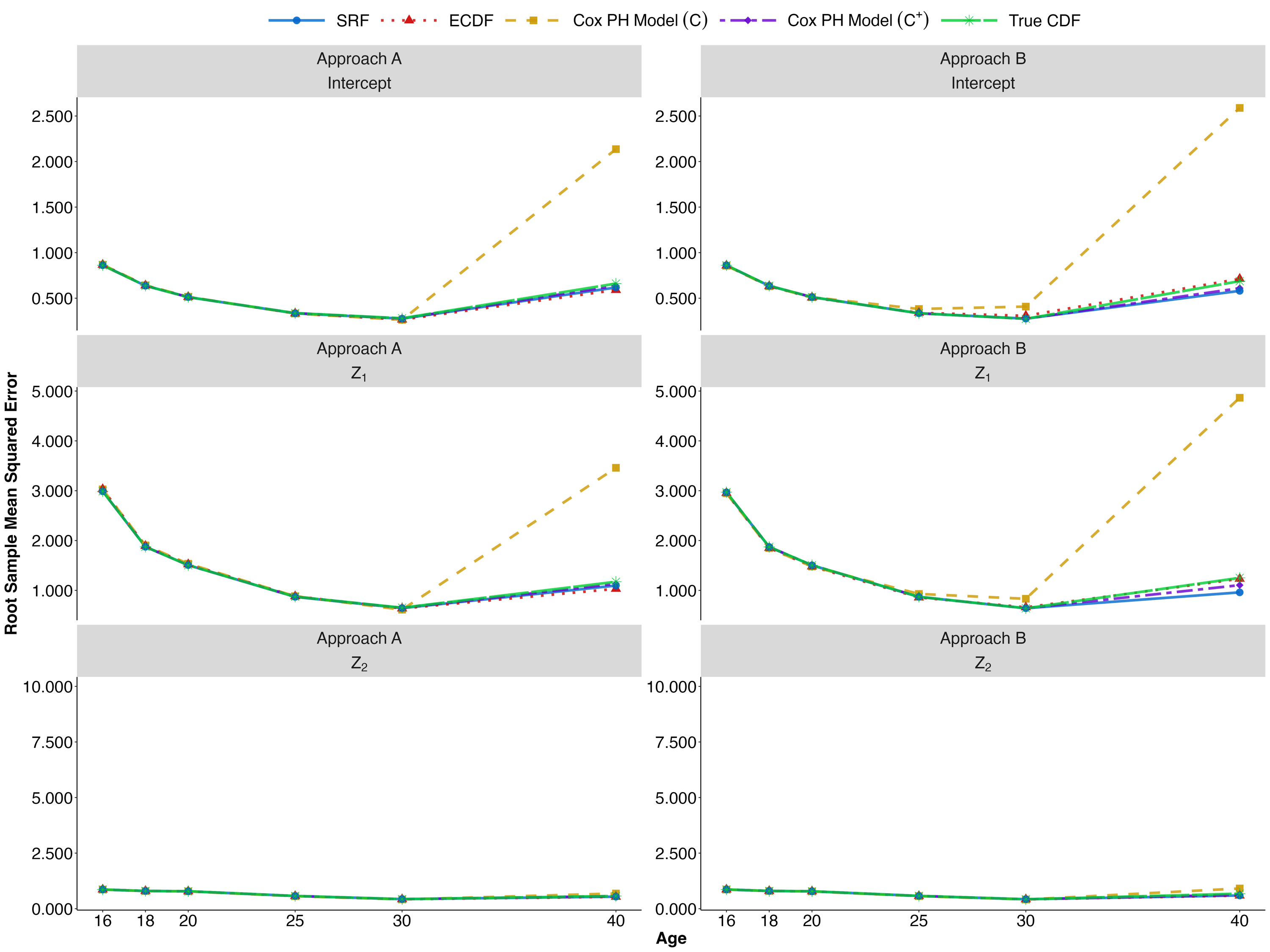}
        \caption{With Firth Correction}
        \label{fig: proj2_sim2_spm_rmse_f}
    \end{subfigure}%
  \caption{The comparison of the root sample mean squared error of the estimates (RSMSE) while outcome is SPM in simulation 2. The $\hat{G}(\cdot \mid \mathbf{Z})$ is obtained using stratified ECDF, Cox PH model, Cox PH model on $C^{*}$, True CDF, and SRF. When using SRF, we choose number of tree equal to 100 and node size equal to 500.}\label{fig: proj2_sim2_spm_rmse_all}
\end{figure}

We also explore the impact of the censoring distribution estimator on coefficient estimation in Simulation~2. Figures~\ref{fig: proj2_sim2_poi_rmse_all} and~\ref{fig: proj2_sim2_spm_rmse_all} present the RSMSE for the POI and SPM outcomes in Simulation~2, respectively. For the POI outcome, the results are consistent with Simulation~1: the SRF, stratified ECDF, and Cox PH model on $C^{*}$ obtain comparable RSMSEs under Approach~A, while the Cox PH model on $C$ obtains substantially higher RSMSE at later ages due to the violation of the proportional hazards assumption. Under Approach~B, the differences among methods are again more pronounced at later ages, with the SRF obtaining the lowest RSMSE. These patterns hold with and without Firth correction. For the SPM outcome without Firth correction, the RSMSE of $\hat{\beta}_{22}$ is extremely large at early ages across all censoring distribution estimators, including the true $G(\cdot \mid \mathbf{Z})$. This confirms that the large RSMSE is driven by the coefficient bias from quasi-complete separation rather than the choice of censoring distribution estimator. With Firth correction, the RSMSE of $\hat{\beta}_{22}$ is substantially reduced and the patterns across censoring estimators resemble those of Simulation~1: the SRF and stratified ECDF obtain the lowest RSMSE, while the Cox PH model on $C$ performs the worst. Overall, the findings from Simulation~1 regarding the choice of censoring distribution estimator remain consistent in the imbalanced data setting of Simulation~2.

\vspace{-10pt}
\begin{figure}[H]
\centering
\scalebox{0.9}{%
\begin{minipage}{\textwidth}
\centering
\begin{subfigure}[b]{0.48\textwidth}
\centering
\includegraphics[width=0.6\textwidth]{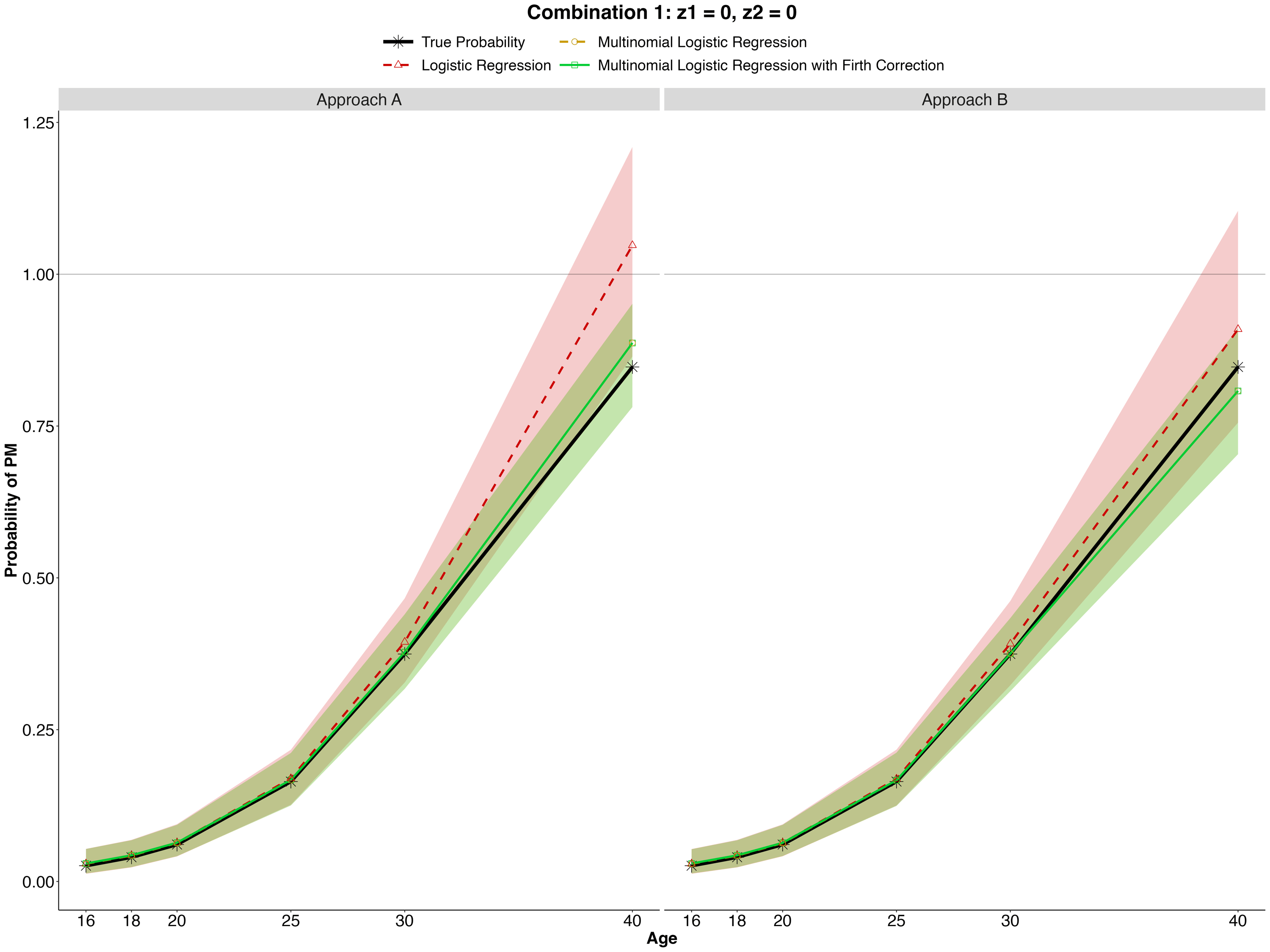}
\caption{}
\end{subfigure}
\hfill
\begin{subfigure}[b]{0.48\textwidth}
\centering
\includegraphics[width=0.6\textwidth]{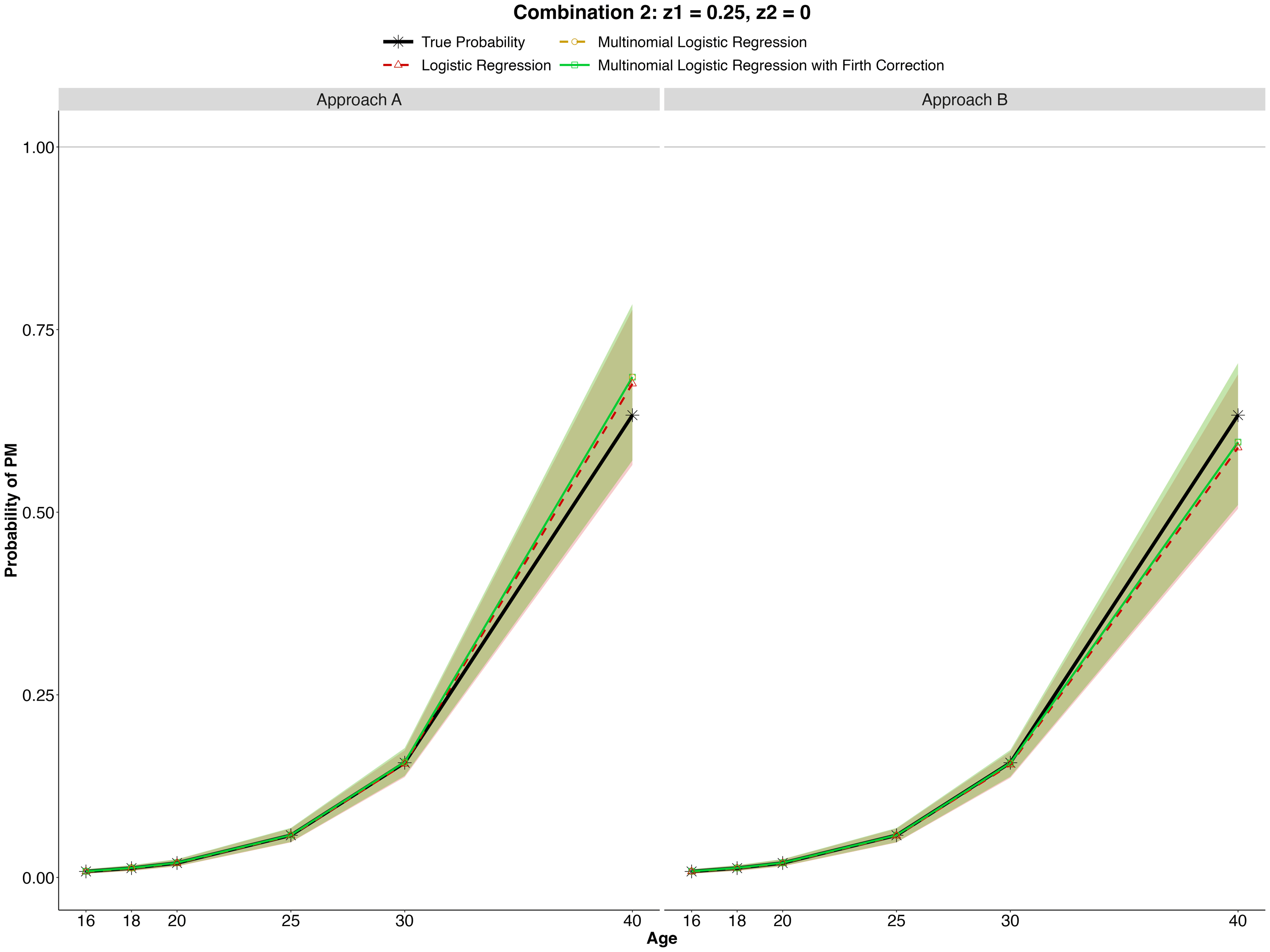}
\caption{}
\end{subfigure}
\begin{subfigure}[b]{0.48\textwidth}
\centering
\includegraphics[width=0.6\textwidth]{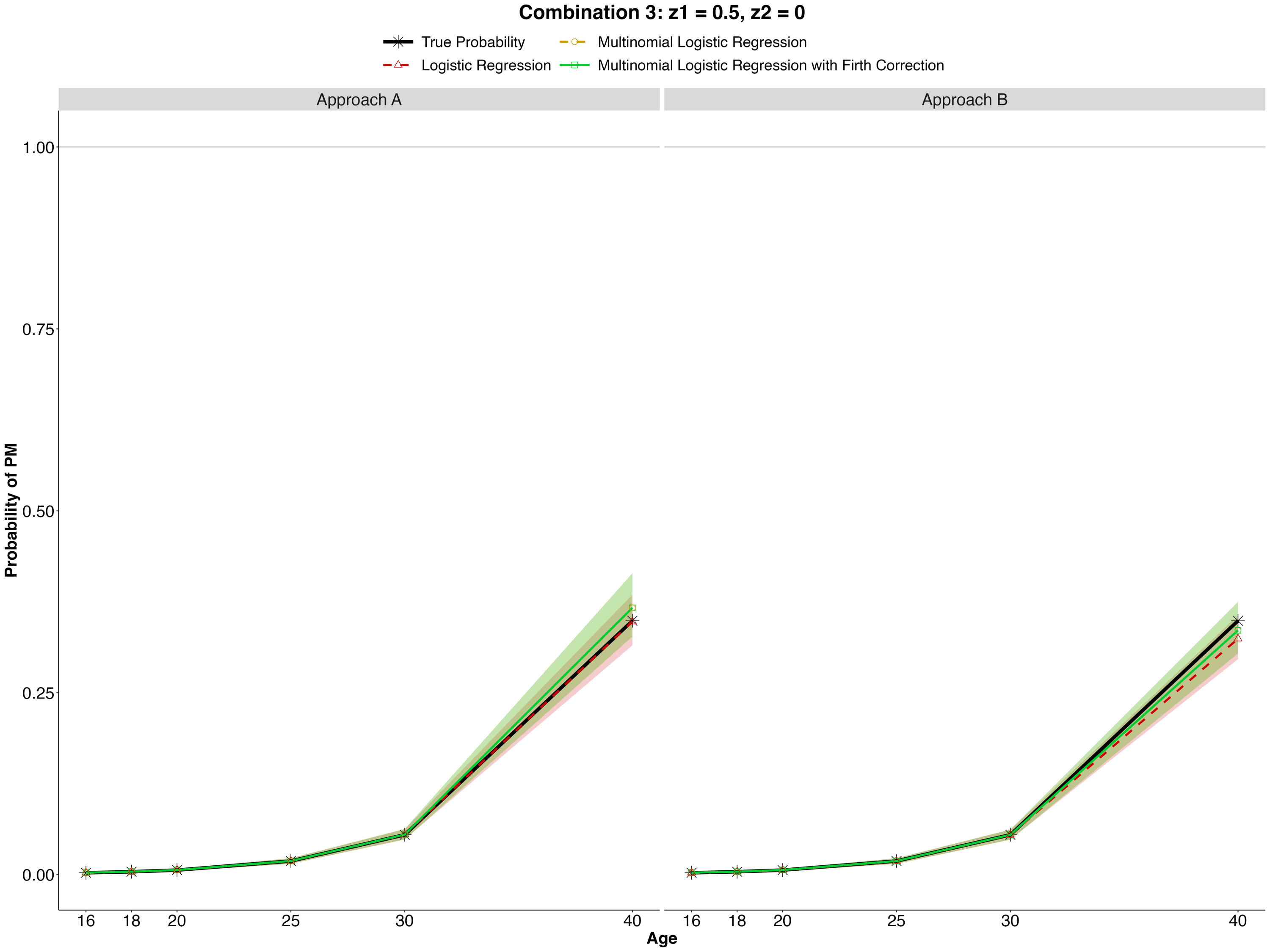}
\caption{}
\end{subfigure}
\hfill
\begin{subfigure}[b]{0.48\textwidth}
\centering
\includegraphics[width=0.6\textwidth]{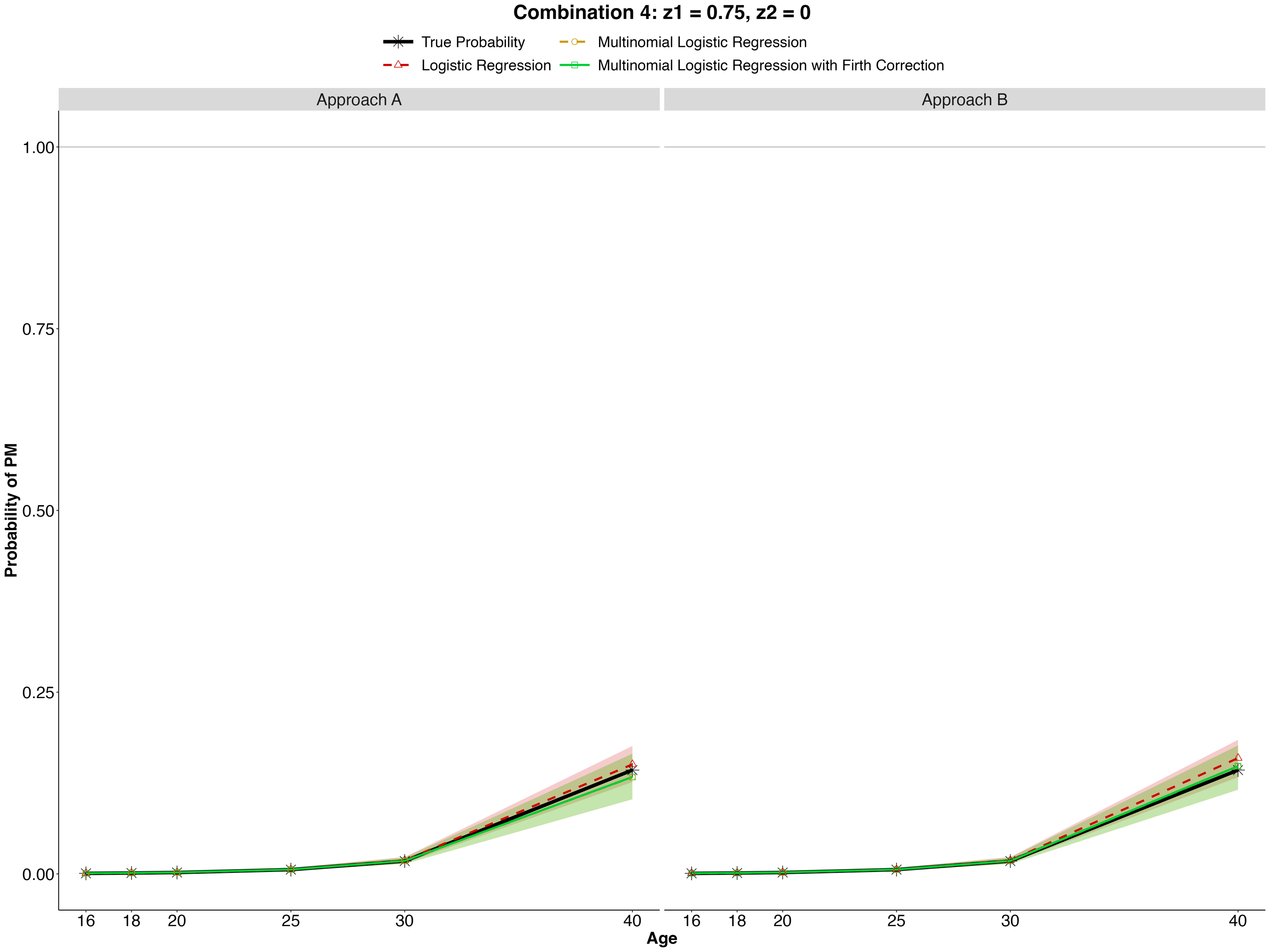}
\caption{}
\end{subfigure}
\begin{subfigure}[b]{0.48\textwidth}
\centering
\includegraphics[width=0.6\textwidth]{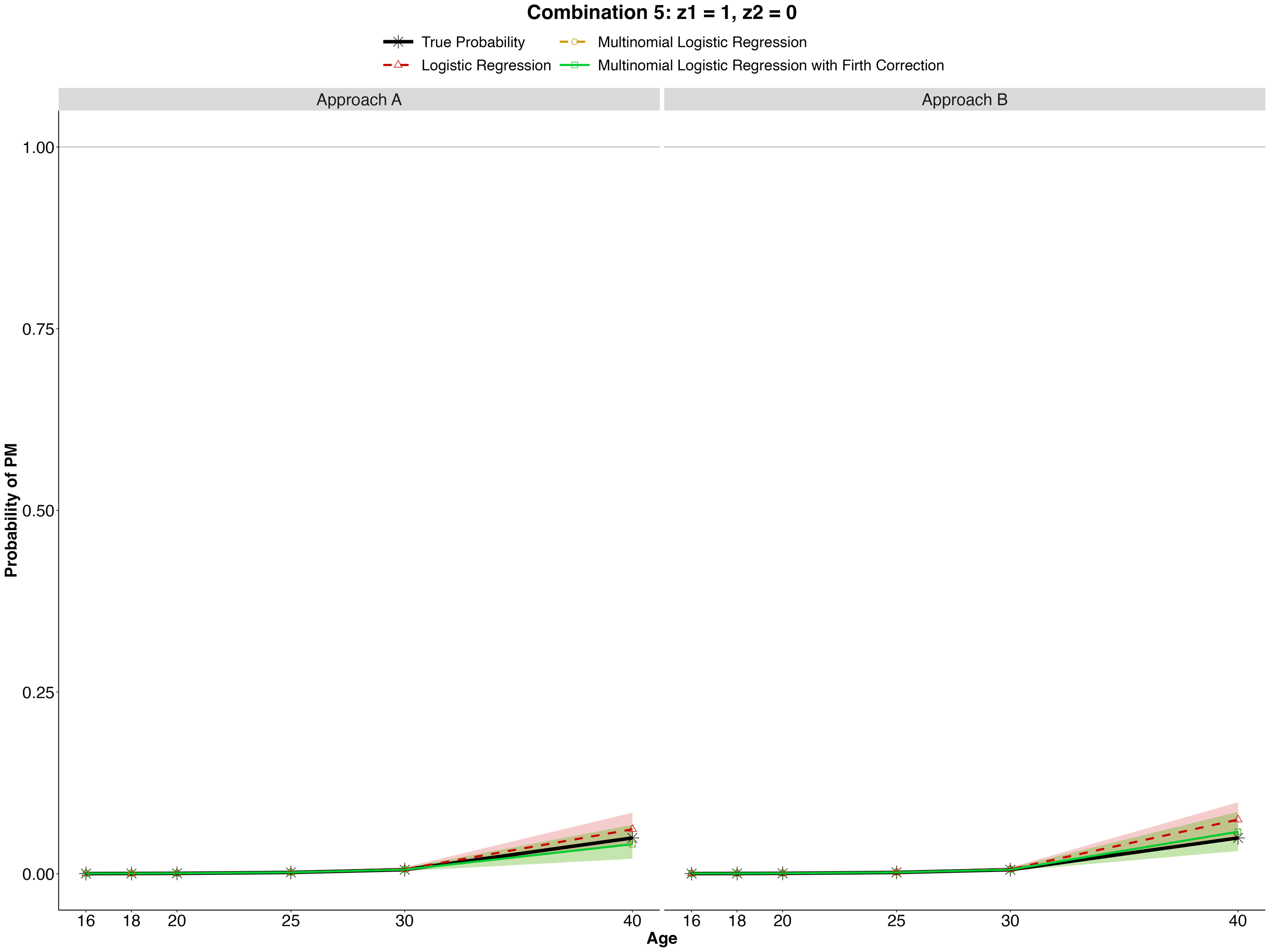}
\caption{}
\end{subfigure}
\hfill
\begin{subfigure}[b]{0.48\textwidth}
\centering
\includegraphics[width=0.6\textwidth]{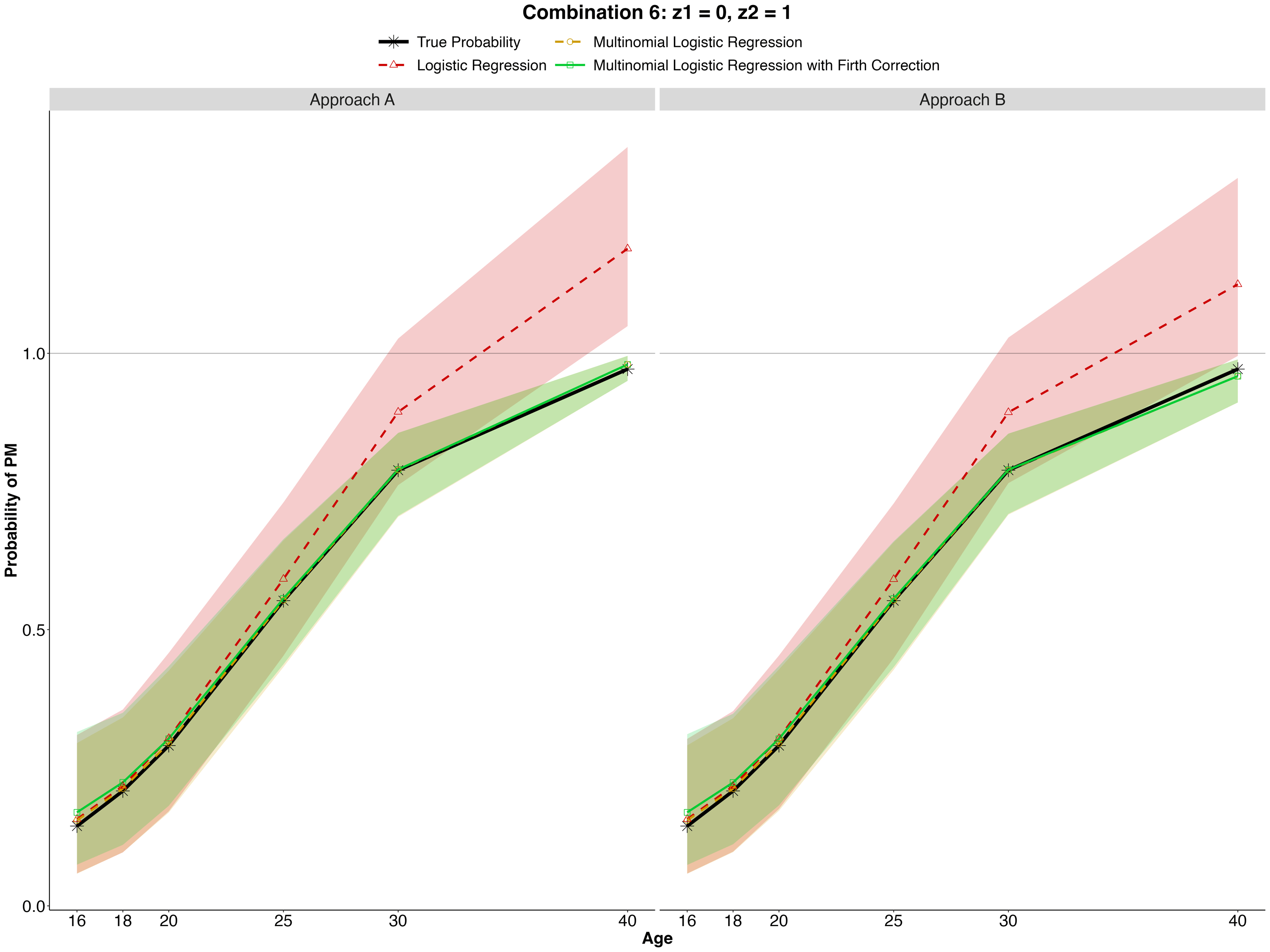}
\caption{}
\end{subfigure}
\begin{subfigure}[b]{0.48\textwidth}
\centering
\includegraphics[width=0.6\textwidth]{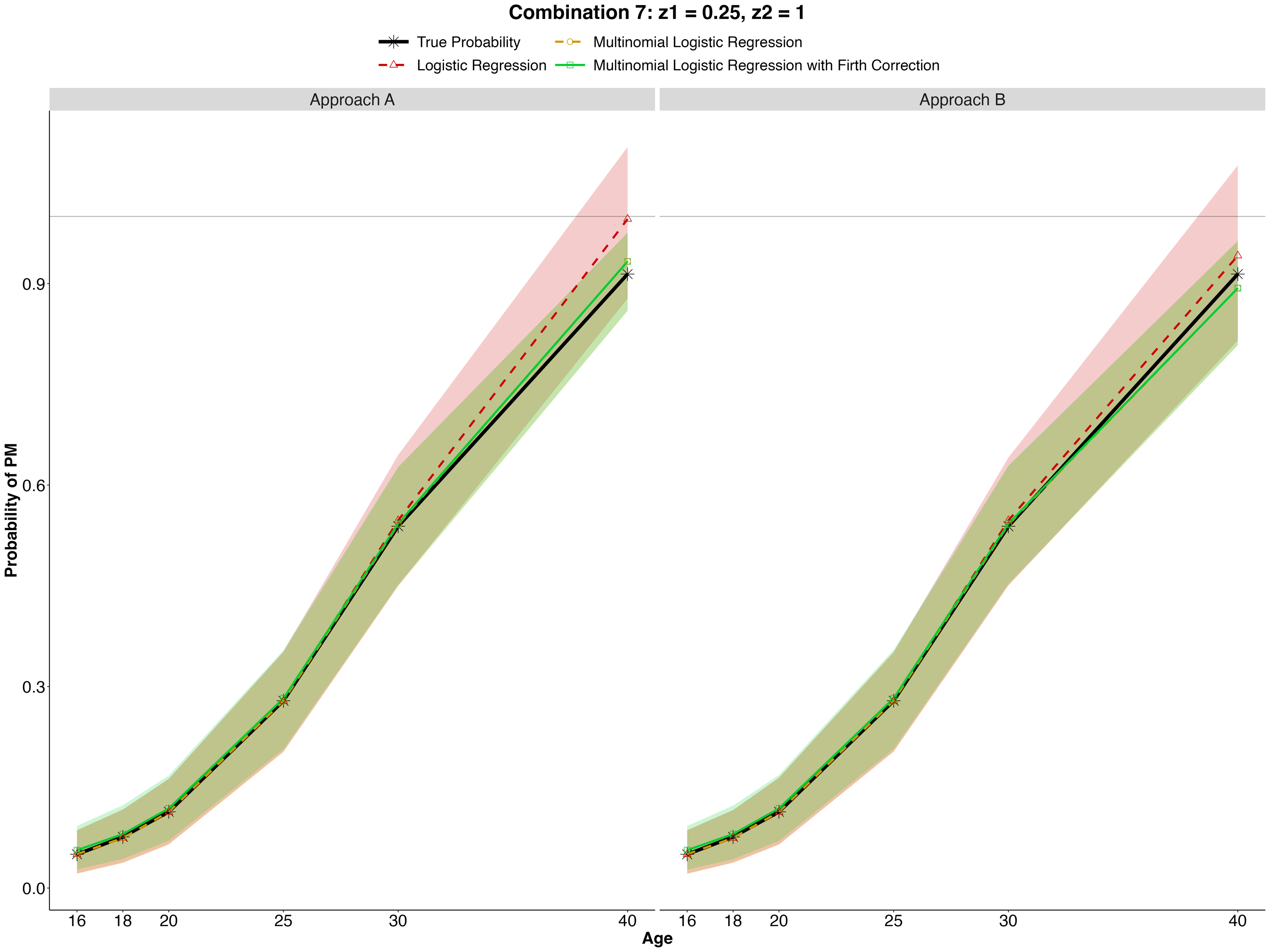}
\caption{}
\end{subfigure}
\hfill
\begin{subfigure}[b]{0.48\textwidth}
\centering
\includegraphics[width=0.6\textwidth]{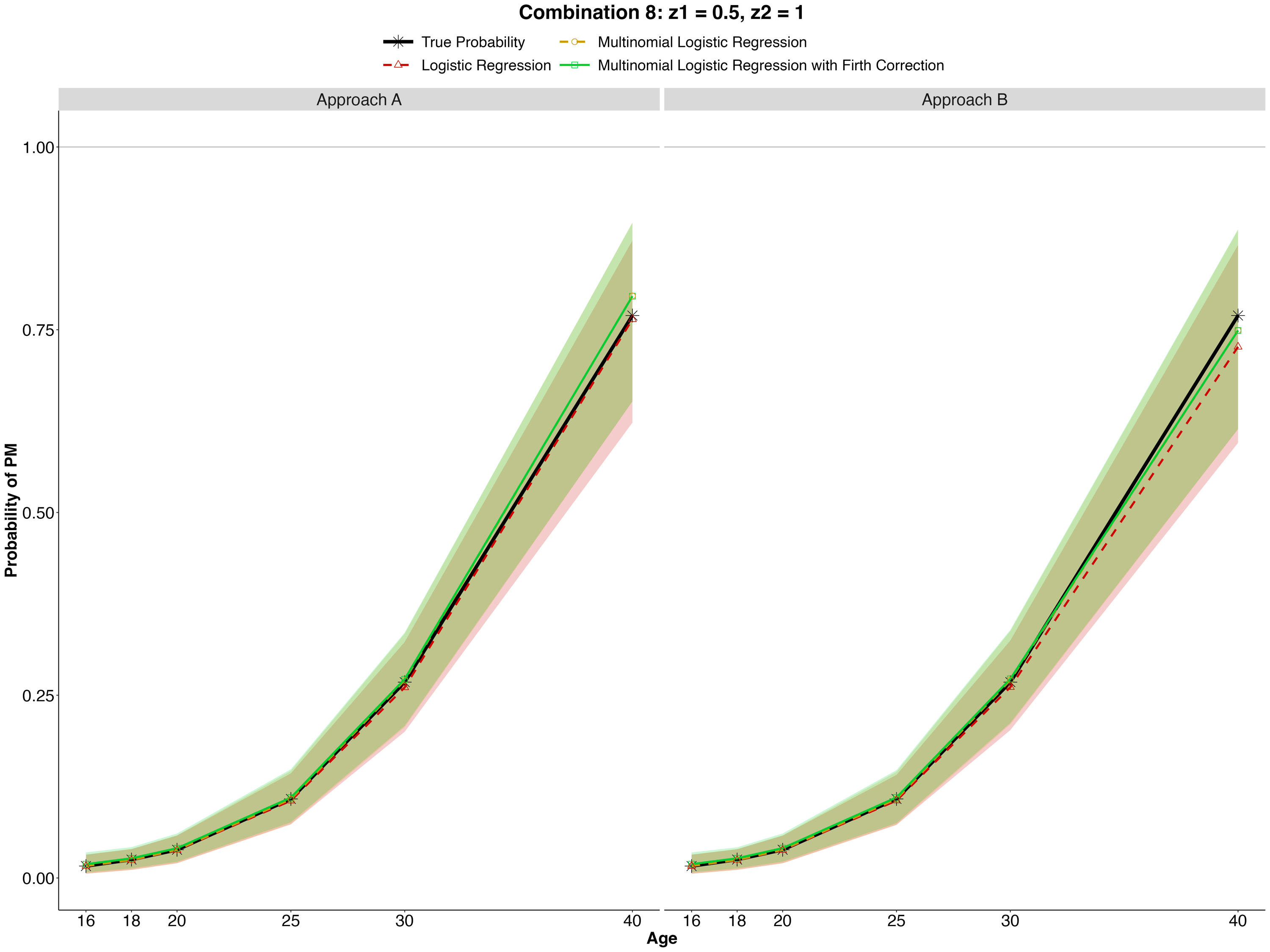}
\caption{}
\end{subfigure}
\begin{subfigure}[b]{0.48\textwidth}
\centering
\includegraphics[width=0.6\textwidth]{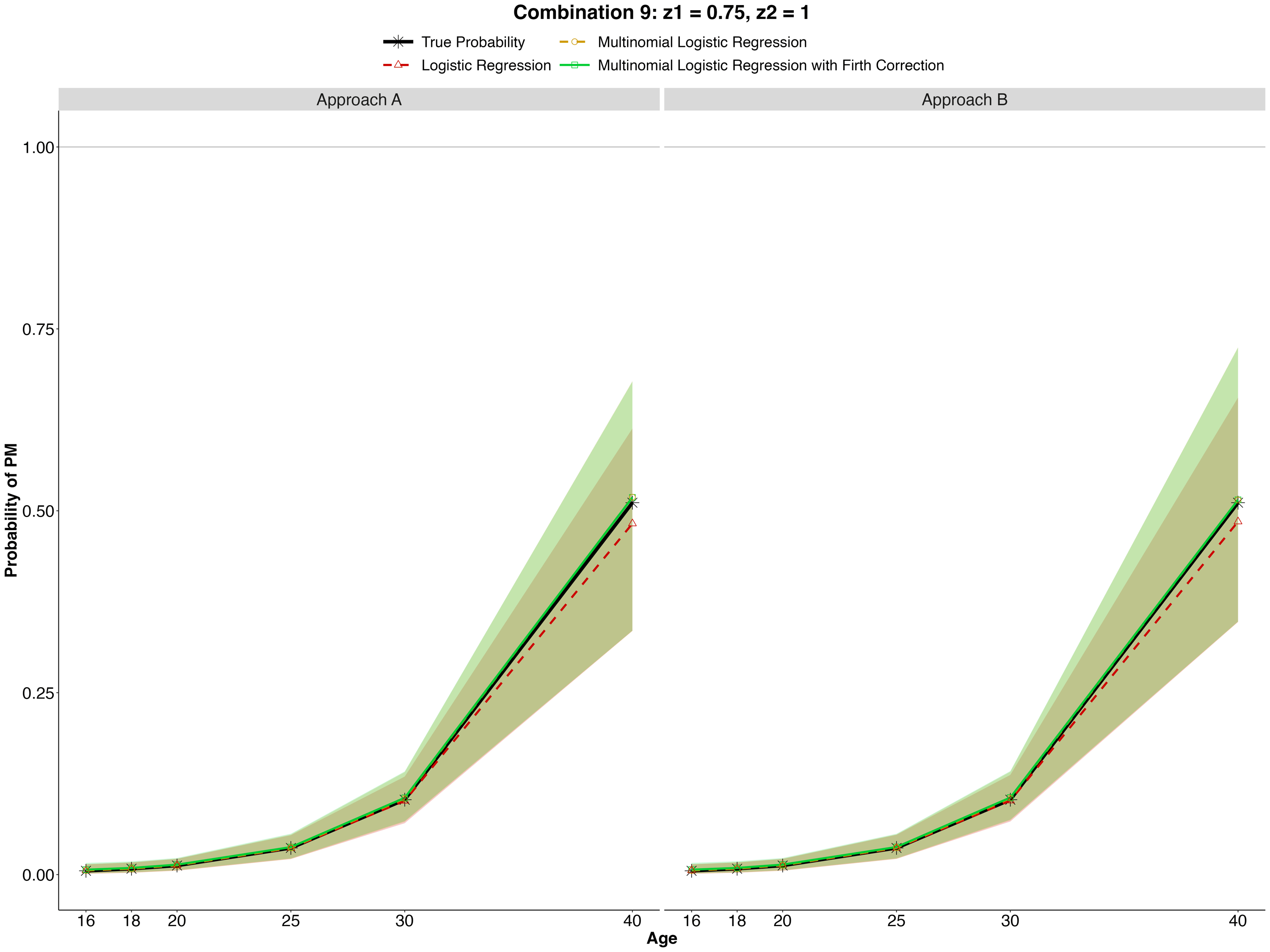}
\caption{}
\end{subfigure}
\hfill
\begin{subfigure}[b]{0.48\textwidth}
\centering
\includegraphics[width=0.6\textwidth]{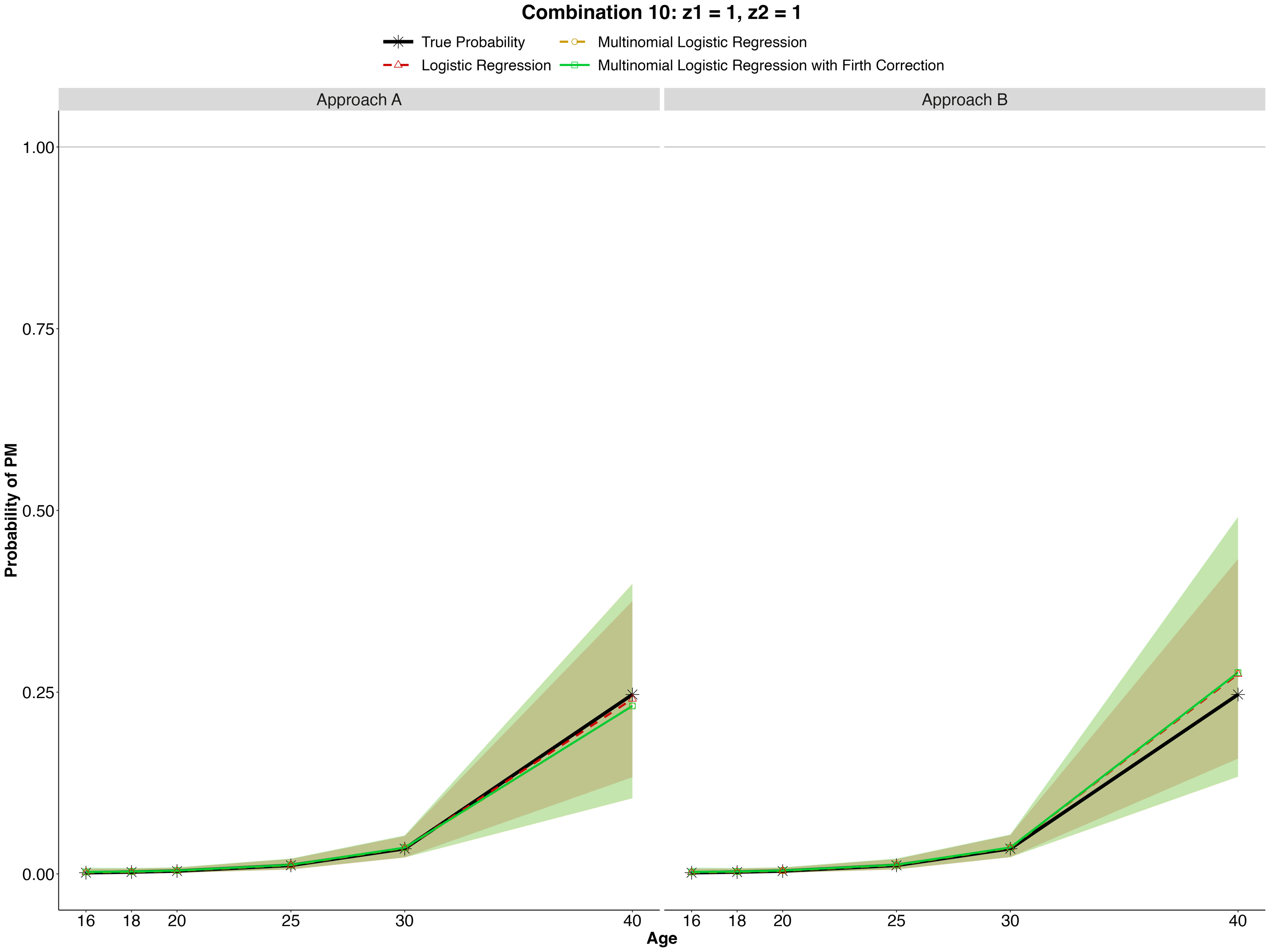}
\caption{}
\end{subfigure}
\end{minipage}%
}
\caption{Estimated probability of PM with 95\% CIs across ages in simulation~2. Red: logistic regression; yellow: multinomial logistic regression without Firth correction; green: multinomial logistic regression with Firth correction. Left and right panels correspond to Approaches~A and~B, respectively. $\hat{G}(\cdot \mid \mathbf{Z})$ is estimated using SRF with the number of trees set to 100 and the node size set to 500.}
\label{fig: sim2_poi_all_prob}
\end{figure}

